\documentclass[12pt,a4paper,pdflatex]{article}%for arXiv (dvipdfmx poses problems)
\usepackage{amsmath,amsfonts,amsthm,amssymb}
\usepackage{tikz}
\usepackage{ytableau}
\allowdisplaybreaks[3]
\numberwithin{equation}{section}

\usepackage[body={16cm,23cm}]{geometry}
\newcommand{\Ds}{{\mathsf D}}

\newcommand{\Qf}{{\mathbf Q}}
\newcommand{\Qb}{{\mathbb Q}}

\newcommand{\Qc}{{\mathcal Q}}

\newcommand{\Ts}{{\mathsf T}}
\newcommand{\Ss}{{\mathsf S}}

\newcommand{\rr}{{\mathfrak r}}
\newcommand{\Bm}{{\mathfrak B}}
\newcommand{\Fm}{{\mathfrak F}}
\newcommand{\Am}{{\mathfrak A}}
\newcommand{\mm}{{\mathtt m}}
\newcommand{\nn}{{\mathtt n}}
\begin{document}
\title{
Folding QQ-relations and transfer matrix eigenvalues:
towards a unified approach to Bethe ansatz for super spin chains
%orthosymplectic superalgebras
}
\author{Zengo Tsuboi
%\footnote{
%E-mail: ztsuboi$\bullet$yahoo.co.jp
%}
\\[8pt]
{\sl
Pacific Quantum Center, Far Eastern Federal University, 
}
\\
{\sl
Sukhanova 8, Vladivostok, 690950, Russia
%
%\footnote{remote contract (work from Japan)}
%
%
}
% \&  
} 
\date{}
\maketitle
%%%%%%%%%%%%%%%%%%%%%%%%%%%%%%%%
\begin{abstract}
 Extending the method proposed in \cite{T11}, we derive QQ-relations (functional relations among Baxter Q-functions)
  and T-functions (eigenvalues of transfer matrices) for fusion vertex models associated with 
 the twisted quantum affine superalgebras 
$U_{q}(gl(2r+1|2s)^{(2)})$, $U_{q}(gl(2r|2s+1)^{(2)})$, $U_{q}(gl(2r|2s)^{(2)})$, $U_{q}(osp(2r|2s)^{(2)})$  and the 
untwisted quantum affine orthosymplectic  superalgebras $U_{q}(osp(2r+1|2s)^{(1)})$ and   
 $U_{q}(osp(2r|2s)^{(1)})$ 
(and their Yangian counterparts,  $Y(osp(2r+1|2s))$ and $Y(osp(2r|2s))$) 
  as reductions (a kind of folding) of those associated with $U_{q}(gl(M|N)^{(1)})$. 
In particular,  we reproduce previously proposed 
generating functions (difference operators) of the T-functions for the symmetric or anti-symmetric 
representations, and 
tableau sum expressions for more general representations for orthosymplectic superalgebras \cite{T99,T99-2}, 
 and obtain Wronskian-type expressions (analogues of Weyl-type character formulas) for them. 
 T-functions for spinorial representations are related to reductions of those for asymptotic limits of
  typical representations of $U_{q}(gl(M|N)^{(1)})$. 
\end{abstract}
%%%%%%%%%%%%%%%%%%%%%%%%%%%%%%%%%%%%
Keywords: Baxter Q-function, QQ-relation, Bethe ansatz, 
orthosymplectic superalgebras, 
Wronskian formula
\\
\\
Journal ref: Nuclear Physics B 1005 (2024) 116607
\\
DOI: https://doi.org/10.1016/j.nuclphysb.2024.116607
\tableofcontents
%%%%%%%%%%%%%%%%%%%%%%%%%%%%%%%%%%%
\section{Introduction}
This paper, together with our recent paper \cite{T21}, is an expansion of section 3.7 in our previous paper \cite{T11}. Namely, we will explain details of it and generalize it further. 

Quantum integrable systems have commuting family of transfer matrices (T-operators). 
Finding eigenvalues of transfer matrices (T-functions) is an important problem in the study of quantum integrable systems. 
For this, the Bethe ansatz is often used. 
The T-functions are expressed in terms of Baxter Q-functions (for short, Q-functions). The Q-functions are eigenvalues of Baxter 
Q-operators. The zeros of the Q-functions give the roots of Bethe ansatz equations.
In general, the Q-functions are not functionally independent and satisfy functional relations, called QQ-relations.  
There are two kinds of QQ-relations for quantum integrable systems associated with $U_{q}(gl(M|N)^{(1)})$ 
(or $U_{q}(sl(M|N)^{(1)})$). 
The bosonic QQ-relations are generalization of the quantum Wronskian condition (cf.\ \cite{BLZ98,PS00}). 
The fermionic QQ-relations came from particle-hole transformations in statistical physics \cite{Wo83}, and are related \cite{T98} 
to odd Weyl reflections \cite{DP85,Se85} of the superalgebra $gl(M|N)$. 

T-functions are generalization of characters of representations of underlying quantum algebras, with a spectral parameter. 
Corresponding to the fact that there are several different expressions of characters, 
there are several different expressions of T-functions: 
 the Cherednik-Bazhanov-Reshetikhin (CBR) determinant formula (an analogue of the Jacobi-Trudi formula) \cite{Ch89,BR90}, 
  Wronskian-like determinant (Casoratian) formulas (analogues of the Weyl character formula) 
  \cite{KLWZ97,BLZ98}, 
 and tableau sum expressions, etc (see \cite{KNS10} for a review).  
 Among them, Wronskian expressions have a merit that the action of the Weyl group on them 
 is manifest. 
Of particular interest is quantum integrable systems associated with superalgebras since  
representations of underlying superalgebras are less well understood and more diverse than those of ordinary (non-super) algebras. 
In view of this, we derived tableau sum and CBR-determinant expressions of T-functions 
by analytic Bethe ansatz \cite{Re83} for fusion vertex models associated with  $U_{q}(gl(M|N)^{(1)})$ 
(or $U_{q}(sl(M|N)^{(1)})$) \cite{T97,T98,T98-2}, 
 $Y(osp(r|2s))$ for $r \ge 3$, $s \ge 1$ \cite{T99},  $U_{q}(osp(2|2s)^{(1)})$ for $s \ge 1$ \cite{T99-2}. 
Establishing Wronskian expressions of T-functions for these is a longstanding problem, and 
in \cite{T09} (together with \cite{BT08}), we  proposed 
Wronskian expressions of T-functions for  the case $U_{q}(gl(M|N)^{(1)})$ (or $U_{q}(sl(M|N)^{(1)})$). 
In this paper we will explain our trial  toward the rest, namely the $U_{q}(osp(r|2s)^{(1)})$ case, 
and also related twisted quantum affine superalgebras cases.

It is known \cite{Z97} that there is a correspondence 
%(some kind of Boson-Fermion correspondence) 
between  representations of superalgebras and ordinary (non-graded) algebras. 
Thus there should be a correspondence between different quantum integrable models 
in accordance with the correspondence between representations of different underlying algebras. 
A relatively well known example for this would be the Izergin-Korepin model \cite{IK81}
 associated with the twisted quantum affine algebra 
$U_{q}(sl(3)^{(2)})$ and a vertex model associated with
 the quantum affine superalgebra $U_{q}(osp(1|2)^{(1)})$ \cite{Kul85,BS87}. 
In the context of the thermodynamic Bethe ansatz, 
coincidence of the Q-system (a system of functional relations among characters of Kirillov-Reshetikhin modules) for $U_{q}(sl(2r+1)^{(2)})$ and $U_{q}(osp(1|2r)^{(1)})$ was pointed out in \cite{T01}. 
Having in mind a correspondence between  the twisted quantum affine superalgebra $U_{q}(gl(2r|1)^{(2)})$ and 
the quantum affine algebra $U_{q}(so(2r+1)^{(1)})$, 
we proposed \cite{T11} a Wronskian solution of the T-system for $U_{q}(so(2r+1)^{(1)})$ 
as a reduction (some kind of folding) of the Wronskian solution for $U_{q}(gl(2r|1)^{(1)})$ \cite{T09}. 
In our recent paper \cite{T21}, we not only explained details of [section 3.7, \cite{T11}], but also gave the QQ-relations for $U_{q}(so(2r+1)^{(1)})$ as a reduction of the QQ-relations for $U_{q}(gl(2r|1)^{(1)})$. 
In this paper, we extend our discussion to more wider algebras, in particular, 
twisted quantum affine superalgebras 
$U_{q}(gl(2r+1|2s)^{(2)})$, $U_{q}(gl(2r|2s+1)^{(2)})$, $U_{q}(gl(2r|2s)^{(2)})$  
(or $U_{q}(sl(2r+1|2s)^{(2)})$, $U_{q}(sl(2r|2s+1)^{(2)})$, $U_{q}(sl(2r|2s)^{(2)})$) and 
quantum affine orthosymplectic  superalgebras $U_{q}(osp(2r+1|2s)^{(1)})$ and $U_{q}(osp(2r|2s)^{(1)})$ 
(and their Yangian counterparts,  $Y(osp(2r+1|2s))$ and $Y(osp(2r|2s))$). 
We will derive T-functions, QQ-relations, Bethe equations for these algebras as reductions of those 
for $U_{q}(gl(M|N)^{(1)})$. 
We have reproduced some of our previous results by analytic Bethe ansatz \cite{T99,T99-2}, 
in particular generating functions of T-functions of fusion vertex models for the symmetric or anti-symmetric representations in the auxiliary space. 

The basic idea on the reduction procedure proposed in \cite{T11} is as follows. 
As remarked in \cite{T09}, there are $2^{M+N}$ kinds of Q-functions $\Qb_{I}(u)$ labeled by $I \subset \{1,2,\dots , M+N \}$ with 
 the spectral parameter $u \in \mathbb{C}$ for quantum integrable systems associated with $U_{q}(gl(M|N)^{(1)})$. 
First we consider a map $\sigma$ which keeps the form of the QQ-relations invariant. 
Then we apply this to the Q-functions $\Qb_{I}(u)$, $I \subset \{1,2,\dots , M+N\}$ and boundary twist parameters $\{z_{a}\}_{a=1}^{M+N}$ and identity the image of them 
with the original ones: $\sigma(\Qb_{I}(u))=\Qb_{I}(u)$, $\sigma(z_{a})=z_{a}$. 
In case we consider reductions to twisted quantum affine superalgebras, we have to make a shift of the spectral parameter: 
$\sigma(\Qb_{I}(u))=\Qb_{I}(u+\eta)$, where $\eta$ is half of the period of the Q-functions. 
The reduction procedure for (non-super) twisted quantum affine algebras for a special class of the index set $I$ was proposed in \cite{KS94-2} (see also \cite{R87}).
The reduction procedure can also be applied to 
the T-functions since the T-functions are expressed in terms of Q-functions.  
In case we consider a reduction to $U_{q}(osp(2r|2s)^{(1)})$, 
 we have to modify the relation $\sigma(z_{a})=z_{a}$ in part and impose additional conditions on Q-functions. 
 This situation is similar to the one in  \cite{KOSY01}, where a reduction of q-characters for $U_{q}(sl(2r+2)^{(1)})$ 
 to q-characters  for $U_{q}(sp(2r)^{(1)})$ was discussed. 
%Some of the reduced Q-functions become square root of the origianl Q-functions. 
 
 %%%
In general, representations of finite Lie algebras other than type A
\footnote{$gl(M|N)$, $sl(M|N)$ or $A(M-1|N-1)$} 
can not be lifted to representations of  Yangians or quantum affine algebras. 
In contrast, evaluation representations based on representations of  $U_{q}(gl(M|N))$ are available 
for  the $U_{q}(gl(M|N)^{(1)})$ case. 
This is a merit to work on the problems as reductions of the $U_{q}(gl(M|N)^{(1)})$  case. 
On the level of supercharacters, one can use various expressions of 
supercharacter formulas of $gl(M|N)$ 
(see for example, \cite{MV03}),  
and consider reductions of them, to get supercharacters of twisted quantum affine superalgebras and 
 quantum affine orthosymplectic superalgebras (or their Yangian counterparts). 
 
 It should be remarked that the Bethe ansatz equations of the Gaudin models associated with 
 $osp(2r+1|2s)$ and $osp(2r|2s)$ were studied in \cite{LM21} in connection to those associated with $gl(r|s)$ 
 (see also the recent paper \cite{Ze23}). 
Although it is not a topic of this paper, our results on XXZ-type models will have some connection to theirs in the Gaudin limit. 

The outline of this paper is as follows. 
In section \ref{preli}, we fix notation and 
summarize preliminaries on Lie superalgebrs in our convention.  
In section \ref{sec:T-fun}, we summarize necessary formulas on T- and Q-functions for $U_{q}(gl(M|N)^{(1)})$, which are taken mainly from \cite{T09,T11,T97,T98}. 
% In section 3.1 and section 3.2, we explain the general procedures of the reduction. 
In section \ref{sec:red}, we explain the general procedure of reductions of the formulas introduced in section \ref{sec:T-fun}. 
%reductions of Q-functions for $U_{q}(gl(M|N)^{(1)})$ to Q-functions for other algebras. 
In section \ref{sec:SN}, we restrict our consideration to reductions along symmetric nesting paths, which correspond to folding with respect to 
symmetric Dynkin diagrams of $gl(M|N)$. 
In subsections \ref{sec:RR} and \ref{sec:SR}, 
the results of the reductions are presented for each value of $(M,N)$:  
QQ-relations, generating functions of the T-functions for the symmetric or anti-symmetric representations, 
Wronskian-type expressions of T-functions, and 
 Bethe ansatz equations are presented. 
 In subsection \ref{sec:QQinv}, the QQ-relations and the Bethe ansatz equations derived in subsections \ref{sec:RR} and \ref{sec:SR} are compactly 
 expressed in terms of simple root systems of underlying algebras.  
We remark that QQ-relations are expressed in terms of root systems 
of underlying (non-super) Lie algebras 
 in connection with discrete Miura opers \cite{MV04} 
and the ODE/IM correspondence \cite{MRV15,MRV15-2}. 
 Our formulation is different from theirs in that
  we use a simple root system of the Lie superalgebra $gl(2r|1)$ for the non-super algebra $U_{q}(so(2r+1)^{(1)})$ case. 
  In subsection \ref{sec:BS}, we derive various T-functions by Bethe strap procedures
   in the analytic Bethe ansatz. 
  One will find similar objects in the context of 
  q-characters in representation theory \cite{FR99,FM99}. 
  We remark that the notion of the Bethe strap  appeared \cite{KS94-1,Su95} 
before the q-characters were introduced. 
 In subsection \ref{sec:SpR}, T-functions for spinorial representations are presented. 
 We remark that T-functions for spinorial representations are obtained as reductions 
 of T-functions of  asymptotic typical representations of $U_{q}(gl(M|N)^{(1)})$, 
 as already demonstrated in \cite{T11,T21} for the $U_{q}(so(2r+1)^{(1)})$ case (a reduction of $(M,N)=(2r,1)$ case). 
 Section \ref{sec:CR} is devoted to concluding remarks. 
 We can consider more reductions to T-functions obtained by reductions. 
 In Appendix A, we consider reductions of $U_{q}(osp(2r|2s)^{(1)})$ case to $U_{q}(osp(2r|2s)^{(2)})$ case. 
In Appendix B, we consider the (super)character limit of T-functions, and their 
decomposition with respect to (super)characters of  
finite Lie superalgebras. We show that 
special cases of them coincide with the characters  
of the Kirillov-Reshetikhin modules of quantum affine algebras (or their Yangian counterparts).

%%%%%%%%%%
\section{Preliminaries}
\label{preli}
%%%%%%%%%%%%%%%%
\subsection{Notation}
For  $M,N \in \mathbb{Z}_{\ge 0}$, we define sets
\begin{align}
\begin{split}
&{\mathfrak B}=\{1,2,\dots, M \},  \\
& {\mathfrak F}=\{M+1,M+2, \dots, M+N \}, \\
&
{\mathfrak I}={\mathfrak B} \sqcup {\mathfrak F},
\end{split}
\end{align}
and an operation
\begin{align}
\begin{split}
&  b^{*}=M+1-b \quad \text{for} \quad b \in {\mathfrak B}, 
 \qquad f^{*}=2M+N+1-f \quad \text{for} \quad f \in {\mathfrak F},
\\
& I^{*}=\{a^{*}| a \in I \} \quad \text{for} \quad I \subset {\mathfrak I}  .
\end{split}
\label{*set} 
\end{align} 
We define the set of acceptable sets
\footnote{This type of sets appeared in the context of  Q-operators associated with $Y(so(2r))$ \cite{FFK20}.}
\begin{align}
\Am =\{ I \subset {\mathfrak I} | a^{*} \notin I  \ \text{for any} \  a \in I  \} .
\end{align}
By definition, $|I | \le |{\mathfrak I} |/2=(M+N)/2 $ if $I \subset \Am$. 
We will use a grading parameter defined on the set $  {\mathfrak B} \sqcup  {\mathfrak F}$:  
\begin{align} 
p_{a}=1 \quad \text{for} \quad a \in {\mathfrak B}, \qquad 
p_{a}=-1 \quad \text{for} \quad a \in {\mathfrak F}. 
\end{align}
We remark that 
$p_{a}=p_{a^{*}}$ holds for any $a \in {\mathfrak I}$. 
We  use the following notation for a matrix:
\begin{align}
(a_{ij})_{i \in B \atop j \in F}=
\begin{pmatrix}
a_{b_{1},f_{1}} & a_{b_{1},f_{2}} & \cdots & a_{b_{1},f_{n}} \\
a_{b_{2},f_{1}} & a_{b_{2},f_{2}} & \cdots & a_{b_{2},f_{n}} \\
\hdotsfor{4} \\
a_{b_{m},f_{1}} & a_{b_{m},f_{2}} & \cdots & a_{b_{m},f_{n}}
\end{pmatrix}
,
\end{align}
where $B=(b_{1},\dots , b_{m} )$, $F=(f_{1},\dots , f_{n} )$. 
In case the tuples $B$ and $F$ are regarded as sets
$B=\{b_{1},\dots , b_{m} \}$, $F=\{f_{1},\dots , f_{n} \}$,  we assume 
 $b_{1}<\dots < b_{m}$, $f_{1}<\dots < f_{n}$.

Consider an arbitrary function $f(u)$ of  $u \in {\mathbb C}$ (the spectral parameter). 
In this paper we use the following notation for a shift of the spectral parameter: 
$f^{[a]}=f(u+a \hbar)$ for an additive shift, and 
$f^{[a]}=f(uq^{a \hbar})$ for a multiplicative shift ($q$-difference), where 
$a \in {\mathbb C}$. 
Here the unit of the shift $\hbar $ is a non-zero fixed complex number.  
If there is no shift ($a=0$), $[0]$ is often omitted: $f^{[0]}=f=f(u)$. 
In the following, we mainly use an additive shift with $\hbar=1$. 
Throughout the paper we assume that the deformation parameter $q$ of the quantum 
affine superalgebras is not a root of unity. 

We denote by $S_{r}$ the symmetric group of order $r$, and by 
$S(I)$ the symmetric group over the elements of a set (or tuple) $I$. 
Let $\tau_{ab} \in S(I)$ be the transposition 
such that $\tau_{ab}(a)=b, \tau_{ab}(b)=a$ and $\tau_{ab}(c)=c$ 
for $c \ne a$, $c \ne b$ ($a,b,c \in I$). 
It is convenient to introduce an operation $\overline{\tau}_{jk}$ that swaps the $j$-th and $k$-th components of a tuple $I$. 
For example, for the tuple $I=(3,1,5,2,4)$, we have $\overline{\tau}_{14}(I)=\tau_{32}(I)=(2,1,5,3,4)$. 

A  partition is a non increasing sequence of positive 
 integers $\mu=(\mu_{1},\mu_{2},\dots) $: 
$\mu_{1} \ge \mu_{2} \ge \dots \ge 0$. 
We often write this in the form $\mu=(l^{m_{r}},(l-1)^{m_{l-1}},\dots,2^{m_{2}},1^{m_{1}})$, 
where $l=\mu_{1}$, and 
 $m_{k}={\rm Card}\{j|\mu_{j}=k \}$. 
We use the same symbol $\mu$ for the Young diagram corresponding to a partition $\mu$. 
The conjugate (transposition) of $\mu$ is defined by $\mu^{\prime}=(\mu_{1}^{\prime},\mu_{2}^{\prime},\dots)$,  
where $\mu_{j}^{\prime}={\rm Card}\{k| \mu_{k} \ge j\}$. 
%%%%%%%%%%%%%%%%%%%%%%%%%
\subsection{Lie superalgebras}
\label{Liesuper}
%In this subsection we briefly summarize Lie superalgebras. 
Although the underlying algebras of the quantum integrable systems in question 
are quantum affine superalgebras (or superYangians), we need simple roots 
and highest weights of (finite) Lie superalgebras for labeling of quantities which 
we discuss in what follows. For details of Lie superalgebras, see for example, \cite{Kac78,FSS89,FSS20,CW12}.

Each simple root $\alpha $ of basic Lie superalgebras carries the grading (parity) $p_{\alpha} \in \{1,-1\}$. 
The root $\alpha$ is called an {\em even root} (bosonic root) if $p_{\alpha}$=1, and an {\em odd root} (fermionic root) if $p_{\alpha}=-1$ . 
Lie superalgebras have several inequivalent simple root systems. 
The simplest one is  the {\em distinguished simple root system}, which contains only one odd root. 
Let $\{\epsilon_{i}\}_{i=1}^{M+N}$ be a basis of the dual space of a Cartan subalgebra of $gl(M|N)$ with 
the bilinear form such that $(\epsilon_{i}| \epsilon_{j})=(\epsilon_{j}| \epsilon_{i})=p_{i}\delta_{ij}$ and  $p_{\epsilon_{i}}=p_{i}$. 
It is convenient  to set $\varepsilon_{i}=\epsilon_{i}$ for $1 \le i  \le M$, $\delta_{i}=\epsilon_{i+M}$ for $1 \le i  \le N$, thus 
$(\varepsilon_{i}| \varepsilon_{j})=\delta_{ij}$, $(\varepsilon_{i}| \delta_{j})=(\delta_{j}|\varepsilon_{i})=0$, 
$(\delta_{i}| \delta_{j})=-\delta_{ij}$. 
We will describe simple root systems of type B, C and D Lie superalgebras  (orthosymplectic Lie superalgebras) 
in terms of subsets 
\footnote{We abuse notation and use the same symbol for different objects.}
of the bases $\{\epsilon_{i}\}_{i=1}^{M+N}$.

%%%%%%%%%%%%%
One can draw a Dynkin diagram for any simple root system. 
To each simple root $\alpha $, one assigns one of the following three types of dots: 

\begin{tikzpicture}[x=1.1pt,y=1.1pt,line width=0.8pt]
\draw (0,0) node[left] {white dot};
\draw (7,0) circle (6);
\draw (15,-1) node[right] {if $(\alpha|\alpha) \ne 0$ and $p_{\alpha}=1,$};
\end{tikzpicture} 

\begin{tikzpicture}[x=1.1pt,y=1.1pt,line width=0.8pt]
\draw (0,0) node[left] {black dot};
\fill (7,0) circle (6);
\draw (16,0) node[right] {if $(\alpha|\alpha) \ne 0$ and $p_{\alpha}=-1,$ and};
\end{tikzpicture}

\begin{tikzpicture}[x=1.1pt,y=1.1pt,line width=0.8pt]
\draw (0,0) node[left] {gray dot};
\draw (7,0) circle (6);
\draw (7-4.24,-4.24)--(7+4.24,4.24);
\draw (7-4.24,4.24)--(7+4.24,-4.24);
\draw (15,0) node[right] {if $(\alpha|\alpha) = 0$.};
\end{tikzpicture}

\begin{tikzpicture}[x=1.1pt,y=1.1pt,line width=0.8pt]
\draw (0,0) node[left] {We also use a symbol};
\fill (7,0) circle (3);
\draw (7,-3) node[below] {$\alpha$};
\draw (16,0) node[right] {to denote one of the above three dots for the root $\alpha $.};
\end{tikzpicture}

The black dot appears in Dynkin diagrams of $osp(2r+1|2s)$.
In \cite{LM21}, the Dynkin diagrams for $osp(2r+1|2s)$ and $osp(2r|2s)$ are defined by attaching one more node to the Dynkin 
diagram for $gl(r|s)$ associated with a ``parity sequence''. 
The parity sequence in \cite{LM21} corresponds to a  subset $(p_{i_{1}},p_{i_{2}},\dots , p_{i_{r+s}})$
\footnote{ or $(p_{i_{M+N-r-s+1}},p_{i_{M+N-r-s+2}},\dots , p_{i_{M+N}})$} 
of the grading parameters of $gl(M|N)$ in this paper. 
Note however that 
we will mainly use tuples (made from ${\mathfrak I}$), rather than the parity sequences, 
to describe the simple root systems of $osp(2r+1|2s)$ and $osp(2r|2s)$. 
%since they are more convenient to realize our proposal [section 3.7, \cite{T11}].  

%%%%%%%%%%%%%%%%%%%%%%%%%%%%%%%
\subsubsection{Simple root systems and highest weights}
\paragraph{Type A}
Let $I_{M+N}=(i_{1},i_{2}, \dots , i_{M+N})$ be any one of the permutations of the tuple $(1,2, \dots, M+N)$. 
The simple root system of  $gl(M|N)$ associated with the tuple $I_{M+N}$ is defined by 
\begin{align}
\alpha_{a}=\epsilon_{i_{M+N+1-a}}-\epsilon_{i_{M+N-a}}.  \label{rootA}
\end{align}

The corresponding Dynkin diagram is given by 

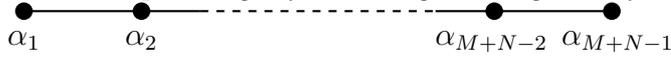
\begin{tikzpicture}[x=1.1pt,y=1.1pt,line width=0.8pt]
\draw (0,0) -- (60,0);
\draw[dashed] (60,0) -- (140,0);
\draw (140,0) -- (200,0);
%\draw[double,double distance=3pt] (200,0) -- (240-6,0);
%
\fill (0,0) circle (3);
\fill (40,0) circle (3);
\fill (160,0) circle (3);
\fill (200,0) circle (3);
%\draw (240,0) circle (6);
%
\draw (0,-3) node[below] {$\alpha_{1}$};
\draw (40,-3) node[below] {$\alpha_{2}$};
\draw (159,-3) node[below] {$\alpha_{M+N-2}$};
\draw (202,-3) node[below] {$\alpha_{M+N-1}$};
%\draw (240,-3) node[below] {$\alpha_{M+N}$};
\end{tikzpicture} 

%There is no black dot for the type A algebra. 
%%%%%%%%%%%%%%%%%%%%%%
In particular for the case $I_{M+N}=(M+N,\dots , 2,1)$, \eqref{rootA} reduces to the distinguished simple root system: 
\begin{align}
\begin{split}
\alpha_{i}&=\varepsilon_{i}-\varepsilon_{i+1} \quad \text{for} \quad i \in \{1,2,\dots , M-1\},
\\
\alpha_{M}&=\varepsilon_{M}-\delta_{1},
\\
\alpha_{i+M}&=\delta_{i}-\delta_{i+1} \quad \text{for} \quad i \in \{1,2,\dots , N-1\},
\end{split}
\label{rootAd}
\end{align}
and the corresponding Dynkin diagram is given by 

%%%
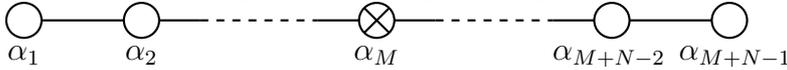
\begin{tikzpicture}[x=1.1pt,y=1.1pt,line width=0.8pt]
\draw  (0,0) circle (6);
\draw (6,0) -- (34,0);
\draw  (40,0) circle (6);
\draw (46,0) -- (60,0);
\draw[dashed] (60,0) -- (100,0);
\draw (100,0) -- (114,0);
\draw  (120,0) circle (6);
\draw (120-4.24,-4.24) -- (120+4.24,4.24);
\draw (120-4.24,4.24) -- (120+4.24,-4.24);
\draw (126,0) -- (140,0);
\draw[dashed] (140,0) -- (180,0);
\draw (180,0) -- (194,0);
\draw (200,0) circle (6);
\draw (206,0) -- (234,0);
\draw (240,0) circle (6);
\draw (0,-5) node[below] {$\alpha_{1}$};
\draw (40,-5) node[below] {$\alpha_{2}$};
\draw (120,-5) node[below] {$\alpha_{M}$};
\draw (199,-5) node[below] {$\alpha_{M+N-2}$};
\draw (242,-5) node[below] {$\alpha_{M+N-1}$};
\end{tikzpicture} 
%

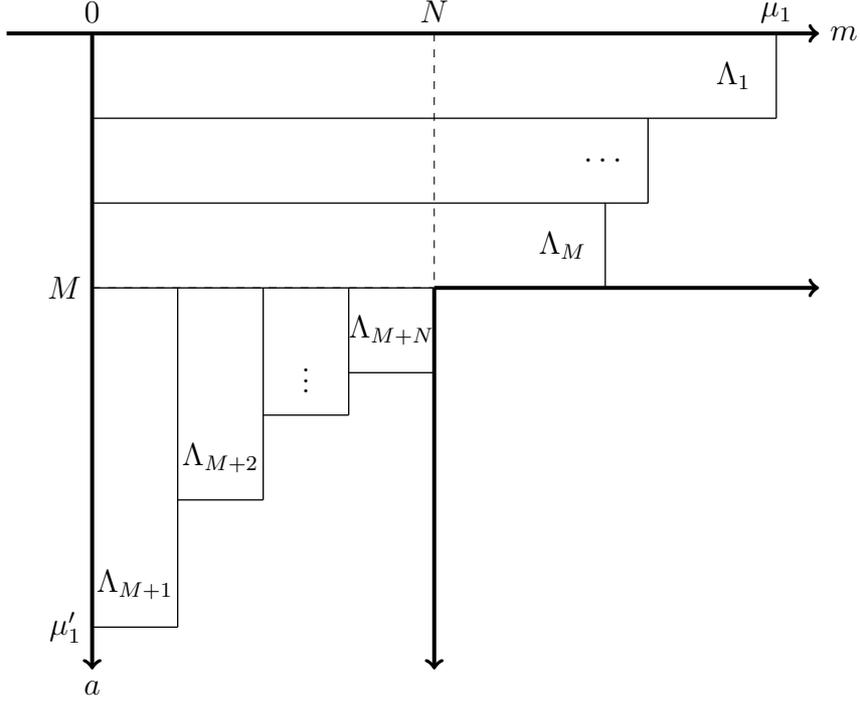
\begin{figure}
\centering
\begin{tikzpicture}[x=1.6pt,y=1.6pt]
%,line width=1.8pt]
\draw[-to,line width=1.5pt]  (-20,0) -- (170,0);
\draw[-to,line width=1.5pt] (80,-60) -- (170,-60);
\draw[-to,line width=1.5pt] (0,0) -- (0,-150);
\draw[-to,line width=1.5pt] (80,-60) -- (80,-150);
\draw[dashed]  (0,-60) -- (80,-60);
\draw[dashed]  (80,0) -- (80,-60);
\draw[line width=0.5pt] (0,-20) -- (160,-20);
\draw[line width=0.5pt] (160,0) -- (160,-20);
\draw[line width=0.5pt] (0,-40) -- (130,-40);
\draw[line width=0.5pt] (130,-20) -- (130,-40);
\draw[line width=0.5pt] (0,-60) -- (120,-60);
\draw[line width=0.5pt] (120,-40) -- (120,-60);
\draw[line width=0.5pt] (60,-80) -- (80,-80);
\draw[line width=0.5pt] (60,-60) -- (60,-90);
\draw[line width=0.5pt] (40,-90) -- (60,-90);
\draw[line width=0.5pt] (40,-60) -- (40,-110);
\draw[line width=0.5pt] (20,-110) -- (40,-110);
\draw[line width=0.5pt] (20,-60) -- (20,-140);
\draw[line width=0.5pt] (0,-140) -- (20,-140);
%\draw[-to,line width=1pt] (32) -- node[midway,left]{$F_{1}^{[2]}$} (22); 
%\draw[dashed] (60,0) -- (140,0);
%\draw[double,double distance=3pt] (200,0) -- (240-6,0);
%
%\draw (240,0) circle (6);
%
\draw (0,0) node[above] {$0$};
\draw (80,0) node[above] {$N$};
\draw (160,0) node[above] {$\mu_{1}$};
\draw (0,-60) node[left] {$M$};
\draw (0,-140) node[left] {$\mu_{1}^{\prime}$};
\draw (150,-10) node {$\Lambda_{1}$};
\draw (120,-30) node {$ \cdots $};
\draw (110,-50) node {$ \Lambda_{M}$};
\draw (10,-130) node {$ \Lambda_{M+1}$};
\draw (30,-100) node {$ \Lambda_{M+2}$};
\draw (50,-80) node {$ \vdots $};
\draw (70,-70) node {$ \Lambda_{M+N}$};
\draw (0,-150) node[below] {$a$};
\draw (170,0) node[right] {$m$};
\end{tikzpicture} 
\caption{$[M,N]$-hook: the Young diagram $\mu $ is related to the highest weight \eqref{HW-A} by \eqref{YW-A}.}
\label{MN-hookA}
\end{figure}

%%%%%%
Let $V(\Lambda )$ be the irreducible representation of $gl(M|N)$ with the highest weight
\begin{align}
\Lambda =\sum_{j=1}^{M} \Lambda_{j} \varepsilon_{j} +\sum_{j=1}^{N} \Lambda_{M+j} \delta_{j} ,
\label{HW-A}
\end{align}
where $\Lambda_{j} \in \mathbb{C}$.
The Kac-Dynkin labels $[b_{1},b_{2},\dots , b_{M+N-1}]$ of  $V(\Lambda )$ are defined by $b_{j}=2(\Lambda | \alpha_{j})/(\alpha_{j} | \alpha_{j})$ 
if $(\alpha_{j} | \alpha_{j}) \ne 0$,  $b_{j}=(\Lambda | \alpha_{j})/(\alpha_{j} |\alpha_{j^{\prime}})$ 
for some $j^{\prime}$ such that $(\alpha_{j} |\alpha_{j^{\prime}}) \ne 0$ if $(\alpha_{j} | \alpha_{j}) = 0$. 
For the distinguished simple roots \eqref{rootAd}, we have
\footnote{Here we set $M^{\prime}=M+1$.}
\begin{align}
b_{j}=\Lambda_{j}-\Lambda_{j+1} \quad \text{for} \quad j \ne M, \qquad
b_{M}=\Lambda_{M}+\Lambda_{M+1}. \label{KD-A}
\end{align}
$V(\Lambda )$ is finite dimensional if $b_{j} \in \mathbb{Z}_{\ge 0}$ for $j \ne M$. 
In case $\Lambda_{j} \in \mathbb{Z}_{\ge 0} $, these parameters are related to 
an $[M,N]$-hook partition $\mu =(\mu_{1},\mu_{2}, \dots )$, $\mu_{1} \ge \mu_{2} \ge \dots \ge 0$, 
$\mu_{M+1} \le N$: 
\begin{align}
\Lambda_{j}=\mu_{j} \quad \text{for} \quad j \in \{1,2,\dots , M \}, \quad 
\Lambda_{M+j}=\max \{ \mu_{j}^{\prime} -M ,0\} 
\quad \text{for} \quad j \in \{1,2,\dots , N \} ,
\label{YW-A}
\end{align}
where $\mu_{k}^{\prime}=\mathrm{Card}\{ j| \mu_{j} \ge k \}$. The $[M,N]$-hook partition describes a Young 
diagram in the $[M,N]$-hook (see Figure \ref{MN-hookA}). 
%%%%%%%%%%%%
%%%%%%%%%%%%%%%%%%%%%%%%%%%%%%%%%%%%%%%%%%%%%

From now on, we  consider the case that the elements of the 
tuple $I_{M+N}$ satisfy $i^{*}_{k}=i_{M+N+1-k}$ for any $k \in \mathfrak{I}$ (for type B, C, D superalgebras). 
We formally set $\epsilon_{k^{*}}= - \epsilon_{k}$. 
%%%%%%%%
\paragraph{Type B}
Set $(M,N)=(2r,2s+1)$, where $r,s \in \mathbb{Z}_{\ge 0}$, $r+s \ge 1$. In this case the tuple has the form $I_{2r+2s+1}=(i_{1},i_{2},\dots , i_{r+s},2r+s+1,i_{r+s}^{*},\dots , i_{2}^{*},i_{1}^{*})$. 
Note that any elements of this tuple are mutually distinct. 
The simple root system of $B(r|s)=osp(2r+1|2s)$ associated with the tuple $I_{2r+2s+1}$ is defined by  
\begin{align}
\beta_{a}=\epsilon_{i^{*}_{a}}-\epsilon_{i^{*}_{a+1}} \quad
 \text{for} \quad a \in \{1,2,\dots , r+s-1 \}, 
 \quad \beta_{r+s}=\epsilon_{i^{*}_{r+s}},
 \label{rootB}
\end{align}

and the corresponding Dynkin diagrams are given by
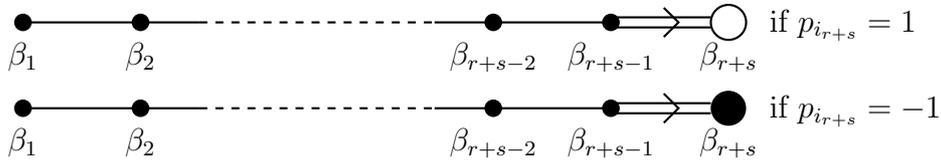
\footnote{We use the tuple $I_{2r+2s+1}$ to emphasize the connection with $gl(2r|2s+1)$, and 
 the last $r+s$ elements for 
labeling of the simple roots. It is possible to use the first $r+s$ elements instead (this looks more standard). In this case, it would be better to reverse the order of the labeling of \eqref{rootA} 
(as $\alpha_{a}=\epsilon_{i_{a}}-\epsilon_{i_{a+1}}$). 
Similar remarks can be applied to the type C and D superalgebras.}

\begin{tikzpicture}[x=1.1pt,y=1.1pt,line width=0.8pt]
\draw (0,0) -- (60,0);
\draw[dashed] (60,0) -- (140,0);
\draw (140,0) -- (200,0);
\draw[double,double distance=3pt] (200,0) -- (240-6,0);
\draw (223-5,5) -- (223,0) -- (223-5,-5);
\fill (0,0) circle (3);
\fill (40,0) circle (3);
\fill (160,0) circle (3);
\fill (200,0) circle (3);
\draw (240,0) circle (6);
\draw (0,-3) node[below] {$\beta_{1}$};
\draw (40,-3) node[below] {$\beta_{2}$};
\draw (160,-3) node[below] {$\beta_{r+s-2}$};
\draw (200,-3) node[below] {$\beta_{r+s-1}$};
\draw (240,-3) node[below] {$\beta_{r+s}$};
\draw (250,-1) node[right] {if $ p_{i_{r+s}}=1$};
\end{tikzpicture} 
%%%%%%%%

\begin{tikzpicture}[x=1.1pt,y=1.1pt,line width=0.8pt]
\draw (0,0) -- (60,0);
\draw[dashed] (60,0) -- (140,0);
\draw (140,0) -- (200,0);
\draw[double,double distance=3pt] (200,0) -- (240-6,0);
\draw (223-5,5) -- (223,0) -- (223-5,-5);
\fill (0,0) circle (3);
\fill (40,0) circle (3);
\fill (160,0) circle (3);
\fill (200,0) circle (3);
\fill (240,0) circle (6);
\draw (0,-3) node[below] {$\beta_{1}$};
\draw (40,-3) node[below] {$\beta_{2}$};
\draw (160,-3) node[below] {$\beta_{r+s-2}$};
\draw (200,-3) node[below] {$\beta_{r+s-1}$};
\draw (240,-3) node[below] {$\beta_{r+s}$};
\draw (250,-1) node[right] {if $ p_{i_{r+s}}=-1$};
\end{tikzpicture} 
%%%%%%%%%%%%%%%%

In particular for the case $I_{2r+2s+1}=(2r+2s+1, 2r+2s, \dots ,2r+s+3, 2 r+s+2,2r,2r-1, 
\dots ,r+2,r+1,2r+s+1 ,r,r-1,\dots, 2,1,2r+s,2r+s-1,\dots ,2r+2,2r+1)$, \eqref{rootB} reduces to 
the distinguished simple root system of $B(r|s)=osp(2r+1|2s)$.
\subparagraph{The case $r>0$:} We have 
\begin{align}
\begin{split}
\beta_{i}&=\delta_{i}-\delta_{i+1} \quad \text{for} \quad i \in \{1,2,\dots , s-1\},
\\
\beta_{s}&=\delta_{s}-\varepsilon_{1},
\\
\beta_{i+s}&=\varepsilon_{i}-\varepsilon_{i+1} \quad \text{for} \quad i \in \{1,2,\dots , r-1\},
\\
\beta_{r+s}&=\varepsilon_{r},
\end{split}
\label{rootBd}
\end{align}

and the corresponding Dynkin diagram is given by 

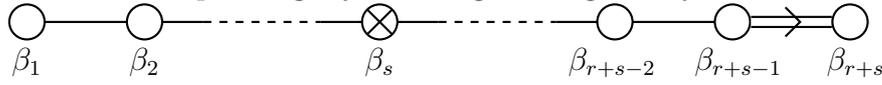
\begin{tikzpicture}[x=1.1pt,y=1.1pt,line width=0.8pt]
\draw  (0,0) circle (6);
\draw (6,0) -- (34,0);
\draw  (40,0) circle (6);
\draw (46,0) -- (60,0);
\draw[dashed] (60,0) -- (100,0);
\draw (100,0) -- (114,0);
\draw  (120,0) circle (6);
\draw (120-4.24,-4.24) -- (120+4.24,4.24);
\draw (120-4.24,4.24) -- (120+4.24,-4.24);
\draw (126,0) -- (140,0);
\draw[dashed] (140,0) -- (180,0);
\draw (180,0) -- (194,0);
\draw (200,0) circle (6);
\draw (206,0) -- (234,0);
\draw (240,0) circle (6);
\draw[double,double distance=3pt] (246.3,0) -- (273.8,0);
\draw (263-5,5) -- (263,0) -- (263-5,-5);
\draw (280,0) circle (6);
\draw (0,-5) node[below] {$\beta_{1}$};
\draw (40,-5) node[below] {$\beta_{2}$};
\draw (120,-5) node[below] {$\beta_{s}$};
\draw (199,-5) node[below] {$\beta_{r+s-2}$};
\draw (242,-5) node[below] {$\beta_{r+s-1}$};
\draw (282,-5) node[below] {$\beta_{r+s}$};
\end{tikzpicture} 
%%%

%%%%%%
Let $V(\Lambda )$ be the irreducible representation of $osp(2r+1|2s)$ with the highest weight
\begin{align}
\Lambda =\sum_{j=1}^{s} \Lambda_{j} \delta_{j} +\sum_{j=1}^{r} \Lambda_{s+j} \varepsilon_{j} ,
\label{HW-B}
\end{align}
where $\Lambda_{j} \in \mathbb{C}$.
The Kac-Dynkin labels $[b_{1},b_{2},\dots , b_{r+s}]$ of  $V(\Lambda )$ are defined by
 $b_{j}=2(\Lambda | \beta_{j})/(\beta_{j} | \beta_{j})$ 
if $(\beta_{j} | \beta_{j}) \ne 0$,  $b_{j}=(\Lambda | \beta_{j}) /(\beta_{j} |\beta_{j^{\prime}})$ 
for some $j^{\prime}$ such that $(\beta_{j} |\beta_{j^{\prime}}) \ne 0$ if $(\beta_{j} | \beta_{j}) = 0$. 
For the distinguished simple roots \eqref{rootBd}, we have
\footnote{Here we set $s^{\prime}=s+1$.}
\begin{align}
b_{j}=\Lambda_{j}-\Lambda_{j+1} \quad \text{for} \quad j \ne s,r+s, \qquad
b_{s}=\Lambda_{s}+\Lambda_{s+1}, \qquad b_{r+s}=2\Lambda_{r+s}. \label{KD-B}
\end{align}
$V(\Lambda )$ is finite dimensional if $b_{j} \in \mathbb{Z}_{\ge 0}$ for $j \ne s$, 
$c=b_{s}-b_{s+1}-b_{s+2}-\cdots -b_{r+s-1}-b_{r+s}/2 \in \mathbb{Z}_{\ge 0}$, and 
$b_{s+c+1}=b_{s+c+2}=\dots =b_{r+s}=0$ if $c < r$. 
In case $\Lambda_{j} \in \mathbb{Z}_{\ge 0} $, these parameters are related to 
an $[r,s]$-hook partition $\mu =(\mu_{1},\mu_{2}, \dots )$, $\mu_{1} \ge \mu_{2} \ge \dots \ge 0$, 
$\mu_{r+1} \le s$: 
\begin{align}
\Lambda_{j}=\mu_{j}^{\prime} \quad \text{for} \quad j \in \{1,2,\dots , s \}, \quad 
\Lambda_{s+j}=\max \{ \mu_{j} -s ,0\} 
\quad \text{for} \quad j \in \{1,2,\dots , r \} ,
\label{YW-B}
\end{align}
where $\mu_{k}^{\prime}=\mathrm{Card}\{ j| \mu_{j} \ge k \}$. The $[r,s]$-hook partition describes a Young 
diagram in the $[r,s]$-hook. This is embedded into the $[2r,2s+1]$-hook of $gl(2r|2s+1)$ (see Figure \ref{MN-hookB}). 
%%%%%%%%%%%%
\begin{figure}
\centering
\begin{tikzpicture}[x=1.6pt,y=1.6pt]
%,line width=1.8pt]
\draw[-to,line width=1.5pt]  (-20,0) -- (230,0);
\draw[-to,dashed, line width=1.5pt] (80,-60) -- (230,-60);
\draw[-to,line width=1.5pt] (180,-120) -- (230,-120);
\draw[-to,line width=1.5pt] (0,0) -- (0,-170);
\draw[-to,dashed,line width=1.5pt] (80,-60) -- (80,-170);
\draw[-to,line width=1.5pt] (180,-120) -- (180,-170);
\draw[dashed]  (0,-60) -- (80,-60);
\draw[dashed]  (0,-120) -- (180,-120);
\draw[dashed]  (80,0) -- (80,-60);
\draw[dashed]  (180,0) -- (180,-120);
\draw[line width=0.5pt] (80,-20) -- (210,-20);
\draw[line width=0.5pt] (210,0) -- (210,-20);
\draw[line width=0.5pt] (80,-40) -- (130,-40);
\draw[line width=0.5pt] (130,-20) -- (130,-40);
\draw[line width=0.5pt] (80,-60) -- (120,-60);
\draw[line width=0.5pt] (120,-40) -- (120,-60);
\draw[line width=0.5pt] (80,0) -- (80,-80);
\draw[line width=0.5pt] (60,-80) -- (80,-80);
\draw[line width=0.5pt] (60,0) -- (60,-90);
\draw[line width=0.5pt] (40,-90) -- (60,-90);
\draw[line width=0.5pt] (40,0) -- (40,-110);
\draw[line width=0.5pt] (20,-110) -- (40,-110);
\draw[line width=0.5pt] (20,0) -- (20,-140);
\draw[line width=0.5pt] (0,-140) -- (20,-140);
%\draw[-to,line width=1pt] (32) -- node[midway,left]{$F_{1}^{[2]}$} (22); 
%\draw[dashed] (60,0) -- (140,0);
%\draw[double,double distance=3pt] (200,0) -- (240-6,0);
%
%\draw (240,0) circle (6);
%
\draw (0,0) node[above] {$0$};
\draw (80,0) node[above] {$s$};
\draw (180,0) node[above] {$2s+1$};
\draw (210,0) node[above] {$\mu_{1}$};
\draw (0,-60) node[left] {$r$};
\draw (0,-120) node[left] {$2r$};
\draw (0,-140) node[left] {$\mu_{1}^{\prime}$};
\draw (200,-10) node {$\Lambda_{s+1}$};
\draw (120,-30) node {$ \cdots $};
\draw (110,-50) node {$ \Lambda_{s+r}$};
\draw (10,-130) node {$ \Lambda_{1}$};
\draw (30,-100) node {$ \Lambda_{2}$};
\draw (50,-80) node {$ \vdots $};
\draw (70,-70) node {$ \Lambda_{s}$};
\draw (0,-170) node[below] {$a$};
\draw (230,0) node[right] {$m$};
\end{tikzpicture} 
\caption{$[r,s]$-hook:in $[2r,2s+1]$-hook: the Young diagram $\mu $ is related to the highest weight \eqref{HW-B} by \eqref{YW-B}.}
\label{MN-hookB}
\end{figure}
%%%%%%%%%%%%%%%%%%%%%%%%%%%%%%%%%%%%%%%%%%%%%

%%%%%%%%%%
%%%%%%%%%%%%%%

\subparagraph{The case $r=0$:} 
The distinguished simple root system of $B(0|s)=osp(1|2s)$ is given by 
\begin{align}
\begin{split}
\beta_{i}&=\delta_{i}-\delta_{i+1} \quad \text{for} \quad i \in \{1,2,\dots , s-1\},
\\
\beta_{s}&=\delta_{s},
\end{split}
\label{rootBd0}
\end{align}

and the corresponding Dynkin diagram has the form

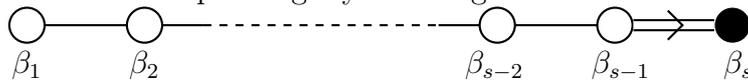
\begin{tikzpicture}[x=1.1pt,y=1.1pt,line width=0.8pt]
\draw  (0,0) circle (6);
\draw (6,0) -- (34,0);
\draw  (40,0) circle (6);
\draw (46,0) -- (60,0);
\draw[dashed] (60,0) -- (140,0);
\draw (140,0) -- (154,0);
\draw (160,0) circle (6);
\draw (166,0) -- (194,0);
\draw (200,0) circle (6);
\draw[double,double distance=3pt] (206.3,0) -- (233.8,0);
\draw (223-5,5) -- (223,0) -- (223-5,-5);
\fill (240,0) circle (6);
\draw (0,-5) node[below] {$\beta_{1}$};
\draw (40,-5) node[below] {$\beta_{2}$};
\draw (159,-5) node[below] {$\beta_{s-2}$};
\draw (202,-5) node[below] {$\beta_{s-1}$};
\draw (242,-5) node[below] {$\beta_{s}$};
\end{tikzpicture} 

%%%%%%
Let $V(\Lambda )$ be the irreducible representation of $osp(1|2s)$ with the highest weight
\begin{align}
\Lambda =\sum_{j=1}^{s} \Lambda_{j} \delta_{j}  ,
\label{HW-B0}
\end{align}
where $\Lambda_{j} \in \mathbb{C}$.
The Kac-Dynkin labels $[b_{1},b_{2},\dots , b_{s}]$ of  $V(\Lambda )$ are given by 
\begin{align}
b_{j}=\Lambda_{j}-\Lambda_{j+1} \quad \text{for} \quad j \ne s,  \qquad b_{s}=2\Lambda_{s}. \label{KD-B0}
\end{align}
$V(\Lambda )$ is finite dimensional if $b_{j} \in \mathbb{Z}_{\ge 0}$ for $j \ne s$, 
$c=b_{s}/2 \in \mathbb{Z}_{\ge 0}$. 
In case $\Lambda_{j} \in \mathbb{Z}_{\ge 0} $, these parameters are related to 
a $[0,s]$-hook partition $\mu =(\mu_{1},\mu_{2}, \dots )$, $\mu_{1} \ge \mu_{2} \ge \dots \ge 0$, 
$\mu_{1} \le s$: 
\begin{align}
\Lambda_{j}=\mu_{j}^{\prime} \quad \text{for} \quad j \in \{1,2,\dots , s \}.
\label{YW-B0}
\end{align}
The $[0,s]$-hook partition describes a Young 
diagram in the $[0,s]$-hook. This is embedded into the $[0,2s+1]$-hook of $gl(0|2s+1)$ (see Figure \ref{MN-hookB0}). 
%%%%%%%%%%
%%%%%%%%%%%%
\begin{figure}
\centering
\begin{tikzpicture}[x=1.6pt,y=1.6pt]
%,line width=1.8pt]
\draw[-to,line width=1.5pt]  (-20,0) -- (220,0);
%\draw[-to,dashed, line width=1.5pt] (80,-60) -- (230,-60);
%\draw[-to,line width=1.5pt] (180,-120) -- (230,-120);
\draw[-to,line width=1.5pt] (0,0) -- (0,-160);
\draw[-to,dashed,line width=1.5pt] (80,0) -- (80,-160);
\draw[-to,line width=1.5pt] (180,0) -- (180,-160);
%
%\draw[dashed]  (0,-60) -- (80,-60);
%\draw[dashed]  (0,-120) -- (180,-120);
\draw[dashed]  (80,0) -- (80,-60);
\draw[dashed]  (180,0) -- (180,-120);
\draw[line width=0.5pt] (80,0) -- (80,-80);
\draw[line width=0.5pt] (60,-80) -- (80,-80);
\draw[line width=0.5pt] (60,0) -- (60,-90);
\draw[line width=0.5pt] (40,-90) -- (60,-90);
\draw[line width=0.5pt] (40,0) -- (40,-110);
\draw[line width=0.5pt] (20,-110) -- (40,-110);
\draw[line width=0.5pt] (20,0) -- (20,-140);
\draw[line width=0.5pt] (0,-140) -- (20,-140);
%\draw[-to,line width=1pt] (32) -- node[midway,left]{$F_{1}^{[2]}$} (22); 
%\draw[dashed] (60,0) -- (140,0);
%\draw[double,double distance=3pt] (200,0) -- (240-6,0);
%
%\draw (240,0) circle (6);
%
\draw (0,0) node[above] {$0$};
\draw (80,0) node[above] {$s$};
\draw (180,0) node[above] {$2s+1$};
\draw (80,7) node[above] {$\mu_{1}$};
\draw (0,-140) node[left] {$\mu_{1}^{\prime}$};
\draw (10,-130) node {$ \Lambda_{1}$};
\draw (30,-100) node {$ \Lambda_{2}$};
\draw (50,-80) node {$ \vdots $};
\draw (70,-70) node {$ \Lambda_{s}$};
\draw (0,-160) node[below] {$a$};
\draw (220,0) node[right] {$m$};
\end{tikzpicture} 
\caption{$[0,s]$-hook:in $[0,2s+1]$-hook: the Young diagram $\mu $ is related to the highest weight \eqref{HW-B0} by \eqref{YW-B0}.}
\label{MN-hookB0}
\end{figure}
%%%%%%%%%%%%%%%%%%%%%%%%%%%%%%%%%%%%%%%%%%%%%

%%%%%%%%%%%%%%%%%%%%%%%%%%%%%%%%%%%%%%%%%%%%%%%

\paragraph{Type C and D}
Set $(M,N)=(2r,2s+2)$, where $r,s \in \mathbb{Z}_{\ge 0}$, $r+s \ge 1$. 
In this case, the tuple has the form $I_{2r+2s+2}=(i_{1},i_{2},\dots , i_{r+s+1},i_{r+s+1}^{*},\dots , i_{2}^{*},i_{1}^{*})$. 
Here we assume $i_{r+s+1}=2r+s+1$ or $2r+s+2$. 
Note that any elements of this tuple are mutually distinct. 
For this tuple $I_{2r+2s+2}$, we define
\begin{align}
\Upsilon=(-1)^{\mathrm{Card} \{ i_{a} \in \{1,2,\dots, r \} |  1 \le a \le r+s \}} 
=(-1)^{\mathrm{Card} \{ i_{a}^{*} \in \{r+1,r+2,\dots, 2r \} |  1 \le a \le r+s \}}.
 \label{sigD}
\end{align}
The simple root systems of $osp(2r|2s)$ ($D(r|s)=osp(2r|2s)$ if $r \ge 2$, $C(s+1)=osp(2|2s)$) associated with the tuple $I_{2r+2s+2}$  are defined as follows.
%
%.  
\subparagraph{The case $i_{r+s} \in \Bm$, $r \ge 1$, $s \ge 0$, $r+s \ge 2$ (type D):} 
We have 
\begin{align}
\begin{split}
\beta_{a}&=\epsilon_{i^{*}_{a}}-\epsilon_{i^{*}_{a+1}} \quad 
\text{for} \quad a \in \{1,2,\dots , r+s-2 \}, 
\\
\beta_{r+s-1}&=\epsilon_{i^{*}_{r+s-1}}-\epsilon_{i^{*}_{r+s}}, 
\\
\beta_{r+s}&=\epsilon_{i^{*}_{r+s-1}}+\epsilon_{i^{*}_{r+s}},
\end{split}
\label{rootD}
\end{align}

and the corresponding Dynkin diagrams are given by 

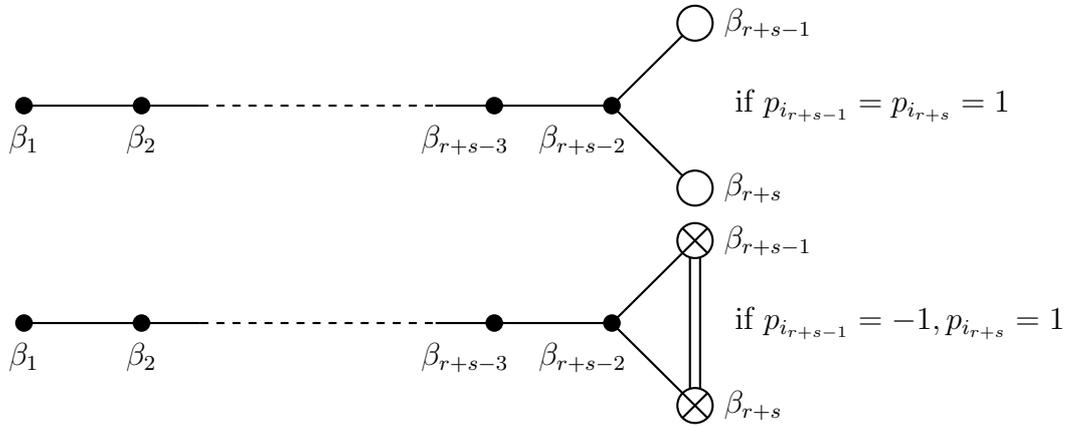
\begin{tikzpicture}[x=1.1pt,y=1.1pt,line width=0.8pt]
\draw (0,0) -- (60,0);
\draw (140,0) -- (200,0) -- (224.04,24.04);
\draw (200,0) -- (224.04,-24.04);
\draw[dashed] (60,0) -- (140,0);
\fill (0,0) circle (3);
\fill (40,0) circle (3);
\fill (160,0) circle (3);
\fill (200,0) circle (3);
\draw (228.28,28.28) circle (6);
\draw (228.28,-28.28) circle (6);
\draw (0,-3) node[below] {$\beta_{1}$};
\draw (40,-3) node[below] {$\beta_{2}$};
\draw (150,-3) node[below] {$\beta_{r+s-3}$};
\draw (190,-3) node[below] {$\beta_{r+s-2}$};
\draw (234.28,28.28) node[right] {$\beta_{r+s-1}$};
\draw (234.28,-28.28) node[right] {$\beta_{r+s}$};
\draw (238,0) node[right] {if $p_{i_{r+s-1}}=p_{i_{r+s}}=1$};
\end{tikzpicture} 
%%%%%%%%%%%%%%

\begin{tikzpicture}[x=1.1pt,y=1.1pt,line width=0.8pt]
\draw (0,0) -- (60,0);
\draw (140,0) -- (200,0) -- (224.04,24.04);
\draw (200,0) -- (224.04,-24.04);
\draw (228.28-4.24,28.28-4.24) -- (228.28+4.24,28.28+4.24);
\draw (228.28-4.24,28.28+4.24) -- (228.28+4.24,28.28-4.24);
\draw (228.28-4.24,-28.28+4.24) -- (228.28+4.24,-28.28-4.24);
\draw (228.28-4.24,-28.28-4.24) -- (228.28+4.24,-28.28+4.24);
\draw[double,double distance=3pt] (228.28,22.28) -- (228.28,-22.28);
\draw[dashed] (60,0) -- (140,0);
\fill (0,0) circle (3);
\fill (40,0) circle (3);
\fill (160,0) circle (3);
\fill (200,0) circle (3);
\draw (228.28,28.28) circle (6);
\draw (228.28,-28.28) circle (6);
\draw (0,-3) node[below] {$\beta_{1}$};
\draw (40,-3) node[below] {$\beta_{2}$};
\draw (150,-3) node[below] {$\beta_{r+s-3}$};
\draw (190,-3) node[below] {$\beta_{r+s-2}$};
\draw (234.28,28.28) node[right] {$\beta_{r+s-1}$};
\draw (234.28,-28.28) node[right] {$\beta_{r+s}$};
\draw (238,0) node[right] {if $p_{i_{r+s-1}}=-1, p_{i_{r+s}}=1$};
\end{tikzpicture} 
%%%%%%

%%%%%%%%

In particular for the case $I_{2r+2s+2}=(2r+2s+2, 2r+2s+1, \dots ,2r+s+4, 2 r+s+3,2r,2r-1, 
\dots ,r+2,r+1,2r+s+2,2r+s+1 ,r,r-1,\dots, 2,1,2r+s,2r+s-1,\dots ,2r+2,2r+1)$, \eqref{rootD} reduces to 
the distinguished simple root system of $D(r|s)=osp(2r|2s)$, $r>1$:
\begin{align}
\begin{split}
\beta_{i}&=\delta_{i}-\delta_{i+1} \quad \text{for} \quad i \in \{1,2,\dots , s-1\},
\\
\beta_{s}&=\delta_{s}-\varepsilon_{1},
\\
\beta_{i+s}&=\varepsilon_{i}-\varepsilon_{i+1} \quad \text{for} \quad i \in \{1,2,\dots , r-1\},
\\
\beta_{r+s}&=\varepsilon_{r-1}+\varepsilon_{r},
\end{split}
\label{rootDd}
\end{align}
and the corresponding Dynkin diagram is given by 

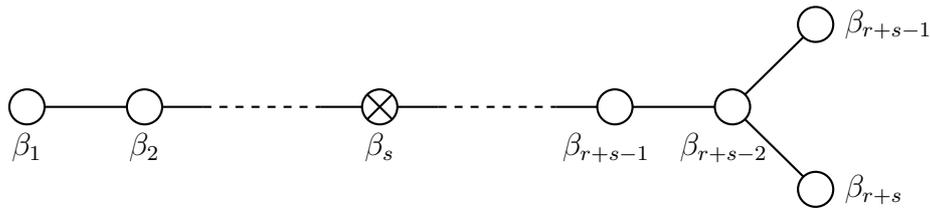
\begin{tikzpicture}[x=1.1pt,y=1.1pt,line width=0.8pt]
\draw  (0,0) circle (6);
\draw (6,0) -- (34,0);
\draw  (40,0) circle (6);
\draw (46,0) -- (60,0);
\draw[dashed] (60,0) -- (100,0);
\draw (100,0) -- (114,0);
\draw  (120,0) circle (6);
\draw (120-4.24,-4.24) -- (120+4.24,4.24);
\draw (120-4.24,4.24) -- (120+4.24,-4.24);
\draw (126,0) -- (140,0);
\draw[dashed] (140,0) -- (180,0);
\draw (180,0) -- (194,0);
\draw (200,0) circle (6);
\draw (206,0) -- (234,0); 
\draw (240,0) circle (6);
\draw (244.24,4.24) -- (264.24,24.04);
\draw (268.28,28.28) circle (6);
\draw (244.24,-4.24) -- (264.24,-24.04);
\draw (268.28,-28.28) circle (6);
\draw (0,-5) node[below] {$\beta_{1}$};
\draw (40,-5) node[below] {$\beta_{2}$};
\draw (120,-5) node[below] {$\beta_{s}$};
\draw (197,-5) node[below] {$\beta_{r+s-1}$};
\draw (237,-5) node[below] {$\beta_{r+s-2}$};;
\draw (274.28,28.28) node[right] {$\beta_{r+s-1}$};
\draw (274.28,-28.28) node[right] {$\beta_{r+s}$};
\end{tikzpicture} 
One can check $\Upsilon =0$ for \eqref{rootDd}. 
We remark that the description of the type D simple root systems here is redundant in that it 
contains extra simple root systems derived by changing the labeling of the $(r+s-1)$-th and $(r+s)$-th nodes of 
the Dynkin diagram of type D (by swapping $i_{r+s}$ and $i_{r+s}^{*}$ for the case $p_{i_{r+s}}=1$; this operation  
 changes the sign of $\Upsilon$). 
 In fact, the type D simple root system $\{ \beta_{a}\}_{a=1}^{r+s}$ with $\Upsilon$ defined by the tuple $I_{2r+2s+2}$ 
 is equivalent to the type D simple root system $\{ \beta_{a}^{\prime} \}_{a=1}^{r+s}$ with $-\Upsilon$ 
 defined by the tuple $\tau_{i_{r+s},i_{r+s}^{*}} (I_{2r+2s+2})$, where $\beta_{a}=:\beta_{a}^{\prime}$
  for $1 \le a \le r+s-2$,  
 $\beta_{r+s-1}=\epsilon_{i^{*}_{r+s-1}}-\epsilon_{i^{*}_{r+s}}=\epsilon_{i^{*}_{r+s-1}}+\epsilon_{i_{r+s}}
 =:\beta_{r+s}^{\prime}$, $\beta_{r+s}=\epsilon_{i^{*}_{r+s-1}}+\epsilon_{i^{*}_{r+s}}=\epsilon_{i^{*}_{r+s-1}}-\epsilon_{i_{r+s}}=:\beta_{r+s-1}^{\prime}$. Thus one can concentrate on the type D simple root systems with $\Upsilon =1$:  
one can start from the distinguished simple root system \eqref{rootDd} and applies Weyl reflections and odd reflections 
to each simple root repeatedly.

%%%%%%

%%%%
Let $V(\Lambda )$ be the irreducible representation of $osp(2r|2s)$ with the highest weight
\begin{align}
\Lambda =\sum_{j=1}^{s} \Lambda_{j} \delta_{j} +\sum_{j=1}^{r} \Lambda_{s+j} \varepsilon_{j} ,
\label{HW-D}
\end{align}
where $\Lambda_{j} \in \mathbb{C}$.
For the distinguished simple roots \eqref{rootDd}, 
the Kac-Dynkin labels $[b_{1},b_{2},\dots , b_{r+s}]$ of  $V(\Lambda )$ is given by
\footnote{Here we set $s^{\prime}=s+1$.}
\begin{align}
b_{j}=\Lambda_{j}-\Lambda_{j+1} \quad \text{for} \quad j \ne s,r+s, \qquad
b_{s}=\Lambda_{s}+\Lambda_{s+1}, \quad b_{r+s}=\Lambda_{r+s-1}+\Lambda_{r+s}. \label{KD-D}
\end{align}
$V(\Lambda )$ is finite dimensional if $b_{j} \in \mathbb{Z}_{\ge 0}$ for $j \ne s$, 
$c=b_{s}-b_{s+1}-b_{s+2}-\cdots -b_{r+s-2}-(b_{r+s-1}+b_{r+s})/2 \in \mathbb{Z}_{\ge 0}$, 
$b_{s+c+1}=b_{s+c+2}=\dots =b_{r+s}=0$ if $c < r-1$, and 
$b_{r+s-1} =b_{r+s}=0$ if $c =r-1$. 
In case $\Lambda_{j} \in \mathbb{Z}_{\ge 0} $, these parameters are related to 
an $[r,s]$-hook partition $\mu =(\mu_{1},\mu_{2}, \dots )$, $\mu_{1} \ge \mu_{2} \ge \dots \ge 0$, 
$\mu_{r+1} \le s$: 
\begin{align}
\Lambda_{j}=\mu_{j}^{\prime} \quad \text{for} \quad j \in \{1,2,\dots , s \}, \quad 
\Lambda_{s+j}=\max \{ \mu_{j} -s ,0\} 
\quad \text{for} \quad j \in \{1,2,\dots , r \} ,
\label{YW-Dp}
\end{align}
or 
\begin{align}
\Lambda_{j}&=\mu_{j}^{\prime} \quad \text{for} \quad j \in \{1,2,\dots , s \}, \quad 
\Lambda_{s+j}=\max \{ \mu_{j} -s ,0\} 
\quad \text{for} \quad j \in \{1,2,\dots , r-1 \}, 
 \nonumber \\
\Lambda_{r+s}&=-\max \{ \mu_{r} -s ,0\}  .
\label{YW-Dm}
\end{align}
 The $[r,s]$-hook partition describes a Young
\footnote{One may consider a ``branch cut'' along the lines $a=r-1$, $m \ge s$ and $r-1 \le a \le r$, $m=s$, to describe \eqref{YW-Dp} and \eqref{YW-Dm} at the same time. The main target domain of the T- and Q-systems 
for tensor representations of $U_{q}(osp(2s|2r)^{(1)})$ would be such an extended $[r,s]$-hook.}
diagram in the $[r,s]$-hook. This is embedded into the $[2r,2s+2]$-hook of $gl(2r|2s+2)$ (see Figure \ref{MN-hookD}). 
%%%%%%%%%%%%
\begin{figure}
\centering
\begin{tikzpicture}[x=1.6pt,y=1.6pt]
%,line width=1.8pt]
\draw[-to,line width=1.5pt]  (-20,0) -- (230,0);
\draw[-to,dashed, line width=1.5pt] (80,-60) -- (230,-60);
\draw[-to,line width=1.5pt] (200,-120) -- (230,-120);
\draw[-to,line width=1.5pt] (0,0) -- (0,-170);
\draw[-to,dashed,line width=1.5pt] (80,-60) -- (80,-170);
\draw[-to,line width=1.5pt] (200,-120) -- (200,-170);
\draw[dashed]  (0,-60) -- (80,-60);
\draw[dashed]  (0,-120) -- (200,-120);
\draw[dashed]  (80,0) -- (80,-60);
\draw[dashed]  (200,0) -- (200,-120);
\draw[line width=0.5pt] (80,-20) -- (220,-20);
\draw[line width=0.5pt] (220,0) -- (220,-20);
\draw[line width=0.5pt] (80,-40) -- (130,-40);
\draw[line width=0.5pt] (130,-20) -- (130,-40);
\draw[line width=0.5pt] (80,-60) -- (120,-60);
\draw[line width=0.5pt] (120,-40) -- (120,-60);
\draw[line width=0.5pt] (80,0) -- (80,-80);
\draw[line width=0.5pt] (60,-80) -- (80,-80);
\draw[line width=0.5pt] (60,0) -- (60,-90);
\draw[line width=0.5pt] (40,-90) -- (60,-90);
\draw[line width=0.5pt] (40,0) -- (40,-110);
\draw[line width=0.5pt] (20,-110) -- (40,-110);
\draw[line width=0.5pt] (20,0) -- (20,-140);
\draw[line width=0.5pt] (0,-140) -- (20,-140);
%\draw[-to,line width=1pt] (32) -- node[midway,left]{$F_{1}^{[2]}$} (22); 
%\draw[dashed] (60,0) -- (140,0);
%\draw[double,double distance=3pt] (200,0) -- (240-6,0);
%
%\draw (240,0) circle (6);
%
\draw (0,0) node[above] {$0$};
\draw (80,0) node[above] {$s$};
\draw (200,0) node[above] {$2s+2$};
\draw (220,0) node[above] {$\mu_{1}$};
\draw (0,-60) node[left] {$r$};
\draw (0,-120) node[left] {$2r$};
\draw (0,-140) node[left] {$\mu_{1}^{\prime}$};
\draw (210,-10) node {$\Lambda_{s+1}$};
\draw (120,-30) node {$ \cdots $};
\draw (110,-50) node {$ |\Lambda_{s+r}|$};
\draw (10,-130) node {$ \Lambda_{1}$};
\draw (30,-100) node {$ \Lambda_{2}$};
\draw (50,-80) node {$ \vdots $};
\draw (70,-70) node {$ \Lambda_{s}$};
\draw (0,-170) node[below] {$a$};
\draw (230,0) node[right] {$m$};
\end{tikzpicture} 
\caption{$[r,s]$-hook:in $[2r,2s+2]$-hook: the Young diagram $\mu $ is related to the highest weight \eqref{HW-D} by \eqref{YW-Dm}.}
\label{MN-hookD}
\end{figure}
%%%%%%%%%%%%%%%%%%%%%%%%%%%%%%%%%%%%%%%%%%%%%
%%%%%%%%%%%%%%%%%%%%%%%%%%

\subparagraph{The case $i_{r+s} \in \Fm$, $r \ge 0$, $s\ge 1$, $r+s \ge 1$ (type C):}
The simple root system is defined by  
\begin{align}
\begin{split}
\beta_{a}&=\epsilon_{i^{*}_{a}}-\epsilon_{i^{*}_{a+1}}, \quad 
 \text{for} \quad a \in \{1,2,\dots , r+s-1 \}, 
 \\
 \beta_{r+s}&=2\epsilon_{i^{*}_{r+s}},
 \end{split}
 \label{rootDC}
 \end{align}
 %%%%%%%
 and the corresponding Dynkin diagram is given by 
 
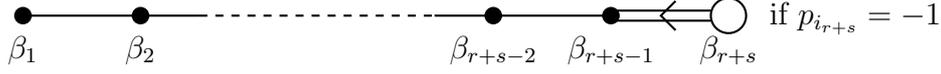
\begin{tikzpicture}[x=1.1pt,y=1.1pt,line width=0.8pt]
\draw (0,0) -- (60,0);
\draw[dashed] (60,0) -- (140,0);
\draw (140,0) -- (200,0);
\draw[double,double distance=3pt] (200,0) -- (240-6,0);
\draw (217+5,-5) -- (217,0) -- (217+5,5);
\fill (0,0) circle (3);
\fill (40,0) circle (3);
\fill (160,0) circle (3);
\fill (200,0) circle (3);
\draw (240,0) circle (6);
\draw (0,-3) node[below] {$\beta_{1}$};
\draw (40,-3) node[below] {$\beta_{2}$};
\draw (160,-3) node[below] {$\beta_{r+s-2}$};
\draw (200,-3) node[below] {$\beta_{r+s-1}$};
\draw (240,-3) node[below] {$\beta_{r+s}$};
\draw (250,-1) node[right] {if $ p_{i_{r+s}}=-1$};
\end{tikzpicture} 
%%%%%%%%%%%%%%%%%%

In particular for the case $r=1$, $s \ge 1$,
 $I_{2s+4}=(2,2s+4, 2s+3,\dots ,4,3,1)$, \eqref{rootDC} reduces to 
the distinguished simple root system of $C(s+1)=osp(2|2s)$:
\begin{align}
\beta_{1}&=\varepsilon_{1}-\delta_{1} , \\
\beta_{i}&=\delta_{i-1}-\delta_{i} \quad \text{for} \quad i \in \{2,3,\dots , s\},
\\
\beta_{s}&=2\delta_{s},
\end{align}
%%%%%
and the corresponding Dynkin diagram is given by

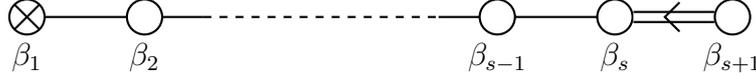
\begin{tikzpicture}[x=1.1pt,y=1.1pt,line width=0.8pt]
\draw  (0,0) circle (6);
\draw (-4.24,-4.24) -- (4.24,4.24);
\draw (-4.24,4.24) -- (4.24,-4.24);
\draw (6,0) -- (34,0);
\draw  (40,0) circle (6);
\draw (46,0) -- (60,0);
\draw[dashed] (60,0) -- (140,0);
\draw (140,0) -- (154,0);
\draw (160,0) circle (6);
\draw (166,0) -- (194,0);
\draw (200,0) circle (6);
\draw[double,double distance=3pt] (206.3,0) -- (233.8,0);
\draw (217+5,-5) -- (217,0) -- (217+5,5);
\draw (240,0) circle (6);
\draw (0,-5) node[below] {$\beta_{1}$};
\draw (40,-5) node[below] {$\beta_{2}$};
\draw (160,-5) node[below] {$\beta_{s-1}$};
\draw (200,-5) node[below] {$\beta_{s}$};
\draw (240,-5) node[below] {$\beta_{s+1}$};
\end{tikzpicture} 
%%%%%%%%%

%%%%
Let $V(\Lambda )$ be the irreducible representation of $osp(2|2s)$ with the highest weight
\begin{align}
\Lambda =\Lambda_{1} \varepsilon_{1} +\sum_{j=1}^{s} \Lambda_{1+j} \delta_{j} ,
\label{HW-Cp}
\end{align}
where $\Lambda_{j} \in \mathbb{C}$.
For the distinguished simple roots \eqref{rootDd}, 
the Kac-Dynkin labels $[b_{1},b_{2},\dots , b_{s+1}]$ of  $V(\Lambda )$ is given by
\footnote{Here we set $1^{\prime}=2$.}
\begin{align}
b_{1}=\Lambda_{1}+\Lambda_{2}, \qquad 
b_{j}=\Lambda_{j}-\Lambda_{j+1} \quad \text{for} \quad j \ne 1,s+1, \qquad
b_{s+1}=\Lambda_{s+1}. \label{KD-C}
\end{align}
$V(\Lambda )$ is finite dimensional if $b_{j} \in \mathbb{Z}_{\ge 0}$ for $j \ne 1$.  
In case $\Lambda_{j} \in \mathbb{Z}_{\ge 0} $, these parameters are related to 
a $[1,s]$-hook partition $\mu =(\mu_{1},\mu_{2}, \dots )$, $\mu_{1} \ge \mu_{2} \ge \dots \ge 0$, 
$\mu_{2} \le s$: 
\begin{align}
\Lambda_{1}= \mu_{1}, \qquad 
\Lambda_{j+1}= \max \{ \mu_{j}^{\prime}-s ,0\},  
\quad \text{for} \quad j \in \{1,2,\dots , s \} .
\label{YW-Cp}
\end{align}
 The $[1,s]$-hook partition describes a Young 
diagram in the $[1,s]$-hook. This is embedded into the $[2,2s+2]$-hook of $gl(2|2s+2)$ (see Figure \ref{MN-hookCp}). 
%%%%%%%%%%%%
\begin{figure}
\centering
\begin{tikzpicture}[x=1.6pt,y=1.6pt]
%,line width=1.8pt]
\draw[-to,line width=1.5pt]  (-20,0) -- (230,0);
\draw[-to,dashed, line width=1.5pt] (80,-20) -- (230,-20);
\draw[-to,line width=1.5pt] (200,-40) -- (230,-40);
\draw[-to,line width=1.5pt] (0,0) -- (0,-170);
\draw[-to,dashed,line width=1.5pt] (80,-40) -- (80,-170);
\draw[-to,line width=1.5pt] (200,-40) -- (200,-170);
\draw[dashed]  (0,-20) -- (80,-20);
\draw[dashed]  (0,-40) -- (200,-40);
\draw[dashed]  (80,0) -- (80,-60);
\draw[dashed]  (200,0) -- (200,-40);
\draw[line width=0.5pt] (0,-20) -- (220,-20);
\draw[line width=0.5pt] (220,0) -- (220,-20);
\draw[line width=0.5pt] (80,-20) -- (80,-80);
\draw[line width=0.5pt] (60,-80) -- (80,-80);
\draw[line width=0.5pt] (60,-20) -- (60,-90);
\draw[line width=0.5pt] (40,-90) -- (60,-90);
\draw[line width=0.5pt] (40,-20) -- (40,-110);
\draw[line width=0.5pt] (20,-110) -- (40,-110);
\draw[line width=0.5pt] (20,-20) -- (20,-140);
\draw[line width=0.5pt] (0,-140) -- (20,-140);
%\draw[-to,line width=1pt] (32) -- node[midway,left]{$F_{1}^{[2]}$} (22); 
%\draw[dashed] (60,0) -- (140,0);
%\draw[double,double distance=3pt] (200,0) -- (240-6,0);
%
%\draw (240,0) circle (6);
%
\draw (0,0) node[above] {$0$};
\draw (80,0) node[above] {$s$};
\draw (200,0) node[above] {$2s+2$};
\draw (220,0) node[above] {$\mu_{1}$};
\draw (0,-20) node[left] {$1$};
\draw (0,-40) node[left] {$2$};
\draw (0,-140) node[left] {$\mu_{1}^{\prime}$};
\draw (210,-10) node {$\Lambda_{1}$};
\draw (10,-130) node {$ \Lambda_{2}$};
\draw (30,-100) node {$ \Lambda_{3}$};
\draw (50,-80) node {$ \vdots $};
\draw (70,-70) node {$ \Lambda_{s+1}$};
\draw (0,-170) node[below] {$a$};
\draw (230,0) node[right] {$m$};
\end{tikzpicture} 
\caption{$[1,s]$-hook:in $[2,2s+2]$-hook: the Young diagram $\mu $ is related to the highest weight \eqref{HW-Cp} by \eqref{YW-Cp}.}
\label{MN-hookCp}
\end{figure}
%%%%%%%%%%%%%%%%%%%%%%%%%%%%%%%%%%%%%%%%%%%%%
%%%%%%%%%%%%%%%%%%%%%%%%%%%%%%%%%%%%%

\subsubsection{Weyl group}
Let $\alpha$ and $\beta$ be roots of a Lie superalgebra. The Weyl group of the Lie superalgebra 
is generated by {\em Weyl reflections}: 
\begin{align}
w_{\alpha}(\beta)=
\beta - \frac{2(\alpha |\beta)}{(\alpha|\alpha)}\alpha  , 
\label{Weven}
\end{align}
where $\alpha$ is an even root. 
The Weyl group is extended by {\em odd reflections} \cite{DP85,Se85}: 
\begin{align}
w_{\alpha}(\beta)=
\begin{cases}
\beta - \frac{2(\alpha |\beta)}{(\alpha|\alpha)}\alpha 
\quad \text{if} \quad (\alpha|\alpha) \ne 0, \\
\beta + \alpha \quad \text{if} \quad (\alpha|\alpha) = 0 \quad \text{and} \quad (\alpha |\beta) \ne 0, \\
\beta  \quad \text{if} \quad (\alpha |\alpha) = 0 \quad \text{and}  \quad (\alpha |\beta) = 0, \\
-\alpha \quad \text{if} \quad \alpha = \beta, 
\end{cases}
\label{Wodd}
\end{align}
where $\alpha$ is an odd root. 
%We remark that \eqref{Weven} and \eqref{Wodd} 
%are realized if we define the action of $w_{\alpha}$ on 
%$\{\epsilon_{i}\}$ as  
%\begin{align}
%w_{\alpha}(\epsilon_{i})=
%\begin{cases}
%\epsilon_{i} - \frac{2(\alpha |\epsilon_{i})}{(\alpha|\alpha)}\alpha 
%\quad \text{if} \quad (\alpha|\alpha) \ne 0, \\
%\epsilon_{i} +  p_{i}\alpha \quad \text{if} \quad (\alpha|\alpha) = 0 \quad \text{and} \quad (\alpha |\epsilon_{i}) \ne 0, \\
%\epsilon_{i}  \quad \text{if}  \quad (\alpha |\epsilon_{i}) = 0.
%\end{cases}
%\label{Wep}
%\end{align}
The Weyl reflections \eqref{Weven} do not change the shape of the Dynkin diagrams, 
while the odd reflections \eqref{Wodd} do.  
Take the type D simple root system \eqref{rootD} for the case $-p_{i_{r+s-1}}=p_{i_{r+s}}=1$, and apply  
 the odd reflection with respect to the $(r+s)$-th odd simple root $\beta_{r+s}$ to this: 
\begin{align}
\begin{split}
w_{\beta_{r+s}} (\beta_{a})&=\beta_{a}=\epsilon_{i^{*}_{a}}-\epsilon_{i^{*}_{a+1}} =:\beta_{a}^{\prime} \quad 
\text{for} \quad a \in \{1,2,\dots , r+s-3 \}, 
\\
w_{\beta_{r+s}} (\beta_{r+s-2})&=\beta_{r+s-2}+\beta_{r+s}=\epsilon_{i^{*}_{r+s-2}}+\epsilon_{i^{*}_{r+s}}
=\epsilon_{i^{*}_{r+s-2}}-\epsilon_{i_{r+s}} =:\beta_{r+s-2}^{\prime},
\\
w_{\beta_{r+s}} (\beta_{r+s-1})&=\beta_{r+s-1}+\beta_{r+s}=2\epsilon_{i^{*}_{r+s-1}} =: \beta_{r+s}^{\prime}, 
\\
w_{\beta_{r+s}} (\beta_{r+s})&=-\beta_{r+s}=-\epsilon_{i^{*}_{r+s-1}}-\epsilon_{i^{*}_{r+s}} 
=\epsilon_{i_{r+s}}-\epsilon_{i^{*}_{r+s-1}} =: \beta_{r+s-1}^{\prime}.
\end{split}
\label{rootDw}
\end{align}
The resultant simple root system $\{ \beta_{a}^{\prime} \}_{a=1}^{r+s}$ is the type C simple root system \eqref{rootDC} 
defined by the tuple $I_{2r+2s+2}^{\prime}=\tau_{i_{r+s-1}^{*},i_{r+s}} \circ \tau_{i_{r+s-1},i_{r+s}^{*}} \circ \tau_{i_{r+s},i_{r+s}^{*}}(I_{2r+2s+2}) =
(i_{1},i_{2},\dots , i_{r+s-2}, i_{r+s}^{*}, i_{r+s-1}, i_{r+s+1},i_{r+s+1}^{*},i_{r+s-1}^{*},
i_{r+s},i_{r+s-2}^{*},\dots , i_{2}^{*},i_{1}^{*})$ with $p_{i_{r+s-1}}=-1$, where $\tau_{a,b}$ permutes $a$ and $b$. 
Note that the labeling of  the $(r+s-1)$-th and $(r+s)$-th nodes of the Dynkin diagram is interchanged, and 
the sign of $\Upsilon$ is changed by the above $w_{\beta_{r+s}}$. 

Suppose we have a  type C simple system \eqref{rootDC} with $p_{i_{r+s}}=-p_{i_{r+s-1}}=-1$. 
In this case, $\beta_{r+s-1}$ is an odd root, and the odd reflection by $\beta_{r+s-1}$ produces
 a type D simple root system $\{\beta_{a}^{\prime} \}_{a=1}^{r+s}$ with $\Upsilon =1$:
\begin{align}
& \beta_{a}^{\prime} =w_{\beta_{r+s-1}}(\beta_{a}) \quad \text{for} \quad 1 \le a \le r+s \quad \text{if} \quad \Upsilon =1;
\label{wCD1}
\\[8pt]
& \beta_{a}^{\prime} =w_{\beta_{r+s-1}}(\beta_{a}) \quad \text{for} \quad 1 \le a \le r+s-2, 
\nonumber \\
&  \beta_{r+s}^{\prime} =w_{\beta_{r+s-1}}(\beta_{r+s-1}), \quad 
\beta_{r+s-1}^{\prime} =w_{\beta_{r+s-1}}(\beta_{r+s})
 \quad \text{if} \quad \Upsilon =-1. 
 \label{wCD2}
\end{align}
If we define the opposite way (that is, \eqref{wCD1} for $\Upsilon =-1$, \eqref{wCD2} for $\Upsilon =1$), 
 the type D simple root system $\{\beta_{a}^{\prime} \}_{a=1}^{r+s}$ should be interpreted as the one with $\Upsilon =-1$. 
Starting form a given simple root system, one can obtain any other simple root systems  
by applying \eqref{Weven} and \eqref{Wodd} repeatedly. 

Let us summarize the relation between the action of the Weyl reflections and odd reflections by simple roots 
and the action of symmetric groups on tuples. 
\paragraph{Type A, \eqref{rootA}:}
\begin{align}
w_{\alpha_{a}}(I_{M+N}) := \tau_{i_{M+N-a},i_{M+N-a+1}}(I_{M+N}) \quad \text{for} \quad 1 \le a \le M+N-1.
 \label{wIA}
\end{align}
%%%%
\paragraph{Type B, \eqref{rootB}:}
\begin{align}
\begin{split}
w_{\beta_{a}}(I_{2r+2s+1}) &:= \tau_{i_{a},i_{a+1}} \circ \tau_{i_{a}^{*},i_{a+1}^{*}}  (I_{2r+2s+1}) \quad \text{for} \quad 1 \le a \le r+s-1,
\\
w_{\beta_{r+s}}(I_{2r+2s+1})& := \tau_{i_{r+s},i_{r+s}^{*}}  (I_{2r+2s+1}) .
\end{split}
 \label{wIB}
\end{align}
%%%%
\paragraph{Type D, \eqref{rootD}, $p_{i_{r+s}}=1$:}
\begin{align}
\begin{split}
w_{\beta_{a}}(I_{2r+2s+2})& := \tau_{i_{a},i_{a+1}} \circ \tau_{i_{a}^{*},i_{a+1}^{*}}  (I_{2r+2s+2}) \quad \text{for} \quad 1 \le a \le r+s-1, 
\\
w_{\beta_{r+s}}(I_{2r+2s+2})& := \tau_{i_{r+s-1},i_{r+s}^{*}} \circ \tau_{i_{r+s-1}^{*},i_{r+s}} (I_{2r+2s+2}) 
\quad \text{if} \quad  p_{\beta_{r+s}}=1 \quad  (p_{i_{r+s-1}}=1),
\\
w_{\beta_{r+s}}(I_{2r+2s+2})& := 
\tau_{i_{r+s-1},i_{r+s-1}^{*}} \circ \tau_{i_{r+s-1},i_{r+s}^{*}} \circ \tau_{i_{r+s-1}^{*},i_{r+s}} (I_{2r+2s+2}) 
\\ 
& \hspace{120pt}  \text{if} \quad  p_{\beta_{r+s}}=-1,
 \quad  (p_{i_{r+s-1}}=-1). 
\end{split}
 \label{wID}
\end{align}
%%%%
\paragraph{Type C, \eqref{rootDC}, $p_{i_{r+s}}=-1$:}
\begin{align}
\begin{split}
w_{\beta_{a}}(I_{2r+2s+2})& := \tau_{i_{a},i_{a+1}} \circ \tau_{i_{a}^{*},i_{a+1}^{*}}  (I_{2r+2s+2}) \quad \text{for} \quad 1 \le a \le r+s-2, 
\\
w_{\beta_{r+s-1}}(I_{2r+2s+2})& := 
\tau_{i_{r+s-1},i_{r+s}} \circ \tau_{i_{r+s-1}^{*},i_{r+s}^{*}} (I_{2r+2s+2}) 
\\ 
\text{if} \quad  p_{\beta_{r+s-1}}=1 &
 \quad  (p_{i_{r+s-1}}=-1) ,\quad \text{or} \quad p_{\beta_{r+s-1}}=-1
 \quad  (p_{i_{r+s-1}}=1),  \quad \Upsilon =1,
 \\
w_{\beta_{r+s-1}}(I_{2r+2s+2})& := 
\tau_{i_{r+s-1},i_{r+s-1}^{*}} \circ \tau_{i_{r+s-1},i_{r+s}} \circ \tau_{i_{r+s-1}^{*},i_{r+s}^{*}} (I_{2r+2s+2}) 
\\ 
& \hspace{80pt}  \text{if} \quad  p_{\beta_{r+s-1}}=-1
 \quad  (p_{i_{r+s-1}}=1),  \quad \Upsilon =-1,
 \\
w_{\beta_{r+s}}(I_{2r+2s+2})& := \tau_{i_{r+s},i_{r+s}^{*}} (I_{2r+2s+2}). 
\end{split}
 \label{wIDC}
\end{align}
%%%%

For any roots $\alpha , \beta , \gamma$ with $(\alpha|\alpha) \ne 0$, one can show
\begin{align}
(w_{\alpha}(\beta)|w_{\alpha}(\gamma)) = (\beta | \gamma ) ,
 \label{IPWb}
\end{align}
and for any roots $\alpha , \beta , \gamma$ with $(\alpha|\alpha)=0$, 
\begin{align}
(w_{\alpha}(\beta)|w_{\alpha}(\gamma)) =
\begin{cases}
(\beta | \gamma )  &\text{if} \quad (\alpha | \beta )(\alpha | \gamma )=0 , \ 
\alpha \ne \beta , \ \alpha \ne \gamma \\
- (\beta | \gamma )   &\text{if}  \quad   \alpha = \beta , \ \alpha \ne \gamma \quad \text{or} \quad 
 \alpha \ne \beta , \ \alpha = \gamma \\
(\beta | \gamma )+ (\alpha | \gamma) + (\beta | \alpha) & \text{if}  \quad 
(\alpha | \beta )(\alpha | \gamma ) \ne 0 , \ \alpha \ne \beta , \ \alpha \ne \gamma \\
0   &\text{if} \quad   \alpha =\beta =\gamma . 
\end{cases}
 \label{IPWf}
\end{align}
%
%\begin{align}
%(w_{\alpha}(\alpha)|w_{\alpha}(\beta)) =
%\begin{cases}
%(\alpha | \beta )  &\text{if} \quad   (\alpha | \alpha ) \ne 0 \\
%- (\alpha | \beta )   &\text{if}  \quad (\alpha | \alpha ) = 0 .
%\end{cases}
% \label{IPW}
%\end{align}

\subsection{Quantum affine superalgebras}
The Dynkin diagrams of 
  affine Lie superalgebras (see \cite{FSS89}) are obtained by extending those of the Lie superalgebras 
 mentioned in the previous subsection. 
Quantum  affine superalgebras are quantization of the universal enveloping algebras of 
 them \cite{Yamane99,KT94}. 
 Rational counterparts of  untwisted quantum affine superalgebras are superYangians. 
 In \cite{AACFR01}, RTT presentation of the superYangian $Y(osp(r|2s))$ is given. 
 A complete classification of representations of quantum affine superalgebras or superYangians, 
 in particular of orthosymplectic type, seems to be still not established, although there are some partial results for the case of super Yangians \cite{Mo21,Mo21-2}. 
%  R-matrices for orthosymplectic superalgebras 
%  appeared in \cite{BS87}. beyond fundamental representations 
%  are \cite{GZ99}. 
%%%%%%%%%%%%%%%%%%%%%%%%%%%%%%%
\section{T- and Q-functions for  $U_{q}(gl(M|N)^{(1)})$}\label{sec:T-fun}
In this section, we will briefly summarize mainly a part of  \cite{T09} (and \cite{T97,T98}), in which 
 miscellaneous formulas on T-and Q-functions 
for quantum integrable models associated with $U_{q}(gl(M|N)^{(1)})$ 
(or $Y(gl(M|N))$) are presented. 
There are some overlaps with the text in \cite{T09,T11,T21}, with respect to review parts of this paper. 
Other references relevant to this section are \cite{KLWZ97,KSZ07,Zabrodin07}, in which 
B\"{a}cklund flows in the context of Bethe ansatz are discussed in detail.

%%%%%%%%%%%%%%%%%%%%%%
 
\subsection{Tableaux sum and CBR-type expressions of T-functions}
Let $I_{M+N}=(i_{1},i_{2},\dots,i_{M+N})$ be 
 any one of the permutations of the tuple $(1,2,\dots, M+N)$, and  
 $I_{a}=(i_{1},i_{2},\dots,i_{a})$ be the first $a$ elements of it, 
where $0 \le a \le M+N$. In particular, $I_{0}$ and $I_{M+N}$ coincide with $\emptyset $ and ${\mathfrak I}$ as sets,  
respectively. 
We use the symbol $\Qb_{I_{a}}(u)$ to denote the Baxter Q-function (for short, Q-function)  labeled by $I_{a}$, 
which is a function of the spectral parameter $u \in {\mathbb C}$.
We suppose that $\Qb_{I_{a}}(u)$ does not depend on the order of the elements of $I_{a}$ and thus 
  $I_{a}$ may be regarded as a subset (rather than a tuple) $\{i_{1},i_{2},\dots, i_{a}\}$ of the full set ${\mathfrak I}$. 
Therefore, as remarked in \cite{T09}, there are $2^{M+N}$ kinds of Q-functions corresponding to 
the number of the subsets of  ${\mathfrak I}$. We use 
the following abbreviation for the labeling of Q-functions: $\Qb_{I_{a},i,j}(u)=\Qb_{I_{a},i,j}=
 \Qb_{\{ i_{1},i_{2},\dots, i_{a},i,j\}}
=\Qb_{i_{1},i_{2},\dots, i_{a},i,j}$ for $i,j \notin I_{a} $, $i \ne j$. 

\paragraph{Fundamental T-function and Bethe ansatz equations}
The eigenvalue formula of the transfer matrix for the fundamental representation of $U_{q}(gl(M|N)^{(1)})$
in the auxiliary space (Perk-Schultz-type model \cite{Perk:1981})   
by  Bethe ansatz has the following form \cite{Sc83,BR08}:
\begin{align}
{\mathsf F}_{(1)}^{I_{M+N}}(u)=
\Qb_{\emptyset}^{[M-N]}
\Qb_{I_{M+N}}
\sum_{a=1}^{M+N}p_{i_{a}}{\mathcal   X}_{I_{a}},
 \label{tab-fund}
\end{align}
\footnote{The summation $\sum_{j \in I_{a}}$ is meant by $\sum_{j \in \{i_{1},i_{2},\dots , i_{a} \} }$.
We also remark that 
${\mathcal   X}_{I_{a}}^{[-M+N]}$ 
corresponds to eq.(2.9) in \cite{T21}, and 
${\mathcal   X}_{I_{a}}^{[-\frac{3(M-N)}{2}]}$ 
corresponds to eq.(2.7) in \cite{T09}. 
Based on the relation $\sum_{j \in \mathfrak{I}}p_{j}=M-N$, one can show 
\begin{align}
{\mathcal   X}_{I_{a}}&=
z_{i_{a}}
\frac{\Qb_{I_{a-1}}^{[\sum_{j \in \overline{I}_{a-1}}p_{j}-2p_{i_{a}}]}
\Qb_{I_{a}}^{[\sum_{j \in \overline{I}_{a}}p_{j}+2p_{i_{a}}]}
}{
\Qb_{I_{a-1}}^{[\sum_{j \in \overline{I}_{a-1}}p_{j}]}
\Qb_{I_{a}}^{[\sum_{j \in \overline{I}_{a}}p_{j}]}
} ,
\label{boxes2} 
\end{align}
where  $\overline{I}_{a}=(i_{a+1},i_{a+2},\dots,i_{M+N})$. 
We will use this to derive \eqref{boxes3}. 
}
where the function ${\mathcal   X}_{I_{a}}={\mathcal   X}_{I_{a}}(u)$ 
of the spectral parameter $u$ is defined by 
\begin{align}
{\mathcal   X}_{I_{a}}(u)=
z_{i_{a}}
\frac{\Qb_{I_{a-1}}^{[M-N-\sum_{j \in I_{a-1}}p_{j}-2p_{i_{a}}]}
\Qb_{I_{a}}^{[M-N-\sum_{j \in I_{a}}p_{j}+2p_{i_{a}}]}
}{
\Qb_{I_{a-1}}^{[M-N-\sum_{j \in I_{a-1}}p_{j}]}
\Qb_{I_{a}}^{[M-N-\sum_{j \in I_{a}}p_{j}]}
} ,
\label{boxes} 
\end{align}
and the complex parameters $\{z_{i} \}_{i \in {\mathfrak I}}$ are boundary twist parameters.

Suppose
\footnote{We normalize the Q-functions so that the zeroth order terms (as series on $q^{-2u}$) 
become $1$. 
In this normalization of the Q-functions, the boundary twist parameters $\{z_{i} \}$ depend on 
$\{n_{I_{a}}\} $. 
One can rewrite \eqref{Q-poly} to a more familiar form: 
 $\Qb_{I_{a}}(u)=(q-q^{-1})^{ n_{I_{a}} } q^{-n_{ I_{a} }u+ \sum_{j=1}^{n_{ I_{a} }} u^{ I_{a} }_{j} }
 \prod_{j=1}^{n_{I_{a}}}[u-u^{ I_{a} }_{j} ]_{q} $, $[u]_{q}=(q^{u}-q^{-u})/(q-q^{-1})$. 
Substituting this into \eqref{boxes}, one finds a factor
$q^{2p_{i_{a}} (n_{I_{a-1}} -n_{I_{a}}) } z_{i_{a}}$, which corresponds to a constant ($\{n_{I_{a}}\}$ independent) 
boundary twist parameter. 
As for the rational case, we redefine  
$\lim_{q \to 1}(q-q^{-1})^{ -n_{I_{a}} } \Qb_{I_{a}}(u) =
\prod_{j=1}^{n_{I_{a}}} (u-u^{I_{a}}_{j})$ as the Q-function $\Qb_{I_{a}}(u)$. 
We also remark that the requirement that the Q-functions have the form \eqref{Q-poly} 
is necessary only when we discuss Bethe ansatz equations, and that 
various functional relations among T-and Q-functions
 are valid irrespective of this requirement.}
 that the Q-functions are finite degree polynomials of $q^{-2u}$ ($u$: the spectral parameter) 
and have zeros at $u=u_{k}^{I_{a}}$: 
\begin{align}
\Qb_{I_{a}}=\Qb_{I_{a}}(u)=\prod_{j=1}^{n_{I_{a}}}(1-q^{-2u+2u^{I_{a}}_{j}}), 
 \label{Q-poly}
\end{align}
where $k \in \{1,2,\dots, n_{I_{a}}\},a \in \{0,1,2,\dots, M+N \} $.  In particular, 
 $u_{k}^{I_{0}}$ and $u_{k}^{I_{M+N}}$ are inhomogeneity 
 of the spectral parameter (known parameters), and $n_{I_{0}}=n_{I_{M+N}}=L$ is (half) of the number of the lattice sites 
 \footnote{Q-functions of an alternating spin chain with $2L$ lattice sites (half of them in the fundamental representation and the other half in the anti-fundamental representation) will have this normalization condition.}.
% 
%\begin{align}
%\Qb_{I_{a}}(x)=\prod_{k=1}^{n_{I_{a}}}\left(1-\frac{x}{x_{k}^{I_{a}}}\right), 
%\label{baxterQ-root}
%\end{align}
The requirement that the T-function \eqref{tab-fund} is free of poles, 
namely,  
\begin{multline}
{\rm Res}_{u=u_{k}^{I_{a}}+\sum_{j \in I_{a}}p_{j}-M+N}
(p_{i_{a}}{\mathcal   X}_{I_{a}}(u)+p_{i_{a+1}}{\mathcal   X}_{I_{a+1}}(u))=0 
\label{chancels}
\\
\text{for} \quad k\in \{1,2,\dots, n_{I_{a}}\} \quad \text{and} \quad a \in \{1,2,\dots, M+N-1 \} 
\end{multline}
produces the following Bethe ansatz equation: 
\begin{multline}
 -1=\frac{p_{i_{a}}z_{i_{a}}}{p_{i_{a+1}}z_{i_{a+1}}}
\frac{\Qb_{I_{a-1}}(u_{k}^{I_{a}}-p_{i_{a}})
\Qb_{I_{a}}(u_{k}^{I_{a}}+2p_{i_{a}})
\Qb_{I_{a+1}}(u_{k}^{I_{a}}-p_{i_{a+1}})} 
{\Qb_{I_{a-1}}(u_{k}^{I_{a}}+p_{i_{a}})
\Qb_{I_{a}}(u_{k}^{I_{a}}-2p_{i_{a+1}})
\Qb_{I_{a+1}}(u_{k}^{I_{a}}+p_{i_{a+1}})}
\\
\text{for} \quad k\in \{1,2,\dots, n_{I_{a}}\} \quad \text{and} \quad a \in \{1,2,\dots, M+N-1 \} .
\label{BAE}
\end{multline}
Here the poles from the known functions $\Qb_{I_{0}} $ and $\Qb_{I_{M+N}} $ are out of the question. 
%
%\begin{multline}
% -1=\frac{p_{i_{a}}z_{i_{a}}}{p_{i_{a+1}}z_{i_{a+1}}}
%\frac{\Qb_{I_{a-1}}(x_{k}^{I_{a}}q^{-p_{i_{a}}})
%\Qb_{I_{a}}(x_{k}^{I_{a}}q^{2p_{i_{a}}})
%\Qb_{I_{a+1}}(x_{k}^{I_{a}}q^{-p_{i_{a+1}}})} 
%{\Qb_{I_{a-1}}(x_{k}^{I_{a}}q^{p_{i_{a}}})
%\Qb_{I_{a}}(x_{k}^{I_{a}}q^{-2p_{i_{a+1}}})
%\Qb_{I_{a+1}}(x_{k}^{I_{a}}q^{p_{i_{a+1}}})}
%\\
%\text{for} \quad k\in \{1,2,\dots, n_{I_{a}}\} \quad \text{and} \quad a \in \{1,2,\dots, M+N-1 \} .
%\label{BAE}
%\end{multline}
%
We assume that the roots of the Q-functions are sufficiently generically distributed. 
We do not go into mathematical rigour on this. 
 %; $u_{k}^{I_{a}} \ne u_{j}^{I_{b}}$ if $(a,k) \ne (b,j)$ and $a,b \in \{1,2,\dots, M+N-1 \} $
%%%%%%%%%%%%%
\paragraph{QQ-relations}
Let $S(I_{M+N})$ be the symmetric group over the components of 
the tuple $I_{M+N}$.  
We assume that $\tau \in S(I_{M+N})$ acts on $I_{a}$ as 
 $\tau(I_{a})=(\tau(i_{1}),\tau(i_{2}),\dots, \tau(i_{a}))$, $0 \le a \le M+N$. 
The action of $\tau \in S(I_{M+N})$ on ${\mathsf F}^{I_{M+N}}_{(1)}$ is defined as 
$\tau({\mathsf F}^{I_{M+N}}_{(1)}):={\mathsf F}^{\tau(I_{M+N})}_{(1)}
={\mathsf F}^{(\tau(i_{1}),\tau(i_{2}),\dots, \tau(i_{M+N}))}_{(1)}$. 
We also set 
$\tau({\mathcal X}_{I_{a}})={\mathcal X}_{\tau(I_{a})}$, 
$\tau (\Qb_{I_{a}})=\Qb_{\tau (I_{a})}$, 
$\tau(z_{a})=z_{\tau(a)},\tau(p_{a})=p_{\tau(a)}$. 
The direct product of 
 the symmetric groups $S({\mathfrak B}) \times S({\mathfrak F})$ 
corresponds to the Weyl group of $gl(M|N)$ (see \eqref{Weven}; we also denote the symmetric group over 
a subset $I$ of ${\mathfrak I}$ as $S(I)$.) 
On the other hand, the elements of $ S(I_{M+N})/(S({\mathfrak B}) \times S({\mathfrak F}))$ 
correspond to the odd reflections of $gl(M|N)$ (see \eqref{Wodd}). 
Consider a permutation $\tau \in S(I_{M+N})=S({\mathfrak I})$ 
such that $\tau(i_{a})=i_{a+1}, \tau(i_{a+1})=i_{a}$ and $\tau(i_{b})=i_{b}$ 
for $b \ne a,a+1$,  
for a fixed $a \in \{1,2,\dots, M+N-1\}$. 
The condition 
$\tau( {\mathsf F}_{(1)}^{I_{M+N}})={\mathsf F}_{(1)}^{I_{M+N}}$ 
is equivalent to 
\begin{align}
p_{i}{\mathcal   X}_{I_{a}}+
p_{j}{\mathcal   X}_{I_{a+1}}=
p_{j}{\mathcal   X}_{\tau(I_{a})}+p_{i}{\mathcal   X}_{\tau(I_{a+1})}, 
\label{Tinv}
\end{align}
where  $i=i_{a},j=i_{a+1}$. 
This implies
\footnote{The 4-term QQ-relation 
\eqref{Tinv} for the case 
$p_{i}=p_{j}$ is equivalent to 
\begin{align}
\left(
\frac{
z_{i}\Qb_{I,i}^{[p_{i}]}
\Qb_{I,j}^{[-p_{i}]}-
z_{j}\Qb_{I,i}^{[-p_{i}]}
\Qb_{I,j}^{[p_{i}]}
}{\Qb_{I}\Qb_{I,i,j}}
\right)^{[2]}
=
\frac{
z_{i}\Qb_{I,i}^{[p_{i}]}
\Qb_{I,j}^{[-p_{i}]}-
z_{j}\Qb_{I,i}^{[-p_{i}]}
\Qb_{I,j}^{[p_{i}]}
}{\Qb_{I}\Qb_{I,i,j}}
 \label{QQratA1}
\end{align}
This means that the right hand side of  \eqref{QQratA1} is a periodic function $\phi $ of 
the spectral parameter: $\phi^{[2]}=\phi$.
The 3-term QQ-relation \eqref{QQb} corresponds to the case that this 
periodic function is a constant $\phi= z_{i}-z_{j}$. This comes from the assumption that 
the Q-functions have the form \eqref{Q-poly}, and the deformation parameter is 
generic. Thus \eqref{QQb} can be a sufficient condition for \eqref{QQratA1} in the general situation. 
The discussion for \eqref{QQf} is parallel to that of  \eqref{QQb}.
}
%%%
 the following functional relations, called QQ-relations
%\footnote{\eqref{QQf} for $p_{i}=-p_{j}=-1$ is the one for $p_{i}=-p_{j}=1$}:
\begin{align}
& (z_{i}-z_{j})\Qb_{I}\Qb_{I,i,j}
=z_{i}\Qb_{I,i}^{[p_{i}]}
\Qb_{I,j}^{[-p_{i}]}-
z_{j}\Qb_{I,i}^{[-p_{i}]}
\Qb_{I,j}^{[p_{i}]}
\qquad 
\text{for} \qquad p_{i}=p_{j},   
\label{QQb}  \\[6pt]
& 
(z_{i}-z_{j})\Qb_{I,i}\Qb_{I,j}=
z_{i}\Qb_{I}^{[-p_{i}]}
\Qb_{I,i,j}^{[p_{i}]}-
z_{j}\Qb_{I}^{[p_{i}]}
\Qb_{I,i,j}^{[-p_{i}]}
\qquad \text{for} \qquad p_{i}=-p_{j}, 
\label{QQf} 
\end{align} 
where $I=I_{a-1}$. 
The 3-term QQ-relations 
\eqref{QQb} and \eqref{QQf} are simplifications
\footnote{Clear the denominators of \eqref{Tinv}. Eqs.\ \eqref{QQb} and \eqref{QQf} 
are ``bi-linearizations'' of this ``multi-linear form''.}
 of the  4-term QQ-relation \eqref{Tinv}. 
The QQ-relations \eqref{QQb} and \eqref{QQf} have a determinant solution \cite{T09}. 
In our convention, it is given by \eqref{emptyT} (\eqref{QQdetsol1}, \eqref{QQdetsol1}). 
That the determinant satisfies the QQ-relations 
was proved in  [Appendix C, \cite{T09}] using the  
Jacobi or Pl\"{u}cker identities. 
From now on, we assume \eqref{QQb} and \eqref{QQf}. 
%$i,j \in {\mathfrak I} \setminus I$ ($i\ne j$).
Various types of 
QQ-relations appeared in the literature (see for example, 
%\cite{BLZ98,PS00,DDT00,BHK02,GS03,MV04,KSZ07} 
\cite{BLZ98,PS00,DDT00,BHK02,MV04,KSZ07}
 and references in \cite{T09}). 
In particular, the form relevant to our discussion
  appeared in \cite{BLZ98}  (\eqref{QQb} for $(M,N)=(2,0)$),  
 \cite{BHK02} (for \eqref{QQb} for $(M,N)=(3,0)$), and \cite{BT08} (\eqref{QQb} and \eqref{QQf} for $(M,N)=(2,1)$). 
In this paper, we use the presentation \cite{T09} of QQ-relations  
 for the whole set of $2^{M+N}$ Q-functions on the Hasse diagram. 
This is necessary for the formulation of Wronskian-type expressions of T-functions. 
The first equations \eqref{QQb} (Bosonic QQ-relations) are generalization of the quantum Wronskian condition. 
 The second equations \eqref{QQf} (Fermionic QQ-relations) came from 
the particle-hole transformations 
%\cite{Wo83,BCFH92,EKS92,T98} 
\cite{Wo83} 
 in statistical mechanics, 
 and are related \cite{T98} to odd Weyl reflections \cite{DP85,Se85} of the superalgebra $gl(M|N)$.  
Fermionic QQ-relations are studied \cite{KSZ07} in relation to B\"{a}cklund transformations in soliton theory.  
In \cite{T09,T11}, we normalized the Q-function for the empty set as $\Qb_{\emptyset}=1$, 
but we do not impose this in this paper. 
%
%We normalize these functions as
%\begin{align}
%\Qb_{\emptyset}=\Qb_{{\mathfrak I}}=1 .
%\label{normali}
%\end{align}
%%

Let $\{\epsilon_{a}\}_{a=1}^{M+N}$ be a basis of the dual space of 
the Cartan subalgebra of $gl(M|N)$  with the bilinear form  
$(\epsilon_{a}|\epsilon_{b})=p_{a}\delta_{ab}$. 
We introduce a map $\sigma$, which is related to an automorphism of $gl(M|N)$ (or  $sl(M|N)$, cf. \cite{FSS89}):
\begin{align}
\sigma(\epsilon_{i})& =-\epsilon_{i^{*}}=
\begin{cases}
-\epsilon_{M+1-i}
\quad \text{for} \quad i \in {\mathfrak B}, \\
-\epsilon_{2M+N+1-i}
\quad \text{for} \quad i \in {\mathfrak F}.
\end{cases}
\label{sigmaep}
\end{align}
Take a Cartan element $h$ of $gl(M|N)$ so that $e^{\epsilon_{a}(h)}=z_{a}$ holds, and define
\begin{align}
\sigma(z_{i})& =z_{i^{*}}^{-1}=
\begin{cases}
z_{M+1-i}^{-1} 
\quad \text{for} \quad i \in {\mathfrak B}, \\
z_{2M+N+1-i}^{-1} 
\quad \text{for} \quad i \in {\mathfrak F}.
\end{cases}
\label{sigmach}
\end{align}
Then we  define the action of this map on the index sets of Q-functions:
\begin{align}
& \sigma (I)
={\mathfrak I} \setminus I^{*} \quad \text{for} \quad I\subset {\mathfrak I} ,
\label{sigma-met} 
\end{align} 
and on the Q-functions: 
\begin{align}
\sigma (\Qb_{I}) 
= \Qb_{\sigma (I)}
\label{sigmaQ}.
\end{align}
We remark that  the form of the QQ-relations \eqref{QQb} and \eqref{QQf} 
are invariant  under the map $\sigma$. 

%%%%%%%%%%%
\paragraph{Root systems and Bethe ansatz}
The Bethe ansatz equation \eqref{BAE} fits into a universal form (cf. \cite{ORW87,RW87,MR97}) 
associated with the root system of the superalgebra,   
supplemented by a sign factor and boundary twist parameters:  
\begin{multline}
 - \frac{\Lambda_{a}(u_{k}^{(a)}-(\omega_{a}|\omega_{a}))}{\Lambda_{a+1}(u_{k}^{(a)}-(\omega_{a}|\omega_{a}))} =
 p_{\alpha_{a}}
 e^{-\alpha_{a}(h)} \hspace{-20pt}
 \prod_{b=1 \atop b \ne a \, \text{if} \, (\alpha_{a}|\alpha_{a}) =0}^{\rr}
 \hspace{-15pt} 
\frac{\Qc_{b}(u_{k}^{(a)}+(\alpha_{a}|\alpha_{b} ))}{\Qc_{b}(u_{k}^{(a)}-(\alpha_{a}|\alpha_{b} ))}
\\
\text{for} \quad k\in \{1,2,\dots, n_{a} \} \quad \text{and} \quad a \in \{1,2,\dots,\rr \} ,
\label{BAEr}
\end{multline}
where  $\rr=M+N-1$ is the rank of $sl(M|N)$; $((\alpha_{a}|\alpha_{b} ))_{1\le a,b \le \rr}$ is the symmetrized Cartan matrix 
with a set of simple roots $\alpha_{a}=\epsilon_{i_{M+N+1-a}}-\epsilon_{i_{M+N-a}}$, 
$\omega_{a}=\sum_{k=1}^{a}\epsilon_{i_{M+N-a+k}}$, 
$(\omega_{a}|\omega_{a})=\sum_{k=1}^{a}p_{i_{M+N-a+k}}$, $p_{\alpha_{a}}=p_{\epsilon_{i_{M+N+1-a}}} p_{\epsilon_{i_{M+N-a}}}$, 
$p_{\epsilon_{a}}=p_{a}$, $u_{k}^{(a)}=u_{k}^{I_{M+N-a}}$. 
We identify 
$\Qc_{a}(u)=\Qb_{I_{M+N-a}}(u)$, 
$n_{a}=n_{I_{M+N-a}}$. 
Left hand side of \eqref{BAEr} depends on the quantum space of the model in question. 
$\{\Lambda_{a}(u)\}_{a=1}^{M+N}$ are the eigenvalues of the diagonal elements of 
a monodromy matrix on the vacuum vector (``vacuum parts'' in the analytic Bethe ansatz). 
 Here we consider the case 
$\Lambda_{1}(u)=\Qb_{\emptyset}(u+M-N) \Qb_{I_{M+N}}(u+2p_{i_{M+N}})$, 
$\Lambda_{a}(u)=\Qb_{\emptyset}(u+M-N) \Qb_{I_{M+N}}(u)$ for $2 \le a \le  M+N-1$, 
$\Lambda_{M+N}(u)=\Qb_{\emptyset}(u+M-N-2p_{i_{1}}) \Qb_{I_{M+N}}(u)$. 
One can rewrite the left hand side of \eqref{BAEr} further: 
\begin{multline}
 - \frac{P_{a}(u_{k}^{(a)}+d_{a})}{P_{a}(u_{k}^{(a)}-d_{a})} =
 p_{\alpha_{a}}
 e^{-\alpha_{a}(h)} \hspace{-20pt}
 \prod_{b=1 \atop b \ne a \, \text{if} \, (\alpha_{a}|\alpha_{a}) =0}^{\rr}
  \hspace{-15pt}
\frac{\Qc_{b}(u_{k}^{(a)}+(\alpha_{a}|\alpha_{b} ))}{\Qc_{b}(u_{k}^{(a)}-(\alpha_{a}|\alpha_{b} ))}
\\
\text{for} \quad k\in \{1,2,\dots, n_{a} \} \quad \text{and} \quad a \in \{1,2,\dots, \rr \} ,
\label{BAEr2}
\end{multline}
where
\begin{align}
P_{a}=
\begin{cases}
\Qc_{0}=\Qb_{I_{M+N}} & \text{if} \quad  a=1, \\
\Qc_{M+N}=\Qb_{I_{0}} & \text{if} \quad  a=M+N-1, \\
1 & \text{otherwise} ,
\end{cases}
\label{vacQQ0}
\end{align}
$d_{a}=(\alpha_{a}|\alpha_{a})/2$ if $(\alpha_{a}|\alpha_{a}) \ne 0$, 
$d_{a}=(\alpha_{a}|\alpha_{a^{\prime}}) \ne 0$ for some simple root $\alpha_{a^{\prime}}$ 
if $(\alpha_{a}|\alpha_{a}) = 0$, in particular 
$d_{1}=p_{i_{M+N}}$, $d_{M+N-1}=p_{i_{1}}$.
\footnote{
$d_{1}=(\alpha_{1}|\alpha_{2})$ if $(\alpha_{1}|\alpha_{1})=0$ since $p_{i_{M+N}}=-p_{i_{M+N-1}}$, 
$d_{1}=(\alpha_{M+N-1}|\alpha_{M+N-2})$ if $(\alpha_{M+N-1}|\alpha_{M+N-1})=0$ since 
$p_{i_{1}}=-p_{i_{2}}$. 
The other option is $d_{a}= - (\alpha_{a}|\alpha_{0})$ for $a=1, M+N-1$, 
where $\alpha_{0}=\epsilon_{i_{1}}-\epsilon_{i_{M+N}}$ is a simple root of the affine Lie
 superalgebra $gl(M|N)^{(1)}$ (the null vector is omitted).} 
The functions $P_{a}(u)$ are related to Drinfeld polynomials, which characterize the quantum space of the model 
(or representation of the underlying algebra) in question 
(cf. \cite{KOS95}). 
%In this case, \eqref{BAEr} has a more compact expression
%\begin{multline}
%- 1=
%p_{\alpha_{a}}
%e^{-\alpha_{a}(h)}
%\prod_{b=0}^{M+N-1}
%\frac{\Qc_{b}(u_{k}^{(a)}+(\alpha_{a}|\alpha_{b} ))}{\Qc_{b}(u_{k}^{(a)}-(\alpha_{a}|\alpha_{b} ))}
%\\
%\text{for} \quad k\in \{1,2,\dots, n_{a} \} \quad \text{and} \quad a \in \{1,2,\dots, M+N-1 \} ,
%\label{BAEr2}
%\end{multline}
%%%%%

The QQ-relations \eqref{QQb} and \eqref{QQf} can also be written in terms of a root system of  $gl(M|N)$. 
For $a \in \{1,2, \dots , M+N-1\}$, they read
\begin{multline}
 (e^{-\alpha_{a}(h)}-1)P_{a} \prod_{b=1 \atop (\alpha_{a}|\alpha_{b}) \ne 0 , b \ne a}^{\rr}\Qc_{b}=
e^{-\alpha_{a}(h)} \Qc_{a}^{[d_{a}]} \widetilde{\Qc}_{a}^{[-d_{a}]} -
\Qc_{a}^{[-d_{a}]} \widetilde{\Qc}_{a}^{[d_{a}]}
\quad
\text{if} \quad (\alpha_{a}|\alpha_{a}) \ne 0,
\label{QQrb0}
\end{multline}
\begin{multline}
 (e^{-\alpha_{a}(h)}-1)  \Qc_{a} \widetilde{\Qc}_{a}
 =
e^{-\alpha_{a}(h)}P_{a}^{[-d_{a}]} \prod_{b=1 \atop (\alpha_{a}|\alpha_{b}) \ne 0,  b \ne a}^{\rr}\Qc_{b}^{[(\alpha_{a}|\alpha_{b})]}
 -
P_{a}^{[d_{a}]} \prod_{b=1 \atop (\alpha_{a}|\alpha_{b}) \ne 0, b \ne a}^{\rr}\Qc_{b}^{[-(\alpha_{a}|\alpha_{b})]}
\\
  \text{if} \quad (\alpha_{a}|\alpha_{a}) = 0,
  \label{QQrf0}
\end{multline}
where
\footnote{One may also write this as $\widetilde{\Qc}_{a}=w_{\alpha_{a}}(\Qc_{a})$ (see subsection \ref{sec:QQinv}).}
$\widetilde{\Qc}_{a}=\Qb_{\widetilde{I}_{M+N-a}}=\Qb_{(i_{1}.i_{2},\dots, i_{M+N-a-1},i_{M+N-a+1})}$. 

We expect that QQ-relations for
 other quantum affine superalgebras or super-Yangians 
associated with simply laced Dynkin diagrams can also be expressed in this form \eqref{QQrb0}-\eqref{QQrf0}. 
We remark that QQ-relations for non-super affine Lie algebras are expressed  in 
terms of root systems 
 in connection with discrete Miura opers \cite{MV04}, and  with the ODE/IM correspondence \cite{MRV15,MRV15-2}. 
 
%%%%%%%%%%%%
\paragraph{T-functions for fusion vertex models}
The Young diagram $\mu$, corresponding to a partition $\mu$, has 
$\mu_{k}$ boxes in the $k$-th row of the plane.
Each box in the Young diagram has the coordinate   
$(i,j)\in {\mathbb Z}_{\ge 1} \times {\mathbb Z}_{\ge 1} $, 
where the row index $i$ increases as one goes down, and the column 
index $j$ increases as one goes from left to right. 
The upper left corner of $\mu$ has the coordinates $(1,1)$. 
Let $\lambda =(\lambda_{1},\lambda_{2},\dots)$ and 
$\mu =(\mu_{1},\mu_{2},\dots)$ be two partitions such that
$\mu_{i} \ge \lambda_{i}: i=1,2,\dots$ and 
$\lambda_{\mu_{1}^{\prime}}=\lambda^{\prime}_{\mu_{1}}=0$. 
We express the skew-Young diagram defined by these two partitions as 
 $\lambda \subset \mu$. 
Each box on the skew-Young diagram $\lambda \subset \mu$ is specified by its coordinate on $\mu$. 
%

%%%%%%%%%%%%%%%
%
We define the space of admissible tableaux  
$\mathsf{Tab}_{I_{K}}(\lambda\subset \mu)$ 
for a tuple $I_{K}=(i_{1},i_{2},\dots , i_{K})$   
on a (skew) Young diagram $\lambda\subset \mu$. 
%Let us take a $K$-tuple ${\mathtt I}=(\gamma(1),\gamma(2),\dots,\gamma(K))$, whose components 
% are mutually distinct elements of ${\mathfrak I}$. 
%Let us write $\chi_{a}(x)=\chi_{I_{a-1},I_{a}}(x)$ \eqref{boxes}. 
%
We assign an integer $t_{ij}$ in each box $(i,j)$ of the diagram. 
An admissible tableau 
$t\in\mathsf{Tab}_{I_{K}}(\lambda\subset \mu)$ 
 is a set of integers $t=\{t_{jk}\}_{(j,k)\in \lambda\subset \mu}$, 
where all $t_{jk} \in \{1, 2, \dots, K \}$  
satisfy the following conditions  
\begin{itemize} 
\item[(i)] $t_{jk}\ge t_{j+1,k},t_{j,k+1}$ \\
\item[(ii)]  $t_{jk} > t_{j,k+1}$ if $i_{t_{jk}}\in {\mathfrak F}$ or
  $i_{t_{j,k+1}} \in {\mathfrak F}$  \\ 
\item[(iii)]  $t_{jk} > t_{j+1,k}$ if $i_{t_{jk}}\in {\mathfrak B}$ or
  $i_{t_{j+1,k}} \in {\mathfrak B}$.
\end{itemize}
% and ${\overline {\mathcal  F}}_{\lambda \subset \mu}^{{\mathtt I}}(x)$. 
We introduce a $\Ts$-function
\footnote{
Here we change the convention of the function ${\mathcal  F}_{\lambda\subset \mu}^{I_{K}}$ in [eq. (2.12), \cite{T21}]. 
 ${\mathcal  F}_{\lambda\subset \mu}^{I_{K}}$ 
in \cite{T21} corresponds to ${\mathcal  F}_{\widetilde{\lambda\subset \mu}}^{I_{K}}$ in 
this paper, where $\widetilde{\lambda\subset \mu}$ is the $180^{\circ}$ rotation of 
$\lambda\subset \mu $ . In particular, both of them coincide if the Young diagram is of rectangular shape.}
 with auxiliary space labeled by 
a skew Young diagram $\lambda\subset \mu$ 
\cite{T97,T98} 
(see \cite{T98-2} for an extension of this T-function, 
\cite{BR90} for $N=0$ case, \cite{LM20} for representation or combinatorial theoretical background and \cite{KOS95} for $U_{q}(B_{r}^{(1)})$ case):  
%\begin{align}
%{\bf [old] }
%{\mathcal  F}_{\lambda\subset \mu}^{I_{K}}=
%\sum_{t \in \mathsf{Tab}_{I_{K}}(\lambda\subset \mu)}
%\prod_{(j,k) \in \lambda\subset \mu}
%p_{\gamma_{t_{j,k}}}
%{\mathcal  X}_{I_{t_{j,k}}}^{[\mu_{1}-\mu_{1}^{\prime}+2j-2k+\mm-\nn-M+N]},
%\label{DVF-tab1} 
%\end{align}
%%
\begin{align}
{\mathcal  F}_{\lambda\subset \mu}^{I_{K}}(u)=
\sum_{t \in \mathsf{Tab}_{I_{K}}(\lambda\subset \mu)}
\prod_{(j,k) \in \lambda\subset \mu}
p_{i_{t_{j,k}}}
{\mathcal  X}_{I_{t_{j,k}}}^{[-\mu_{1}+\mu_{1}^{\prime}-2j+2k+\mm-\nn-M+N]},
\label{DVF-tab1} 
\end{align}
where the summation is taken over all the admissible tableaux, and 
the products are taken over all the boxes of the 
Young diagram $\lambda\subset \mu$;   
$\mm:={\rm card} (I_{K} \cap {\mathfrak B})$, 
$\nn:={\rm card} (I_{K} \cap {\mathfrak F})$. 
We also set ${\mathcal  F}_{\emptyset}^{I_{K}}=1$, 
%${\overline {\mathcal  F}}_{\emptyset}^{I}=1$, 
and 
${\mathcal  F}_{\mu}^{\emptyset}=0$ and 
for the non-empty Young diagram $\mu $.
Note that the admissible tableaux $\mathsf{  Tab}_{I_{K}}(\lambda \subset \mu)$ 
 becomes an empty set if the Young diagram $ \lambda \subset \mu$ 
contains a rectangular sub-diagram of a height of $\mm+1$ and a width of $\nn+1$,
 and consequently
(\ref{DVF-tab1})
%W and (\ref{DVF-tab2})
 vanishes for such Young diagram. 
 Thus the T-functions for $\lambda = \emptyset $ are defined on the 
$[\mm,\nn]$-hook (L-hook; cf.\ Figure \ref{MN-hookA}). 
Let us give examples of \eqref{DVF-tab1} for the cases $(M,N)=(2,1)$, $K=3$, 
$I_{3}=(i_{1},i_{2},i_{3})=(2,3,1)$, 
$\lambda =\emptyset$, $\mu=(1)$ and  $\mu=(1^{2})$. In these cases, we have 
$\Bm=\{1,2\}$, $\Fm=\{3\}$, $\mm=2$,  $\nn=1$,  
 $I_{2}=(i_{1},i_{2})=(2,3)$, $I_{1}=(i_{1})=(2)$, $I_{0}=\emptyset$, $p_{i_{1}}=p_{2}=1$, $p_{i_{2}}=p_{3}=-1$, 
 $p_{i_{3}}=p_{1}=1$. 
  Thus \eqref{DVF-tab1} reduces to
\begin{align}
{\mathcal  F}_{(1)}^{I_{3}}&=  {\mathcal  X}_{I_{3}} - {\mathcal  X}_{I_{2}}+ {\mathcal  X}_{I_{1}}
=
z_{1} \frac{\Qb_{231}^{[2]} \Qb_{23}^{[-1]} }{ \Qb_{231} \Qb_{23}^{[1]} } 
-z_{3} \frac{\Qb_{23}^{[-1]} \Qb_{2}^{[2]} }{  \Qb_{23}^{[1]} \Qb_{2} }
+z_{2} \frac{ \Qb_{2}^{[2]} \Qb_{\emptyset }^{[-1]} }{  \Qb_{2} \Qb_{\emptyset}^{[1]}  } ,
\\
{\mathcal  F}_{(1^{2})}^{I_{3}}&=
- {\mathcal  X}_{I_{3}}^{[1]} {\mathcal  X}_{I_{2}}^{[-1]} +{\mathcal  X}_{I_{3}}^{[1]} {\mathcal  X}_{I_{1}}^{[-1]} 
+{\mathcal  X}_{I_{2}}^{[1]} {\mathcal  X}_{I_{2}}^{[-1]}-{\mathcal  X}_{I_{2}}^{[1]} {\mathcal  X}_{I_{1}}^{[-1]}
\nonumber \\
&=-z_{1}z_{3} \frac{\Qb_{231}^{[3]} \Qb_{23}^{[-2]} \Qb_{2}^{[1]} }{ \Qb_{231}^{[1]} \Qb_{23}^{[2]} \Qb_{2}^{[-1]} }
+z_{1}z_{2} \frac{\Qb_{231}^{[3]} \Qb_{23} \Qb_{2}^{[1]} \Qb_{\emptyset}^{[-2]} }{ \Qb_{231}^{[1]} \Qb_{23}^{[2]} \Qb_{2}^{[-1]} \Qb_{\emptyset}} 
+(z_{3})^{2} \frac{\Qb_{23}^{[-2]} \Qb_{2}^{[3]} }{ \Qb_{23}^{[2]} \Qb_{2}^{[-1]} } -
z_{3}z_{2} \frac{ \Qb_{23} \Qb_{2}^{[3]} \Qb_{\emptyset}^{[-2]} }{ \Qb_{23}^{[2]} \Qb_{2}^{[-1]} \Qb_{\emptyset}}  .
\label{DVF-ex1} 
\end{align}
%%%%%%

\paragraph{(Super)character limit} 
We define the {\em (super)character limit} by the operation:
\begin{align}
\zeta : \Qb_{I} \to 1 \quad  \text{for all} \quad  I \subset \mathfrak{I}. 
\label{ch-limit}
\end{align}
In this limit, we have $\zeta (\mathfrak{\chi}_{I_{a}})=z_{i_{a}}$ for \eqref{boxes}. 
In general, T-functions reduce to the (super)characters of representations of underlying algebras. 
In particular, $\zeta ({\mathcal  F}_{ \mu}^{I_{M+N}})$ coincides with the
 supercharacter of the highest weight 
 representation of $gl(M|N)$ with the highest weight \eqref{HW-A} for \eqref{YW-A} 
 (in case $I_{M+N}=(M+N,\dots, 2,1)$). 

%%%%%%%%%
\paragraph{Generating functions of T-functions}
For Young diagrams with one rows or columns,  there are 
generating functions for \eqref{DVF-tab1}: 
\begin{align}
{\mathbf W}_{I_{K}}({\mathbf X})&=\overleftarrow{\prod_{k=1}^{K}}
 (1-{\mathcal X}_{I_{k}}{\mathbf X})^{-p_{i_{k}}} 
= \sum_{a=0}^{\infty} {\mathcal F}_{(a)}^{I_{K}
[a -1-\mm+\nn+M-N]}{\mathbf X}^{a }, 
\label{gene1}
\\
{\mathbf W}_{I_{K}}({\mathbf X})^{-1}&=\overrightarrow{\prod_{k=1}^{K}}
 (1-{\mathcal X}_{I_{k}}{\mathbf X})^{p_{i_{k}}} 
= \sum_{a=0}^{\infty}(-1)^{a} {\mathcal F}_{(1^{a})}^{I_{K}
[a -1-\mm+\nn+M-N]}{\mathbf X}^{a }, 
\label{gene2}
\end{align}
where ${\mathbf X}$ is a shift operator ${\mathbf X}f=f^{[2]}{\mathbf X}$ for any function $f$ of the spectral parameter. 
This type of generating functions for $K=M+N$ appeared in \cite{T97,T98} 
(\cite{KSZ07,Zabrodin07,T09} for $0<K<M+N $ case; \cite{KLWZ97} for $N=0$ case). 
The supercharacter limit of these at $K=M+N$, namely $\zeta({\mathbf W}_{I_{K}}({\mathbf X}))$ and 
$\zeta({\mathbf W}_{I_{K}}({\mathbf X})^{-1})$  are the generating functions of 
 the symmetric and anti-symmetric representations of $gl(M|N)$, respectively (see \eqref{gench-A}). 
The condition that the generating function ${\mathbf W}_{I_{K}}({\mathbf X})$ 
for $K \ge a+1$ is invariant under the transposition $\tau \in S(I_{M+N})$ of $i_{a}$ 
and $i_{a+1}$, namely 
${\tau(\mathbf W}_{I_{K}}({\mathbf X}))={\mathbf W}_{\tau(I_{K})}({\mathbf X})={\mathbf W}_{I_{K}}({\mathbf X})$ 
is equivalent to the {\em discrete zero curvature condition} (cf. \cite{KSZ07,Zabrodin07}):
\begin{align}
 (1-{\mathcal X}_{I_{a}}{\mathbf X})^{p_{i_{a}}} (1-{\mathcal X}_{I_{a+1}}{\mathbf X})^{p_{i_{a+1}}} 
=
 (1-{\mathcal X}_{\tau(I_{a}) }{\mathbf X})^{p_{\tau(i_{a})}} (1-{\mathcal X}_{\tau(I_{a+1})}{\mathbf X})^{p_{\tau(i_{a+1})}} ,
\label{ZCC}
\end{align}
where $\tau(i_{a})=i_{a+1}, \tau(i_{a+1})=i_{a}$. 
This relation \eqref{ZCC} boils down to \eqref{Tinv} and an identity.  
\begin{multline}
\left({\mathcal X}^{[-p_{i_{a+1} }]}_{I_{a} } \right)^{p_{i_{a}}} 
\left({\mathcal X}^{[p_{i_{a} }]}_{I_{a+1} } \right)^{p_{i_{a+1}}}
=
\left({\mathcal X}^{[-p_{\tau(i_{a+1}) }]}_{\tau(I_{a}) } \right)^{p_{\tau(i_{a})}} 
\left({\mathcal X}^{[p_{\tau(i_{a}) }]}_{\tau(I_{a+1}) } \right)^{p_{\tau(i_{a+1})}}
\\
=(z_{i_{a}})^{p_{i_{a}}} (z_{i_{a+1}})^{p_{i_{a+1}}}
\frac{\Qb_{I_{a-1}}^{[M-N-\sum_{j \in I_{a+1}}p_{j}-1]}
\Qb_{I_{a+1}}^{[M-N-\sum_{j \in I_{a-1}}p_{j}+1]}
}{
\Qb_{I_{a-1}}^{[M-N-\sum_{j \in I_{a+1}}p_{j} +1]}
\Qb_{I_{a+1}}^{[M-N-\sum_{j \in I_{a-1}}p_{j}-1]}
} .
\label{id-A}
\end{multline}
The invariance for the case $K \le a-1$ is trivial. 
Thus the T-functions $ {\mathcal F}_{(b)}^{I_{K}}$ and $ {\mathcal F}_{(1^{b})}^{I_{K}}$ for 
$b \ge 0$, $K \ne a$ are invariant under the transposition $\tau$. 
Therefore, these functions are invariant under $S(I_{K}) \times S(\overline{I}_{K})$,  
where $\overline{I}_{K}=(i_{K+1},i_{K+2},\dots , i_{M+N})$, 
since the symmetric group is generated by transpositions. 
In particular,  the T-functions $ {\mathcal F}_{(b)}^{I_{M+N}}$ and $ {\mathcal F}_{(1^{b})}^{I_{M+N}}$ 
are invariant under $S(I_{M+N})$. This means that these are independent of 
the order of the elements of tuple $I_{M+N}$ under the QQ-relations \eqref{QQb} and \eqref{QQf}. 
%They obey the following recursion relations (cf.\ \cite{KSZ07,Zabrodin07}): 
%\begin{align}
%& (1-{\mathcal X}_{I_{K+1}}{\mathbf X})^{-p_{i_{K+1}}} {\mathbf W}_{I_{K}}({\mathbf X})={\mathbf W}_{I_{K+1}}({\mathbf X}).
%\label{rec1}
%\end{align}

One can derive Baxter type equations from the kernels of  \eqref{gene1} and \eqref{gene2}. 
\begin{align}
0={\mathbf W}_{I_{K}}({\mathbf X}) \cdot {\mathfrak Q}_{I_{1}}&=
\sum_{a=0}^{\infty} {\mathcal F}_{(a)}^{I_{K}
[a -1-\mm+\nn+M-N]}{\mathfrak Q}_{I_{1}}^{[2a] } \quad \text{if} \quad p_{i_{1}}=-1, 
\label{Bax1}
\\
0={\mathbf W}_{I_{K}}({\mathbf X})^{-1} \cdot  {\mathfrak Q}_{I_{K}}&=
 \sum_{a=0}^{\infty}(-1)^{a} {\mathcal F}_{(1^{a})}^{I_{K}
[a -1-\mm+\nn+M-N]}{\mathfrak Q}_{I_{K}}^{[2a] } \quad \text{if} \quad p_{i_{K}}=1, 
\label{Bax2}
\end{align}
where 
\begin{align}
{\mathfrak Q}_{I_{a}}&=
e^{-\frac{u}{2}\log z_{i_{a}}}
\left(
\frac{\Qb_{I_{a-1}}^{[M-N-\sum_{j \in I_{a-1}}p_{j}-p_{i_{a}}-1]}
}{
\Qb_{I_{a}}^{[M-N-\sum_{j \in I_{a}}p_{j}+p_{i_{a}}-1]}
}
\right)^{p_{i_{a}}} ,
\end{align}
and ${\mathbf X} \cdot {\mathfrak Q}_{I_{a}}={\mathfrak Q}_{I_{a}}^{[2]}$. 
%%%%%%%%%%%%%%%%%%%%%%%%%%%%%%%%%
It is possible to consider reverse order version of \eqref{gene1} and \eqref{gene2}:
\begin{align}
{\mathbf W}^{\prime}_{I_{K}}({\mathbf X}^{-1})&=\overrightarrow{\prod_{k=1}^{K}}
 (1-{\mathcal X}_{I_{k}}{\mathbf X}^{-1})^{-p_{i_{k}}} 
= \sum_{a=0}^{\infty} {\mathcal F}_{(a)}^{I_{K}
[-a +1-\mm+\nn+M-N]}{\mathbf X}^{-a }, 
\label{gene3}
\\
{\mathbf W}^{\prime}_{I_{K}}({\mathbf X}^{-1})^{-1}&=\overleftarrow{\prod_{k=1}^{K}}
 (1-{\mathcal X}_{I_{k}}{\mathbf X}^{-1})^{p_{i_{k}}} 
= \sum_{a=0}^{\infty}(-1)^{a} {\mathcal F}_{(1^{a})}^{I_{K}
[-a +1-\mm+\nn+M-N]}{\mathbf X}^{-a }.
\label{gene4}
\end{align}
These are invariant under the action of $S(I_{K})\times S(\overline{I}_{K})$. 
%
%They obey the following recursion relations (cf.\ \cite{KSZ07,Zabrodin07}): 
%\begin{align}
%& (1-{\mathcal X}_{I_{K+1}}{\mathbf X})^{-p_{i_{K+1}}} {\mathbf W}_{I_{K}}({\mathbf X})={\mathbf W}_{I_{K+1}}({\mathbf X}).
%\label{rec1}
%\end{align}
One can derive Baxter type equations from the kernels of  \eqref{gene3} and \eqref{gene4}. 
\begin{align}
0={\mathbf W}^{\prime}_{I_{K}}({\mathbf X}^{-1}) \cdot {\mathfrak Q}^{\prime}_{I_{K}}&=
\sum_{a=0}^{\infty} {\mathcal F}_{(a)}^{I_{K}
[-a +1-\mm+\nn+M-N]}{\mathfrak Q}_{I_{K}}^{\prime [-2a] } \quad \text{if} \quad p_{i_{K}}=-1, 
\label{Bax3}
\\
0={\mathbf W}^{\prime}_{I_{K}}({\mathbf X}^{-1})^{-1} \cdot  {\mathfrak Q}^{\prime}_{I_{1}}&=
 \sum_{a=0}^{\infty}(-1)^{a} {\mathcal F}_{(1^{a})}^{I_{K}
[-a +1-\mm+\nn+M-N]}{\mathfrak Q}_{I_{1}}^{\prime [-2a] } \quad \text{if} \quad p_{i_{1}}=1, 
\label{Bax4}
\end{align}
where 
\begin{align}
{\mathfrak Q}^{\prime}_{I_{a}}&=
e^{\frac{u}{2}\log z_{i_{a}}}
\left(
\frac{
\Qb_{I_{a}}^{[M-N-\sum_{j \in I_{a}}p_{j}+p_{i_{a}}+1]}
}{\Qb_{I_{a-1}}^{[M-N-\sum_{j \in I_{a-1}}p_{j}-p_{i_{a}}+1]}
}
\right)^{p_{i_{a}}} ,
\end{align}
and ${\mathbf X}^{-1} \cdot {\mathfrak Q}^{\prime}_{I_{a}}={\mathfrak Q}_{I_{a}}^{\prime [-2]}$. 

%%%%%%%%%%%
\paragraph{Supersymmetric Cherednik-Bazhanov-Reshetikhin formula}
The tableau sum formula \eqref{DVF-tab1} has  determinant expressions
\footnote{
For the 180 degree rotated Young diagram $\widetilde{\lambda \subset \mu}$, one can show 
\begin{align}
{\mathcal F}_{\widetilde{\lambda \subset \mu}}^{I_{K}}&= 
\begin{vmatrix}
    \left(
    {\mathcal F}^{I_{K} \, [\mu_{1}-\mu_{1}^{\prime}+\mu_{i}^{\prime}+\lambda_{j}^{\prime}-i-j+1]}_{(1^{\mu_{i}^{\prime}-\lambda_{j}^{\prime}-i+j})} \
    \right)_{1 \le i \le \mu_{1} \atop
    1 \le j \le \mu_{1}} 
 \end{vmatrix}   
\end{align}
based on the identity $|A_{ij}|_{1 \le i,j \le K}=|A_{K+1-j,K+1-i}|_{1 \le i,j \le K}$ for a matrix $(A_{ij})_{1 \le i,j \le K}$. 
This corresponds to [(2.13), \cite{T21}].}
%%%%%%%%%%%%%%%%%%%%%%%%%%%%%%%%%%%
\begin{align}
{\mathcal F}_{\lambda \subset \mu}^{I_{K}}&= 
\begin{vmatrix}
    \left(
    {\mathcal F}^{I_{K} \, [-\mu_{1}+\mu_{1}^{\prime}-\mu_{i}^{\prime}-\lambda_{j}^{\prime}+i+j-1]}_{(1^{\mu_{i}^{\prime}-\lambda_{j}^{\prime}-i+j})} \
    \right)_{1 \le i \le \mu_{1} \atop
    1 \le j \le \mu_{1}}
 \end{vmatrix}   
 \label{superJT1}
 \\
&= 
\begin{vmatrix}
    \left(
    {\mathcal F}^{I_{K} \, [-\mu_{1}+\mu_{1}^{\prime}+\mu_{i}^{\prime}+\lambda_{j}^{\prime}-i-j+1]}_{(\mu_{i}-\lambda_{j}-i+j)} \
    \right)_{1 \le i \le \mu_{1}^{\prime} \atop
    1 \le j \le \mu_{1}^{\prime}}
 \end{vmatrix}   ,
 \label{superJT2}
\end{align}
where  ${\mathcal F}_{(1^{0})}^{I_{K}}={\mathcal F}_{(0)}^{I_{K}}=1$ and 
${\mathcal F}_{(1^{a})}^{I_{K}}={\mathcal F}_{(a)}^{I_{K}} = 0$ for $a <0$. 
These determinant expressions for $K=M+N$ correspond to 
the supersymmetric Cherednik-Bazhanov-Reshetikhin formulas (supersymmetric CBR formulas or quantum supersymmetric Jacobi-Trudi formulas) \cite{T97,T98} 
(see also \cite{KV07,LM20,KOS95}), which are 
supersymmetric extensions of the CBR formula (quantum Jacobi-Trudi formula) \cite{Ch89,BR90}. 
%%%
%To kill this kind of pole, 
In order to cancel the poles by the functions $\Qb_{\emptyset}$ and $\Qb_{I_{K}}$, 
we introduce the following transformation
\footnote{In the right hand side, we do not need the 180 degrees rotated Young diagram (see [(2.14), \cite{T21}])  
since we have changed the convention of the function \eqref{DVF-tab1}.}
 for any skew Young diagram $\lambda \subset \mu$:
\begin{align}
 {\mathsf F}_{\lambda \subset \mu}^{I_{K}}
 =
\Phi_{\lambda \subset \mu}^{I_{K}} 
{\mathcal  F}_{\lambda \subset \mu}^{I_{K}} 
\label{TF-rel03}
\end{align}
with the overall factor defined by 
\begin{multline}
\Phi_{\lambda \subset \mu}^{I_{K}}=
\Qb_{\emptyset}^{[-\mu_{1}-\mu_{1}^{\prime}+2\mu_{\mu^{\prime}_{1} }+\mm-\nn]}
\Qb_{I_{K}}^{[-\mu_{1}+\mu_{1}^{\prime}+2\lambda_{1}]} 
\times 
\\
\times
\prod_{j=1}^{\mu^{\prime}_{1}-1} 
\left(\Qb_{\emptyset}^{[-\mu_{1}+\mu_{1}^{\prime}-2j+2\mu_{j}+\mm-\nn]} \right)^{\theta ((\mu_{j}-\mu_{j+1})(\mu_{j}-\lambda_{j}) >0)}
\times 
\\
\times 
\prod_{j=2}^{\min(\lambda^{\prime}_{1}+1,\mu^{\prime}_{1}) } 
\left(\Qb_{I_{K}}^{[-\mu_{1}+\mu_{1}^{\prime}-2j+2\lambda_{j}+2]} \right)^{\theta ((\lambda_{j-1}-\lambda_{j})(\mu_{j}-\lambda_{j}) >0)} 
 ,
\label{overallF}
\end{multline}
where $\lambda_{\lambda_{1}^{\prime} +1 }=0$, $\theta(\text{True})=1$, $\theta(\text{False})=0$, 
$\prod_{j=c_{1}}^{c_{2}} (\dots ) =1$ if $c_{1}>c_{2}$. 
In case the Young diagram $\lambda \subset \mu $ is of rectangular shape, \eqref{overallF} reduces to 
\begin{align}
\Phi_{(\mu_{1}^{\mu^{\prime}_{1}})}^{I_{K}}=\Qb_{\emptyset}^{[\mm-\nn+\mu_{1}-\mu_{1}^{\prime}]} 
\Qb_{I_{K}}^{[-\mu_{1}+\mu_{1}^{\prime}]} .
 \label{overallFreq}
\end{align}
The normalization by the function \eqref{overallFreq} was used in the previous paper [eq.\ (2.14), \cite{T21}]. 
%We change the normalization of the T-function $ {\mathsf F}_{\lambda \subset \mu}^{I_{K}}$ 
%for non-rectangular Young diagrams.  

The T-function ${\mathcal F}_{\lambda \subset \mu}^{I_{K}}$ is 
 invariant under the action of $S(I_{K}) \times S(\overline{I}_{K})$ since the matrix elements 
of \eqref{superJT1} are. In particular, the T-function ${\mathcal F}_{\lambda \subset \mu}^{I_{M+N}}$ is 
 invariant under the action of $S(I_{M+N}) $ under  the QQ-relations \eqref{QQb} and \eqref{QQf}. 
This is also the case with $ {\mathsf F}_{\lambda \subset \mu}^{I_{K}}$.  
 Then we define ${\mathcal F}_{\lambda \subset \mu}^{B,F}:={\mathcal F}_{\lambda \subset \mu}^{I_{K}}$ and 
 ${\mathsf F}_{\lambda \subset \mu}^{B,F}:={\mathsf F}_{\lambda \subset \mu}^{I_{K}}$, where 
 $B=I_{K} \cap \Bm$ and $F=I_{K} \cap \Fm$ as sets. 
 The function \eqref{overallF} does not depend on the order of the elements of $I_{K}$ since $\Qb_{I_{K}}=\Qb_{B,F}$. Thus we may set 
$ \Phi_{\lambda \subset \mu}^{B,F} := \Phi_{\lambda \subset \mu}^{I_{K}}$. 
 %%%%%%%%%%%%
 \paragraph{Bethe strap}
 T-functions obtained by analytic Bethe ansatz have {\em Bethe strap} structure \cite{KS94-1,Su95}. 
T-functions are generalization of (super)characters. One sees supercharacters by setting all the Q-functions 
to $1$ as in \eqref{ch-limit}. 
Thus each term of a T-function carries a weight of 
a representation of  an underlying algebra. The term which carries the highest weight is called the 
 {\em top term}. 

%%%%%
Let us explain the Bethe strap procedure for the $U_{q}(gl(M|N)^{(1)})$ case. 
We introduce the following function 
\begin{align}
F_{a}(u)=
 p_{\alpha_{a}}
 e^{-\alpha_{a}(h)}
 \frac{P_{a}(u-d_{a} )}{P_{a}(u+d_{a})}
 \prod_{b=1}^{\rr}
\frac{\Qc_{b}(u+(\alpha_{a}| \alpha_{b})) }{\Qc_{b}(u-(\alpha_{a}|\alpha_{b} ) )}
\quad
\text{for}  \quad a \in \{1,2,\dots, \rr \} .
\label{BAEr2F}
\end{align}
The Bethe ansatz equation \eqref{BAEr2} is equivalent to  $F_{a}(u_{k}^{(a)})=-1$, $k\in \{1,2,\dots, n_{a} \} $. 
The adjacent terms in \eqref{boxes} are related to each other as
\begin{align}
& {\mathcal   X}_{I_{M+N+1-a}} F_{a}^{[M-N-\sum_{j \in I_{M+N-a}} p_{j} ]}= p_{\alpha_{a}} {\mathcal   X}_{I_{M+N-a}}, 
\quad \text{for} \quad 1 \le a \le M+N-1 . 
\label{acF}
\end{align}
This means that the common poles  of $ {\mathcal   X}_{I_{M+N+1-a}} $ and $ {\mathcal   X}_{I_{M+N-a}}$ 
at $u=u_{k}^{(a)}-(M-N-\sum_{j \in I_{M+N-a}} p_{j} )$ ($k\in \{1,2,\dots, n_{a} \} $) 
cancel with each other under the Bethe ansatz equation \eqref{BAEr2} (see \eqref{chancels} for $a \to M+N-a$). 
 The T-function \eqref{tab-fund}  is an analogue of the supercharacter of the 
 highest weight representation 
  $V(\epsilon_{i_{M+N}})$ of $gl(M|N)$ with the highest weight $\epsilon_{i_{M+N}}$. 
The term ${\mathcal   X}_{I_{M+N}}$ in \eqref{tab-fund} is the top term, which carries the highest weight $\epsilon_{i_{M+N}}$ 
of $gl(M|N)$. 
$F_{a}$ looks like the root vector corresponding to the root $-\alpha_{a}$. 
The term ${\mathcal   X}_{I_{M+N-a}}$ carries the weight $\epsilon_{i_{M+N-a}}=\epsilon_{i_{M+N}} - \sum_{b=1}^{a}\alpha_{b}$ 
because of the action of $F_{1},F_{2},\dots , F_{\rr}$. 
The whole set of the terms of the T-function \eqref{tab-fund} forms a connected graph by the relation \eqref{acF}. 
The Bethe strap is a procedure to find the minimal pole-free set by repeatedly multiplying the top term by the
 function \eqref{BAEr2F}. The T-functions obtained by the Bethe strap procedures are expected to form 
  connected graphs if the auxiliary spaces are irreducible representations of an underlying algebra. 
  In particular, the top term of the T-function $\mathcal{F}^{I_{M+N}}_{\mu}$ 
for $I_{M+N}=(M+N,\dots , 2,1)$ is considered (cf.\ eqs.\ (2.28), (2.29) in \cite{T97})  to be 
\begin{align}
\mathsf{hw}_{\mu}(u)=
\prod_{j=1}^{M }
\prod_{k=1}^{\mu_{j}}
{\mathcal  X}_{I_{M+N+1-j}}^{[-\mu_{1}+\mu_{1}^{\prime}-2j+2k]} 
\prod_{k=1}^{ N }
\prod_{j=M+1}^{\mu^{\prime}_{k}}
(-1)
{\mathcal  X}_{I_{N+1-k}}^{[-\mu_{1}+\mu_{1}^{\prime}-2j+2k]} 
\label{topA}
\end{align}
since it carries the $gl(M|N)$ highest weight \eqref{HW-A} for \eqref{YW-A}. 
In fact, we have $\zeta(\mathsf{hw}_{\mu}(u))=(-1)^{\sum_{k=1}^{N} \max\{\mu_{k}^{\prime}-M,0 \}}e^{\Lambda (h)}$, where $e^{\epsilon_{a}(h)}=z_{a}$, $a \in \mathfrak{I}$. 
Here we set $\mu_{j}=0$ if $j>\mu^{\prime}_{1}$, $\mu^{\prime}_{k}=0$ if $k>\mu_{1}$, 
$\prod_{j=a}^{b}(\cdots)=1$ if $a>b$.

\begin{figure}
 \begin{tikzpicture}[x=10mm,y=10mm]
  \node (3) at (-2.5,0) {$ {\mathcal  X}_{I_{3}}$};
  \node (2) at (0,0) {$-{\mathcal  X}_{I_{2}}$};
  \node (1) at (2.5,0) {${\mathcal  X}_{I_{1}} $};
  \node[below of = 2]  {${\mathcal F}_{(1)}^{(2,3,1)}$};
  \draw[-to,line width=1pt] (3) -- node[midway,above]{$F_{1}^{[1]}$} (2); 
  \draw[-to,line width=1pt] (2) -- node[midway,above]{$F_{2}$} (1); 
\end{tikzpicture}
%%%%%
\begin{tikzpicture}[x=10mm,y=10mm]
  \node (32) at (0,5) {$- {\mathcal  X}_{I_{3}}^{[1]} {\mathcal  X}_{I_{2}}^{[-1]}$};
  \node (22) at (-2.5,2.5) {${\mathcal  X}_{I_{2}}^{[1]} {\mathcal  X}_{I_{2}}^{[-1]}$};
  \node (31) at (2.5,2.5) {${\mathcal  X}_{I_{3}}^{[1]} {\mathcal  X}_{I_{1}}^{[-1]}  $};
  \node (21) at (0,0) {$-{\mathcal  X}_{I_{2}}^{[1]} {\mathcal  X}_{I_{1}}^{[-1]}$};
  \node[below of = 21]  {${\mathcal F}_{(1^{2})}^{(2,3,1)}$};
  \draw[-to,line width=1pt] (32) -- node[midway,left]{$F_{1}^{[2]}$} (22); 
  \draw[-to,line width=1pt] (32) -- node[midway,right]{$F_{2}^{[-1]}$} (31); 
  \draw[-to,line width=1pt] (22) -- node[midway,left]{$F_{2}^{[-1]}$} (21); 
  \draw[-to,line width=1pt] (31) -- node[midway,right]{$F_{1}^{[2]}$}  (21); 
\end{tikzpicture}
\caption{Bethe strap structures of T-functions for $U_{q}(gl(2|1)^{(1)})$.}
\label{BS1}
\end{figure}
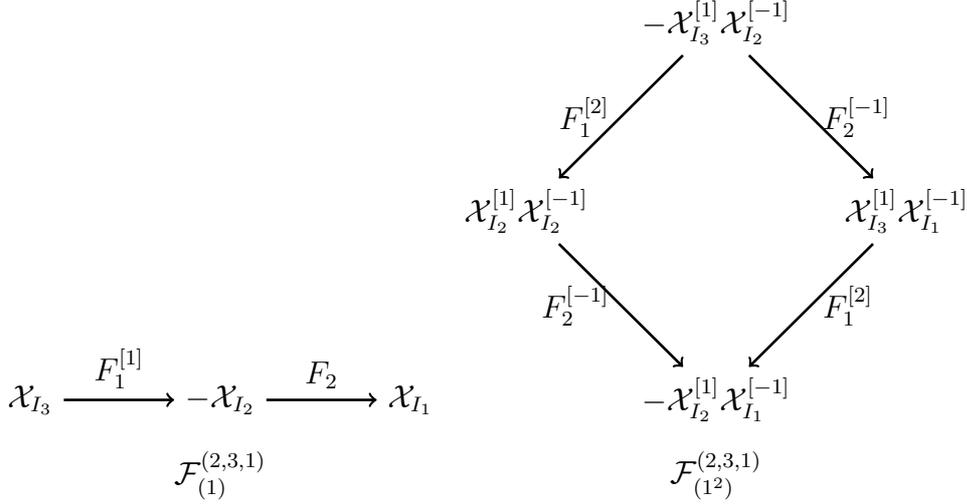

Let us give examples for the case $U_{q}(gl(2|1)^{(1)})$. 
In case $I_{3}=(2,3,1)$, \eqref{BAEr2F} reduces to 
 \begin{align}
F_{1} &= -\frac{z_{3}}{z_{1}}
 \frac{\Qb_{231}^{[-1]} \Qb_{2}^{[1]} }{ \Qb_{231}^{[1]} \Qb_{2}^{[-1]} } ,
 \qquad 
 F_{2} = -\frac{z_{2}}{z_{3}}
 \frac{\Qb_{23}^{[1]} \Qb_{\emptyset}^{[-1]} }{ \Qb_{23}^{[-1]} \Qb_{\emptyset}^{[1]} } .
\end{align}
The Bethe straps for the T-functions \eqref{DVF-ex1} form connected graphs described in Figure \ref{BS1}.
 %%%
%%%
   See \cite{T98} for more examples of Bethe straps for other representations.  
  The notion of the Bethe strap in the analytic Bethe ansatz appeared \cite{KS94-1,Su95} 
before q-characters in representation theory \cite{FR99,FM99} were introduced. 
In the theory of q-characters, the parameters $z_{j}$ are usually included in Q-functions 
and ratios of Q-functions are used as variables. In addition, the Q-functions $\Qb_{\emptyset}$ and
 $\Qb_{\mathfrak{I}}$, which are related to  ``vacuum parts''  in the analytic Bethe ansatz,
   are normalized to $1$,  
and  the top term \eqref{topA} is called 
the ``highest weight monominal''. 
The Bethe strap procedures may be mathematically justified by the theory of q-characters, but this is beyond 
the scope of this paper. 
 %%%%%%%%%%%%%%%%%%%%
 \subsection{Wronskian-type expressions of T-functions}
Let us introduce a determinant 
\footnote{
This is related to the sparse determinant in [eq.(3.24) in \cite{T11}] as
$
\Delta^{B,R
,[\xi ]}_  
{F,S  
} 
=
\Delta^{B,\emptyset,R
,[0; \xi ]}_  
{F,S, \emptyset, \emptyset  
} 
$,
where  we set $B_{1}=B, B_{2}=T_{1}=T_{2}=\emptyset$, $\eta =0$.
Moreover, this is related to the determinant in [eq.(3.4) in \cite{T09}] 
as 
$\Delta^{B,R
,[\xi ]}_  
{F,S  
} 
=\Delta^{B,R}_  
{F,S  
}(x q^{\xi}) $ in case q-difference is adopted. 
}
over a block matrix 
labeled by sets $B,R,F,S$ ($B \subset \Bm$, $|B|=\mm $; $F \subset \Fm$, $|F|=\nn$; $R,S \subset \mathbb{Z}$): 
\begin{align}  
\Delta^{B,R,[\xi ]}_  
{F,S  
} 
&=   
\begin{vmatrix}  
\left( \frac{\Qb_{b,f}^{[\xi ]}}{z_{b}- z_{f} } 
\right)_{  
\genfrac{}{}{0pt}{}{b\in B, }{f \in F} } 
&   
 \left(
z_{b }^{j-1} \Qb_{b}^{[\xi+2j-1]}
\right)_{  
\genfrac{}{}{0pt}{}{b\in B, }{j \in S} }    
  \\[6pt]  
\left(
(-z_{f})^{i-1}\Qb_{f}^{[\xi-2i+1 ]}
\right)_{  
\genfrac{}{}{0pt}{}{i \in R, }{f \in F} }  
& (0)_{|R|\times |S|}    
\end{vmatrix}  
,  \label{sparcedetbb}
\end{align}  
where  $\xi \in \mathbb{C}$,  and $(0)_{|R|\times |S|} $ is $|R|$ by $ |S|$ zero matrix.  
The number of the elements of the sets must satisfy 
$|B|+|R|=|F| + |S| $.
For any Young diagram $\mu$, we introduce a 
number, called $(\mm,\nn)$-index \cite{MV03}:
\begin{align}
\xi_{\mm,\nn}(\mu):={\rm min}\{j \in {\mathbb Z}_{\ge 1}|\mu_{j}+\mm-j \le \nn-1\}.
\label{mn-index}
\end{align}
In particular, we have $1 \le \xi_{\mm,\nn}(\mu) \le \mm+1$, $\xi_{\mm,0}(\mu)=\mm+1$ and $\xi_{0,\nn}(\mu)=1$
 for $\mu_{\mm+1} \le \nn$, and 
$\xi_{\mm,\nn}(\mu)=\mm+1$ for $\mu_{\mm+1} \le n \le \mu_{\mm}$; 
$\xi_{\mm,\nn}(\emptyset)=\max\{\mm-\nn+1,1\}$. 
We often abbreviate $\xi_{\mm,\nn}(\mu)$ as $\xi_{\mm,\nn}$. 
The denominator formula of the supercharacter of $gl(\mm|\nn)$ can be written as
 \cite{MV03}: 
\begin{align}
 \Ds( B|F)
&= 
\frac{\prod_{ b,b^{\prime } \in B, \atop b < b^{\prime } }(z_{b} - z_{b^{\prime}})
\prod_{f,f^{\prime } \in F, \atop f < f^{\prime } }(z_{f^{\prime} } - z_{f})
}
{\prod_{ (b,f) \in B \times F } (z_{b} - z_{f})}.
\end{align} 

For any Young diagram $\mu$, 
we introduce the following function
\footnote{$ \Ts_{\mu}^{B,F [-\frac{\mm-\nn}{2}]}$ 
corresponds to eq.(3.15) in \cite{T09} (in a different normalization).}
%. 
%\begin{align} 
\begin{multline}
 \Ts_{\mu}^{B,F}
:= (-1)^{(\mm+\nn+1)(\xi_{\mm,\nn}(\mu ) +1) +\frac{(\mm-\nn)(\mm+\nn-1)}{2}} \times 
 \\ 
 \times 
 \frac{
 \Phi_{\mu}^{B,F}
 }
{ \Phi_{(\mu_{1}^{\mu^{\prime}_{1}})}^{B,F} }
\, 
 \Psi^{(\mm,\nn)}_{\mu}
\Delta^{B, (r_{1},r_{2},\dots,r_{\nn-\mm+\xi_{\mm,\nn}-1}), [-\mm+\nn +\mu_{1}^{\prime}-\mu_{1}]}
_{F,(s_{1},s_{2},\dots,s_{\xi_{\mm,\nn}-1} )} 
 \Ds( B|F)^{-1},
\label{unnor-t1}
\end{multline}
where $s_{l}=\mu_{\xi_{\mm,\nn} -l}+\mm-\nn-\xi_{\mm,\nn}(\mu)+l+1$,
$r_{k}=\mu_{\nn-\mm+\xi_{\mm,\nn}-k}^{\prime}+k-\xi_{\mm,\nn}(\mu)+1$ and 
\begin{align}
& \Psi^{(\mm,\nn)}_{\mu}=
 \frac{\Qb_{\emptyset}^{[\mm-\nn+\mu_{1}-\mu_{1}^{\prime}]} \Qb^{[-\mm+\nn-\mu_{1}+\mu_{1}^{\prime}]}_{\emptyset}
 \left( \Qb_{\emptyset}^{[-\mm+\nn-\mu_{1}+\mu_{1}^{\prime}]}\right)^{-\mm-1+\xi_{\mm,\nn}}}
{
\prod^{\nn-\mm+\xi_{\mm,\nn}-1}_{i=1} \Qb_{\emptyset}^{[-\mm+\nn-\mu_{1}+\mu_{1}^{\prime}-2r_{i}+2]}
\prod^{\xi_{\mm,\nn}-1}_{j=1} \Qb_{\emptyset}^{[-\mm+\nn-\mu_{1}+\mu_{1}^{\prime}+2s_{j}-2]} 
 } ,
 \label{Del-nor}
\\
& \frac{
 \Phi_{\mu}^{B,F}
 }
{ \Phi_{(\mu_{1}^{\mu^{\prime}_{1}})}^{B,F} }=
\Qb_{\emptyset}^{[-\mu_{1}-\mu_{1}^{\prime}+2\mu_{\mu^{\prime}_{1} }+\mm-\nn]}
( \Qb_{\emptyset}^{[\mu_{1}-\mu_{1}^{\prime}+\mm-\nn]} )^{-1}
\prod_{j=1}^{\mu^{\prime}_{1}-1} 
\left(\Qb_{\emptyset}^{[-\mu_{1}+\mu_{1}^{\prime}-2j+2\mu_{j}+\mm-\nn]} \right)^{\theta (\mu_{j}-\mu_{j+1} >0)}
 ,
\label{overallnor}
\end{align}
If the Young diagram $\mu$ is of rectangular shape, the factor \eqref{overallnor} becomes $1$. 
In this case, the normalization of \eqref{unnor-t1} reduces to the one in the previous paper \cite{T21}.
%Note that right hand side of \eqref{unnor-t1} is invariant under any permutation of the elements in each of the tuples $B$ and $F$, 
%and thus $B$ and $F$ in the left hand side can be interpreted as just sets: 
%$B=\{b_{1},b_{2},\dots, b_{m}\}$, 
%$F=\{f_{1},f_{2},\dots, f_{n} \}$. 
We remark that the (super)character limit of \eqref{unnor-t1} at $(\mm,\nn)=(M,N)$, 
namely $\zeta( \Ts_{\mu}^{\Bm,\Fm})$
 coincides with the determinant formula \cite{MV03} of the supercharacter of 
 the highest weight representation $V(\Lambda)$ of $gl(M|N)$ with the highest weight 
 \eqref{HW-A} (with \eqref{YW-A}). It is a natural generalization of the Schur function 
 (Weyl-type formula). 
Specializing \eqref{unnor-t1} for the empty diagram, we obtain 
a Wronskian-like determinant solution [Theorem 3.2 in \cite{T09}] of the QQ-relations \eqref{QQb} and \eqref{QQf}: 
\begin{align}
 \Ts_{\emptyset}^{B,F}
 &=
(-1)^{\frac{(\mm-\nn)(\mm+\nn-1)}{2}}  \Psi^{(\mm,\nn)}_{\emptyset}
\Delta^{B, \langle 1, \nn-\mm \rangle , [-\mm+\nn ]}
_{F, \langle 1, \mm-\nn \rangle } 
 \Ds( B|F)^{-1}
 \nonumber 
 \\
 &=\Qb_{B,F} \Qb_{\emptyset}^{[\mm-\nn]},
 \label{emptyT}
\end{align}
where we introduce notation
\begin{align}
\langle a,b \rangle =
\begin{cases}
\{a,a+1,a+2, \dots, b \} & \text{for} \quad b -a \in {\mathbb Z}_{\ge 0}, \\[3pt]
\emptyset & \text{for}  \quad  b -a  \notin {\mathbb Z}_{\ge 0} .
\end{cases}
\end{align}
Explicitly, \eqref{emptyT} reads
\begin{multline}  
\Qb_{B,F} =   
 \frac{(-1)^{\frac{(\mm-\nn)(\mm+\nn-1)}{2}} }
{
 (\Qb_{\emptyset}^{[\nn-\mm]})^{\nn-1}
\prod^{\mm-\nn}_{j=1} \Qb_{\emptyset}^{[\nn-\mm+2j-2]} 
 \Ds( B|F)
  }
\begin{vmatrix}  
\left( \frac{\Qb_{b,f}^{[\nn-\mm ]}}{z_{b}- z_{f} } 
\right)_{  
\genfrac{}{}{0pt}{}{b\in B, }{f \in F} } 
&   
 \left(
z_{b }^{j-1} \Qb_{b}^{[\nn-\mm+2j-1]}
\right)_{  
\genfrac{}{}{0pt}{}{b\in B, }{j \in \langle 1,\mm-\nn \rangle} }      
\end{vmatrix}  
\\
\text{for} \quad \mm \ge \nn ,
\label{QQdetsol1}
\end{multline}
%%%
\begin{multline}
\Qb_{B,F} =    
 \frac{(-1)^{\frac{(\mm-\nn)(\mm+\nn-1)}{2}} }
{
 (\Qb_{\emptyset}^{[\nn-\mm]})^{\mm-1}
\prod^{\nn-\mm}_{j=1} \Qb_{\emptyset}^{[\nn-\mm-2j+2]} 
 \Ds( B|F)
  }
\begin{vmatrix}  
\left( \frac{\Qb_{b,f}^{[\nn-\mm ]}}{z_{b}- z_{f} } 
\right)_{  
\genfrac{}{}{0pt}{}{b\in B, }{f \in F} }  
  \\[6pt]  
\left(
(-z_{f})^{i-1}\Qb_{f}^{[\nn-\mm-2i+1 ]}
\right)_{  
\genfrac{}{}{0pt}{}{i \in \langle 1,\nn-\mm \rangle, }{f \in F} }    
\end{vmatrix}  
\\
\text{for} \quad \mm \le \nn .
 \label{QQdetsol2}
\end{multline}  
For any rectangular Young diagram, we set
\begin{align}
\Ts_{a,m}^{B,F}&=
\begin{cases}
\Ts^{B,F}_{(m^{a})} & \text{for} \quad a, m \in \mathbb{Z}_{\ge 1}
\\
\Qb_{\emptyset}^{[\mm-\nn-a]} (\Qb_{\emptyset}^{[\mm-\nn+a]})^{-1} \Ts^{B,F[a]}_{\emptyset} & \text{for} \quad a \in \mathbb{Z}_{\ge 0},
 \quad m  =0
\\
\Qb_{\emptyset}^{[\mm-\nn+m]} (\Qb_{\emptyset}^{[\mm-\nn-m]})^{-1} 
\Ts^{B,F[-m]}_{\emptyset} & \text{for} \quad  a=0 , \quad m \in \mathbb{Z}
\\
0  & \text{otherwise}  .
\end{cases}
 \label{Trec0}
\end{align}
%%%
More explicitly, we have:
\\
\noindent 
for $a \le \mm-\nn$, we have $\xi_{\mm,\nn}((m^{a}))=\mm-\nn+1$ and 
\begin{align}
 \Ts_{a,m}^{{B,F}}
= 
(-1)^{\frac{(\mm-\nn)(\mm+\nn-1)}{2}} \Psi^{(\mm,\nn)}_{a,m}\Delta^{B, \emptyset , [-\mm+\nn+a-m]}
_{F,\langle 1,\mm-\nn-a \rangle  \sqcup \langle \mm-\nn-a+m+1,\mm-\nn+m \rangle}
 \Ds( B|F)^{-1},
  \label{Trec1}
\end{align}
for $a-m \le \mm-\nn \le a $, we have $\xi_{\mm,\nn}((m^{a}))=a+1$ and 
\begin{align}
\Ts_{a,m}^{B,F}
= (-1)^{(\mm+\nn+1)a+\frac{(\mm-\nn)(\mm+\nn-1)}{2}} \Psi^{(\mm,\nn)}_{a,m}
\Delta^{B, \langle 1, \nn-\mm+a \rangle , [-\mm+\nn+a-m]}
_{F,\langle \mm-\nn-a+m+1 , \mm-\nn+m \rangle }
 \Ds( B|F)^{-1},
  \label{Trec2}
\end{align}
for $-m \le \mm-\nn \le a-m $, we have $\xi_{\mm,\nn}((m^{a}))=\mm-\nn+m+1$ and 
\begin{align}
\Ts_{a,m}^{B,F}
= (-1)^{(\mm+\nn+1)m+\frac{(\mm-\nn)(\mm+\nn-1)}{2}} \Psi^{(\mm,\nn)}_{a,m}
\Delta^{B,\langle \nn-\mm-m+a+1,\nn-\mm+a \rangle , [-\mm+\nn+a-m]}
_{F,\langle 1, \mm-\nn+m \rangle}
 \Ds( B|F)^{-1},
 \label{Trec3}
\end{align}
for $\mm-\nn \le -m$, we have $\xi_{\mm,\nn}((m^{a}))=1$ and 
\begin{align}
\Ts_{a,m}^{B,F}
= (-1)^{\frac{(\mm-\nn)(\mm+\nn-1)}{2}} \Psi^{(\mm,\nn)}_{a,m}
\Delta^{B,\langle 1,\nn-\mm-m \rangle \sqcup \langle \nn-\mm-m+a+1,\nn-\mm+a \rangle , [-\mm+\nn+a-m]}
_{F, \emptyset}
 \Ds( B|F)^{-1},
  \label{Trec4}
\end{align}
where we set
\begin{align}
\Psi^{(\mm,\nn)}_{a,m}&=
\begin{cases}
\Psi^{(\mm,\nn)}_{(m^{a})} & \text{for} \quad a, m \in \mathbb{Z}_{\ge 1}
\\
\Qb_{\emptyset}^{[\mm-\nn-a]} (\Qb_{\emptyset}^{[\mm-\nn+a]})^{-1} \Psi^{(\mm,\nn) [a]}_{\emptyset} & \text{for} \quad a \in \mathbb{Z}_{\ge 0}, \quad m  =0
\\
\Qb_{\emptyset}^{[\mm-\nn+m]} (\Qb_{\emptyset}^{[\mm-\nn-m]})^{-1} 
\Psi^{(\mm,\nn) [-m]}_{\emptyset} & \text{for} \quad  a=0 , \quad m \in \mathbb{Z}
\\
1  & \text{otherwise}  .
\end{cases}
 \label{Tas-nor}
\end{align}
Applying the Laplace expansion formula to \eqref{Trec0}-\eqref{Trec4}, we obtain useful expressions \cite{T09}
\begin{multline}    
\Ts^{B,F}_{a,m} =   
\sum_{J \subset F,  \atop
|J|=m}  
\frac{\prod_{j \in J}(-z_{j})^{a-m-\mm+\nn} \prod_{(b,j) \in B \times J } (z_{b}-z_{j}) }
{\prod_{(i,j) \in (F\setminus J) \times J  } (z_{i}-z_{j}) }
\Qb^{[a]}_{B,F\setminus J}  
\Qb^{[-a+\mm-\nn]}_{J}  
\\
 \qquad \qquad  \text{for} \quad a\ge m+\mm-\nn,  
\label{soQQb-1}  
\end{multline}
\begin{multline}
\Ts^{B,F}_{a,m} =   
\sum_{I \subset B, \atop |I|=a} 
\frac{ \prod_{i \in I} z_{i}^{m-a+\mm-\nn}
\prod_{(i,f) \in I \times F } (z_{i}-z_{f})}
{\prod_{(i,j) \in I \times (B\setminus I)  } (z_{i}-z_{j}) }  
\Qb^{[m+\mm-\nn]}_{I}  
\Qb^{[-m]}_{B\setminus I, F}  
\quad 
\text{for} \quad  a \le m + \mm-\nn, \label{soQQb-2}  
\end{multline} 
where the summation is taken over any possible subsets.  
The T-functions $\Ts^{B,F}_{a,m}$ solve \cite{T09} the T-system for $U_{q}(gl(M|N)^{(1)})$ (or $U_{q}(sl(M|N)^{(1)})$,  
or its Yangian counterpart) \cite{T97,T98}. 
The T-functions \eqref{soQQb-1} and \eqref{soQQb-2} are defined for the integer parameters $a$ and $m$. 
One can consider analytic continuation of \eqref{soQQb-1} with respect to $a$ and  \eqref{soQQb-2} with respect to $m$ 
 to the whole complex plane. 
However, in most cases, the analytically continued T-functions for the generic complex parameters ($a$ or $m$)
do not give T-functions for irreducible representations of the underlying algebra. 
Exceptions are T-functions for one parameter families of finite dimensional typical  irreducible representations of superalgebras, 
which correspond to \eqref{soQQb-1} for $m=\nn =N$ and \eqref{soQQb-2} for $a=\mm =M $ 
(analytic continuation of T-functions were considered in \cite{T98-2,T99-2}. 

%%%
We also have [eq.(3.67) in  \cite{T09}].
\begin{align}
\Ts^{B,F}_{\mu}={\mathsf F}^{B,F}_{\mu} 
 \label{T=F}
\end{align}
This is proven for general non-rectangular Young diagrams for $|F| =0$ case and rectangular Young diagrams for $|B||F| \ne 0$ case, 
and is a conjecture for general non-rectangular Young diagrams for $|B||F| \ne 0$ case (see page 426 in \cite{T09}).

\paragraph{T-functions for asymptotic representations}
Let $\mu=(\mu_{1},\mu_{2},\dots, \mu_{\mu_{1}^{\prime}})$ be a partition with $\mu_{1}^{\prime} \le \mm $. 
For an integer $c \ge \mm$, we define a partition 
$\mu +(\nn^{c})=(\mu_{1} +\nn,\mu_{2} +\nn,\dots, \mu_{\mu_{1}^{\prime}} +\nn, \underbrace{\nn,\dots ,\nn}_{c-\mu_{1}^{\prime}})$. We can show [cf.\ eq.\ (4.6) in \cite{T09}; eq.\ (3.52) in \cite{T21}]
\begin{align}
\Ts^{B,F [\mu_{1}^{\prime}-c] }_{\mu +(\nn^{c})}  =   
\prod_{ f \in F } (-z_{f})^{c-\mm }
\prod_{ (b,f) \in B \times F } (z_{b}-z_{f}) 
\left( \Qb^{[\mm -\mu_{1}-\mu_{1}^{\prime} +2\mu_{\mu^{\prime}_{1}}]}_{\emptyset }   \right)^{-1}
 \Qb^{[\mm -\nn -\mu_{1}+\mu_{1}^{\prime} -2c]}_{F } 
 \Ts^{B,\emptyset }_{\mu } ,
 \label{typ-sh}  
\end{align} 
where we use the property of the determinant 
\begin{align}
\begin{vmatrix}
A & B \\
C & \mathbf{0}
\end{vmatrix}
=(-1)^{\mm \nn} |C| |B|
\end{align}
for $\mm \times \nn $ matrix $A$,  $\mm \times \mm $ matrix $B$,  $\nn \times \nn $ matrix $C$ and 
 $\nn \times \mm $ zero matrix $\mathbf{0}$, 
and  the relation
\begin{align}
\frac{ \mathsf{D}(B |\emptyset) \mathsf{D}(\emptyset | F)}{\mathsf{D}(B|F)}=
\prod_{ (b,f) \in B \times F } (z_{b}-z_{f}) .
\end{align}
We consider the following limit of \eqref{typ-sh} with respect to the parameter $c$ 
[cf.\ eq.\ (3.53) in \cite{T21}]:
\begin{align}
 \Qb^{[\mm -\mu_{1}-\mu_{1}^{\prime} +2\mu_{\mu^{\prime}_{1}}]}_{\emptyset }  
\lim_{c}
\prod_{ f \in F } (-z_{f})^{-c }
\Ts^{B,F [\mu_{1}^{\prime}-c] }_{\mu +(\nn^{c})}  =   
\prod_{ f \in F } (-z_{f})^{-\mm }
\prod_{ (b,f) \in B \times F } (z_{b}-z_{f}) 
 \Ts^{B,\emptyset }_{\mu } 
 \label{typ-lim}  
\end{align} 
where we assume $\lim_{c} \Qb_{\emptyset}^{[-2c]}=\lim_{c} \Qb_{F}^{[-2c]}=1$. 
We will use reductions of \eqref{typ-lim} to describe T-functions for spinorial representations of 
orthosymplectic superalgebras. One can also use $ {\mathsf F}^{B,\emptyset }_{\mu } $ instead
 of $ \Ts^{B,\emptyset }_{\mu } $ in \eqref{typ-lim}. 
%%%%%%%%%%%%%%%%%%%%%%%%%%%%%%%%%%%%%%%%%%%%%%%%%
\section{Reductions of T- and Q-functions}
Now we would like to explain details of our proposal [section 3.7 in \cite{T11}] and its extension. 
In this section we consider reductions of T- and Q-functions introduced in the previous section. 
The reduction procedures in this section are an extension of the methods to 
derive the Bethe ansatz equations and T-functions 
for twisted quantum affine algebras from the ones for untwisted quantum 
affine algebras \cite{KS94-2} (see also \cite{R87}). 
We will also consider an extension of \cite{KOSY01}, in which a reduction 
from the $U_{q}(sl(2r+2)^{(1)})$ case to the $U_{q}(sp(2r)^{(1)})$ case was discussed.  
Besides these one will find new features that are not present in \cite{KS94-2,R87,KOSY01}. 
%%%%%%%%%%
\subsection{Reductions of Q-functions by automorphisms}
\label{sec:red}
We find that reductions on the  QQ-relations by the map $\sigma$ 
 (and some dualities among different superalgebras  \cite{Z97}) 
 produce QQ-relations (and from zeros of Q-functions,  
Bethe equations) and T-functions (and in particular, solutions of the T-systems)
 associated with algebras different from the original ones. 
The reductions here are basically accomplished by identifying the image of 
the Q-functions and the parameters $\{z_{a}\}$ by the map $\sigma$ 
with the original ones (up to overall factors and manipulations on the spectral parameter in some cases). 
Let ${\mathfrak D}$ be a subset of the sets $\Bm$ or $\Fm$ such that ${\mathfrak D}^{*}={\mathfrak D}$ and $|{\mathfrak D}|=2$, 
or ${\mathfrak D}=\emptyset$. 
Let us consider ``$gl(M|N)^{(2)}$ type reduction'' by $\sigma$
\footnote{It may be possible to generalize this by considering 
compositions of $\sigma$ 
and the $GL(M) \times GL(N)$-symmetry \cite{T11,GKLT10} of the QQ-relations. 
Here we consider only the simplest case.}
: 
\begin{align}
\begin{split}
&\sigma (\Qb_{I})=\Qb_{\mathfrak{I} \setminus I^{*}}=\Qb_{I}^{[\eta]} \quad \text{for} \quad I \subset {\mathfrak I}, 
\\
&
\sigma (z_{a})=z_{a^{*}}^{-1}=z_{a} \quad \text{for} \quad a \in {\mathfrak I} \setminus {\mathfrak D} ,
\\
&
\sigma (z_{a})=z_{a^{*}}^{-1}= - z_{a}=1  \quad  \text{or} \ -1  
 \quad \text{for} \quad a \in {\mathfrak D} ,
\end{split}
\label{reduction-sigma}
\end{align} 
where $\eta =0$, or $2\eta \in {\mathbb C}^{\times}$ is the common period
\footnote{Depending on the normalization of the Q-functions, a sign factor may appear $\Qb_{J}^{[2\eta]}= \pm \Qb_{J}$. 
We normalize the Q-functions so that the sign factor is always $1$.
Let  $\sigma $ be an automorphism of order $\kappa $ ($\sigma^{\kappa}=1$). 
In general, $\kappa \eta $ corresponds to the common period of the Q-functions in case $\eta \ne 0$.
Here we consider only the case $\kappa=2$.}
 of the Q-functions: $|\eta | $ is the minimal non-zero number such that 
 $\Qb_{J}^{[2\eta]}=\Qb_{J}$ for all $J \subset {\mathfrak I}$. 
 In particular, we have
 %
%\footnote{Eq. \eqref{Q=Q} represents a kind of `boson-fermion correspondence',
%which means that the Q-functions with a `bosonic' index and a `fermionic' index are equivalent. 
%This is different from the boson-fermion correspondence in soliton theory.}
% 
\begin{align}
&\Qb_{\Bm}=\Qb_{\Fm}^{[\eta]} , \label{Q=Q}
\\
&\Qb_{\mathfrak{I}}=\Qb_{\Bm,\Fm}=\Qb_{\emptyset}^{[\eta]} . 
%\label{Q=Q2}
\end{align}
We remark that 
$\sigma (I)=I$ holds if and only if $|I|=(M+N)/2$ and $I \cap I^{*}=\emptyset $.  This is possible only if both $M$ and $N$ 
are even numbers. For this index set $I$, the Q-function satisfies $\Qb_{I}^{[\eta]}=\Qb_{I}$. 
In case $\eta \ne 0$, 
this suggests a factorization $\Qb_{I}=\Qf_{I}\Qf_{I}^{[\eta]}=:\Qf^{2}_{I}$, where $\Qf_{I}^{[2\eta]}=\Qf_{I}$.  
In case the Q-function $\Qb_{I}$ has the form \eqref{Q-poly}, this means that 
\begin{align}
\Qb_{I}=\Qb_{I}(u)&=\prod_{j=1}^{n_{I}}(1-q^{-2u+2u^{I}_{j}})
\nonumber 
\\
&=\prod_{j=1}^{m_{I}}(1-q^{-2u+2v^{I}_{j}})(1+q^{-2u+2v^{I}_{j}})
=\prod_{j=1}^{m_{I}}(1-q^{-4u+4v^{I}_{j}}), 
 \label{Q-polytw1}
\end{align}
where 
\begin{gather}
\Qf_{I}=\Qf_{I}(u)=\prod_{j=1}^{m_{I}}(1-q^{-2u+2v^{I}_{j}}), 
\notag
\\
 n_{I}=2m_{I}, \qquad \{u_{j}\}_{j=1}^{n_{I}} = \{v_{j}\}_{j=1}^{m_{I}} \sqcup \{v_{j}+\eta \}_{j=1}^{m_{I}}, 
\qquad \eta =\frac{\pi i }{2 \log q } .
 \label{Q-polytw2}
\end{gather}
If $M$ or $N$ are odd, fixed points  $\mathfrak{f} \in {\mathfrak I}$  by $^{*}$ appear: $\mathfrak{f}^{*}=\mathfrak{f}$, 
 $\sigma(z_{\mathfrak{f}})=z_{\mathfrak{f}}^{-1}=z_{\mathfrak{f}}$. Thus we have $z_{\mathfrak{f}}=\pm 1$. 
 The minus sign $z_{\mathfrak{f}}=-1$ effectively changes the sign of $p_{\mathfrak{f}}$ from the grading of the superalgebra 
(see \eqref{boxes} and \eqref{tab-fund}), 
which induces a duality among superalgebras (correspondence between 
a twisted quantum affine (super)algebra and an untwisted quantum affine (super)algebra). 
 We will set $z_{\mathfrak{f}}=-1$ (resp. $z_{\mathfrak{f}}=1$) when we consider reductions to untwisted quantum affine superalgebras 
  (resp.\ twisted quantum affine superalgebras). 
In case we consider reductions to twisted quantum affine superalgebras, we assume $\eta \ne 0$. 
In case we consider reductions to quantum affine superalgebras of type A and B, we assume ${\mathfrak D} = \emptyset$ (regular reduction). 
  We need more reductions on Q-functions in addition to \eqref{reduction-sigma} for reduction to quantum affine superalgebras of type C and D, where we  assume
   ${\mathfrak D} \ne \emptyset$ (singular reduction).

In case $B=\Bm$, $F=\Fm$, 
$\Qb_{\Bm,\Fm \setminus J}=\Qb_{J^{*}}^{[\eta]}$ and
 $\Qb_{\Bm \setminus I,\Fm}=\Qb_{I^{*}}^{[\eta]}$ hold in 
\eqref{soQQb-1} and \eqref{soQQb-2}. Thus they reduce to 
\begin{multline}    
\Ts^{\Bm,\Fm}_{a,m} =   
\sum_{J \subset \Fm,  \atop
|J|=m}  
\frac{\prod_{j \in J}(-z_{j})^{a-m-M+N} \prod_{(b,j) \in \Bm \times J } (z_{b}-z_{j}) }
{\prod_{(i,j) \in (\Fm \setminus J) \times J  } (z_{i}-z_{j}) }
\Qb^{[a+\eta]}_{J^{*}}  
\Qb^{[-a+M-N]}_{J}  
\\
 \qquad \qquad  \text{for} \quad a\ge m+M-N,  
\label{re1}  
\end{multline}
\begin{multline}
\Ts^{\Bm,\Fm}_{a,m} =   
\sum_{I \subset \Bm, \atop |I|=a} 
\frac{ \prod_{i \in I} z_{i}^{m-a+M-N}
\prod_{(i,f) \in I \times \Fm } (z_{i}-z_{f})}
{\prod_{(i,j) \in I \times (\Bm \setminus I)  } (z_{i}-z_{j}) }  
\Qb^{[m+M-N]}_{I}  
\Qb^{[-m+\eta]}_{I^{*}}  
\quad 
\text{for} \quad  a \le m + M-N \label{re2}  .
\end{multline} 
In case $\mathfrak{D}=\emptyset$, one can show
\begin{align}
\Ts^{\emptyset ,\Fm [\eta]}_{a,N-m}&=\left(\prod_{f \in \Fm}(- z_{f})\right)^{a}\Ts^{\emptyset,\Fm}_{a,m}
\quad \text{for} \quad a \ge 0, \quad 0 \le m \le N,
\label{re1-du}
\\
\Ts^{\Bm,\emptyset [\eta]}_{M-a, m}&=\left(\prod_{b \in \Bm} z_{b} \right)^{m}\Ts^{\Bm,\emptyset}_{a,m}
\quad \text{for}  \quad 1 \le a \le M, \quad m \ge 0,
\label{re2-du}
\end{align}
where the prefactors of \eqref{re1-du} and \eqref{re2-du} take $1$ or $-1$. 
%%%
Let $\hat{\Ts}^{\Bm,\Fm}_{a,m}$ be an 
analytic continuation of the right hand side of \eqref{re1} with respect to $a$ and  that of \eqref{re2} with respect to $m$ 
to the whole complex plane.
One can show
\begin{align}
\hat{\Ts}^{\Bm,\Fm [\eta]}_{-a+M-N,m}&=\left(-\frac{\prod_{i \in \Fm}z_{i}}{\prod_{b \in \Bm} z_{b}}\right)^{m}\hat{\Ts}^{\Bm,\Fm}_{a,m}
\quad \text{for} 
\ a \in \mathbb{C} \ \text{and} \ \mathfrak{D} \cap \Fm =\emptyset  \ \text{for} \ \eqref{re1},
\label{re1-con}
\\
\hat{\Ts}^{\Bm,\Fm [\eta]}_{a,-m-M+N}&=\left((-1)^{N-M+a}\frac{\prod_{j \in \Bm}z_{j}}{\prod_{f \in \Fm} z_{f}}\right)^{a}\hat{\Ts}^{\Bm,\Fm}_{a,m}
\quad \text{for} 
\ m \in \mathbb{C} \ \text{and} \ \mathfrak{D} \cap \Bm =\emptyset  \ \text{for} \ \eqref{re2},
\label{re2-con}
\end{align}
where the prefactors of \eqref{re1-con} and \eqref{re2-con} take $1$ or $-1$. 
%%%%%%%%

%%%%%%%%%%%%%%%%%%
\subsection{Symmetric nesting path}
\label{sec:SN}
Let $I_{M+N}=(i_{1},i_{2},\dots,i_{M+N})$ be 
 any one of the permutations of $(1,2,\dots, M+N)$, and  
 $I_{a}=(i_{1},i_{2},\dots,i_{a})$ be the first $a$ elements of it, 
where $0 \le a \le M+N$. In particular, $I_{0}$ and $I_{M+N}$ coincide with $\emptyset $ and ${\mathfrak I}$ as sets,  
respectively.  
The set of the tuples $\{I_{a}\}_{a=0}^{M+N}$ is called the {\em nesting path} defined by the tuple $I_{M+N}$. 
For $1 \le a \le M+N-1$, we define $\widetilde{I}_{a}=(i_{1},i_{2},\dots,i_{a-1},i_{a+1})$. 
In this subsection, we consider a special class of nesting paths.  

Suppose $i_{k}^{*}=i_{M+N+1-k}$ holds for any $k \in {\mathfrak I}$. In this case, 
 $I_{M+N}=(i_{1},i_{2},\dots, i_{\frac{M+N}{2}},i_{\frac{M+N}{2}}^{*},\dots , i_{2}^{*}, i_{1}^{*})$ 
if both $M+N$ and  $MN$ are even, and 
$I_{M+N}=(i_{1},i_{2},\dots, i_{\frac{M+N-1}{2}}, {\mathfrak f}, i_{\frac{M+N-1}{2}}^{*},\dots , i_{2}^{*}, i_{1}^{*})$ 
if $M+N$ is odd and $MN$ is even (there is a fixed point ${\mathfrak f}:=i_{\frac{M+N+1}{2}}^{*}=i_{\frac{M+N+1}{2}}$; 
${\mathfrak f}=(M+1)/2$  if $M$ is odd, and ${\mathfrak f}=M+(N+1)/2$ if $N$ is odd).  
We may say that the nesting path is {\em symmetric} in the sense that 
the Q-functions along this nesting path are symmetric up to the half period:  
$\Qb_{I_{a}}^{[\eta]}=\Qb_{I_{M+N-a}}$ for any $a \in \{0,1,\dots M+N\}$. 
Note that this is impossible if $MN$ is odd since two fixed points $((M+1)/2)^{*}=(M+1)/2$, 
$(M+(N+1)/2)^{*}=M+(N+1)/2$ appear. In this case, we propose to 
consider{\em almost symmetric nesting path} defined by 
 $i_{k}^{*}=i_{M+N+1-k}$  for any $k \in {\mathfrak I}\setminus \{(M+N)/2,(M+N)/2+1\}$ and 
 $(\mathfrak{f}_{1},\mathfrak{f}_{2}):=
 (i_{\frac{M+N}{2}},i_{\frac{M+N}{2}+1})=((M+1)/2,M+(N+1)/2)$ or $(M+(N+1)/2,(M+1)/2)$, 
 namely $I_{M+N}=(i_{1},i_{2},\dots, i_{\frac{M+N-2}{2}},{\mathfrak f}_{1},{\mathfrak f}_{2} , i_{\frac{M+N-2}{2}}^{*},\dots , i_{2}^{*}, i_{1}^{*})$.  
Along this almost symmetric nesting path, we have   
$\Qb_{I_{a}}^{[\eta]}=\Qb_{I_{M+N-a}}$ for any $a \in \{0,1,\dots M+N\}\setminus \{(M+N)/2\}$, and 
$\Qb_{I_{\frac{M+N}{2}}}^{[\eta]}=\Qb_{I_{\frac{M+N-2}{2}},\mathfrak{f}_{1}}^{[\eta]}=\Qb_{I_{\frac{M+N-2}{2}},\mathfrak{f}_{2}}$. 
See Figures \ref{HasseQ1},\ref{HasseQ2},\ref{HasseQ3} for examples of symmetric nesting paths.  
We remark that $I_{a} \in \Am$ holds as set if $a \le (M+N)/2$ and $MN$ is even, or 
 if $a \le (M+N-2)/2$ and $MN$ is odd. 

The functions ${\mathcal   X}_{I_{a}}$ on any symmetric (or almost symmetric) nesting path can be expressed in terms of the Q-functions 
$\Qb_{I_{a}}$ for $a \le \frac{M+N}{2}$. 
In fact, \eqref{boxes} for $a >  \frac{M+N}{2} $ can be rewritten as  
%We also introduce $(K-1)$-tuples $ \hat{I}=(\gamma_{1},\gamma_{2},\dots,\gamma_{K-1})$. 
% and $ \check{\mathtt I}=(\gamma_{2},\gamma_{3},\dots,\gamma_{K})$.  
\begin{align}
{\mathcal   X}_{I_{M+N+1-a}}&=
z_{i_{a}^{*}}
\frac{\Qb_{I_{a-1}}^{[\sum_{j \in I_{a}}p_{j}+p_{i_{a}}+\eta]}
\Qb_{I_{a}}^{[\sum_{j \in I_{a}}p_{j}-2p_{i_{a}}+\eta ]}
}{
\Qb_{I_{a-1}}^{[\sum_{j \in I_{a}}p_{j}-p_{i_{a}}+\eta]}
\Qb_{I_{a}}^{[\sum_{j \in I_{a}}p_{j}+\eta ]}
} 
\quad \text{for} \quad a \le \frac{M+N}{2}.
\label{boxes3} 
\end{align}
Although the T-function \eqref{DVF-tab1} can be non-zero on the $[M,N]$-hook
 (see Figures \ref{MN-hookB}, \ref{MN-hookB0}, \ref{MN-hookD}, \ref{MN-hookCp}), 
 the main target domain after the reduction 
 is the $[r,s]$-hook, in which we have observations on the meaning (relationship with the labels of  representations, irreducibility) of T-functions 
  through the Bethe strap (see subsection \ref{sec:BS}). 
  The T-system (for tensor representations; cf. \cite{T99,T99-2}) will be defined on the $[r,s]$-hook 
  (or its extension). 
  We expect that the T-functions outside of the $[r,s]$-hook are described in terms of 
  the ones in the $[r,s]$-hook (see \cite{KOS95,T21} for the case $U_{q}(so(2r+1)^{(1)})$). 
 The main target domain of the Wronskian-type expression of the T-function \eqref{unnor-t1} 
 (for $(M,N)=(\mm,\nn))$) is also  $[r,s]$-hook after the reduction. 
  The identity \eqref{T=F} (with \eqref{TF-rel03}) assumes the QQ-relations 
  \eqref{QQb} and \eqref{QQf}, so further investigation is needed to see how far it holds when the consideration is restricted to the symmetric nesting paths. Note, however, that the identity 
   \eqref{T=F} always holds under the (super)character limit \eqref{ch-limit}. 
  After the reduction, $\zeta(\Ts^{\Bm,\Fm}_{\mu})$ gives a Weyl-type (super)character formula for  
   finite dimensional representations of quantum affine superalgebras (or super-Yangians). 
   This does not always give irreducible (super)characters, especially for type C or D superalgebras. 
   We expect that criteria for irreducibility and 
   cues for extracting irreducible components are suggested by the Bethe strap (see subsection \ref{sec:BS}).

The simple root system $\{\alpha_{a}=\epsilon_{i_{M+N+1-a}}-\epsilon_{i_{M+N-a}}=
\epsilon_{i^{*}_{a}}-\epsilon_{i^{*}_{a+1}} \}_{a=1}^{M+N-1}$ on the symmetric nesting path is symmetric in the sense that 
$\sigma(\alpha_{a})=-\epsilon_{i^{*}_{M+N+1-a}}+\epsilon_{i^{*}_{M+N-a}}=
-\epsilon_{i_{a}}+\epsilon_{i_{a+1}} =\alpha_{M+N-a}$, and thus the associated Dynkin diagram is symmetric with respect to the map $\sigma$. 

 Let $\mathfrak{W}$ be  a subgroup of the 
permutation group $ S(I_{M+N})=S({\mathfrak I})$, 
which preserves the entire set of symmetric nesting paths. 
$\mathfrak{W}$ preserves the whole set of the symmetric Dynkin diagrams of $gl(M|N)$. 
We will discuss the invariance of T-functions under this ($\mathfrak{W}$-symmetry). 
%%%%%
\begin{figure}
\centering
\begin{tikzpicture}[x=8mm,y=8mm]
  \node (123) at (0,6) {$\Qb_{123}$};
  \node (12) at (-2.5,4) {$\Qb_{12}$};
  \node (13) at (0,4) {$\Qb_{13}$};
  \node (23) at (2.5,4) {$\Qb_{23}$};
  \node (1) at (-2.5,2) {$\Qb_{1}$};
  \node (2) at (0,2) {$\Qb_{2}$};
  \node (3) at (2.5,2) {$\Qb_{3} $};
  \node (em) at (0,0) {$\Qb_{\emptyset}$};
  \node[below of = em]  {$gl(3|0)$ or $gl(0|3)$};
  \draw[line width=1.5pt] (em) -- (1) -- (12) -- (123) -- (23) -- (3) -- (em);
  \draw (1) -- (13) -- (123)
  (3) -- (13)
  (em) -- (2);
  \draw[preaction={draw=white, -,line width=7pt}] (12) -- (2) -- (23);
\end{tikzpicture}
%%%
\begin{tikzpicture}[x=8mm,y=8mm]
  \node (123) at (0,6) {$\Qb_{123}$};
  \node (12) at (-2.5,4) {$\Qb_{12}$};
  \node (13) at (0,4) {$\Qb_{13}$};
  \node (23) at (2.5,4) {$\Qb_{23}$};
  \node (1) at (-2.5,2) {$\Qb_{1}$};
  \node (2) at (0,2) {$\Qb_{2}$};
  \node (3) at (2.5,2) {$\Qb_{3} $};
  \node (em) at (0,0) {$\Qb_{\emptyset}$};
  \node[below of = em]  {$gl(2|1)$};
  \draw[line width=1.5pt] (em) -- (1) -- (13) -- (123) -- (23) 
 (2) -- (em);
  \draw (1) -- (12) -- (123)
  (3) -- (13)
  (em) -- (3) -- (23);
  \draw[preaction={draw=white, -,line width=7pt}] (12) -- (2);
\draw[line width=1.5pt,preaction={draw=white, -,line width=7pt}] (2) -- (23);
\end{tikzpicture}
%%%
\begin{tikzpicture}[x=8mm,y=8mm]
  \node (123) at (0,6) {$\Qb_{123}$};
  \node (12) at (-2.5,4) {$\Qb_{12}$};
  \node (13) at (0,4) {$\Qb_{13}$};
  \node (23) at (2.5,4) {$\Qb_{23}$};
  \node (1) at (-2.5,2) {$\Qb_{1}$};
  \node (2) at (0,2) {$\Qb_{2}$};
  \node (3) at (2.5,2) {$\Qb_{3} $};
  \node (em) at (0,0) {$\Qb_{\emptyset}$};
  \node[below of = em]  {$gl(1|2)$};
  \draw[line width=1.5pt] (em) -- (3) -- (13) -- (123) -- (12) 
 (2) -- (em);
  \draw (3) -- (23) -- (123)
  (1) -- (13)
  (em) -- (1) -- (12);
  \draw[preaction={draw=white, -,line width=7pt}] (23) -- (2);
\draw[line width=1.5pt,preaction={draw=white, -,line width=7pt}] (2) -- (12);
\end{tikzpicture}
\caption{Hasse diagrams for Q-functions: The thick lines denote symmetric nesting paths. 
There are two symmetric nesting paths for each algebra.}
\label{HasseQ1}
\end{figure}
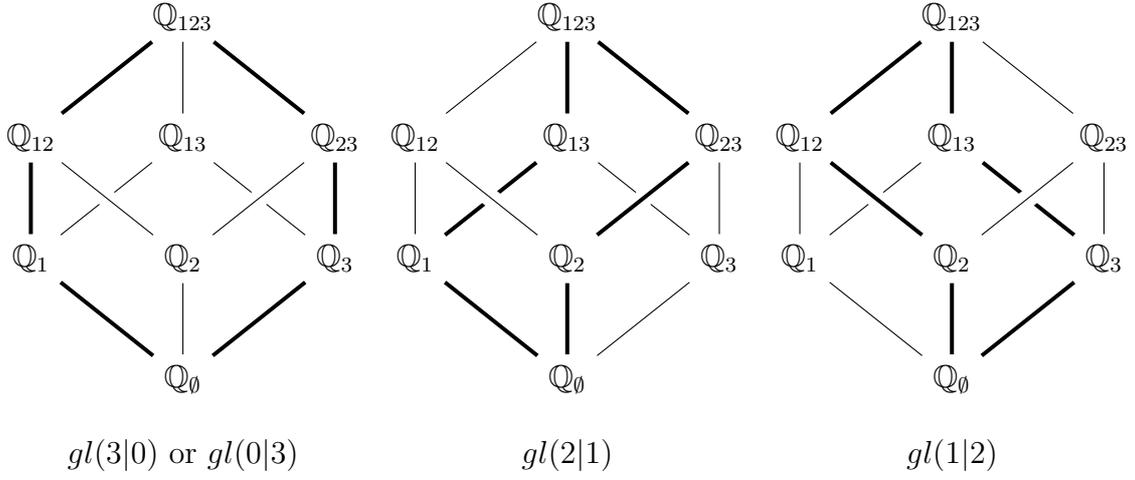
%%%
%%%
\begin{figure}
\centering
\begin{tikzpicture}[x=10mm,y=10mm]
\node (1234) at (0,8) {$\Qb_{1234}$};
\node (123) at (-3.9,6) {$\Qb_{123}$};
\node (124) at (-1.3,6) {$\Qb_{124}$};
\node (134) at (1.3,6) {$\Qb_{134} $};
\node (234) at (3.9,6) {$\Qb_{234} $};
\node (12) at (-5,4) {$\Qb_{12}$};
\node (13) at (-3,4) {$\Qb_{13}$};
\node (14) at (-1,4) {$\Qb_{14}$};
\node (23) at (1,4) {$\Qb_{23}$};
\node (24) at (3,4) {$\Qb_{24}$};
\node (34) at (5,4) {$\Qb_{34}$};
\node (1) at (-3.9,2) {$\Qb_{1}$};
\node (2) at (-1.3,2) {$\Qb_{2}$};
\node (3) at (1.3,2) {$\Qb_{3} $};
\node (4) at (3.9,2) {$\Qb_{4} $};
\node (em) at (0,0) {$\Qb_{\emptyset}$};
  \node[below of = em]  {$gl(4|0)$ or $gl(0|4)$};
  \draw[line width=1.5pt] (em) -- (1) -- (12) -- (123) -- (1234) 
 (1) -- (13)-- (123);
  \draw[line width=1.5pt,densely dotted] (em) -- (2) -- (12) -- (124) -- (1234) 
 (2) -- (24)-- (124);
  \draw[line width=1.5pt,dashed] (em) -- (3) -- (13) -- (134) -- (1234) 
 (3) -- (34)-- (134);
  \draw[line width=1.5pt,loosely dotted] (em) -- (4) -- (24) -- (234) -- (1234) 
 (4) -- (34)-- (234);
\draw (1) -- (14) -- (124)
 (2) -- (23) -- (123) 
(3) -- (23) -- (234) 
 (4) -- (14) -- (134);
%\draw[preaction={draw=white, -,line width=7pt}] (23) -- (2);
%\draw[line width=1.5pt,preaction={draw=white, -,line width=7pt}] (2) -- (12);
\end{tikzpicture}
\caption{Hasse diagrams for Q-functions: The thick or dotted lines denote symmetric nesting paths. 
There are eight symmetric nesting paths for each algebra.}
\label{HasseQ2}
\end{figure}
%%%
\begin{figure}
\centering
\begin{tikzpicture}[x=10mm,y=10mm]
\node (1234) at (0,8) {$\Qb_{1234}$};
\node (123) at (-3.9,6) {$\Qb_{123}$};
\node (124) at (-1.3,6) {$\Qb_{124}$};
\node (134) at (1.3,6) {$\Qb_{134} $};
\node (234) at (3.9,6) {$\Qb_{234} $};
\node (12) at (-5,4) {$\Qb_{12}$};
\node (13) at (-3,4) {$\Qb_{13}$};
\node (14) at (-1,4) {$\Qb_{14}$};
\node (23) at (1,4) {$\Qb_{23}$};
\node (24) at (3,4) {$\Qb_{24}$};
\node (34) at (5,4) {$\Qb_{34}$};
\node (1) at (-3.9,2) {$\Qb_{1}$};
\node (2) at (-1.3,2) {$\Qb_{2}$};
\node (3) at (1.3,2) {$\Qb_{3} $};
\node (4) at (3.9,2) {$\Qb_{4} $};
\node (em) at (0,0) {$\Qb_{\emptyset}$};
  \node[below of = em]  {$gl(2|2)$};
  \draw[line width=1.5pt] (em) -- (1) -- (13) -- (134) -- (1234) 
 (1) -- (14)-- (134);
  \draw[line width=1.5pt,densely dotted] (em) -- (2) -- (23) -- (234) -- (1234) 
 (2) -- (24)-- (234);
  \draw[line width=1.5pt,dashed] (em) -- (3) -- (13) -- (123) -- (1234) 
 (3) -- (23)-- (123);
  \draw[line width=1.5pt,loosely dotted] (em) -- (4) -- (14) -- (124) -- (1234) 
 (4) -- (24)-- (124);
\draw (1) -- (12) -- (123)
 (2) -- (12) -- (124) 
(3) -- (34) -- (134) 
 (4) -- (34) -- (234);
%\draw[preaction={draw=white, -,line width=7pt}] (23) -- (2);
%\draw[line width=1.5pt,preaction={draw=white, -,line width=7pt}] (2) -- (12);
\end{tikzpicture}
\caption{Hasse diagrams for Q-functions: The thick or dotted lines denote symmetric nesting paths. 
There are eight symmetric nesting paths.}
\label{HasseQ3}
\end{figure}
%%%%%%%%%%%%%%%%%%%%%%%%%%%%%%%%%%  
\subsection{Regular reductions}
\label{sec:RR}
In this subsection, we consider reductions for the case $\mathfrak{D}=\emptyset $. 
In this case the resultant Bethe ansatz equations are always reductions of 
the ones for $U_{q}(gl(M|N)^{(1)})$. 
%%%%%%
\subsubsection{$U_{q}(osp(2r+1|2s)^{(1)})$ case}
We assume $r,s \in \mathbb{Z}_{\ge 0}$, $r+s \ge 1$, and set
\begin{multline}
(M,N)=(2r,2s+1), \quad \Bm=\{1,2,\dots , 2r\}, \quad  \Fm=\{2r+1,2r+2,\dots ,2r+2s+1\}, 
\\
\mathfrak{D}=\emptyset, \quad \eta=0, \quad  
z_{2r+s+1}=-1 .
\label{zz-b}
\end{multline}
In particular for $s=0$,  this reduces to the case  $U_{q}(so(2r+1)^{(1)})$, which is already 
explained in \cite{T21}. 

%%%%%%%%%%%
\paragraph{QQ-relations}
For a symmetric nesting path defined by 
$I_{2r+2s+1}=(i_{1},i_{2},\dots, i_{r+s}, 2r+s+1 , i_{r+s}^{*} \dots , i_{2}^{*}, i_{1}^{*})$, 
 the QQ-relations \eqref{QQb} reduce to
\begin{align}
& (z_{i_{a}}-z_{i_{a+1}})\Qb_{I_{a-1}}\Qb_{I_{a+1}}
=z_{i_{a}}\Qb_{I_{a}}^{[p_{i_{a}}]}
\Qb_{\widetilde{I}_{a}}^{[-p_{i_{a}}]}-
z_{i_{a+1}}\Qb_{I_{a}}^{[-p_{i_{a}}]}
\Qb_{\widetilde{I}_{a}}^{[p_{i_{a}}]}
\nonumber \\
&
\qquad \text{for} \quad a \in \{1,2,\dots , r+s-1\},   \quad 
p_{i_{a}}=p_{i_{a+1}}, 
\label{QQb1}  \\
& (z_{i_{a}}-z_{i_{a+1}})\Qb_{I_{a}}\Qb_{\widetilde{I}_{a}}
=z_{i_{a}}\Qb_{I_{a-1}}^{[-p_{i_{a}}]}
\Qb_{I_{a+1}}^{[p_{i_{a}}]}-
z_{i_{a+1}}\Qb_{I_{a-1}}^{[p_{i_{a}}]}
\Qb_{I_{a+1}}^{[-p_{i_{a}}]}
\nonumber \\
&
\qquad \text{for} \quad a \in \{1,2,\dots , r+s-1\},   \quad 
p_{i_{a}}=-p_{i_{a+1}}, 
\label{QQb2}  \\
& (z_{i_{r+s}}+1)\Qb_{I_{r+s-1}}\Qb_{I_{r+s}}
=z_{i_{r+s}}\Qb_{I_{r+s}}^{[-1]}
\Qb_{\widetilde{I}_{r+s}}^{[1]}
+\Qb_{I_{r+s}}^{[1]}
\Qb_{\widetilde{I}_{r+s}}^{[-1]}
\quad \text{for} \quad 
p_{i_{r+s}}=-1, 
\label{QQb3}  
  \\
& (z_{i_{r+s}}+1)\Qb_{I_{r+s}}\Qb_{\widetilde{I}_{r+s}}
=z_{i_{r+s}}\Qb_{I_{r+s-1}}^{[-1]}
\Qb_{I_{r+s}}^{[1]}
+\Qb_{I_{r+s-1}}^{[1]}
\Qb_{I_{r+s}}^{[-1]}
\quad \text{for} \quad 
p_{i_{r+s}}=1,
\label{QQb4}  
\end{align}
where $\widetilde{I}_{a}=(i_{1},i_{2},\dots, i_{a-1},i_{a+1})$, $i_{r+s+1}=2r+s+1$.  
Eqs.\ \eqref{QQb1} and  \eqref{QQb3} are reductions of \eqref{QQb} for 
$I=I_{a-1}$, $(i,j)=(i_{a},i_{a+1})$ for $1 \le a \le r+s-1$  and $a=r+s$, respectively
\footnote{The QQ-relations for $a>r+s$ reduce to the ones for $a \le r+s$. 
For example,  \eqref{QQb3} is also a reduction of \eqref{QQb} for 
$I=I_{r+s}$, $(i,j)=(2r+s+1,i_{r+s}^{*})$. 
% since $\Qb_{I_{a}}=\Qb_{I_{2r+2s+1-a}}$ holds on the symmetric nesting path.
 }.
Eqs.\ \eqref{QQb2} and  \eqref{QQb4} are reductions of \eqref{QQf} for 
$I=I_{a-1}$, $(i,j)=(i_{a},i_{a+1})$ for $1 \le a \le r+s-1$  and $a=r+s$, respectively. 
Instead of \eqref{QQb4}, one may use 
\begin{align}
& (z_{i_{r+s}}-1)\Qb^{[1]}_{I_{r+s-1}}\Qb_{I_{r+s-1}}
=z_{i_{r+s}}\Qb_{I_{r+s}}^{[1]}
\Qb_{\breve{I}_{r+s}}
-\Qb_{I_{r+s}}
\Qb_{\breve{I}_{r+s}}^{[1]}
\quad \text{for} \quad 
p_{i_{r+s}}=1,
\label{QQb5}  
\end{align}
where $\breve{I}_{r+s}=(i_{1},i_{2},\dots , i_{r+s-1},i^{*}_{r+s})$. 
One can derive \eqref{QQb5} in the same way as explained in [section 3.2, \cite{T21}] for $s=0$ case. 
%%%%%%%%%%%%
We will use the following QQ-relations:
\begin{align}
& (z_{i_{r+s}}-z_{i_{r+s}}^{-1})\Qb_{I_{r+s-1}}\Qb_{\widetilde{I}_{r+s}}
=z_{i_{r+s}} \Qb_{I_{r+s}}^{[1]} \Qb_{\breve{I}_{r+s}}^{[-1]} - z^{-1}_{i_{r+s}} \Qb_{I_{r+s}}^{[-1]} \Qb_{\breve{I}_{r+s}}^{[1]} 
\quad \text{for} \quad 
p_{i_{r+s}}=1,
 \label{QQb5-2}
\\
& (z_{i_{r+s}}^{-1}+1)\Qb_{\breve{I}_{r+s}}\Qb_{\widetilde{I}_{r+s}}
=z_{i_{r+s}}^{-1} \Qb_{I_{r+s-1}}^{[-1]} \Qb_{\breve{I}_{r+s}}^{[1]} + \Qb_{I_{r+s-1}}^{[1]} \Qb_{\breve{I}_{r+s}}^{[-1]}
\quad \text{for} \quad 
p_{i_{r+s}}=1.
 \label{QQb5-3}
\end{align}
Eqs. \eqref{QQb5-2} and  \eqref{QQb5-3} are reductions of  \eqref{QQb} for $I=I_{r+s-1}$, $(i,j)=(i_{r+s},i^{*}_{r+s})$
 and  \eqref{QQf} for $I=I_{r+s-1}$, $(i,j)=(i^{*}_{r+s},2r+s+1)$, respectively.  
Now we prove \eqref{QQb5} step by step as follows. 
\begin{multline}
[\text{left hand side of \eqref{QQb5}}] \times 
(z_{i_{r+s}}-z_{i_{r+s}}^{-1})\Qb_{\widetilde{I}_{r+s}} =
\\
=(z_{i_{r+s}}-1)\Qb^{[1]}_{I_{r+s-1}}
\underbrace{(z_{i_{r+s}}-z_{i_{r+s}}^{-1}) \Qb_{I_{r+s-1}}\Qb_{\widetilde{I}_{r+s}}}_{\text{apply \eqref{QQb5-2}}}
  \\
=(z_{i_{r+1}}-1)\Qb^{[1]}_{I_{r+s-1}}
(z_{i_{r+s}} \Qb_{I_{r+s}}^{[1]} \Qb_{\breve{I}_{r+s}}^{[-1]} - z^{-1}_{i_{r+s}} \Qb_{I_{r+s}}^{[-1]} \Qb_{\breve{I}_{r+s}}^{[1]}), 
\label{proofQQb}
\end{multline}
\begin{multline}
[\text{right hand side of \eqref{QQb5}}] \times 
(z_{i_{r+s}}-z_{i_{r+s}}^{-1})\Qb_{\widetilde{I}_{r+s}}=
\\
=(z_{i_{r+s}}-z_{i_{r+s}}^{-1})
 (z_{i_{r+s}} \Qb_{I_{r+s}}^{[1]} 
\underbrace{ \Qb_{\breve{I}_{r+s}}\Qb_{\widetilde{I}_{r+s}}}_{\text{apply \eqref{QQb5-3}}}
-
\underbrace{\Qb_{I_{r+s}} \Qb_{\widetilde{I}_{r+s}}}_{\text{apply \eqref{QQb4}}} \Qb_{\breve{I}_{r+s}}^{[1]})
  \\
=(z_{i_{r+s}}-z_{i_{r+s}}^{-1})
 \bigl(z_{i_{r+s}} \Qb_{I_{r+s}}^{[1]} 
(z_{i_{r+s}}^{-1}+1)^{-1} (z_{i_{r+s}}^{-1} \Qb_{I_{r+s-1}}^{[-1]} \Qb_{\breve{I}_{r+s}}^{[1]} +
 \Qb_{I_{r+s-1}}^{[1]}\Qb_{\breve{I}_{r+s}}^{[-1]})
\\
-
(z_{i_{r+s}}+1)^{-1}
(z_{i_{r+s}} \Qb_{I_{r+s-1}}^{[-1]} \Qb_{I_{r+s}}^{[1]} +
 \Qb_{I_{r+s-1}}^{[1]}\Qb_{I_{r+s}}^{[-1]})
\Qb_{\breve{I}_{r+s}}^{[1]}\bigr)
\\
=[\text{right hand side of \eqref{proofQQb}}] .
\end{multline}
Hence \eqref{QQb5} holds since 
$(z_{i_{r+s}}-z_{i_{r+s}}^{-1})\Qb_{\widetilde{I}_{r+s}}$ is not identically zero.  
One can also show that 
\eqref{QQb4} and \eqref{QQb5-3} follow from \eqref{QQb5} and \eqref{QQb5-2}. 
 Eqs.\ \eqref{QQb1} and \eqref{QQb5} for 
the case $s=0$, $(i_{1},i_{2},\dots , i_{r})=(1,2,\dots , r)$ correspond to [eqs.\ (6.11) and (6.14) in \cite{DDMST06}]. 

%%%%%%
\paragraph{T-functions and Bethe ansatz equations}
Eq.\ \eqref{boxes} reduces to 
\begin{align}
\begin{split}
{\mathcal   X}_{I_{a}}&=
z_{i_{a}}
\frac{\Qb_{I_{a-1}}^{[2r-2s-1-\sum_{j \in I_{a}}p_{j}-p_{i_{a}}]}
\Qb_{I_{a}}^{[2r-2s-1-\sum_{j \in I_{a}}p_{j}+2p_{i_{a}}]}
}{
\Qb_{I_{a-1}}^{[2r-2s-1-\sum_{j \in I_{a}}p_{j}+p_{i_{a}}]}
\Qb_{I_{a}}^{[2r-2s-1-\sum_{j \in I_{a}}p_{j}]}
} 
\quad \text{for} \quad 1 \le a \le r+s ,
\\
{\mathcal   X}_{I_{r+s+1}}&=
-
\frac{\Qb_{I_{r+s}}^{[r-s+1]}
\Qb_{I_{r+s}}^{[r-s-2]}
}{
\Qb_{I_{r+s}}^{[r-s-1]}
\Qb_{I_{r+s}}^{[r-s]}
}  ,
\\
{\mathcal   X}_{I_{2r+2s+2-a}}&=
z_{i_{a}}^{-1}
\frac{\Qb_{I_{a-1}}^{[\sum_{j \in I_{a}}p_{j}+p_{i_{a}}]}
\Qb_{I_{a}}^{[\sum_{j \in I_{a}}p_{j}-2p_{i_{a}}]}
}{
\Qb_{I_{a-1}}^{[\sum_{j \in I_{a}}p_{j}-p_{i_{a}}]}
\Qb_{I_{a}}^{[\sum_{j \in I_{a}}p_{j} ]}
} 
\quad \text{for} \quad 1 \le a \le r+s .
\end{split}
\label{boxes-b} 
\end{align}
The T-function \eqref{tab-fund} reduces to 
\begin{align}
{\mathsf F}_{(1)}^{I_{2r+2s+1}}=
\Qb_{\emptyset}^{[2r-2s-1]}
\Qb_{\emptyset}
\left(\sum_{a=1}^{r+s}p_{i_{a}}({\mathcal   X}_{I_{a}}+{\mathcal   X}_{I_{2r+2s+2-a}}) -{\mathcal   X}_{I_{r+s+1}} \right).
 \label{tab-fund-b}
\end{align}
The pole-free condition of the T-function \eqref{tab-fund-b} 
namely,  
\begin{align}
&{\rm Res}_{u=u_{k}^{I_{a}}+\sum_{j \in I_{a}}p_{j}-2r+2s+1}
(p_{i_{a}}{\mathcal   X}_{I_{a}}(u)+p_{i_{a+1}}{\mathcal   X}_{I_{a+1}}(u))=0 ,
\label{chancels-b1}
\\
&{\rm Res}_{u=u_{k}^{I_{a}}-\sum_{j \in I_{a}}p_{j}}
(p_{i_{a}}{\mathcal   X}_{I_{2r+2s+2-a}}(u)+p_{i_{a+1}}{\mathcal   X}_{I_{2r+2s+1-a}}(u))=0 
\label{chancels-b2}
\\
&\qquad \qquad \text{for} \quad k\in \{1,2,\dots, n_{I_{a}}\} \quad \text{and} \quad a \in \{1,2,\dots, r+s \} 
\nonumber 
\end{align}
produces the following Bethe ansatz equations: 
\begin{align}
\begin{split}
 -1&=\frac{p_{i_{a}}z_{i_{a}}}{p_{i_{a+1}}z_{i_{a+1}}}
\frac{\Qb_{I_{a-1}}(u_{k}^{I_{a}}-p_{i_{a}})
\Qb_{I_{a}}(u_{k}^{I_{a}}+2p_{i_{a}})
\Qb_{I_{a+1}}(u_{k}^{I_{a}}-p_{i_{a+1}})} 
{\Qb_{I_{a-1}}(u_{k}^{I_{a}}+p_{i_{a}})
\Qb_{I_{a}}(u_{k}^{I_{a}}-2p_{i_{a+1}})
\Qb_{I_{a+1}}(u_{k}^{I_{a}}+p_{i_{a+1}})}
\\
& \hspace{50pt} \text{for} \quad k\in \{1,2,\dots, n_{I_{a}}\} \quad \text{and} \quad a \in \{1,2,\dots, r+s-1 \} ,
\\
 -1&=p_{i_{r+s}}z_{i_{r+s}}
\frac{\Qb_{I_{r+s-1}}(u_{k}^{I_{r+s}}-p_{i_{r+s}})
\Qb_{I_{r+s}}(u_{k}^{I_{r+s}}+2p_{i_{r+s}})
\Qb_{I_{r+s}}(u_{k}^{I_{r+s}}+1)} 
{\Qb_{I_{r+s-1}}(u_{k}^{I_{r+s}}+p_{i_{r+s}})
\Qb_{I_{r+s}}(u_{k}^{I_{r+s}}+2)
\Qb_{I_{r+s}}(u_{k}^{I_{r+s}}-1)}
\\
&\hspace{50pt}  \text{for} \quad k\in \{1,2,\dots, n_{I_{r+s}}\}  .
\end{split}
\label{BAEb}
\end{align}
This is a reduction of \eqref{BAE} on the symmetric nesting path. 
Note that \eqref{chancels-b1} and \eqref{chancels-b2} produce the 
same Bethe equations.  
Eqs.\ \eqref{tab-fund-b} and \eqref{BAEb} agree with  the known results by algebraic Bethe ansatz \cite{GM04} in case $i_{k} \in \Fm$ for $1 \le k \le s$ 
and $i_{k} \in \Bm$ for $s+1 \le k \le r+s$. 
One can also derive the Bethe ansatz equations \eqref{BAEb} from the QQ-relations \eqref{QQb1}-\eqref{QQb4} 
by considering the zeros of the Q-functions. One can also use  \eqref{QQb5} instead of \eqref{QQb4}.
The tableaux sum expression of the T-function \eqref{DVF-tab1} reproduces [eq.\ (3.38), \cite{T99}]
\footnote{
The functions $\Qb_{\emptyset}$, 
$\Qb_{I_{a}}$, 
$\Qb_{\emptyset}^{[2r-2s-1]}\Qb_{\emptyset}{\mathcal   X}_{I_{a}}$, 
$-\Qb_{\emptyset}^{[2r-2s-1]}\Qb_{\emptyset}{\mathcal   X}_{I_{r+s+1}}$, and 
$\Qb_{\emptyset}^{[2r-2s-1]}\Qb_{\emptyset}{\mathcal   X}_{I_{2r+2s+2-a}}$ 
correspond to $\phi(u)$, $Q_{a}(u)$, $\boxed{\overline{a}}_{u}$, $\boxed{0}_{u}$ and $\boxed{a}_{u}$ in [eq.\ (3.13), \cite{T99}], 
where $1 \le a \le r+s$. 
To be precise, T-functions in \cite{T99} do not contain boundary twist parameters explicitly. 
The boundary twist parameters do not affect the rule of the tableaux sum, and thus 
 can be recovered easily. 
}
under the reduction. 
Moreover,  $\mathsf{T}^{\Bm,\Fm}_{\mu}$ (from \eqref{unnor-t1}) 
and its (super)character limit $\zeta(\mathsf{T}^{\Bm,\Fm}_{\mu})$ 
give a Wronskian expression of the T-function and 
a Weyl-type supercharacter formula respectively after the reduction. 
The Young diagram $\mu$ is related to the labels of the representation through 
\eqref{HW-B}-\eqref{YW-B}. 
%%%%%%%%%%%%%%%%%%

The generating functions \eqref{gene1} and \eqref{gene2} reduce to  
\begin{align}
{\mathbf W}_{I_{2r+2s+1}}({\mathbf X})&=
\overrightarrow{\prod_{a=1}^{r+s}} (1-{\mathcal X}_{I_{2r+2s+2-a}}{\mathbf X})^{-p_{i_{a}}} 
 (1-{\mathcal X}_{I_{r+s+1}}{\mathbf X} )
\overleftarrow{\prod_{a=1}^{r+s}} (1-{\mathcal X}_{I_{a}}{\mathbf X})^{-p_{i_{a}}} 
\nonumber \\
&= \sum_{a=0}^{\infty} {\mathcal F}_{(a)}^{I_{2r+2s+1}
[a -1]}{\mathbf X}^{a }, 
\label{gene1B}
\\
{\mathbf W}_{I_{2r+2s+1}}({\mathbf X})^{-1}&=
\overrightarrow{\prod_{a=1}^{r+s}} (1-{\mathcal X}_{I_{a}}{\mathbf X})^{p_{i_{a}}} 
 (1-{\mathcal X}_{I_{r+s+1}}{\mathbf X} )^{-1}
\overleftarrow{\prod_{a=1}^{r+s}} (1-{\mathcal X}_{I_{2r+2s+2-a}}{\mathbf X})^{p_{i_{a}}} 
\nonumber \\
&= \sum_{a=0}^{\infty}(-1)^{a} {\mathcal F}_{(1^{a})}^{I_{2r+2s+1}
[a -1]}{\mathbf X}^{a }.
\label{gene2B}
\end{align}
In this way, we recover [eqs. (B.2) and (B.1) in \cite{T99}] (see also [eqs. (2.7a) and (2.7b) in \cite{KOS95}] for the case $s=0$)
\footnote{In \cite{T99}, we considered only the formulas for the distinguished Dynkin diagram, while the formulas here 
are the ones for general Dynkin diagrams.}. 
Baxter type equations follow from the kernels of  \eqref{gene1B} and \eqref{gene2B}, which are 
reductions of \eqref{Bax3} and \eqref{Bax4}. 
%%%%%%%%

The relations \eqref{re1-con} and \eqref{re2-con} reduce to 
\begin{align}
\hat{\Ts}^{\Bm,\Fm}_{-a+2r-2s-1,m}&=\hat{\Ts}^{\Bm,\Fm}_{a,m}
\quad \text{for any} \ a \in \mathbb{C} \quad \text{for} \quad \eqref{re1},
\label{re1-con-b}
\\
\hat{\Ts}^{\Bm,\Fm}_{a,-m-2r+2s+1}&=(-1)^{a} \hat{\Ts}^{\Bm,\Fm}_{a,m}
\quad \text{for any} \ m \in \mathbb{C} \quad \text{for} \quad \eqref{re2}.
\label{re2-con-b}
\end{align}
We remark that \eqref{re2-con-b} for $a=1$ and $s=0$ corresponds to the Yangian $Y(so(2r+1))$ case of [Proposition 7.58 in \cite{FKT21}]. 
Let us write down $a=1$ case of \eqref{re2}. 
\begin{align}
\Ts^{\Bm,\Fm}_{1,m} &=   
\sum_{i =1}^{2r} 
\frac{ z_{i}^{m-2+2r-2s}
\prod_{f =2r+1 }^{2r+2s+1} (z_{i}-z_{f})}
{\prod_{j=1  \atop j \ne i}^{2r} (z_{i}-z_{j}) }  
\Qb^{[m+2r-2s-1]}_{i}  
\Qb^{[-m]}_{i^{*}}  
\nonumber \\
&=   
\sum_{i =1}^{r}
\left(
\chi_{i}^{+} 
\Qb^{[m+2r-2s-1]}_{i}  
\Qb^{[-m]}_{i^{*}}  
+
\chi_{i^{*}}^{+} 
\Qb^{[m+2r-2s-1]}_{i^{*}}  
\Qb^{[-m]}_{i}  
\right) 
\quad 
\text{for} \quad  2s-2r+2 \le m  \label{re2a=1}  .
\end{align} 
where the character parts are given by
\begin{align}
\chi_{i}^{+} &=
\frac{ z_{i}^{m-2+2r-2s}
(z_{i}+1)\prod_{f =2r+1 }^{2r+s} (z_{i}-z_{f})(z_{i}-z_{f}^{-1})}
{\prod_{j=1}^{i-1} (z_{i}-z_{j}) \prod_{j=i+1}^{r} (z_{i}-z_{j})
\prod_{j=1}^{r} (z_{i}-z_{j}^{-1})} 
\nonumber
\\
&=
\frac{(-1)^{i-1} z_{i}^{m+i-1}\prod_{j =1 }^{i-1}z_{j}^{-1}
\prod_{f =2r+1 }^{2r+s} (1-\frac{z_{f}}{z_{i}})(1-\frac{1}{z_{i}z_{f}})}
{(1-\frac{1}{z_{i}})\prod_{j=1}^{i-1} (1-\frac{z_{i}}{z_{j}}) \prod_{j=i+1}^{r} (1-\frac{z_{j}}{z_{i}})
\prod_{j=1 \atop j \ne i}^{r} (1-\frac{1}{z_{i}z_{j}})} 
\quad \text{for} \quad 1 \le i \le r,
\label{chp1}
\\
\chi_{i^{*}}^{+} &=
\frac{(-1)^{i} z_{i}^{-m+i-2r+2s}\prod_{j =1 }^{i-1}z_{j}^{-1}
\prod_{f =2r+1 }^{2r+s} (1-\frac{z_{f}}{z_{i}})(1-\frac{1}{z_{i}z_{f}})}
{(1-\frac{1}{z_{i}})\prod_{j=1}^{i-1} (1-\frac{z_{i}}{z_{j}}) \prod_{j=i+1}^{r} (1-\frac{z_{j}}{z_{i}})
\prod_{j=1 \atop j \ne i}^{r} (1-\frac{1}{z_{i}z_{j}})} 
\quad \text{for} \quad 1 \le i \le r.
\label{chp2}
\end{align}
We remark that \eqref{chp1} and \eqref{chp2} for $s=0$ coincide with 
 the Yangian $Y(so(2r+1))$ case of [eq.\ (9.25) in \cite{FKT21}] and that \eqref{re2a=1} for $s=0$, 
 which is [eq.\ (3.26) for $a=1$ in \cite{T21}] (up to an overall factor), corresponds
\footnote{Compare $\Ts^{\Bm,\Fm [-r-\frac{1}{2}]}_{1,m}$ for $s=0$ with [eq. (9.28) in \cite{FKT21}]. 
In our convention, the unit of shift of the spectral parameter is twice as large as theirs. Their parameters $\tau_{j}$ 
correspond to our parameters $z_{j}$. 
The sign factors $(-1)^{i-1}$ and $(-1)^{i}$ are included in the character parts \eqref{chp1} and \eqref{chp2}. If we understand correctly, \cite{FKT21} mainly focuses on specific
 representations of the Yangians $Y(\mathfrak{g})$ that are lift of those of 
 $\mathfrak{g}=B_{r},C_{r},D_{r}$. Note that the evaluation map is not available for general representations for these cases. Thus, it will be an interesting problem to apply their method to evaluation representations of 
 $Y(gl(M|N))$ (or $U_{q}(gl(M|N)^{(1)})$) and investigate their reductions.}
  to [eq. (9.28) in \cite{FKT21}]. 
%%%%%%%%%%%%%%%%

\paragraph{ $\mathfrak{W}$-symmetry}
%Consider a permutation $\tau_{ab} \in S(I_{M+N})=S(I_{2r+2s+1})=S({\mathfrak I})$ 
%such that $\tau_{ab}(a)=b, \tau_{ab}(b)=a$ and $\tau_{ab}(c)=c$ 
%for $c \ne a,b$,  
%for a fixed $a,b \in \{1,2,\dots, M+N-1\}$. 
%Here we assume that $i_{a}$ and $i_{a}^{*}=i_{2r+2s+2-a}$ are independent with respect to this operation.  
%Thus in order to keep stay on the symmetric nesting path, we have to apply 
%two independent operations $\tau_{i_{a}i_{a+1}}$ and $\tau_{i_{a}^{*} i_{a+1}^{*}}=\tau_{i_{2r+2s+2-a}i_{2r+2s+1-a}}$ for $a \in %\{1,2,\dots, r+s-1 \}$ to the T-function 
% at the same time. 
%
%Note that $\tau_{i_{a}i_{a+1}}$ and $\tau_{i_{a}^{*} i_{a+1}^{*}}=\tau_{i_{2r+2s+2-a}i_{2r+2s+1-a}}$ are not independent 
%operation since we are considering  symmetric nesting path 
%$i_{a}^{*}=i_{2r+2s+2-a}$. 
We would like to consider a subgroup $\mathfrak{W}= \mathbb{Z}^{r+s}_{2} \rtimes S_{r+s}$ of the 
permutation group $ S(I_{M+N})=S(I_{2r+2s+1})=S({\mathfrak I})$, 
which preserves the entire  set of  symmetric nesting paths, and discuss the invariance of
 the T-function ${\mathsf F}_{(1)}^{I_{2r+2s+1}}$ under it. 
$\mathfrak{W}$ is generated by two kinds of operations of the form: 
%combination of the transposition of $i_{a}$ and $i_{a+1}$, and transposition of $i^{*}_{a}$ and $i^{*}_{a+1}$ 
$\mathfrak{s}=\overline{\tau}_{a,a+1} \circ \overline{\tau}_{2r+2s+1-a,2r+2s+2-a} $,
%$\mathfrak{s}=\tau_{i_{a}i_{a+1}} \circ \tau_{i_{a}^{*} i_{a+1}^{*}} $, 
$\mathfrak{s} ( I_{2r+2s+1} )=\tau_{i_{a}i_{a+1}} \circ \tau_{i_{a}^{*} i_{a+1}^{*}}  ( I_{2r+2s+1} )=(i_{1},i_{2},\dots, i_{a-1},i_{a+1},i_{a},i_{a+2}, \dots, i_{r+s}, 2r+s+1, i_{r+s}^{*}, \dots ,
i_{a+2}^{*},i_{a}^{*},i_{a+1}^{*},i_{a-1}^{*},\dots , i_{2}^{*}, i_{1}^{*})$ 
for $a \in \{1,2,\dots , r+s-1 \}$,  
and 
$\mathfrak{k}=\overline{\tau}_{r+s,r+s+2} $,  
%$\mathfrak{k}=\tau_{i_{r+s}i^{*}_{r+s}} $,  
$\mathfrak{k} ( I_{2r+2s+1})=\tau_{i_{r+s}i^{*}_{r+s}}( I_{2r+2s+1})=(i_{1},i_{2},\dots, i_{r+s-1},i^{*}_{r+s}, 2r+s+1 , i_{r+s}, i_{r+s-1}^{*}, \dots , i_{2}^{*}, i_{1}^{*})$. 
Let $\mathfrak{s} ( I_{a} )$ denotes the sub-tuple consisting of the first
 $a$-components of $\mathfrak{s} ( I_{2r+2s+1} )$. 
Similarly, let $\mathfrak{k} ( I_{a} )$ denotes the sub-tuple consisting of the first
 $a$-components of $\mathfrak{k} ( I_{2r+2s+1} )$. 
The condition 
$\mathfrak{s} ( {\mathsf F}_{(1)}^{I_{2r+2s+1}})= {\mathsf F}_{(1)}^{\mathfrak{s} (I_{2r+2s+1})}={\mathsf F}_{(1)}^{I_{2r+2s+1}}$ 
is equivalent to the  8-term QQ-relations
\begin{multline}
p_{i_{a}}{\mathcal   X}_{I_{a}}+
p_{i_{a+1}}{\mathcal   X}_{I_{a+1}} +
p_{i_{a}}{\mathcal   X}_{I_{2r+2s+2-a}}+
p_{i_{a+1}}{\mathcal   X}_{I_{2r+2s+1-a}} =
\\
p_{i_{a+1}}{\mathcal   X}_{\mathfrak{s}(I_{a})}+p_{i_{a}}{\mathcal   X}_{\mathfrak{s}(I_{a+1})} 
+
p_{i_{a+1}}{\mathcal   X}_{\mathfrak{s}(I_{2r+2s+2-a})}+p_{i_{a}}{\mathcal   X}_{\mathfrak{s}(I_{2r+2s+1-a})}.
\label{TinvB0}
\end{multline}
The following set of 4-term QQ-relations is a 
 sufficient condition
%
%\footnote{We expect that \eqref{TinvB1} and  \eqref{TinvB2} are
% also necessary condition for \eqref{TinvB0}.} 
%
for \eqref{TinvB0}:
\begin{align}
p_{i_{a}}{\mathcal   X}_{I_{a}}+
p_{i_{a+1}}{\mathcal   X}_{I_{a+1}} &=
p_{i_{a+1}}{\mathcal   X}_{\mathfrak{s}(I_{a})}+p_{i_{a}}{\mathcal   X}_{\mathfrak{s}(I_{a+1})}, 
\label{TinvB1}
\\
p_{i_{a}}{\mathcal   X}_{I_{2r+2s+2-a}}+
p_{i_{a+1}}{\mathcal   X}_{I_{2r+2s+1-a}} &=
p_{i_{a+1}}{\mathcal   X}_{\mathfrak{s}(I_{2r+2s+2-a})}+p_{i_{a}}{\mathcal   X}_{\mathfrak{s}(I_{2r+2s+1-a})}.
\label{TinvB2}
\end{align}
The relation \eqref{TinvB2} is equivalent to \eqref{TinvB1}, which 
follows from the 3-term simplified QQ-relations \eqref{QQb1} and \eqref{QQb2}. 
Considering this and the fact that  \eqref{tab-fund-b} involves two copies of the same Bethe equations through \eqref{chancels-b1} and \eqref{chancels-b2}, 
it would be reasonable to interpret 
 \eqref{TinvB0} as a composite of \eqref{TinvB1} and \eqref{TinvB2}. 
The condition 
$\mathfrak{k} ( {\mathsf F}_{(1)}^{I_{2r+2s+1}})= {\mathsf F}_{(1)}^{\mathfrak{k} (I_{2r+2s+1})}= {\mathsf F}_{(1)}^{I_{2r+2s+1}}$ is equivalent to the following 6-term QQ-relation
\begin{align}
p_{i_{r+s}}{\mathcal   X}_{I_{r+s}} -{\mathcal   X}_{I_{r+s+1}} +p_{i_{r+s}}{\mathcal   X}_{I_{r+s+2}} 
=
p_{i_{r+s}}{\mathcal   X}_{\mathfrak{k} (I_{r+s})} - {\mathcal   X}_{\mathfrak{k} (I_{r+s+1})}
+p_{i_{r+s}}{\mathcal   X}_{\mathfrak{k} (I_{r+s+2}) }  .
\label{TinvB3}
\end{align}
%where $p_{b}=p_{b^{*}}, z_{b^{*}}=z_{b}^{-1}$. 
One can show
\footnote{Taking ratio of both sides of \eqref{QQb5} and the ones shifted by $u \to u-1$, we obtain 
\begin{align}
& \frac{\Qb^{[-1]}_{I_{r+s-1}}}{\Qb^{[1]}_{I_{r+s-1}}}
=
\frac{z_{i_{r+s}}\Qb_{I_{r+s}} \Qb_{\breve{I}_{r+s}}^{[-1]} -\Qb_{I_{r+s}}^{[-1]} \Qb_{\breve{I}_{r+s}} }
{z_{i_{r+s}}\Qb_{I_{r+s}}^{[1]} \Qb_{\breve{I}_{r+s}} -\Qb_{I_{r+s}} \Qb_{\breve{I}_{r+s}}^{[1]} }.
\label{QQb5rat}  
\end{align}
Then we eliminate the Q-function $\Qb_{I_{r+s-1}}$ from \eqref{TinvB3} by \eqref{QQb5rat}.
}
 that \eqref{TinvB3} holds under the 3-term QQ-relation \eqref{QQb5} in case $p_{i_{r+s}}=1$. 
Thus one  can forget about \eqref{QQb4}, which deviates from the  symmetric nesting paths.   
At the moment, we do not have a simple analogue 
 of \eqref{QQb5} for the case  $p_{i_{r+s}}=-1$, 
and thus have to use 3-term QQ-relations,
\footnote{
The following relation follows from  \eqref{QQb3}, \eqref{QQb} and \eqref{QQb7}:
\begin{align}
& (z_{i_{r+s}}-1)\Qb_{\widetilde{I}_{r+s}}\Qb_{\widetilde{I}_{r+s}} ^{[1]}
=z_{i_{r+s}}\Qb_{I_{r+s}} \Qb_{\breve{I}_{r+s}}^{[1]}
-\Qb_{I_{r+s}}^{[1]} \Qb_{\breve{I}_{r+s}}
\quad \text{for} \quad 
p_{i_{r+s}}=-1,
\label{QQb8}  
\end{align}
In a sense, this is an analogue of  \eqref{QQb5} for the case $p_{i_{r+s}}=-1$. However, what we need is 
not this but a QQ-relation among $\Qb_{I_{r+s-1}}$, $\Qb_{I_{r+s}}$, $\Qb_{\breve{I}_{r+s}}$. 
It is possible to eliminate $\Qb_{\widetilde{I}_{r+s}}$ from \eqref{QQb3}, \eqref{QQb} and \eqref{QQb7}, 
and derive a relation among $\Qb_{I_{r+s-1}}$, $\Qb_{I_{r+s}}$, $\Qb_{\breve{I}_{r+s}}$. 
However, both the final expression and the 6-term QQ-relation \eqref{TinvB3} are too cumbersome to use. 
%show \eqref{TinvB3}. 
%\begin{multline}
% (z_{i_{r+s}}+1) (z_{i_{r+s}}-z_{i_{r+s}}^{-1}) \Qb_{\widetilde{I}_{r+s-1}}\Qb_{\widetilde{I}_{r+s-1}} ^{[1]}
%=z_{i_{r+s}}\Qb_{I_{r+s}} \Qb_{\breve{I}_{r+s}}^{[1]}
%-\Qb_{I_{r+s}}^{[1]} \Qb_{\breve{I}_{r+s}}
%\quad \text{for} \quad 
%p_{i_{r+s}}=-1,
%\label{QQb9}  
%\end{multline}
 }
which deviate from the symmetric nesting paths to show \eqref{TinvB3}. 
Split the operation $\mathfrak{k}$ into three steps
\footnote{Instead of this, one may take 
$\mathfrak{k} ( I_{2r+2s+1})=\tau_{2r+s+1,i^{*}_{r+s}} \circ \tau_{i_{r+s},i^{*}_{r+s}} \circ \tau_{i_{r+s},2r+s+1}
(I_{2r+2s+1})$.}
: 
$\mathfrak{k} ( I_{2r+2s+1})=\tau_{i_{r+s},2r+s+1} \circ \tau_{i_{r+s},i^{*}_{r+s}} \circ \tau_{2r+s+1,i^{*}_{r+s}} 
(I_{2r+2s+1})=
\tau_{i_{r+s},2r+s+1}  \circ \tau_{i_{r+s},i^{*}_{r+s}}(i_{1},i_{2},
\dots, i_{r+s-1} , i_{r+s}, i^{*}_{r+s}, 2r+s+1,i_{r+s-1}^{*}, \dots , i_{2}^{*}, i_{1}^{*})
=\tau_{i_{r+s},2r+s+1} (i_{1},i_{2},\dots, i_{r+s-1}, i^{*}_{r+s}, i_{r+s},  2r+s+1,i_{r+s-1}^{*}, \dots , i_{2}^{*}, i_{1}^{*})
=(i_{1},i_{2},\dots, i_{r+s-1},i^{*}_{r+s}, 2r+s+1 , i_{r+s}, i_{r+s-1}^{*}, \dots , i_{2}^{*}, i_{1}^{*})$. 
One sees non-symmetric nesting paths in the intermediate steps. 
We have to use  \eqref{QQb3} and 
\begin{align}
& (z_{i_{r+s}}-z_{i_{r+s}}^{-1})\Qb_{I_{r+s-1}}\Qb_{\widetilde{I}_{r+s}}
=z_{i_{r+s}} \Qb_{I_{r+s}}^{[-1]}
\Qb_{\breve{I}_{r+s}}^{[1]}
-z_{i_{r+s}}^{-1}\Qb_{I_{r+s}}^{[1]}
\Qb_{\breve{I}_{r+s}}^{[-1]}
\quad \text{for} \quad 
p_{i_{r+s}}=-1.
\label{QQb7} 
\\
& (z_{i_{r+s}}^{-1}+1)\Qb_{I_{r+s-1}}\Qb_{\breve{I}_{r+s}}
=z_{i_{r+s}}^{-1} \Qb_{\breve{I}_{r+s}}^{[-1]}
\Qb_{\widetilde{I}_{r+s}}^{[1]}
+\Qb_{\breve{I}_{r+s}}^{[1]}
\Qb_{\widetilde{I}_{r+s}}^{[-1]}
\quad \text{for} \quad 
p_{i_{r+s}}=-1, 
\label{QQb6}  
\end{align}
for each step.  
Eqs. \eqref{QQb7} is a reduction of \eqref{QQb} for $I=I_{r+s-1}$, $(i,j)=(i_{r+s},i_{r+s}^{*})$. 
Eqs. \eqref{QQb6} is a reduction of \eqref{QQb} for 
 $I=I_{r+s-1} \sqcup (i_{r+s}^{*})$, $(i,j)=(i_{r+s},2r+s+1)$ [or  $I=I_{r+s-1}$, $(i,j)=(i_{r+s}^{*},2r+s+1)$]. 
In short, we have to use reductions of \eqref{Tinv} three times to show \eqref{TinvB3}. 
T-functions and Bethe ansatz equations on non-symmetric nesting paths do not  necessarily have 
 the standard form given in \eqref{boxes-b}-\eqref{BAEb}. 
At the moment, the T-functions for the symmetric nesting paths form a closed system 
under $\mathfrak{W}$ only for the case $s=0$ (the $U_{q}(so^{(1)}(2r+1))$ case) 
if we stick to the three-term QQ-relations (instead of 4- or 6-term QQ-relations). 
It remains to be seen whether we can exclude T-and Q-functions on non-symmetric nesting paths. 

The condition that the generating function ${\mathbf W}_{I_{2r+2s+1}}({\mathbf X})$ 
 is invariant under $\mathfrak{s}$, namely 
${\mathfrak{s}(\mathbf W}_{I_{2r+2s+1}}({\mathbf X}))={\mathbf W}_{\mathfrak{s}(I_{2r+2s+1})}({\mathbf X})={\mathbf W}_{I_{2r+2s+1}}({\mathbf X})$ 
is equivalent to the discrete zero curvature condition (a reduction of \eqref{ZCC}):
\begin{align}
& (1-{\mathcal X}_{I_{a}}{\mathbf X})^{p_{i_{a}}} (1-{\mathcal X}_{I_{a+1}}{\mathbf X})^{p_{i_{a+1}}} 
=
 (1-{\mathcal X}_{\mathfrak{s}(I_{a}) }{\mathbf X})^{p_{i_{a+1}}} 
(1-{\mathcal X}_{\mathfrak{s}(I_{a+1})}{\mathbf X})^{p_{i_{a}}} ,
\\
 &(1-{\mathcal X}_{I_{2r+2s+1-a}}{\mathbf X})^{p_{i_{a+1}}} (1-{\mathcal X}_{I_{2r+2s+2-a}}{\mathbf X})^{p_{i_{a}}} 
=
\nonumber 
\\
& \hspace{100pt} =
 (1-{\mathcal X}_{\mathfrak{s}(I_{2r+2s+1-a}) }{\mathbf X})^{p_{i_{a}}} 
(1-{\mathcal X}_{\mathfrak{s}(I_{2r+2s+2-a})}{\mathbf X})^{p_{i_{a+1}}} ,
\label{ZCCB}
\end{align}
where $a \in \{1,2,\dots , r+s-1 \}$. 
These relations \eqref{ZCCB} boil down to \eqref{TinvB1}, \eqref{TinvB2}  and a reduction of the identity \eqref{id-A}.  
 
The condition that the generating function ${\mathbf W}_{I_{2r+2s+1}}({\mathbf X})$ 
 is invariant under $\mathfrak{k}$, namely 
${\mathfrak{k}(\mathbf W}_{I_{2r+2s+1}}({\mathbf X}))={\mathbf W}_{\mathfrak{k}(I_{2r+2s+1})}({\mathbf X})={\mathbf W}_{I_{2r+2s+1}}({\mathbf X})$ 
is equivalent to the following form of discrete zero curvature condition:
\begin{multline}
 (1-{\mathcal X}_{I_{r+s}}{\mathbf X})^{p_{i_{r+s}}}  
(1-{\mathcal X}_{I_{r+s+1}}{\mathbf X})^{-1} 
 (1-{\mathcal X}_{I_{r+s+2}}{\mathbf X})^{p_{i_{r+s}}} 
=
\\
=
 (1-{\mathcal X}_{\mathfrak{k}(I_{r+s})}{\mathbf X})^{p_{i_{r+s}}}  (1-{\mathcal X}_{\mathfrak{k}(I_{r+s+1})}{\mathbf X})^{-1} 
 (1-{\mathcal X}_{\mathfrak{k}(I_{r+s+2})}{\mathbf X})^{p_{i_{r+s}}} 
\label{ZCCB2}
\end{multline}
Consider the expansion of \eqref{ZCCB2} with respect to the non-negative powers of ${\mathbf X}$. 
The coefficient of ${\mathbf X}$ in  \eqref{ZCCB2} is equivalent to \eqref{TinvB3}. 
At the moment we have a proof of \eqref{ZCCB2} which uses reductions of \eqref{ZCC} three times, 
namely a proof by way of non-symmetric nesting paths. 
The T-functions $ {\mathcal F}_{(b)}^{I_{2r+2s+1}}$ and $ {\mathcal F}_{(1^{b})}^{I_{2r+2s+1}}$ 
are invariant under $S(I_{2r+2s+1})$ if reductions of the QQ-relations \eqref{QQb} and \eqref{QQf} 
on non-symmetric nesting paths as well as on symmetric nesting paths are imposed.  
Whether it is possible to restrict the reduction procedures 
to the symmetric nesting paths and construct  T-functions which are  
 $\mathfrak{W}$-invariant  under the 3-term QQ-relations 
is an open question.

%%%%%%%%%%%%%%%%%%%%%%%%%%%%%%%%%%%%%%%%%%%  
\subsubsection{$U_{q}(gl(2r|2s+1)^{(2)})$ case}
This case
\footnote{We remark that R-matrices for a class of representations of  $U_{q}(gl(m|n)^{(2)})$ are available in \cite{GZ99}.}
 is parallel to the case  $U_{q}(osp(2r+1|2s)^{(1)})$. 
We set
\begin{multline}
(M,N)=(2r,2s+1), \quad \Bm=\{1,2,\dots , 2r\}, \quad  \Fm=\{2r+1,2r+2,\dots ,2r+2s+1\}, 
\\
\mathfrak{D}=\emptyset, \quad \eta \ne 0, \quad  
z_{2r+s+1}=1 .
\label{zz-t1}
\end{multline}

%%%%%%%%%%%
\paragraph{QQ-relations}
For a symmetric nesting path defined by 
$I_{2r+2s+1}=(i_{1},i_{2},\dots, i_{r+s}, 2r+s+1 , i_{r+s}^{*} \dots , i_{2}^{*}, i_{1}^{*})$, 
the QQ-relations \eqref{QQb} and \eqref{QQf} reduce to
\begin{align}
& (z_{i_{a}}-z_{i_{a+1}})\Qb_{I_{a-1}}\Qb_{I_{a+1}}
=z_{i_{a}}\Qb_{I_{a}}^{[p_{i_{a}}]}
\Qb_{\widetilde{I}_{a}}^{[-p_{i_{a}}]}-
z_{i_{a+1}}\Qb_{I_{a}}^{[-p_{i_{a}}]}
\Qb_{\widetilde{I}_{a}}^{[p_{i_{a}}]}
\nonumber \\
&
\qquad \text{for} \quad a \in \{1,2,\dots , r+s-1\},   \quad 
p_{i_{a}}=p_{i_{a+1}}, 
\label{QQt11}  \\
& (z_{i_{a}}-z_{i_{a+1}})\Qb_{I_{a}}\Qb_{\widetilde{I}_{a}}
=z_{i_{a}}\Qb_{I_{a-1}}^{[-p_{i_{a}}]}
\Qb_{I_{a+1}}^{[p_{i_{a}}]}-
z_{i_{a+1}}\Qb_{I_{a-1}}^{[p_{i_{a}}]}
\Qb_{I_{a+1}}^{[-p_{i_{a}}]}
\nonumber \\
&
\qquad \text{for} \quad a \in \{1,2,\dots , r+s-1\},   \quad 
p_{i_{a}}=-p_{i_{a+1}}, 
\label{QQt12}  \\
& (z_{i_{r+s}}-1)\Qb_{I_{r+s-1}}\Qb^{[\eta]}_{I_{r+s}}
=z_{i_{r+s}}\Qb_{I_{r+s}}^{[-1]}
\Qb_{\widetilde{I}_{r+s}}^{[1]}
-\Qb_{I_{r+s}}^{[1]}
\Qb_{\widetilde{I}_{r+s}}^{[-1]}
\quad \text{for} \quad 
p_{i_{r+s}}=-1, 
\label{QQt13}  
  \\
& (z_{i_{r+s}}-1)\Qb_{I_{r+s}}\Qb_{\widetilde{I}_{r+s}}
=z_{i_{r+s}}\Qb_{I_{r+s-1}}^{[-1]} \Qb_{I_{r+s}}^{[\eta+1]}
-\Qb_{I_{r+s-1}}^{[1]} \Qb_{I_{r+s}}^{[\eta-1]}
\quad \text{for} \quad 
p_{i_{r+s}}=1.
\label{QQt14}  
\end{align}
Eqs.\ \eqref{QQt11} and  \eqref{QQt13} are reductions of \eqref{QQb} for 
$I=I_{a-1}$, $(i,j)=(i_{a},i_{a+1})$ for $1 \le a \le r+s-1$  and $a=r+s$, respectively
\footnote{The QQ-relations for $a>r+s$ reduce to the ones for $a \le r+s$.}.
Eqs.\ \eqref{QQt12} and  \eqref{QQt14} are reductions of \eqref{QQf} for 
$I=I_{a-1}$, $(i,j)=(i_{a},i_{a+1})$ for $1 \le a \le r+s-1$  and $a=r+s$, respectively. 
Instead of \eqref{QQt14}, one may use 
\begin{align}
& (z_{i_{r+s}}+1)\Qb^{[1]}_{I_{r+s-1}}\Qb^{[\eta]}_{I_{r+s-1}}
=z_{i_{r+s}}\Qb_{I_{r+s}}^{[\eta+1]}
\Qb_{\breve{I}_{r+s}}
+\Qb_{I_{r+s}}
\Qb_{\breve{I}_{r+s}}^{[\eta+1]}
\quad \text{for} \quad 
p_{i_{r+s}}=1,
\label{QQt15}  
\end{align}
where $\breve{I}_{r+s}=(i_{1},i_{2},\dots , i_{r+s-1},i^{*}_{r+s})$. 
One can derive \eqref{QQt15} in the same way as \eqref{QQb5}.  
We will use the following QQ-relations:
\begin{align}
& (z_{i_{r+s}}-z_{i_{r+s}}^{-1})\Qb_{I_{r+s-1}}\Qb^{[\eta]}_{\widetilde{I}_{r+s}}
=z_{i_{r+s}} \Qb_{I_{r+s}}^{[1]} \Qb_{\breve{I}_{r+s}}^{[-1]} - z^{-1}_{i_{r+s}} \Qb_{I_{r+s}}^{[-1]} \Qb_{\breve{I}_{r+s}}^{[1]} 
\quad \text{for} \quad 
p_{i_{r+s}}=1,
 \label{QQt16}
\\
& (z_{i_{r+s}}^{-1}-1)\Qb_{\breve{I}_{r+s}}\Qb_{\widetilde{I}_{r+s}}
=z_{i_{r+s}}^{-1} \Qb_{I_{r+s-1}}^{[-1]} \Qb_{\breve{I}_{r+s}}^{[\eta+1]} - \Qb_{I_{r+s-1}}^{[1]} \Qb_{\breve{I}_{r+s}}^{[\eta-1]}
\quad \text{for} \quad 
p_{i_{r+s}}=1.
 \label{QQt17}
\end{align}
Eqs. \eqref{QQt16} and  \eqref{QQt17} are reductions of  \eqref{QQb} for $I=I_{r+s-1}$, $(i,j)=(i_{r+s},i^{*}_{r+s})$
 and  \eqref{QQf} for $I=I_{r+s-1}$, $(i,j)=(i^{*}_{r+s},2r+s+1)$, respectively.  
Now we prove \eqref{QQt15} step by step as follows. 
\begin{multline}
[\text{left hand side of \eqref{QQt15}}] \times 
(z_{i_{r+s}}-z_{i_{r+s}}^{-1})\Qb_{\widetilde{I}_{r+s}} =
\\
=(z_{i_{r+s}}+1)\Qb^{[1]}_{I_{r+s-1}}
\underbrace{(z_{i_{r+s}}-z_{i_{r+s}}^{-1}) \Qb_{I_{r+s-1}}^{[\eta]} \Qb_{\widetilde{I}_{r+s}}}_{\text{apply \eqref{QQt16}}}
  \\
=(z_{i_{r+1}}+1)\Qb^{[1]}_{I_{r+s-1}}
(z_{i_{r+s}} \Qb_{I_{r+s}}^{[\eta+1]} \Qb_{\breve{I}_{r+s}}^{[\eta-1]} - z^{-1}_{i_{r+s}} \Qb_{I_{r+s}}^{[\eta-1]} \Qb_{\breve{I}_{r+s}}^{[\eta+1]}), 
\label{proofQQ}
\end{multline}
\begin{multline}
[\text{right hand side of \eqref{QQt15}}] \times 
(z_{i_{r+s}}-z_{i_{r+s}}^{-1})\Qb_{\widetilde{I}_{r+s}}=
\\
=(z_{i_{r+s}}-z_{i_{r+s}}^{-1})
 (z_{i_{r+s}} \Qb_{I_{r+s}}^{[\eta+1]} 
\underbrace{ \Qb_{\breve{I}_{r+s}}\Qb_{\widetilde{I}_{r+s}}}_{\text{apply \eqref{QQt17}}}
+ 
\underbrace{\Qb_{I_{r+s}} \Qb_{\widetilde{I}_{r+s}}}_{\text{apply \eqref{QQt14}}} \Qb_{\breve{I}_{r+s}}^{[\eta+1]})
  \\
=(z_{i_{r+s}}-z_{i_{r+s}}^{-1})
 \bigl(z_{i_{r+s}} \Qb_{I_{r+s}}^{[\eta+1]} 
(z_{i_{r+s}}^{-1}-1)^{-1} (z_{i_{r+s}}^{-1} \Qb_{I_{r+s-1}}^{[-1]} \Qb_{\breve{I}_{r+s}}^{[\eta+1]} - 
 \Qb_{I_{r+s-1}}^{[1]}\Qb_{\breve{I}_{r+s}}^{[\eta-1]})
\\
+
(z_{i_{r+s}}-1)^{-1}
(z_{i_{r+s}} \Qb_{I_{r+s-1}}^{[-1]} \Qb_{I_{r+s}}^{[\eta+1]} - 
 \Qb_{I_{r+s-1}}^{[1]}\Qb_{I_{r+s}}^{[\eta-1]})
\Qb_{\breve{I}_{r+s}}^{[\eta+1]}\bigr)
\\
=[\text{right hand side of \eqref{proofQQ}}] .
\end{multline}
Hence \eqref{QQt15} holds since 
$(z_{i_{r+s}}-z_{i_{r+s}}^{-1})\Qb_{\widetilde{I}_{r+s}}$ is not identically zero.  
One can also show that 
\eqref{QQt14} and \eqref{QQt17} follow from \eqref{QQt15} and \eqref{QQt16}. 

%%%%%%%%%%
\paragraph{T-functions and Bethe ansatz equations}
Under the reduction, \eqref{boxes} reduces to 
\begin{align}
\begin{split}
{\mathcal   X}_{I_{a}}&=
z_{i_{a}}
\frac{\Qb_{I_{a-1}}^{[2r-2s-1-\sum_{j \in I_{a}}p_{j}-p_{i_{a}}]}
\Qb_{I_{a}}^{[2r-2s-1-\sum_{j \in I_{a}}p_{j}+2p_{i_{a}}]}
}{
\Qb_{I_{a-1}}^{[2r-2s-1-\sum_{j \in I_{a}}p_{j}+p_{i_{a}}]}
\Qb_{I_{a}}^{[2r-2s-1-\sum_{j \in I_{a}}p_{j}]}
} 
\quad \text{for} \quad 1 \le a \le r+s ,
\\
{\mathcal   X}_{I_{r+s+1}}&=
\frac{\Qb_{I_{r+s}}^{[r-s+1]}
\Qb_{I_{r+s}}^{[r-s-2+\eta]}
}{
\Qb_{I_{r+s}}^{[r-s-1]}
\Qb_{I_{r+s}}^{[r-s+\eta]}
}  ,
\\
{\mathcal   X}_{I_{2r+2s+2-a}}&=
z_{i_{a}}^{-1}
\frac{\Qb_{I_{a-1}}^{[\sum_{j \in I_{a}}p_{j}+p_{i_{a}}+\eta]}
\Qb_{I_{a}}^{[\sum_{j \in I_{a}}p_{j}-2p_{i_{a}}+\eta]}
}{
\Qb_{I_{a-1}}^{[\sum_{j \in I_{a}}p_{j}-p_{i_{a}}+\eta]}
\Qb_{I_{a}}^{[\sum_{j \in I_{a}}p_{j} +\eta]}
} 
\quad \text{for} \quad 1 \le a \le r+s .
\end{split}
\label{boxes-t1} 
\end{align}
The T-function \eqref{tab-fund} reduces to 
\begin{align}
{\mathsf F}_{(1)}^{I_{2r+2s+1}}=
\Qb_{\emptyset}^{[2r-2s-1]}
\Qb_{\emptyset}^{[\eta ]}
\left(\sum_{a=1}^{r+s}p_{i_{a}}({\mathcal   X}_{I_{a}}+{\mathcal   X}_{I_{2r+2s+2-a}}) -{\mathcal   X}_{I_{r+s+1}} \right),
 \label{tab-fund-t1}
\end{align}
The pole-free condition of the T-function \eqref{tab-fund-t1} 
produces the following Bethe ansatz equations: 
\begin{align}
\begin{split}
 -1&=\frac{p_{i_{a}}z_{i_{a}}}{p_{i_{a+1}}z_{i_{a+1}}}
\frac{\Qb_{I_{a-1}}(u_{k}^{I_{a}}-p_{i_{a}})
\Qb_{I_{a}}(u_{k}^{I_{a}}+2p_{i_{a}})
\Qb_{I_{a+1}}(u_{k}^{I_{a}}-p_{i_{a+1}})} 
{\Qb_{I_{a-1}}(u_{k}^{I_{a}}+p_{i_{a}})
\Qb_{I_{a}}(u_{k}^{I_{a}}-2p_{i_{a+1}})
\Qb_{I_{a+1}}(u_{k}^{I_{a}}+p_{i_{a+1}})}
\\
& \hspace{50pt} \text{for} \quad k\in \{1,2,\dots, n_{I_{a}}\} \quad \text{and} \quad a \in \{1,2,\dots, r+s-1 \} ,
\\
 1&=p_{i_{r+s}}z_{i_{r+s}}
\frac{\Qb_{I_{r+s-1}}(u_{k}^{I_{r+s}}-p_{i_{r+s}})
\Qb_{I_{r+s}}(u_{k}^{I_{r+s}}+2p_{i_{r+s}})
\Qb_{I_{r+s}}(u_{k}^{I_{r+s}}+1+\eta)} 
{\Qb_{I_{r+s-1}}(u_{k}^{I_{r+s}}+p_{i_{r+s}})
\Qb_{I_{r+s}}(u_{k}^{I_{r+s}}+2)
\Qb_{I_{r+s}}(u_{k}^{I_{r+s}}-1+\eta)}
\\
&\hspace{50pt}  \text{for} \quad k\in \{1,2,\dots, n_{I_{r+s}}\}  .
\end{split}
\label{BAEt1}
\end{align}
This is a reduction of \eqref{BAE} on the symmetric nesting path. 
One can also derive the Bethe ansatz equations \eqref{BAEt1} from the QQ-relations \eqref{QQt11}-\eqref{QQt14} 
by considering the zeros of the Q-functions. One may use  \eqref{QQt15} instead of \eqref{QQt14}. 
A tableaux sum expression of the T-function is provided by \eqref{DVF-tab1}
 after reduction. 
Moreover,  $\mathsf{T}^{\Bm,\Fm}_{\mu}$ (from \eqref{unnor-t1}) 
and its (super)character limit $\zeta(\mathsf{T}^{\Bm,\Fm}_{\mu})$ 
give a Wronskian expression of the T-function and 
a Weyl-type supercharacter formula respectively after  reduction. 
%The Young diagram $\mu$ is related to the labels of the representation through 
%\eqref{HW-B}-\eqref{YW-B}. 

%%
The generating functions \eqref{gene1} and \eqref{gene2} in this case have the same structure as \eqref{gene1B} and \eqref{gene2B}
\begin{align}
{\mathbf W}_{I_{2r+2s+1}}({\mathbf X})&=
\overrightarrow{\prod_{a=1}^{r+s}} (1-{\mathcal X}_{I_{2r+2s+2-a}}{\mathbf X})^{-p_{i_{a}}} 
 (1-{\mathcal X}_{I_{r+s+1}}{\mathbf X} )
\overleftarrow{\prod_{a=1}^{r+s}} (1-{\mathcal X}_{I_{a}}{\mathbf X})^{-p_{i_{a}}} 
\nonumber \\
&= \sum_{a=0}^{\infty} {\mathcal F}_{(a)}^{I_{2r+2s+1}
[a -1]}{\mathbf X}^{a }, 
\label{gene1t1}
\\
{\mathbf W}_{I_{2r+2s+1}}({\mathbf X})^{-1}&=
\overrightarrow{\prod_{a=1}^{r+s}} (1-{\mathcal X}_{I_{a}}{\mathbf X})^{p_{i_{a}}} 
 (1-{\mathcal X}_{I_{r+s+1}}{\mathbf X} )^{-1}
\overleftarrow{\prod_{a=1}^{r+s}} (1-{\mathcal X}_{I_{2r+2s+2-a}}{\mathbf X})^{p_{i_{a}}} 
\nonumber \\
&= \sum_{a=0}^{\infty}(-1)^{a} {\mathcal F}_{(1^{a})}^{I_{2r+2s+1}
[a -1]}{\mathbf X}^{a }.
\label{gene2t1}
\end{align}
%%%%%%%%
Baxter type equations follow from the kernels of  \eqref{gene1t1} and \eqref{gene2t1}, which are 
reductions of \eqref{Bax3} and \eqref{Bax4}. 
One can also discuss the $\mathfrak{W}$-symmetry in the same way as the 
$U_{q}(osp(2r+1|2s)^{(1)})$ case. 

%%%%%%%%%%%%%%%%%%%%%%%%%%%%%%%%%%  
\subsubsection{$U_{q}(gl(2r+1|2s)^{(2)})$ case}
This case is similar to the cases $U_{q}(osp(2r+1|2s)^{(1)})$ and $U_{q}(gl(2r|2s+1)^{(2)})$. 
We set
\begin{multline}
(M,N)=(2r+1,2s), \quad \Bm=\{1,2,\dots , 2r+1\}, \quad  \Fm=\{2r+2,2r+3,\dots ,2r+2s+1\}, 
\\
\mathfrak{D}=\emptyset, \quad \eta \ne 0, \quad  
z_{r+1}=1 .
\label{zz-t2}
\end{multline}

%%%%%%%%%%%
\paragraph{QQ-relations}
For a symmetric nesting path defined by 
$I_{2r+2s+1}=(i_{1},i_{2},\dots, i_{r+s}, r+1 , i_{r+s}^{*} \dots , i_{2}^{*}, i_{1}^{*})$, 
the QQ-relations \eqref{QQb} and \eqref{QQf} reduce to
\begin{align}
& (z_{i_{a}}-z_{i_{a+1}})\Qb_{I_{a-1}}\Qb_{I_{a+1}}
=z_{i_{a}}\Qb_{I_{a}}^{[p_{i_{a}}]}
\Qb_{\widetilde{I}_{a}}^{[-p_{i_{a}}]}-
z_{i_{a+1}}\Qb_{I_{a}}^{[-p_{i_{a}}]}
\Qb_{\widetilde{I}_{a}}^{[p_{i_{a}}]}
\nonumber \\
&
\qquad \text{for} \quad a \in \{1,2,\dots , r+s-1\},   \quad 
p_{i_{a}}=p_{i_{a+1}}, 
\label{QQt21}  \\
& (z_{i_{a}}-z_{i_{a+1}})\Qb_{I_{a}}\Qb_{\widetilde{I}_{a}}
=z_{i_{a}}\Qb_{I_{a-1}}^{[-p_{i_{a}}]}
\Qb_{I_{a+1}}^{[p_{i_{a}}]}-
z_{i_{a+1}}\Qb_{I_{a-1}}^{[p_{i_{a}}]}
\Qb_{I_{a+1}}^{[-p_{i_{a}}]}
\nonumber \\
&
\qquad \text{for} \quad a \in \{1,2,\dots , r+s-1\},   \quad 
p_{i_{a}}=-p_{i_{a+1}}, 
\label{QQt22}  \\
& (z_{i_{r+s}}-1)\Qb_{I_{r+s-1}}\Qb^{[\eta]}_{I_{r+s}}
=z_{i_{r+s}}\Qb_{I_{r+s}}^{[1]}
\Qb_{\widetilde{I}_{r+s}}^{[-1]}
-\Qb_{I_{r+s}}^{[-1]}
\Qb_{\widetilde{I}_{r+s}}^{[1]}
\quad \text{if} \quad 
p_{i_{r+s}}=1, 
\label{QQt23}  
  \\
& (z_{i_{r+s}}-1)\Qb_{I_{r+s}}\Qb_{\widetilde{I}_{r+s}}
=z_{i_{r+s}}\Qb_{I_{r+s-1}}^{[1]} \Qb_{I_{r+s}}^{[\eta-1]}
-\Qb_{I_{r+s-1}}^{[-1]} \Qb_{I_{r+s}}^{[\eta+1]}
\quad \text{if} \quad 
p_{i_{r+s}}=-1.
\label{QQt24}  
\end{align}
Eqs.\ \eqref{QQt21} and  \eqref{QQt23} are reductions of \eqref{QQb} for 
$I=I_{a-1}$, $(i,j)=(i_{a},i_{a+1})$ for $1 \le a \le r+s-1$  and $a=r+s$, respectively
\footnote{The QQ-relations for $a>r+s$ reduce to the ones for $a \le r+s$.}.
Eqs.\ \eqref{QQt22} and  \eqref{QQt24} are reductions of \eqref{QQf} for 
$I=I_{a-1}$, $(i,j)=(i_{a},i_{a+1})$ for $1 \le a \le r+s-1$  and $a=r+s$, respectively. 
Instead of \eqref{QQt24}, one may use 
\begin{align}
& (z_{i_{r+s}}+1)\Qb^{[-1]}_{I_{r+s-1}}\Qb^{[\eta]}_{I_{r+s-1}}
=z_{i_{r+s}}\Qb_{I_{r+s}}^{[\eta-1]}
\Qb_{\breve{I}_{r+s}}
+\Qb_{I_{r+s}}
\Qb_{\breve{I}_{r+s}}^{[\eta-1]}
\quad \text{if} \quad 
p_{i_{r+s}}=-1,
\label{QQt25}  
\end{align}
where $\breve{I}_{r+s}=(i_{1},i_{2},\dots , i_{r+s-1},i^{*}_{r+s})$. 
One can derive \eqref{QQt15} in the same way as \eqref{QQb5}.  

In the context of representation theory, 
QQ-relations for twisted quantum affine (non-super) algebras ($s=0$ case) 
appeared in \cite{FH16} and were proved in \cite{W22}. 
These papers confirm
\footnote{As a matter of fact, we have a difficulty in comparing \eqref{QQt21} (for $s=0$, $a=r-1$) 
and \eqref{QQt23} (for $s=0$) with 
[eq.\ (3.9), \cite{FH16}] for the last two elements of $I_{\sigma}$. 
On the other hand, \eqref{QQt21} and \eqref{QQt23} for $s=0$ appear to agree with 
[eqs.\ (5.4), (5.5), \cite{W22}]. In any case, what is important for us is that \eqref{QQt21}-\eqref{QQt25} produce the real Bethe ansatz equations \eqref{BAEt2} derived by algebraic Bethe ansatz. 
%Comment on [Remark 3.4, \cite{FH16}]: the QQ-relations on a Hasse diagram introduced in [BFLMS] 
%were known before [BFLMS] ([BFLMS] constructed Q-operators that 
%satisfy the QQ-relations.) 
%The formulation based on the Hasse diagram appeared in \cite{T09} before [BFLMS].
}
 our proposal [section 3.7, \cite{T11}] on reductions of the 
QQ-relations for the case $(M,N)=(2r+1,0)$, 
which was inspired by \cite{KS94-2}. 
%%%%%%%%%%%

%%%%%%
\paragraph{T-functions and Bethe ansatz equations}
Under the reduction, \eqref{boxes} reduces to 
\begin{align}
\begin{split}
{\mathcal   X}_{I_{a}}&=
z_{i_{a}}
\frac{\Qb_{I_{a-1}}^{[2r-2s+1-\sum_{j \in I_{a}}p_{j}-p_{i_{a}}]}
\Qb_{I_{a}}^{[2r-2s+1-\sum_{j \in I_{a}}p_{j}+2p_{i_{a}}]}
}{
\Qb_{I_{a-1}}^{[2r-2s+1-\sum_{j \in I_{a}}p_{j}+p_{i_{a}}]}
\Qb_{I_{a}}^{[2r-2s+1-\sum_{j \in I_{a}}p_{j}]}
} 
\quad \text{for} \quad 1 \le a \le r+s ,
\\
{\mathcal   X}_{I_{r+s+1}}&=
\frac{\Qb_{I_{r+s}}^{[r-s-1]}
\Qb_{I_{r+s}}^{[r-s+2+\eta]}
}{
\Qb_{I_{r+s}}^{[r-s+1]}
\Qb_{I_{r+s}}^{[r-s+\eta]}
}  ,
\\
{\mathcal   X}_{I_{2r+2s+2-a}}&=
z_{i_{a}}^{-1}
\frac{\Qb_{I_{a-1}}^{[\sum_{j \in I_{a}}p_{j}+p_{i_{a}}+\eta]}
\Qb_{I_{a}}^{[\sum_{j \in I_{a}}p_{j}-2p_{i_{a}}+\eta]}
}{
\Qb_{I_{a-1}}^{[\sum_{j \in I_{a}}p_{j}-p_{i_{a}}+\eta]}
\Qb_{I_{a}}^{[\sum_{j \in I_{a}}p_{j} +\eta]}
} 
\quad \text{for} \quad 1 \le a \le r+s .
\end{split}
\label{boxes-t2} 
\end{align}
The T-function \eqref{tab-fund} reduces to 
\begin{align}
{\mathsf F}_{(1)}^{I_{2r+2s+1}}=
\Qb_{\emptyset}^{[2r-2s+1]}
\Qb_{\emptyset}^{[\eta]}
\left(\sum_{a=1}^{r+s}p_{i_{a}}({\mathcal   X}_{I_{a}}+{\mathcal   X}_{I_{2r+2s+2-a}}) +{\mathcal   X}_{I_{r+s+1}} \right).
 \label{tab-fund-t2}
\end{align}
The pole-free condition of the T-function \eqref{tab-fund-t2} 
produces the following Bethe ansatz equations: 
\begin{align}
\begin{split}
 -1&=\frac{p_{i_{a}}z_{i_{a}}}{p_{i_{a+1}}z_{i_{a+1}}}
\frac{\Qb_{I_{a-1}}(u_{k}^{I_{a}}-p_{i_{a}})
\Qb_{I_{a}}(u_{k}^{I_{a}}+2p_{i_{a}})
\Qb_{I_{a+1}}(u_{k}^{I_{a}}-p_{i_{a+1}})} 
{\Qb_{I_{a-1}}(u_{k}^{I_{a}}+p_{i_{a}})
\Qb_{I_{a}}(u_{k}^{I_{a}}-2p_{i_{a+1}})
\Qb_{I_{a+1}}(u_{k}^{I_{a}}+p_{i_{a+1}})}
\\
& \hspace{50pt} \text{for} \quad k\in \{1,2,\dots, n_{I_{a}}\} \quad \text{and} \quad a \in \{1,2,\dots, r+s-1 \} ,
\\
- 1&=p_{i_{r+s}}z_{i_{r+s}}
\frac{\Qb_{I_{r+s-1}}(u_{k}^{I_{r+s}}-p_{i_{r+s}})
\Qb_{I_{r+s}}(u_{k}^{I_{r+s}}+2p_{i_{r+s}})
\Qb_{I_{r+s}}(u_{k}^{I_{r+s}}-1+\eta)} 
{\Qb_{I_{r+s-1}}(u_{k}^{I_{r+s}}+p_{i_{r+s}})
\Qb_{I_{r+s}}(u_{k}^{I_{r+s}}-2)
\Qb_{I_{r+s}}(u_{k}^{I_{r+s}}+1+\eta)}
\\
&\hspace{50pt}  \text{for} \quad k\in \{1,2,\dots, n_{I_{r+s}}\}  .
\end{split}
\label{BAEt2}
\end{align}
This is a reduction of \eqref{BAE} on the symmetric nesting path. 
One can also derive the Bethe ansatz equations \eqref{BAEt2} from the QQ-relations \eqref{QQt21}-\eqref{QQt24} 
by considering the zeros of the Q-functions. One can also use  \eqref{QQt25} instead of \eqref{QQt24}.
Eqs.\ \eqref{tab-fund-t2} and \eqref{BAEt2} agree with the known results by algebraic Bethe ansatz \cite{GM04} in case $i_{k} \in \Fm$ for $1 \le k \le s$ 
and $i_{k} \in \Bm$ for $s+1 \le k \le r+s$. 
We remark that  this reduces to the case  $U_{q}(gl(2r+1)^{(2)})$ \cite{R87,KS94-2} for $s=0$.
A tableaux sum expression of the T-function is provided by \eqref{DVF-tab1}
 after reduction. 
Moreover,  $\mathsf{T}^{\Bm,\Fm}_{\mu}$ (from \eqref{unnor-t1}) 
and its (super)character limit $\zeta(\mathsf{T}^{\Bm,\Fm}_{\mu})$ 
give a Wronskian expression of the T-function and 
a Weyl-type supercharacter formula respectively after  reduction. 
%The Young diagram $\mu$ is related to the labels of the representation through 
%\eqref{HW-B}-\eqref{YW-B}. 
%%%%%%%%%

The generating functions \eqref{gene1} and \eqref{gene2} reduce to  
\begin{align}
{\mathbf W}_{I_{2r+2s+1}}({\mathbf X})&=
\overrightarrow{\prod_{a=1}^{r+s}} (1-{\mathcal X}_{I_{2r+2s+2-a}}{\mathbf X})^{-p_{i_{a}}} 
 (1-{\mathcal X}_{I_{r+s+1}}{\mathbf X} )^{-1}
\overleftarrow{\prod_{a=1}^{r+s}} (1-{\mathcal X}_{I_{a}}{\mathbf X})^{-p_{i_{a}}} 
\nonumber \\
&= \sum_{a=0}^{\infty} {\mathcal F}_{(a)}^{I_{2r+2s+1}
[a -1]}{\mathbf X}^{a }, 
\label{gene1t2}
\\
{\mathbf W}_{I_{2r+2s+1}}({\mathbf X})^{-1}&=
\overrightarrow{\prod_{a=1}^{r+s}} (1-{\mathcal X}_{I_{a}}{\mathbf X})^{p_{i_{a}}} 
 (1-{\mathcal X}_{I_{r+s+1}}{\mathbf X} )
\overleftarrow{\prod_{a=1}^{r+s}} (1-{\mathcal X}_{I_{2r+2s+2-a}}{\mathbf X})^{p_{i_{a}}} 
\nonumber \\
&= \sum_{a=0}^{\infty}(-1)^{a} {\mathcal F}_{(1^{a})}^{I_{2r+2s+1}
[a -1]}{\mathbf X}^{a }.
\label{gene2t2}
\end{align}
We remark that  \eqref{gene2t2} for $s=0$, corresponds to [eq. (2.12) in \cite{T02}]. 
Baxter type equations follow from the kernels of  \eqref{gene1t2} and \eqref{gene2t2}, which are 
reductions of \eqref{Bax3} and \eqref{Bax4}. 
%%%%%%%%
One can also discuss the $\mathfrak{W}$-symmetry in the same way as the 
$U_{q}(osp(2r+1|2s)^{(1)})$ case. 
%%%%%%%%%%%%%%%%%%%%%%%%%%%%%%%%%%  
\subsubsection{$U_{q}(gl(2r|2s)^{(2)})$ case}
We set
\begin{multline}
(M,N)=(2r,2s), \quad \Bm=\{1,2,\dots , 2r\}, \quad  \Fm=\{2r+1,2r+2,\dots ,2r+2s\}, 
\\
\mathfrak{D}=\emptyset, \quad \eta \ne 0
\label{zz-t3}
\end{multline}

%%%%%%%%%%%
\paragraph{QQ-relations}
For a symmetric nesting path defined by 
$I_{2r+2s}=(i_{1},i_{2},\dots, i_{r+s},  i_{r+s}^{*} \dots , i_{2}^{*}, i_{1}^{*})$, 
the QQ-relations \eqref{QQb} and \eqref{QQf} reduce to
\begin{align}
& (z_{i_{a}}-z_{i_{a+1}})\Qb_{I_{a-1}}\Qb_{I_{a+1}}
=z_{i_{a}}\Qb_{I_{a}}^{[p_{i_{a}}]}
\Qb_{\widetilde{I}_{a}}^{[-p_{i_{a}}]}-
z_{i_{a+1}}\Qb_{I_{a}}^{[-p_{i_{a}}]}
\Qb_{\widetilde{I}_{a}}^{[p_{i_{a}}]}
\nonumber \\
&
\qquad \text{for} \quad a \in \{1,2,\dots , r+s-2\},   \quad 
p_{i_{a}}=p_{i_{a+1}}, 
\label{QQt31}  \\
& (z_{i_{a}}-z_{i_{a+1}})\Qb_{I_{a}}\Qb_{\widetilde{I}_{a}}
=z_{i_{a}}\Qb_{I_{a-1}}^{[-p_{i_{a}}]}
\Qb_{I_{a+1}}^{[p_{i_{a}}]}-
z_{i_{a+1}}\Qb_{I_{a-1}}^{[p_{i_{a}}]}
\Qb_{I_{a+1}}^{[-p_{i_{a}}]}
\nonumber \\
&
\qquad \text{for} \quad a \in \{1,2,\dots , r+s-2\},   \quad 
p_{i_{a}}=-p_{i_{a+1}}, 
\label{QQt32}  \\
& (z_{i_{r+s-1}}-z_{i_{r+s}})\Qb_{I_{r+s-2}}\Qf^{2}_{I_{r+s}}
=z_{i_{r+s-1}}\Qb_{I_{r+s-1}}^{[p_{i_{r+s-1}}]} \Qb_{\widetilde{I}_{r+s-1}}^{[-p_{i_{r+s-1}}]}
-z_{i_{r+s}}\Qb_{I_{r+s-1}}^{[-p_{i_{r+s-1}}]} \Qb_{\widetilde{I}_{r+s-1}}^{[p_{i_{r+s-1}}]}
\nonumber \\
&
\qquad \text{if} \quad p_{i_{r+s-1}}=p_{i_{r+s}}, 
\label{QQt33}  
  \\
& (z_{i_{r+s-1}}-z_{i_{r+s}})\Qb_{I_{r+s-1}}\Qb_{\widetilde{I}_{r+s-1}}
=z_{i_{r+s-1}}\Qb_{I_{r+s-2}}^{[-p_{i_{r+s-1}}]} \Qf_{I_{r+s}}^{2[p_{i_{r+s-1}}]}
- z_{i_{r+s}}\Qb_{I_{r+s-2}}^{[p_{i_{r+s-1}}]} \Qf_{I_{r+s}}^{2[-p_{i_{r+s-1}}]}
\nonumber \\
&
\qquad \text{if} \quad  p_{i_{r+s-1}}=-p_{i_{r+s}},
\label{QQt34}   \\
& (z_{i_{r+s}}-z_{i_{r+s}}^{-1})\Qb^{2}_{I_{r+s-1}}
=z_{i_{r+s}}\Qf_{I_{r+s}}^{2[p_{i_{r+s}}]} \Qf_{\widetilde{I}_{r+s}}^{2[-p_{i_{r+s}}]}
-z_{i_{r+s}}^{-1}\Qf_{I_{r+s}}^{2[-p_{i_{r+s}}]} \Qf_{\widetilde{I}_{r+s}}^{2[p_{i_{r+s}}]} ,
\label{QQt35} 
\end{align}
where $\Qb^{2}_{I_{r+s-1}}=\Qb_{I_{r+s-1}}\Qb^{[\eta]}_{I_{r+s-1}}$, 
$\Qb_{I_{r+s}}=\Qf^{2}_{I_{r+s}}=\Qf_{I_{r+s}}\Qf^{[\eta]}_{I_{r+s}}$. 
Eqs.\ \eqref{QQt31}, \eqref{QQt33} and \eqref{QQt35} are reductions of \eqref{QQb} for 
$I=I_{a-1}$, $(i,j)=(i_{a},i_{a+1})$ for $1 \le a \le r+s-2$, $a=r+s-1$  and $a=r+s$, respectively
\footnote{The QQ-relations for $a>r+s$ reduce to the ones for $a \le r+s$.}.
Eqs.\ \eqref{QQt32} and  \eqref{QQt34} are reductions of \eqref{QQf} for 
$I=I_{a-1}$, $(i,j)=(i_{a},i_{a+1})$ for $1 \le a \le r+s-2$  and $a=r+s-1$, respectively. 
Let $\{v^{I_{r+s}}_{k}\}_{k=1}^{m_{I_{k}}}$  be the zeros of the Q-function $\Qf_{I_{r+s}}$. 
$\Qb_{I_{r+s}}=\Qf_{I_{r+s}}\Qf^{[\eta]}_{I_{r+s}}$  means that 
$\{u^{I_{r+s}}_{k}\}_{k=1}^{n_{I_{k}}}=\{v^{I_{r+s}}_{k}\}_{k=1}^{m_{I_{k}}} \sqcup \{v^{I_{r+s}}_{k} +\eta \}_{k=1}^{m_{I_{k}}}$, 
 $n_{I_{k}}=2m_{I_{k}}$ holds. 

In the context of representation theory, 
QQ-relations for twisted quantum affine (non-super) algebras ($s=0$ case) 
appeared in \cite{FH16} and were proved in \cite{W22}. 
These papers, at least partially
\footnote{As a matter of fact, we have a difficulty in comparing \eqref{QQt33} and \eqref{QQt35} for $s=0$ with 
[eq.\ (3.9), \cite{FH16}] for the last two elements of $I_{\sigma}$. 
Moreover, \eqref{QQt35} for $s=0$  looks different from 
[eq.\ (5.3) or eq.\ (5.11), \cite{W22}] for $i=n$. This difference might be superficial, but what is important for us is that \eqref{QQt31}-\eqref{QQt35} produce the real Bethe ansatz equations \eqref{BAEt3} derived by algebraic Bethe ansatz.}, 
confirm our proposal [section 3.7, \cite{T11}] on reductions of the 
QQ-relations for the case $(M,N)=(2r,0)$, 
which was inspired by \cite{KS94-2}. 

%%%%%%%%%%%%
\paragraph{T-functions and Bethe ansatz equations}
Under the reduction, 
\eqref{boxes} reduces to 
\begin{align}
\begin{split}
{\mathcal   X}_{I_{a}}&=
z_{i_{a}}
\frac{\Qb_{I_{a-1}}^{[2r-2s-\sum_{j \in I_{a}}p_{j}-p_{i_{a}}]}
\Qb_{I_{a}}^{[2r-2s-\sum_{j \in I_{a}}p_{j}+2p_{i_{a}}]}
}{
\Qb_{I_{a-1}}^{[2r-2s-\sum_{j \in I_{a}}p_{j}+p_{i_{a}}]}
\Qb_{I_{a}}^{[2r-2s-\sum_{j \in I_{a}}p_{j}]}
} 
\quad \text{for} \quad 1 \le a \le r+s-1 ,
\\
{\mathcal   X}_{I_{r+s}}&=
z_{i_{r+s}}
\frac{ \Qb_{I_{r+s-1}}^{[r-s-p_{i_{r+s}}]} 
\Qf_{I_{r+s}}^{2[r-s+2p_{i_{r+s}}]} 
}{
\Qb_{I_{r+s-1}}^{[r-s+p_{i_{r+s}}]}
\Qf_{I_{r+s}}^{2[r-s]} 
} ,
\\
{\mathcal   X}_{I_{r+s+1}}&=
z_{i_{r+s}}^{-1}
\frac{\Qb_{I_{r+s-1}}^{[r-s+p_{i_{r+s}}+\eta]}
\Qf_{I_{r+s}}^{2[r-s-2p_{i_{r+s}}]} 
}{
\Qb_{I_{r+s-1}}^{[r-s-p_{i_{r+s}}+\eta]}
\Qf_{I_{r+s}}^{2[r-s]}
} ,
\\
{\mathcal   X}_{I_{2r+2s+1-a}}&=
z_{i_{a}}^{-1}
\frac{\Qb_{I_{a-1}}^{[\sum_{j \in I_{a}}p_{j}+p_{i_{a}}+\eta]}
\Qb_{I_{a}}^{[\sum_{j \in I_{a}}p_{j}-2p_{i_{a}}+\eta]}
}{
\Qb_{I_{a-1}}^{[\sum_{j \in I_{a}}p_{j}-p_{i_{a}}+\eta]}
\Qb_{I_{a}}^{[\sum_{j \in I_{a}}p_{j} +\eta]}
} 
\quad \text{for} \quad 1 \le a \le r+s-1 ,
\end{split}
\label{boxes-t3} 
\end{align}
where $\Qf_{I_{r+s}}^{2}=\Qf_{I_{r+s}} \Qf_{I_{r+s}}^{[\eta]}  $.
The T-function \eqref{tab-fund} reduces to 
\begin{align}
{\mathsf F}_{(1)}^{I_{2r+2s}}=
\Qb_{\emptyset}^{[2r-2s]}
\Qb_{\emptyset}^{[\eta]}
\sum_{a=1}^{r+s}p_{i_{a}}({\mathcal   X}_{I_{a}}+{\mathcal   X}_{I_{2r+2s+1-a}})  ,
 \label{tab-fund-t3}
\end{align}
The pole-free condition of the T-function \eqref{tab-fund-t3} 
produces the following Bethe ansatz equations: 
\begin{align}
\begin{split}
 -1&=\frac{p_{i_{a}}z_{i_{a}}}{p_{i_{a+1}}z_{i_{a+1}}}
\frac{\Qb_{I_{a-1}}(u_{k}^{I_{a}}-p_{i_{a}})
\Qb_{I_{a}}(u_{k}^{I_{a}}+2p_{i_{a}})
\Qb_{I_{a+1}}(u_{k}^{I_{a}}-p_{i_{a+1}})} 
{\Qb_{I_{a-1}}(u_{k}^{I_{a}}+p_{i_{a}})
\Qb_{I_{a}}(u_{k}^{I_{a}}-2p_{i_{a+1}})
\Qb_{I_{a+1}}(u_{k}^{I_{a}}+p_{i_{a+1}})}
\\
& \hspace{50pt} \text{for} \quad k\in \{1,2,\dots, n_{I_{a}}\} \quad \text{and} \quad a \in \{1,2,\dots, r+s-2 \} ,
\\
- 1&= \frac{p_{i_{r+s-1}}z_{i_{r+s-1}}}{p_{i_{r+s}}z_{i_{r+s}}} 
\frac{\Qb_{I_{r+s-2}}(u_{k}^{I_{r+s-1}}-p_{i_{r+s-1}})  \Qb_{I_{r+s-1}}(u_{k}^{I_{r+s-1}}+2p_{i_{r+s-1}}) 
\Qf_{I_{r+s}}^{2}(u_{k}^{I_{r+s-1}}-p_{i_{r+s}})} 
{\Qb_{I_{r+s-2}}(u_{k}^{I_{r+s-1}}+p_{i_{r+s-1}})  \Qb_{I_{r+s-1}}(u_{k}^{I_{r+s-1}}-2p_{i_{r+s}})
\Qf_{I_{r+s}}^{2}(u_{k}^{I_{r+s-1}}+p_{i_{r+s}})} 
\\
&\hspace{50pt}  \text{for} \quad k\in \{1,2,\dots, n_{I_{r+s-1}}\}  ,
\\
- 1&=z_{i_{r+s}}^{2}
\frac{\Qb^{2}_{I_{r+s-1}}(v_{k}^{I_{r+s}}-p_{i_{r+s}}) 
\Qf_{I_{r+s}}^{2}(v_{k}^{I_{r+s}}+2p_{i_{r+s}})} 
{\Qb^{2}_{I_{r+s-1}}(v_{k}^{I_{r+s}}+p_{i_{r+s}}) 
\Qf_{I_{r+s}}^{2}(v_{k}^{I_{r+s}}-2p_{i_{r+s}})} 
%\\
%&\hspace{50pt} 
\quad 
 \text{for} \quad k\in \{1,2,\dots, m_{I_{r+s}}\}  ,
\end{split}
\label{BAEt3}
\end{align}
where $\Qb^{2}_{I_{r+s-1}}(u)=\Qb_{I_{r+s-1}}(u)\Qb_{I_{r+s-1}}(u+\eta)$, 
$\Qf^{2}_{I_{r+s}}(u)=\Qf_{I_{r+s}}(u)\Qf_{I_{r+s}}(u+\eta)$. 
Note that $\{v_{k}^{I_{r+s}} + \eta \}_{k=1}^{m_{I_{r+s}}}$  also satisfies the last equation of \eqref{BAEt3}. 
 \eqref{BAEt3} is a reduction of \eqref{BAE} on the symmetric nesting path. 
One can also derive the Bethe ansatz equations \eqref{BAEt3} from the QQ-relations \eqref{QQt31}-\eqref{QQt35} 
by considering the zeros of the Q-functions. 
Eqs. \eqref{tab-fund-t3} and \eqref{BAEt3} 
agree with  the known results by algebraic Bethe ansatz \cite{GM04} in case $i_{k} \in \Fm$ for $1 \le k \le s$ 
and $i_{k} \in \Bm$ for $s+1 \le k \le r+s$. 
We remark that  this reduces to the case  $U_{q}(gl(2r)^{(2)})$ \cite{R87,KS94-2} for $s=0$. 
A tableaux sum expression of the T-function is provided by \eqref{DVF-tab1}
 after reduction. 
Moreover,  $\mathsf{T}^{\Bm,\Fm}_{\mu}$ (from \eqref{unnor-t1}) 
and its (super)character limit $\zeta(\mathsf{T}^{\Bm,\Fm}_{\mu})$ 
give a Wronskian expression of the T-function and 
a Weyl-type supercharacter formula respectively after  reduction. 
%The Young diagram $\mu$ is related to the labels of the representation through 
%\eqref{HW-B}-\eqref{YW-B}. 
%%%%%%%%%%%%%%%%%

The generating functions \eqref{gene1} and \eqref{gene2} reduce to  
\begin{align}
{\mathbf W}_{I_{2r+2s}}({\mathbf X})&=
\overrightarrow{\prod_{a=1}^{r+s}} (1-{\mathcal X}_{I_{2r+2s+1-a}}{\mathbf X})^{-p_{i_{a}}} 
\overleftarrow{\prod_{a=1}^{r+s}} (1-{\mathcal X}_{I_{a}}{\mathbf X})^{-p_{i_{a}}} 
\nonumber \\
&= \sum_{a=0}^{\infty} {\mathcal F}_{(a)}^{I_{2r+2s}
[a -1]}{\mathbf X}^{a }, 
\label{gene1t3}
\\
{\mathbf W}_{I_{2r+2s}}({\mathbf X})^{-1}&=
\overrightarrow{\prod_{a=1}^{r+s}} (1-{\mathcal X}_{I_{a}}{\mathbf X})^{p_{i_{a}}} 
\overleftarrow{\prod_{a=1}^{r+s}} (1-{\mathcal X}_{I_{2r+2s+1-a}}{\mathbf X})^{p_{i_{a}}} 
\nonumber \\
&= \sum_{a=0}^{\infty}(-1)^{a} {\mathcal F}_{(1^{a})}^{I_{2r+2s}
[a -1]}{\mathbf X}^{a }.
\label{gene2t3}
\end{align}
We remark that  \eqref{gene2t3} for $s=0$ corresponds to [eq. (2.13) in \cite{T02}]. 
Baxter type equations follow from the kernels of  \eqref{gene1t3} and \eqref{gene2t3}, which are 
reductions of \eqref{Bax3} and \eqref{Bax4}. 
%%%%%%%%%%%%%%%%

\paragraph{ $\mathfrak{W}$-symmetry}
We would like to consider a subgroup $\mathfrak{W}= \mathbb{Z}^{r+s}_{2} \rtimes S_{r+s}$ of the 
permutation group $ S(I_{M+N})=S(I_{2r+2s})=S({\mathfrak I})$, 
which preserves the entire set of symmetric nesting paths, and discuss the invariance of
 the T-function ${\mathsf F}_{(1)}^{I_{2r+2s}}$ under it. 
$\mathfrak{W}$ is generated by two kinds of operations of the form: 
$\mathfrak{s}=\overline{\tau}_{a,a+1} \circ \overline{\tau}_{2r+2s-a,2r+2s+1-a} $, 
$\mathfrak{s} ( I_{2r+2s} )=\tau_{i_{a}i_{a+1}} \circ \tau_{i_{a}^{*} i_{a+1}^{*}} ( I_{2r+2s} )=(i_{1},i_{2},\dots, i_{a-1},i_{a+1},i_{a},i_{a+2}, \dots, i_{r+s},  i_{r+s}^{*}, \dots ,
i_{a+2}^{*},i_{a}^{*},i_{a+1}^{*},i_{a-1}^{*},\dots , i_{2}^{*}, i_{1}^{*})$ 
for $a \in \{1,2,\dots , r+s-1 \}$,  
and 
$\mathfrak{k}=\overline{\tau}_{r+s,r+s+1} $, 
$\mathfrak{k} ( I_{2r+2s})=\tau_{i_{r+s}i^{*}_{r+s}}( I_{2r+2s})= (i_{1},i_{2},\dots, i_{r+s-1},i^{*}_{r+s}, i_{r+s}, i_{r+s-1}^{*}, \dots , i_{2}^{*}, i_{1}^{*})$.
The condition 
$\mathfrak{s} ( {\mathsf F}_{(1)}^{I_{2r+2s}})= {\mathsf F}_{(1)}^{\mathfrak{s} (I_{2r+2s})}={\mathsf F}_{(1)}^{I_{2r+2s}}$ 
is equivalent to the  8-term QQ-relations
\begin{multline}
p_{i_{a}}{\mathcal   X}_{I_{a}}+
p_{i_{a+1}}{\mathcal   X}_{I_{a+1}} +
p_{i_{a}}{\mathcal   X}_{I_{2r+2s+1-a}}+
p_{i_{a+1}}{\mathcal   X}_{I_{2r+2s-a}} =
\\
p_{i_{a+1}}{\mathcal   X}_{\mathfrak{s}(I_{a})}+p_{i_{a}}{\mathcal   X}_{\mathfrak{s}(I_{a+1})} 
+
p_{i_{a+1}}{\mathcal   X}_{\mathfrak{s}(I_{2r+2s+1-a})}+p_{i_{a}}{\mathcal   X}_{\mathfrak{s}(I_{2r+2s-a})}.
\label{Tinvt30}
\end{multline}
The following set of 4-term QQ-relations is a
 sufficient condition for \eqref{Tinvt30}:
\begin{align}
p_{i_{a}}{\mathcal   X}_{I_{a}}+
p_{i_{a+1}}{\mathcal   X}_{I_{a+1}} &=
p_{i_{a+1}}{\mathcal   X}_{\mathfrak{s}(I_{a})}+p_{i_{a}}{\mathcal   X}_{\mathfrak{s}(I_{a+1})}, 
\label{Tinvt31}
\\
p_{i_{a}}{\mathcal   X}_{I_{2r+2s+1-a}}+
p_{i_{a+1}}{\mathcal   X}_{I_{2r+2s-a}} &=
p_{i_{a+1}}{\mathcal   X}_{\mathfrak{s}(I_{2r+2s+1-a})}+p_{i_{a}}{\mathcal   X}_{\mathfrak{s}(I_{2r+2s-a})}.
\label{Tinvt32} 
\end{align}
The relation \eqref{Tinvt32} is equivalent to \eqref{Tinvt31}, 
which follows from the 3-term QQ-relations \eqref{QQt31}-\eqref{QQt34}.
The condition 
$\mathfrak{k} ( {\mathsf F}_{(1)}^{I_{2r+2s}})= {\mathsf F}_{(1)}^{\mathfrak{k} (I_{2r+2s})}= {\mathsf F}_{(1)}^{I_{2r+2s}}$ 
is equivalent to  the following 4-term QQ-relations
\begin{align}
{\mathcal   X}_{I_{r+s}} +{\mathcal   X}_{I_{r+s+1}} 
&=
{\mathcal   X}_{\mathfrak{k} (I_{r+s})} + {\mathcal   X}_{\mathfrak{k} (I_{r+s+1})}, 
\label{Tinvt33}
\end{align}
which follows from the 3-term QQ-relation \eqref{QQt35}. 
All these relations  \eqref{Tinvt31}-\eqref{Tinvt33} 
are reductions of \eqref{Tinv}. Thus the T-function ${\mathsf F}_{(1)}^{I_{2r+2s}}$ 
on the symmetric nesting path 
is  $\mathfrak{W}$-invariant under the QQ-relations \eqref{QQt31}-\eqref{QQt35}. 
%%%%%%%%%

The condition that the generating function ${\mathbf W}_{I_{2r+2s}}({\mathbf X})$ 
 is invariant under $\mathfrak{s}$ and $\mathfrak{k}$, namely 
${\mathfrak{s}(\mathbf W}_{I_{2r+2s}}({\mathbf X}))={\mathbf W}_{\mathfrak{s}(I_{2r+2s})}({\mathbf X})={\mathbf W}_{I_{2r+2s}}({\mathbf X})$ and 
${\mathfrak{k}(\mathbf W}_{I_{2r+2s}}({\mathbf X}))={\mathbf W}_{\mathfrak{k}(I_{2r+2s})}({\mathbf X})={\mathbf W}_{I_{2r+2s}}({\mathbf X})$ 
is equivalent to the discrete zero curvature condition (a reduction of \eqref{ZCC}):
\begin{align}
\begin{split}
& (1-{\mathcal X}_{I_{a}}{\mathbf X})^{p_{i_{a}}} (1-{\mathcal X}_{I_{a+1}}{\mathbf X})^{p_{i_{a+1}}} 
=
 (1-{\mathcal X}_{\mathfrak{s}(I_{a}) }{\mathbf X})^{p_{i_{a+1}}} 
(1-{\mathcal X}_{\mathfrak{s}(I_{a+1})}{\mathbf X})^{p_{i_{a}}} ,
\\
 &(1-{\mathcal X}_{I_{2r+2s-a}}{\mathbf X})^{p_{i_{a+1}}} (1-{\mathcal X}_{I_{2r+2s+1-a}}{\mathbf X})^{p_{i_{a}}} 
=
\\
& \hspace{100pt} =
 (1-{\mathcal X}_{\mathfrak{s}(I_{2r+2s-a}) }{\mathbf X})^{p_{i_{a}}} 
(1-{\mathcal X}_{\mathfrak{s}(I_{2r+2s+1-a})}{\mathbf X})^{p_{i_{a+1}}} ,
\\
& (1-{\mathcal X}_{I_{r+s}}{\mathbf X})^{p_{i_{r+s}}}  
(1-{\mathcal X}_{I_{r+s+1}}{\mathbf X})^{p_{i_{r+s}}}
=
 (1-{\mathcal X}_{\mathfrak{k}(I_{r+s})}{\mathbf X})^{p_{i_{r+s}}}  (1-{\mathcal X}_{\mathfrak{k}(I_{r+s+1})}{\mathbf X})^{p_{i_{r+s}}} ,
\end{split}
\label{ZCCt3}
\end{align}
where $a \in \{1,2,\dots , r+s-1 \}$. 
These relations \eqref{ZCCB} boil down to \eqref{Tinvt31}-\eqref{Tinvt33}  and reductions
 of the identity \eqref{id-A}.  
Therefore the T-functions $ {\mathcal F}_{(b)}^{I_{2r+2s}}$ and $ {\mathcal F}_{(1^{b})}^{I_{2r+2s}}$ 
on the symmetric nesting paths 
are invariant under ${\mathfrak W}$ if the QQ-relations \eqref{QQt31}-\eqref{QQt35} are imposed.   
%The T-functions for the symmetric nesting paths form a closed system under $\mathfrak{W}$. 
One may be able to exclude the T- and Q-functions on the non-symmetric nesting paths 
from  consideration.
%%%%%%%%%%%%%%%%%
\subsection{Singular reductions}
\label{sec:SR}
In this subsection, we consider reductions for the case $\mathfrak{D} \ne \emptyset $. 
This case is more hypothetical than the regular reduction because it requires additional non-trivial ansatzes.  
In particular, the resultant Bethe ansatz equations are not always reductions of 
the ones for $U_{q}(gl(2r|2s+2)^{(1)})$. A part of \eqref{BAE} becomes singular under reductions and 
has to be excluded from our consideration. 
This suggests that 
%the naive correspondence 
not all the representations of  $U_{q}(osp(2r|2s)^{(1)})$  
are naive reductions of those of $U_{q}(gl(M|N)^{(1)})$.
\footnote{The relations  \eqref{facd1}-\eqref{facd3} 
(and also \eqref{facc01}, \eqref{facc03}) suggest that 
 the reductions of some (asymptotic) representations $W$ of  $U_{q}(gl(2r|2s+2)^{(1)})$  
 decompose into those $W_{1},W_{2}$ of  $U_{q}(osp(2r|2s)^{(1)})$: 
 $W \simeq  W_{1} \otimes W_{2}$ (on the level of the trace). 
%the correspondence may not be one to one.
In addition, there is a freedom to permute the elements of the set $\mathfrak{D}$.  
In this sense, the original system for $U_{q}(gl(2r|2s+2)^{(1)})$ splits into two systems for $U_{q}(osp(2r|2s)^{(1)})$ after reductions. 
We also remark that 
two copies of T-functions appear after reductions (see [Theorem 2.5, \cite{KOSY01}]).} 
%between the general linear superalgebra and orthosymplectic superalgebras is partially broken. 
%To what extent the correspondence holds is still not fully understood. 
To what extent the reductions work is still not fully understood. 
This subsection is our trial to understand this 
in the context of Bethe ansatz. We will present candidates of QQ-relations, 
which produce Bethe ansatz equations known in the literature. 
%%%%%%%%%%%
\subsubsection{$U_{q}(sp(2r)^{(1)})$ case}
Before we consider the $U_{q}(osp(2r|2s)^{(1)})$ case, we summarize the $U_{q}(sp(2r)^{(1)})$ case \cite{KOSY01} in our terminology, as warm-up. 
After reading the next subsection, one will realize that \cite{KOSY01} is 
  the tip of the iceberg.

We assume $r \in \mathbb{Z}_{\ge 1}$, and set
\begin{multline}
(M,N)=(2r+2,0), \quad \Bm=\{1,2,\dots , 2r+2 \}, \quad  \Fm=\emptyset, 
\\
\mathfrak{D}=\{r+1,r+2 \}, \quad \eta=0, \quad  
z_{r+1}=-z_{r+2}=1 .
\label{zz-c0}
\end{multline}

%%%%%%%%%%%
\paragraph{QQ-relations}
For a symmetric nesting path defined by 
$I_{2r+2}=(i_{1},i_{2},\dots, i_{r+1},  i_{r+1}^{*} \dots , i_{2}^{*}, i_{1}^{*})$,  $i_{r+1} \in \mathfrak{D}$, 
we consider additional reductions
\begin{align} 
%\begin{split}
\Qb_{I_{r}}&=\Qf_{I_{r}}^{[-1]}\Qf_{I_{r}}^{[1]}, 
\label{facc01}
\\
 (z_{i_{r}}+z_{i_{r+1}})\Qb_{\widetilde{I}_{r}}
&=z_{i_{r}}\Qf_{I_{r}}^{[1]}
\Qf_{\breve{I}_{r}}^{[-1]}+
z_{i_{r+1}} \Qf_{I_{r}}^{[-1]}
\Qf_{\breve{I}_{r}}^{[1]}
\label{facc02}
\\
\Qb_{I_{r+1}}&=(\Qf_{I_{r}})^{2},
%\end{split}
\label{facc03}
\end{align}
where $\breve{I}_{r}=(i_{1},i_{2},\dots , i_{r-1},i^{*}_{r}) $.  Eq.\ \eqref{facc01} and \eqref{facc03} for 
the case $(i_{1},i_{2},\dots , i_{r+1})=(1,2,\dots , r+1)$ correspond to [eq.\ (B.4) in \cite{KOSY01}]. 
Then the QQ-relations \eqref{QQb} along this symmetric nesting path reduce to
\begin{align}
& (z_{i_{a}}-z_{i_{a+1}})\Qb_{I_{a-1}}\Qb_{I_{a+1}}
=z_{i_{a}}\Qb_{I_{a}}^{[1]}
\Qb_{\widetilde{I}_{a}}^{[-1]}-
z_{i_{a+1}}\Qb_{I_{a}}^{[-1]}
\Qb_{\widetilde{I}_{a}}^{[1]}
\quad 
\text{for} \quad a \in \{1,2,\dots , r-2\},   
\label{QQsp1}  \\
& (z_{i_{r-1}}-z_{i_{r}})\Qb_{I_{r-2}}\Qf^{[-1]}_{I_{r}}\Qf^{[1]}_{I_{r}}
=z_{i_{r-1}}\Qb_{I_{r-1}}^{[1]}
\Qb_{\widetilde{I}_{r-1}}^{[-1]}-
z_{i_{r}}\Qb_{I_{r-1}}^{[-1]}
\Qb_{\widetilde{I}_{r-1}}^{[1]}
 .
\label{QQsp2} 
\\
& (z_{i_{r}}^{2}-1)\Qb_{I_{r-1}}
=z_{i_{r}}^{2}\Qf_{I_{r}}^{[2]}
\Qf_{\breve{I}_{r}}^{[-2]}-
\Qf_{I_{r}}^{[-2]}
\Qf_{\breve{I}_{r}}^{[2]}
 .
\label{QQsp3}  
\end{align}
Eqs.\ \eqref{QQsp1}, \eqref{QQsp2} and \eqref{QQsp3} are reductions of \eqref{QQb} for 
$I=I_{a-1}$, $(i,j)=(i_{a},i_{a+1})$ for $1 \le a \le r-2$, $a=r-1$ and $a=r$, respectively. In \eqref{QQsp3}, 
 $z_{i_{r+1}}^{2}=1$ is used. 
 Eqs.\ \eqref{facc02} and \eqref{QQsp1}, \eqref{QQsp2} and \eqref{QQsp3} for 
the case $(i_{1},i_{2},\dots , i_{r+1})=(1,2,\dots , r+1)$ correspond to [eqs.\ (7.26), (7.27) and (7.29) in \cite{DDMST06}]. 
Let $\{v^{I_{r}}_{k}\}_{k=1}^{m_{I_{k}}}$  be the zeros of the Q-function $\Qf_{I_{r}}$. Eq. \eqref{facc01}  means that 
$\{u^{I_{r}}_{k}\}_{k=1}^{n_{I_{k}}}=\{v^{I_{r}}_{k}-1\}_{k=1}^{m_{I_{k}}} \sqcup \{v^{I_{r}}_{k} +1 \}_{k=1}^{m_{I_{k}}}$, 
 $n_{I_{k}}=2m_{I_{k}}$ holds. 
 
%%%%%%
\paragraph{T-functions and Bethe ansatz equations} 
Under the reduction, \eqref{boxes} reduces to 
\begin{align}
\begin{split}
{\mathcal   X}_{I_{a}}&=
z_{i_{a}}
\frac{\Qb_{I_{a-1}}^{[2r+1-a]}
\Qb_{I_{a}}^{[2r+4-a]}
}{
\Qb_{I_{a-1}}^{[2r+3-a]}
\Qb_{I_{a}}^{[2r+2-a]}
} 
\quad \text{for} \quad 1 \le a \le r-1 ,
\\
{\mathcal   X}_{I_{r}}&=
z_{i_{r}}
\frac{ \Qb_{I_{r-1}}^{[r+1]} \Qf_{I_{r}}^{[r+5]}  
}{
\Qb_{I_{r-1}}^{[r+3]}
\Qf_{I_{r}}^{[r+1]} 
} ,
\\
{\mathcal   X}_{I_{r+1}}&=-{\mathcal   X}_{I_{r+2}}=
z_{i_{r+1}}
\frac{ \Qf_{I_{r}}^{[r-1]} \Qf_{I_{r}}^{[r+3]}  
}{
(\Qf_{I_{r}}^{[r+1]} )^{2}
} ,
\\
{\mathcal   X}_{I_{r+3}}&=
z_{i_{r}}^{-1}
\frac{ \Qb_{I_{r-1}}^{[r+1]} \Qf_{I_{r}}^{[r-3]}  
}{
\Qb_{I_{r-1}}^{[r-1]}
\Qf_{I_{r}}^{[r+1]} 
} ,
\\
{\mathcal   X}_{I_{2r+3-a}}&=
z_{i_{a}}^{-1}
\frac{\Qb_{I_{a-1}}^{[a+1]}
\Qb_{I_{a}}^{[a-2]}
}{
\Qb_{I_{a-1}}^{[a-1]}
\Qb_{I_{a}}^{[a]}
} 
\quad \text{for} \quad 1 \le a \le r-1 .
\end{split}
\label{boxes-zz-c0} 
\end{align}
The T-function \eqref{tab-fund} reduces to 
\begin{align}
{\mathsf F}_{(1)}^{I_{2r+2}}=
\Qb_{\emptyset}^{[2r+2]}
\Qb_{\emptyset}
\sum_{a=1}^{r}({\mathcal   X}_{I_{a}}+{\mathcal   X}_{I_{2r+3-a}})
 \label{tab-fund-c0}
\end{align}
Note that the terms ${\mathcal   X}_{I_{r+1}}$ and ${\mathcal   X}_{I_{r+2}}$ are missing in \eqref{tab-fund-c0} because of cancellation. 
The pole-free condition of the T-function \eqref{tab-fund-c0} 
produces the following Bethe ansatz equations: 
\begin{align}
%\begin{split}
 -1&=\frac{z_{i_{a}}}{z_{i_{a+1}}}
\frac{\Qb_{I_{a-1}}(u_{k}^{I_{a}}-1)
\Qb_{I_{a}}(u_{k}^{I_{a}}+2)
\Qb_{I_{a+1}}(u_{k}^{I_{a}}-1)} 
{\Qb_{I_{a-1}}(u_{k}^{I_{a}}+1)
\Qb_{I_{a}}(u_{k}^{I_{a}}-2)
\Qb_{I_{a+1}}(u_{k}^{I_{a}}+1)}
\nonumber \\
& \hspace{50pt} \text{for} \quad k\in \{1,2,\dots, n_{I_{a}}\} \quad \text{and} \quad a \in \{1,2,\dots, r-2 \} , \label{BAEc01}
\\
- 1&= \frac{z_{i_{r-1}}}{z_{i_{r}}} 
\frac{\Qb_{I_{r-2}}(u_{k}^{I_{r-1}}-1)  \Qb_{I_{r-1}}(u_{k}^{I_{r-1}}+2) 
\Qf_{I_{r}}(u_{k}^{I_{r-1}}-2)} 
{\Qb_{I_{r-2}}(u_{k}^{I_{r-1}}+1)  \Qb_{I_{r-1}}(u_{k}^{I_{r-1}}-2)
\Qf_{I_{r}}(u_{k}^{I_{r-1}}+2)} 
\nonumber \\
&\hspace{50pt}  \text{for} \quad k\in \{1,2,\dots, n_{I_{r-1}}\}  ,\label{BAEc02}
\\
- 1&=z_{i_{r}}^{2}
\frac{\Qb_{I_{r-1}}(v_{k}^{I_{r}}-2) 
\Qf_{I_{r}}(v_{k}^{I_{r}}+4)} 
{\Qb_{I_{r-1}}(v_{k}^{I_{r}}+2) 
\Qf_{I_{r}}(v_{k}^{I_{r}}-4)} 
%\\
%&\hspace{50pt} 
\quad 
 \text{for} \quad k\in \{1,2,\dots, m_{I_{r}}\}  .
%\end{split}
\label{BAEc03}
\end{align}
Eqs. \eqref{BAEc01} and \eqref{BAEc02} are reductions of \eqref{BAE} on the symmetric nesting path, while \eqref{BAEc03} 
is not. 
Eqs. \eqref{tab-fund-c0}-\eqref{BAEc03} agree with  the known results by analytic Bethe ansatz \cite{R87}. 
One can also derive the Bethe ansatz equations \eqref{BAEc01}-\eqref{BAEc03} from the QQ-relations \eqref{QQsp1}-\eqref{QQsp3} 
by considering the zeros of the Q-functions. 
The tableaux sum expression of the T-function \eqref{DVF-tab1} (for one row Young diagrams and one column Young diagrams) 
reproduces [eqs. (3.9), (3.17), \cite{KS94-1}]
\footnote{
The functions $\Qb_{\emptyset}$, 
$\Qb_{I_{b}}$ ($1 \le b \le r-1$), $\Qf_{I_{r}}$, 
$\Qb_{\emptyset}^{[2r+2]}\Qb_{\emptyset}{\mathcal   X}_{I_{a}}$  and 
$\Qb_{\emptyset}^{[2r+2]}\Qb_{\emptyset}{\mathcal   X}_{I_{2r+3-a}}$ 
correspond to $\phi(u)$, $Q_{b}(u)$ ($1 \le b \le r-1$),  $Q_{r}(u)$, $\boxed{\overline{a}}_{u}$ and $\boxed{a}_{u}$ in [eqs.\ (3.4a), (3.4b) for $p=1$, \cite{KS94-1}], 
where $1 \le a \le r$ (the unit of the shift of the spectral parameter in \cite{KS94-1} is half of the one in this paper). 
}
under the reduction. 

The generating functions \eqref{gene1} and \eqref{gene2} reduce to  the ones in  \cite{KOSY01}: 
\begin{align}
{\mathbf W}_{I_{2r+2}}({\mathbf X})&=
\overrightarrow{\prod_{a=1}^{r}} (1-{\mathcal X}_{I_{2r+3-a}}{\mathbf X})^{-1} 
 (1-{\mathcal X}_{I_{r}} \, {\mathcal X}^{[2]}_{I_{r+3}}{\mathbf X}^{2})^{-1}
\overleftarrow{\prod_{a=1}^{r}} (1-{\mathcal X}_{I_{a}}{\mathbf X})^{-1} 
\nonumber \\
&= \sum_{a=0}^{\infty} {\mathcal F}_{(a)}^{I_{2r+2}
[a -1]}{\mathbf X}^{a }, 
\label{gene1C}
\\
{\mathbf W}_{I_{2r+2}}({\mathbf X})^{-1}&=
\overrightarrow{\prod_{a=1}^{r}} (1-{\mathcal X}_{I_{a}}{\mathbf X}) 
 (1-{\mathcal X}_{I_{r}} \, {\mathcal X}^{[2]}_{I_{r+3}}{\mathbf X}^{2})
\overleftarrow{\prod_{a=1}^{r}} (1-{\mathcal X}_{I_{2r+3-a}}{\mathbf X})
\nonumber \\
&= \sum_{a=0}^{2r+2}(-1)^{a} {\mathcal F}_{(1^{a})}^{I_{2r+2}
[a -1]}{\mathbf X}^{a }. 
\label{gene2C}
\end{align}
Note that the terms ${\mathcal X}_{I_{r+2}}$ and ${\mathcal X}_{I_{r+1}}$ disappear from the 
formula  because of cancellation. By \eqref{DVF-tab1}, ${\mathcal F}_{(1^{a})}^{I_{2r+2}}=0$ if $a>2r+2$. 
Baxter type equations follow from the kernels of  \eqref{gene1C} and \eqref{gene2C}, which are 
reductions of \eqref{Bax3} and \eqref{Bax4}. 
%%%%%%%%%%%%%%

\paragraph{ $\mathfrak{W}$-symmetry}
We would like to consider a subgroup $\mathfrak{W}= \mathbb{Z}^{r}_{2} \rtimes S_{r}$ of the 
permutation group $ S(I_{M+N})=S(I_{2r+2})=S({\mathfrak I})$, 
which preserves the  entire set\footnote{We fix the elements of $\mathfrak{D}$.} of  
symmetric nesting paths, and discuss the invariance of
 the T-function ${\mathsf F}_{(1)}^{I_{2r+2}}$ under it. 
$\mathfrak{W}$ is generated by two kinds of operations of the form: 
$\mathfrak{s}=\overline{\tau}_{a,a+1} \circ \overline{\tau}_{2r+2-a,2r+3-a} $, 
$\mathfrak{s} ( I_{2r+2} )=\tau_{i_{a}i_{a+1}} \circ \tau_{i_{a}^{*} i_{a+1}^{*}} ( I_{2r+2} )=(i_{1},i_{2},\dots, i_{a-1},i_{a+1},i_{a},i_{a+2}, \dots, i_{r},i_{r+1},i_{r+1}^{*},  i_{r}^{*}, \dots ,
i_{a+2}^{*},i_{a}^{*},i_{a+1}^{*},i_{a-1}^{*},\dots , i_{2}^{*}, i_{1}^{*})$ 
for $a \in \{1,2,\dots , r-1 \}$,  
and 
$\mathfrak{k}=\overline{\tau}_{r,r+3} $, 
$\mathfrak{k} ( I_{2r+2})= \tau_{i_{r}i^{*}_{r}}( I_{2r+2})=(i_{1},i_{2},\dots, i_{r-1},i^{*}_{r}, i_{r+1}, i^{*}_{r+1},i_{r}, i_{r-1}^{*}, \dots , i_{2}^{*}, i_{1}^{*})$.
The condition 
$\mathfrak{s} ( {\mathsf F}_{(1)}^{I_{2r+2}})= {\mathsf F}_{(1)}^{\mathfrak{s} (I_{2r+2})}={\mathsf F}_{(1)}^{I_{2r+2}}$ 
is equivalent to the  8-term QQ-relations
\begin{align}
{\mathcal   X}_{I_{a}}+
{\mathcal   X}_{I_{a+1}} +
{\mathcal   X}_{I_{2r+3-a}}+
{\mathcal   X}_{I_{2r+2-a}} =
{\mathcal   X}_{\mathfrak{s}(I_{a})}+{\mathcal   X}_{\mathfrak{s}(I_{a+1})} 
+
{\mathcal   X}_{\mathfrak{s}(I_{2r+3-a})}
+{\mathcal   X}_{\mathfrak{s}(I_{2r+2-a})}.
\label{Tinvsp0}
\end{align}
The following set of 4-term QQ-relations is a
 sufficient condition for \eqref{Tinvsp0}:
\begin{align}
{\mathcal   X}_{I_{a}}+ {\mathcal   X}_{I_{a+1}} &=
{\mathcal   X}_{\mathfrak{s}(I_{a})}+{\mathcal   X}_{\mathfrak{s}(I_{a+1})}, 
\label{Tinvsp1}
\\
{\mathcal   X}_{I_{2r+3-a}}+{\mathcal   X}_{I_{2r+2-a}} &=
{\mathcal   X}_{\mathfrak{s}(I_{2r+3-a})}+{\mathcal   X}_{\mathfrak{s}(I_{2r+2-a})}.
\label{Tinvsp2} 
\end{align}
The relation \eqref{Tinvsp2} is equivalent to \eqref{Tinvsp1}, 
which follows from the 3-term QQ-relations \eqref{QQsp1} and \eqref{QQsp2}.
The condition 
$\mathfrak{k} ( {\mathsf F}_{(1)}^{I_{2r+2}})= {\mathsf F}_{(1)}^{\mathfrak{k} (I_{2r+2})}= {\mathsf F}_{(1)}^{I_{2r+2}}$ 
is equivalent to the following 4-term  QQ-relations 
\begin{align}
{\mathcal   X}_{I_{r}} +{\mathcal   X}_{I_{r+3}} 
&=
{\mathcal   X}_{\mathfrak{k} (I_{r})} + {\mathcal   X}_{\mathfrak{k} (I_{r+3})}, 
\label{Tinvsp3}
\end{align}
which follows from the 3-term QQ-relation \eqref{QQsp3}. The relations  \eqref{Tinvsp1} and \eqref{Tinvsp2} 
are reductions of \eqref{Tinv}, while the relation \eqref{Tinvsp3} is not. Thus the T-function ${\mathsf F}_{(1)}^{I_{2r+2}}$ 
on the symmetric nesting path 
is  $\mathfrak{W}$-invariant under the 3-term QQ-relations \eqref{QQsp1}-\eqref{QQsp3}. 
%%%%%%%%%

The condition that the generating function ${\mathbf W}_{I_{2r+2}}({\mathbf X})$ 
 is invariant under $\mathfrak{s}$, namely 
${\mathfrak{s}(\mathbf W}_{I_{2r+2}}({\mathbf X}))={\mathbf W}_{\mathfrak{s}(I_{2r+2})}({\mathbf X})={\mathbf W}_{I_{2r+2}}({\mathbf X})$,  
is equivalent to the discrete zero curvature condition (a reduction of \eqref{ZCC}):
\begin{align}
\begin{split}
& (1-{\mathcal X}_{I_{a}}{\mathbf X}) (1-{\mathcal X}_{I_{a+1}}{\mathbf X})
=
 (1-{\mathcal X}_{\mathfrak{s}(I_{a}) }{\mathbf X})
(1-{\mathcal X}_{\mathfrak{s}(I_{a+1})}{\mathbf X}),
\\
 &(1-{\mathcal X}_{I_{2r+2-a}}{\mathbf X}) (1-{\mathcal X}_{I_{2r+3-a}}{\mathbf X})
 =
 (1-{\mathcal X}_{\mathfrak{s}(I_{2r+2-a}) }{\mathbf X})
(1-{\mathcal X}_{\mathfrak{s}(I_{2r+3-a})}{\mathbf X}) ,
\end{split}
\label{ZCCsp1}
\end{align}
where $a \in \{1,2,\dots , r-1 \}$. 
These relations \eqref{ZCCsp1} boil down to \eqref{Tinvsp1} and \eqref{Tinvsp2}  and a reduction of the identity \eqref{id-A}.  
The condition that the generating function ${\mathbf W}_{I_{2r+2}}({\mathbf X})$ 
 is invariant under $\mathfrak{k}$, namely 
${\mathfrak{k}(\mathbf W}_{I_{2r+2}}({\mathbf X}))={\mathbf W}_{\mathfrak{k}(I_{2r+2})}({\mathbf X})={\mathbf W}_{I_{2r+2}}({\mathbf X})$, 
is equivalent to the following discrete zero curvature condition:
\begin{multline}
(1-{\mathcal X}_{I_{r}}{\mathbf X}) (1-{\mathcal X}_{I_{r}} \, {\mathcal X}^{[2]}_{I_{r+3}}{\mathbf X}^{2})
(1-{\mathcal X}_{I_{r+3}} {\mathbf X})
=
\\
=
 (1-{\mathcal X}_{\mathfrak{k}(I_{r})}{\mathbf X})
 (1-{\mathcal X}_{\mathfrak{k}(I_{r}) } \, {\mathcal X}^{[2]}_{\mathfrak{k}(I_{r+3})}{\mathbf X}^{2})
 (1-{\mathcal X}_{\mathfrak{k}(I_{r+3})}{\mathbf X}).
\label{ZCCsp2}
\end{multline}
Consider the expansion of \eqref{ZCCsp2} with respect to the non-negative powers of $ {\mathbf X}$.  
The coefficients of $ {\mathbf X}$ on both sides of \eqref{ZCCsp2} give the relation \eqref{Tinvsp3}, 
which follows from the QQ-relation \eqref{QQsp3}. The relation derived from the coefficients of $ {\mathbf X}^{3}$ 
also  follows from the QQ-relation \eqref{QQsp3}. The  relation derived from  $ {\mathbf X}^{4}$ 
is trivially valid.  
The other coefficients are $0$ or $1$. 
Therefore the T-functions $ {\mathcal F}_{(b)}^{I_{2r+2}}$ and $ {\mathcal F}_{(1^{b})}^{I_{2r+2}}$ 
on the symmetric nesting paths 
are invariant under ${\mathfrak W}$ if the QQ-relations \eqref{QQsp1}-\eqref{QQsp3} are imposed.   
%The T-functions for the symmetric nesting paths form a closed system under $\mathfrak{W}$. 
One may be able to exclude the T- and Q-functions on the non-symmetric nesting paths 
from consideration.
%%%%%%%%%%%%%%%%%
\subsubsection{$U_{q}(osp(2r|2s)^{(1)})$ case}\label{sec:QQD}
We assume $r,s \in \mathbb{Z}_{\ge 0}$, $r+s \ge 2$ or $(r,s)=(0,1)$, and set
\begin{multline}
(M,N)=(2r,2s+2), \quad \Bm=\{1,2,\dots , 2r \}, \quad  \Fm=\{2r+1,2r+2,\dots ,2r+2s+2\}, 
\\
\mathfrak{D}=\{2r+s+1,2r+s+2 \}, \quad \eta=0, \quad  
z_{2r+s+1}=-z_{2r+s+2}=1 .
\label{zz-d}
\end{multline}
In particular for $s=0$,  this reduces to the case  $U_{q}(so(2r)^{(1)})$. 
The case $r=0$ is parallel to the $U_{q}(sp(2s)^{(1)})$ case, but the role of row and column of 
Young diagrams has to be interchanged.
%%%%%%%%%%%
\paragraph{QQ-relations}
For a symmetric nesting path defined by 
$I_{2r+2s+2}=(i_{1},i_{2},\dots, i_{r+s+1},  i_{r+s+1}^{*} \dots , i_{2}^{*}, i_{1}^{*})$, 
$i_{r+s+1} \in \mathfrak{D}$, 
we consider additional reductions 
\begin{align} 
\Qb_{I_{r+s}}&=\Qf_{I_{r+s}}^{[-1]}\Qf_{I_{r+s}}^{[1]}, 
\label{facd1}
\\
\Qb_{I_{r+s+1}}&=(\Qf_{I_{r+s}})^{2} 
\label{facd1b} 
\end{align}
and
\begin{align} 
\Qb_{I_{r+s-1}}&=\Qf_{\breve{I}_{r+s}}\Qf_{I_{r+s}} \quad \text{if} \quad i_{r+s} \in \mathfrak{B}, 
\label{facd2}
\\
\Qb_{\widetilde{I}_{r+s-1}}&=\Qf_{\acute{I}_{r+s}}\Qf_{I_{r+s}} \quad \text{if} \quad i_{r+s-1} \in \mathfrak{B}, 
\label{facd3}
\\
 (z_{i_{r+s}}+z_{i_{r+s+1}})\Qb_{\widetilde{I}_{r+s}}
&=z_{i_{r+s}} \Qf_{I_{r+s}}^{[-1]} \Qf_{\breve{I}_{r+s}}^{[1]}+
z_{i_{r+s+1}} \Qf_{I_{r+s}}^{[1]} \Qf_{\breve{I}_{r+s}}^{[-1]} 
 \quad \text{if} \quad i_{r+s} \in \mathfrak{F},
\label{facd4}  
\end{align}
where $\widetilde{I}_{r+s}=(i_{1},i_{2},\dots, i_{r+s-1},i_{r+s+1})$, 
$\breve{I}_{r+s}=(i_{1},i_{2},\dots, i_{r+s-1},i^{*}_{r+s})$,  $\acute{I}_{r+s}=(i_{1},i_{2},\dots, i_{r+s-2},i^{*}_{r+s-1},i_{r+s})$. 
The relation \eqref{facd3}  is the same type relation as \eqref{facd2}
  on another symmetric nesting path defied by 
$\tau_{i_{r+s-1},i_{r+s}} \circ \tau_{i_{r+s-1}^{*},i_{r+s}^{*}} ( I_{2r+2s+2} )=(i_{1},i_{2},\dots, i_{r+s-2},i_{r+s},i_{r+s-1}, i_{r+s+1}, 
i_{r+s+1}^{*},i_{r+s-1}^{*},i_{r+s}^{*},i_{r+s-2}^{*}, \dots , i_{2}^{*}, i_{1}^{*})$, 
$i_{r+s+1} \in \mathfrak{D}$.
%, which follows from \eqref{facd2} by permuting $i_{r+s-1}$ and $i_{r+s}$ in case $i_{r+s-1} \in \Bm$.
QQ-relations can be interpreted as functional relations associated with Dynkin diagrams (see subsection \ref{Liesuper} and 
subsection \ref{sec:QQinv}). 
The QQ-relations \eqref{QQb} and \eqref{QQf} reduce to the following functional relations:
\\ \noindent
{\bf for $a$-th node ($1 \le a \le r+s-2$ for type C, $1 \le a \le r+s-3$ for type D):}
\begin{align}
& (z_{i_{a}}-z_{i_{a+1}})\Qb_{I_{a-1}}\Qb_{I_{a+1}}
=z_{i_{a}}\Qb_{I_{a}}^{[p_{i_{a}}]}
\Qb_{\widetilde{I}_{a}}^{[-p_{i_{a}}]}-
z_{i_{a+1}}\Qb_{I_{a}}^{[-p_{i_{a}}]}
\Qb_{\widetilde{I}_{a}}^{[p_{i_{a}}]}
\quad \text{if} \quad p_{i_{a}}=p_{i_{a+1}},  \quad \text{and} 
\nonumber \\
&
 \text{for} \; \; a \in \{1,2,\dots , r+s-3\} \; \; \text{if} \; \; i_{r+s} \in \Bm,   \quad 
 \text{for} \; \; a \in \{1,2,\dots , r+s-2\} \; \; \text{if} \; \; i_{r+s} \in \Fm
, 
\label{QQd1}  \\
& (z_{i_{a}}-z_{i_{a+1}})\Qb_{I_{a}}\Qb_{\widetilde{I}_{a}}
=z_{i_{a}}\Qb_{I_{a-1}}^{[-p_{i_{a}}]}
\Qb_{I_{a+1}}^{[p_{i_{a}}]}-
z_{i_{a+1}}\Qb_{I_{a-1}}^{[p_{i_{a}}]}
\Qb_{I_{a+1}}^{[-p_{i_{a}}]}
\quad \text{if} \quad p_{i_{a}}=-p_{i_{a+1}},  \quad \text{and} 
\nonumber \\
&
 \text{for} \; \; a \in \{1,2,\dots , r+s-3\} \; \; \text{if} \; \; i_{r+s} \in \Bm,   \quad 
 \text{for} \; \; a \in \{1,2,\dots , r+s-2\} \; \; \text{if} \; \; i_{r+s} \in \Fm ,
\label{QQd2}  
\end{align}
%%%%%%%%
{\bf for $(r+s-1)$-th node of type C:}
\begin{align}
& (z_{i_{r+s-1}}-z_{i_{r+s}})\Qb_{I_{r+s-2}}\Qf^{[-1]}_{I_{r+s}} \Qf^{[1]}_{I_{r+s}}
=
 z_{i_{r+s-1}}\Qb_{I_{r+s-1}}^{[-1]} \Qb_{\widetilde{I}_{r+s-1}}^{[1]}
 -
z_{i_{r+s}}\Qb_{I_{r+s-1}}^{[1]} \Qb_{\widetilde{I}_{r+s-1}}^{[-1]}
\nonumber \\
& \hspace{270pt} \text{if} \quad  i_{r+s-1},  i_{r+s} \in \Fm,  
\label{QQd7} 
\\
& (z_{i_{r+s-1}}-z_{i_{r+s}})\Qb_{I_{r+s-1}}\Qf_{\acute{I}_{r+s}}
=
 z_{i_{r+s-1}}\Qb_{I_{r+s-2}}^{[-1]} \Qf_{I_{r+s}}^{[2]}
 -
z_{i_{r+s}} \Qb_{I_{r+s-2}}^{[1]} \Qf_{I_{r+s}}^{[-2]}
\nonumber \\
& \hspace{270pt} \text{if} \quad  i_{r+s-1} \in \Bm,  \quad i_{r+s} \in \Fm,  
\label{QQd9}  
\end{align}
%%%%
{\bf for $(r+s)$-th node of type C:}
\begin{align}
& (z_{i_{r+s}}^{2}-1)\Qb_{I_{r+s-1}}
=
 z_{i_{r+s}}^{2}\Qf_{I_{r+s}}^{[-2]} \Qf_{\breve{I}_{r+s}}^{[2]}
 -
 \Qf_{I_{r+s}}^{[2]} \Qf_{\breve{I}_{r+s}}^{[-2]}
\quad  \text{if} \quad   i_{r+s} \in \Fm,  
\label{QQd11} 
\end{align}
%%%
{\bf for $(r+s-2)$-th node of type D:}
\begin{align}
& (z_{i_{r+s-2}}-z_{i_{r+s-1}})\Qb_{I_{r+s-3}} \Qf_{\breve{I}_{r+s}}\Qf_{I_{r+s}}
=
 z_{i_{r+s-2}}\Qb_{I_{r+s-2}}^{[p_{i_{r+s-2}}]}
\Qb_{\widetilde{I}_{r+s-2}}^{[-p_{i_{r+s-2}}]}
\nonumber \\
&
\hspace{70pt} -
z_{i_{r+s-1}}\Qb_{I_{r+s-2}}^{[-p_{i_{r+s-2}}]}
\Qb_{\widetilde{I}_{r+s-2}}^{[p_{i_{r+s-2}}]}
\quad \text{if} \quad p_{i_{r+s-2}}=p_{i_{r+s-1}}  \quad \text{and} \quad i_{r+s} \in \Bm,  
\label{QQd3}  \\
& (z_{i_{r+s-2}}-z_{i_{r+s-1}})\Qb_{I_{r+s-2}} \Qb_{\widetilde{I}_{r+s-2}} 
=
 z_{i_{r+s-2}}\Qb_{I_{r+s-3}}^{[-p_{i_{r+s-2}}]}
\Qf_{\breve{I}_{r+s}}^{[p_{i_{r+s-2}}]} \Qf_{I_{r+s}}^{[p_{i_{r+s-2}}]}
\nonumber \\
&
\hspace{10pt} -
z_{i_{r+s-1}}\Qb_{I_{r+s-3}}^{[p_{i_{r+s-2}}]}
 \Qf_{\breve{I}_{r+s}}^{[-p_{i_{r+s-2}}]} \Qf_{I_{r+s}}^{[-p_{i_{r+s-2}}]}
\quad \text{if} \quad p_{i_{r+s-2}}=-p_{i_{r+s-1}}  \; \; \;  \text{and} \; \; \;  i_{r+s} \in \Bm,  
\label{QQd4} 
\end{align}
%%%%
{\bf for $(r+s-1)$-th and $(r+s)$-th nodes of type D (simply laced):}
\begin{align}
& (z_{i_{r+s-1}}-z_{i_{r+s}})\Qb_{I_{r+s-2}}
=
 z_{i_{r+s-1}}\Qf_{\breve{I}_{r+s}}^{[1]} \Qf_{\acute{I}_{r+s}}^{[-1]}
 -
z_{i_{r+s}}\Qf_{\breve{I}_{r+s}}^{[-1]} \Qf_{\acute{I}_{r+s}}^{[1]}
\quad \text{if} \quad  i_{r+s-1}, i_{r+s} \in \Bm,  
\label{QQd5}  \\
& (z_{i_{r+s-1}}-z_{i_{r+s}}^{-1})\Qb_{I_{r+s-2}}
=
 z_{i_{r+s-1}}\Qf_{I_{r+s}}^{[1]} \Qf_{\grave{I}_{r+s}}^{[-1]}
 -
z_{i_{r+s}}^{-1}\Qf_{I_{r+s}}^{[-1]} \Qf_{\grave{I}_{r+s}}^{[1]}
\quad \text{if} \quad  i_{r+s-1}, i_{r+s} \in \Bm,  
\label{QQd6}  
\end{align}
%%%%%%%%
{\bf for $(r+s-1)$-th and $(r+s)$-th nodes of type D (non-simply laced):}
\begin{align}
& (z_{i_{r+s-1}}-z_{i_{r+s}})\Qb_{\widetilde{I}_{r+s-1}}\Qf_{\breve{I}_{r+s}}
=
 z_{i_{r+s-1}}\Qb_{I_{r+s-2}}^{[1]} \Qf_{I_{r+s}}^{[-2]}
 -
z_{i_{r+s}} \Qb_{I_{r+s-2}}^{[-1]} \Qf_{I_{r+s}}^{[2]}
\nonumber \\
& \hspace{270pt} \text{if} \quad  i_{r+s-1} \in \Fm, \quad i_{r+s} \in \Bm,  
\label{QQd8}  \\
& (z_{i_{r+s-1}}-z_{i_{r+s}}^{-1}) \Qb_{\dot{I}_{r+s-1}} \Qf_{I_{r+s}}
=
 z_{i_{r+s-1}}\Qb_{I_{r+s-2}}^{[1]} \Qf_{\breve{I}_{r+s}}^{[-2]}
 -
z_{i_{r+s}}^{-1} \Qb_{I_{r+s-2}}^{[-1]} \Qf_{\breve{I}_{r+s}}^{[2]}
\nonumber \\
& \hspace{270pt} \text{if} \quad  i_{r+s-1} \in \Fm, \quad  i_{r+s} \in \Bm,  
\label{QQd10}  
\end{align} 
where $\grave{I}_{r+s}=(i_{1},i_{2},\dots , i_{r+s-2},i^{*}_{r+s-1},i^{*}_{r+s})$, 
$\dot{I}_{r+s-1}=(i_{1},i_{2},\dots , i_{r+s-2},i^{*}_{r+s})$.
Eqs.\ \eqref{QQd1}, \eqref{QQd7}, \eqref{QQd11}, \eqref{QQd3} and \eqref{QQd5}  are reductions of \eqref{QQb} for 
$I=I_{a-1}$, $(i,j)=(i_{a},i_{a+1})$ for $1 \le a \le r+s-3$ (or $r+s-2$), $a=r+s-1$, $a=r+s$, $a=r+s-2$
 and $a=r+s-1$ respectively
\footnote{
The QQ-relations from \eqref{QQb} and \eqref{QQf} for $I=I_{a-1}$, $a\ge r+s+2$ basically
 reduce to the ones for $a \le r+s$. 
However, care must be taken if $i$ or $j$ is an element of $\mathfrak{D}$. 
This may lead to contradictions or over-constraints to the system. 
A remedy for this would be to set 
$(z_{2r+s+1},z_{2r+s+2})=(\sqrt{-1},-\sqrt{-1})$ or $(-\sqrt{-1},\sqrt{-1})$ 
since this satisfies $\sigma(z_{2r+s+1})=z_{2r+s+2}^{-1}=z_{2r+s+1}$. 
However this does not necessary give desirable T-functions beyond fundamental representations 
in the auxiliary space (the factor $1-{\mathcal X}_{I_{r+s+3}}{\mathbf X} \, {\mathcal X}_{I_{r+s}}{\mathbf X}$ 
in \eqref{gene1D}-\eqref{can-gen} becomes
 $1+{\mathcal X}_{I_{r+s+3}}{\mathbf X} \, {\mathcal X}_{I_{r+s}}{\mathbf X}$). 
The QQ-relation from \eqref{QQb} for $I=I_{r+s}$, $(i,j)=(i_{r+s+1},i_{r+s+1}^{*})$, namely 
$2( \Qb_{I_{r+s}})^{2} = \Qb_{I_{r+s}}^{[-1]} \Qb_{\widetilde{I}_{r+s}}^{[1]} + 
\Qb_{I_{r+s}}^{[1]} \Qb_{\widetilde{I}_{r+s}}^{[-1]}$  may also 
 over-constrain the system (in fact, $\{ i_{r+s+1},i_{r+s+1}^{*}\} \notin \mathcal{A}$). 
 %This is related to how much the correspondence between the general linear superalgebras 
%and the ortho-symplectic  superalgebras holds and at which point the correspondence breaks down. 
So we have to exclude unnecessary Q-functions and QQ-relations from the system. 
%We may also have to modify the reduction procedures. 
This point requires further research. 
}. 
Eq.  \eqref{QQd6} is a reduction of \eqref{QQb} for 
$I=I_{r+s-2}$, $(i,j)=(i_{r+s-1},i_{r+s}^{*})$. 
Eqs.\ \eqref{QQd2}, \eqref{QQd9}, \eqref{QQd4} and \eqref{QQd8} are reductions of \eqref{QQf} for 
$I=I_{a-1}$, $(i,j)=(i_{a},i_{a+1})$ for $1 \le a \le r+s-3$ (or $ r+s-2$), 
$a=r+s-1$, $a=r+s-2$, and $a=r+s-1$, respectively. 
Eq.  \eqref{QQd10} is a reduction of \eqref{QQf} for 
$I=I_{r+s-2}$, $(i,j)=(i_{r+s-1},i_{r+s}^{*})$. 
Eqs. \eqref{QQd5} and \eqref{QQd6} are of the same type. 
Eqs. \eqref{QQd8} and \eqref{QQd10} are also of the same type. 
They are related to each other by the permutation of $i_{r+s}$ and $i^{*}_{r+s}$, which corresponds to 
the symmetry of the Dynkin diagrams of type D with respect to their final pair of nodes. 
%%%%%%

For $ I \in \Am$ and mutually distinct 
$i,i^{*} \in \mathfrak{I}$ such that  
$I \sqcup I^{*} \sqcup \{i,i^{*}\} \sqcup \mathfrak{D}= \mathfrak{I}$, 
 \eqref{facd2} and \eqref{facd3} are summarized as
\begin{align}
 \Qb_{I}
&=
 \Qf_{I,i} \Qf_{I,i^{*}} \quad  \text{if} \quad  i \in \Bm .
\label{facd2s}  
\end{align}
Note that $|I|=r+s-1$ holds.
For $ I \in \Am$ and mutually distinct 
$ i,j,i^{*},j^{*} \in \mathfrak{I}$ such that  
$I \sqcup I^{*} \sqcup \{i,i^{*},j,j^{*}\} \sqcup \mathfrak{D}= \mathfrak{I}$, 
 \eqref{QQd5} and \eqref{QQd6} are summarized as
\begin{align}
 (z_{i}-z_{j})\Qb_{I }
&=
 z_{i}\Qf_{I,i,j^{*}}^{[1]} \Qf_{I,i^{*},j}^{[-1]}
 -
z_{j} \Qf_{I,i,j^{*}}^{[-1]} \Qf_{I,i^{*},j}^{[1]}\quad  \text{if}  \quad p_{i}=p_{j}=1,
\label{QQd5s}  
\end{align}
and \eqref{QQd9}, \eqref{QQd8} and \eqref{QQd10} are summarized as
\begin{align}
 (z_{i}-z_{j})\Qb_{I,i }\Qf_{I,i^{*},j}
&=
 z_{i}\Qb_{I}^{[-1]} \Qf_{I,i,j}^{[2]}
 -
z_{j} \Qb_{I}^{[1]} \Qf_{I,i,j}^{[-2]} \quad  \text{if}  \quad p_{i}=-p_{j}.
\label{QQd9s}  
\end{align}
Note that $|I|=r+s-2$ holds. Eqs. \eqref{facd2s} and \eqref{QQd5s} for $s=0$ corresponds to 
eqs. (5.4) and (5.8) in \cite{FFK20}, respectively.
%%%%
Let $\{v^{I_{r+s}}_{k}\}_{k=1}^{m_{I_{k}}}$  and 
$\{v^{\breve{I}_{r+s}}_{k}\}_{k=1}^{m_{\breve{I}_{k}}}$ be the zeros of the Q-functions $\Qf_{I_{r+s}}$ 
and $\Qf_{\breve{I}_{r+s}}$, respectively. Eq. \eqref{facd2}  means that 
$\{u^{I_{r+s-1}}_{k}\}_{k=1}^{n_{I_{k}}}=\{v^{I_{r+s}}_{k}\}_{k=1}^{m_{I_{k}}} \sqcup \{v^{\breve{I}_{r+s}}_{k}\}_{k=1}^{m_{\breve{I}_{k}}}$ holds if $i_{r+s} \in \mathfrak{B}$. 
We remark that QQ-relations related to  $osp(4|6)$ symmetry were discussed
\footnote{Eq. (7.9) in \cite{BCFGT17} looks like a reduction of \eqref{QQf} for $I=\emptyset $, 
 $i \in \Bm $, $j \in \mathfrak{D}$.}
 in \cite{BCFGT17} 
in the context of the quantum spectral curve for $AdS_{4}/CFT_{3}$.  
In particular, there is a room to reconsider the reduction procedures and 
 find QQ-relations corresponding to [eqs.\ (7.51), (7.52), \cite{BCFGT17}] 
 as substitutes of the QQ-relations \eqref{QQd9}, \eqref{QQd8}, \eqref{QQd10} and \eqref{QQd9s} (see also the last paragraph of this subsection). 

%%%%%%
\paragraph{T-functions and Bethe ansatz equations}
Under the reduction, \eqref{boxes} reduces to 
\begin{align}
%\begin{split}
{\mathcal   X}_{I_{a}}&=
z_{i_{a}}
\frac{\Qb_{I_{a-1}}^{[2r-2s-2-\sum_{j \in I_{a}}p_{j}-p_{i_{a}}]}
\Qb_{I_{a}}^{[2r-2s-2-\sum_{j \in I_{a}}p_{j}+2p_{i_{a}}]}
}{
\Qb_{I_{a-1}}^{[2r-2s-2-\sum_{j \in I_{a}}p_{j}+p_{i_{a}}]}
\Qb_{I_{a}}^{[2r-2s-2-\sum_{j \in I_{a}}p_{j}]}
}
\nonumber \\ 
& \qquad \qquad \text{for} \quad 
\begin{cases}
 1 \le a \le r+s-2 & \text{if} \quad i_{r+s} \in \mathfrak{B} 
\\
1 \le a \le r+s-1 & \text{if} \quad i_{r+s} \in \mathfrak{F},
\end{cases}
\nonumber  \\
{\mathcal   X}_{I_{r+s-1}}&=
z_{i_{r+s-1}}
\frac{ \Qb_{I_{r+s-2}}^{[r-s-1-p_{i_{r+s-1}}]}
\Qf_{\breve{I}_{r+s}}^{[r-s-1+2p_{i_{r+s-1}}]}    \Qf_{I_{r+s}}^{[r-s-1+2p_{i_{r+s-1}}]}
}{
\Qb_{I_{r+s-2}}^{[r-s-1+p_{i_{r+s-1}}]}
\Qf_{\breve{I}_{r+s}}^{[r-s-1]} \Qf_{I_{r+s}}^{[r-s-1]} 
} \quad \text{if} \quad i_{r+s} \in \mathfrak{B} ,
\nonumber  \\
{\mathcal   X}_{I_{r+s}}&=
\begin{cases}
z_{i_{r+s}}
\frac{ \Qf_{\breve{I}_{r+s}}^{[r-s-3]}
  \Qf_{I_{r+s}}^{[r-s+1]}   
}{
\Qf_{\breve{I}_{r+s}}^{[r-s-1]}
\Qf_{I_{r+s}}^{[r-s-1]} 
} 
& \text{if} \quad i_{r+s} \in \mathfrak{B} 
\\[12pt]
z_{i_{r+s}}
\frac{ \Qb_{I_{r+s-1}}^{[r-s-1]}
  \Qf_{I_{r+s}}^{[r-s-5]}   
}{
\Qb_{I_{r+s-1}}^{[r-s-3]}
\Qf_{I_{r+s}}^{[r-s-1]} 
} 
& \text{if} \quad i_{r+s} \in \mathfrak{F} ,
\end{cases}
\nonumber  \\
{\mathcal   X}_{I_{r+s+1}}&=-{\mathcal   X}_{I_{r+s+2}}=
z_{i_{r+s+1}}
\frac{ \Qf_{I_{r+s}}^{[r-s+1]} \Qf_{I_{r+s}}^{[r-s-3]}  
}{
(\Qf_{I_{r+s}}^{[r-s-1]} )^{2}
} ,
\nonumber \\
{\mathcal   X}_{I_{r+s+3}}&=
\begin{cases}
z_{i_{r+s}}^{-1}
\frac{ \Qf_{\breve{I}_{r+s}}^{[r-s+1]}
  \Qf_{I_{r+s}}^{[r-s-3]}   
}{
\Qf_{\breve{I}_{r+s}}^{[r-s-1]}
\Qf_{I_{r+s}}^{[r-s-1]} 
}
& \text{if} \quad i_{r+s} \in \mathfrak{B} 
\\[12pt] 
z_{i_{r+s}}^{-1}
\frac{ \Qb_{I_{r+s-1}}^{[r-s-1]}
  \Qf_{I_{r+s}}^{[r-s+3]}   
}{
\Qb_{I_{r+s-1}}^{[r-s+1]}
\Qf_{I_{r+s}}^{[r-s-1]} 
}
& \text{if} \quad i_{r+s} \in \mathfrak{F} ,
\end{cases}
\nonumber \\
{\mathcal   X}_{I_{r+s+4}}&=
z_{i_{r+s-1}}^{-1}
\frac{ \Qb_{I_{r+s-2}}^{[r-s-1+p_{i_{r+s-1}}]}
  \Qf_{\breve{I}_{r+s}}^{[r-s-1-2p_{i_{r+s-1}}]}    \Qf_{I_{r+s}}^{[r-s-1-2p_{i_{r+s-1}}]}
}{
\Qb_{I_{r+s-2}}^{[r-s-1-p_{i_{r+s-1}}]}
\Qf_{\breve{I}_{r+s}}^{[r-s-1]} \Qf_{I_{r+s}}^{[r-s-1]} 
} \quad \text{if} \quad i_{r+s} \in \mathfrak{B}  ,
\nonumber \\
{\mathcal   X}_{I_{2r+2s+3-a}}&=
z_{i_{a}}^{-1}
\frac{\Qb_{I_{a-1}}^{[\sum_{j \in I_{a}}p_{j}+p_{i_{a}}]}
\Qb_{I_{a}}^{[\sum_{j \in I_{a}}p_{j}-2p_{i_{a}}]}
}{
\Qb_{I_{a-1}}^{[\sum_{j \in I_{a}}p_{j}-p_{i_{a}}]}
\Qb_{I_{a}}^{[\sum_{j \in I_{a}}p_{j} ]}
} 
\nonumber \\ 
& \qquad \qquad \text{for} \quad 
\begin{cases}
 1 \le a \le r+s-2 & \text{if} \quad i_{r+s} \in \mathfrak{B} 
\\
1 \le a \le r+s-1 & \text{if} \quad i_{r+s} \in \mathfrak{F}.
\end{cases}
%\end{split}
\label{boxes-zz-d} 
\end{align}
The T-function \eqref{tab-fund} reduces to 
\begin{align}
{\mathsf F}_{(1)}^{I_{2r+2s+2}}=
\Qb_{\emptyset}^{[2r-2s-2]}
\Qb_{\emptyset}
\sum_{a=1}^{r+s}p_{i_{a}}({\mathcal   X}_{I_{a}}+{\mathcal   X}_{I_{2r+2s+3-a}}) .
 \label{tab-fund-d}
\end{align}
The pole-free condition of the T-function \eqref{tab-fund-d} 
produces the following Bethe ansatz equations: 
\\
\noindent
{\bf for $a$-th node ($1 \le a \le r+s-2$ for type C, $1 \le a \le r+s-3$ for type D):}
\begin{align}
 -1&=\frac{p_{i_{a}}z_{i_{a}}}{p_{i_{a+1}}z_{i_{a+1}}}
\frac{\Qb_{I_{a-1}}(u_{k}^{I_{a}}-p_{i_{a}})
\Qb_{I_{a}}(u_{k}^{I_{a}}+2p_{i_{a}})
\Qb_{I_{a+1}}(u_{k}^{I_{a}}-p_{i_{a+1}})} 
{\Qb_{I_{a-1}}(u_{k}^{I_{a}}+p_{i_{a}})
\Qb_{I_{a}}(u_{k}^{I_{a}}-2p_{i_{a+1}})
\Qb_{I_{a+1}}(u_{k}^{I_{a}}+p_{i_{a+1}})}
 \quad \text{for} \ k\in \{1,2,\dots, n_{I_{a}}\} 
\nonumber \\
&\ \text{and} \quad a \in \{1,2,\dots, r+s-3 \}
\ \  \text{if} \ i_{r+s} \in \Bm  ,
\quad a \in \{1,2,\dots, r+s-2 \}
\ \ \text{if} \ i_{r+s} \in \Fm  ;
\label{BAEd1}
\end{align}
{\bf for $(r+s-1)$-th node of type C:}
\begin{align}
 1&= \frac{p_{i_{r+s-1}}z_{i_{r+s-1}}}{z_{i_{r+s}}} 
\frac{\Qb_{I_{r+s-2}}(u_{k}^{I_{r+s-1}}-p_{i_{r+s-1}})  \Qb_{I_{r+s-1}}(u_{k}^{I_{r+s-1}}+2p_{i_{r+s-1}}) 
\Qf_{I_{r+s}}(u_{k}^{I_{r+s-1}}+2)} 
{\Qb_{I_{r+s-2}}(u_{k}^{I_{r+s-1}}+p_{i_{r+s-1}})  \Qb_{I_{r+s-1}}(u_{k}^{I_{r+s-1}}+2)
\Qf_{I_{r+s}}(u_{k}^{I_{r+s-1}}-2)} 
\nonumber  \\
&\hspace{50pt}  \text{for} \quad k\in \{1,2,\dots, n_{I_{r+s-1}}\} \quad \text{if} \quad i_{r+s} \in \Fm ,
\end{align}
{\bf for $(r+s)$-th node of type C:}
\begin{align}
- 1&=z_{i_{r+s}}^{2}
\frac{\Qb_{I_{r+s-1}}(v_{k}^{I_{r+s}}+2) 
\Qf_{I_{r+s}}(v_{k}^{I_{r+s}}-4)} 
{\Qb_{I_{r+s-1}}(v_{k}^{I_{r+s}}-2) 
\Qf_{I_{r+s}}(v_{k}^{I_{r+s}}+4)} 
%\\
%&\hspace{50pt} 
\quad 
 \text{for} \quad k\in \{1,2,\dots, m_{I_{r+s}}\} \quad \text{if} \quad i_{r+s} \in \Fm  ,
\end{align}
%\nonumber \\
{\bf for $(r+s-2)$-th node of type D:}
\begin{align}
- 1&= \frac{p_{i_{r+s-2}}z_{i_{r+s-2}}}{p_{i_{r+s-1}}z_{i_{r+s-1}}} 
\frac{\Qb_{I_{r+s-3}}(u_{k}^{I_{r+s-2}}-p_{i_{r+s-2}}) \Qb_{I_{r+s-2}}(u_{k}^{I_{r+s-2}}+2p_{i_{r+s-2}}) } 
{\Qb_{I_{r+s-3}}(u_{k}^{I_{r+s-2}}+p_{i_{r+s-2}}) \Qb_{I_{r+s-2}}(u_{k}^{I_{r+s-2}}-2p_{i_{r+s-1}})} \times
\nonumber \\
& \hspace{50pt} \times \frac{
 \Qf_{\breve{I}_{r+s}}(u_{k}^{I_{r+s-2}}-p_{i_{r+s-1}}) 
\Qf_{I_{r+s}}(u_{k}^{I_{r+s-2}}-p_{i_{r+s-1}})} 
{ \Qf_{\breve{I}_{r+s}}(u_{k}^{I_{r+s-2}}+p_{i_{r+s-1}})
\Qf_{I_{r+s}}(u_{k}^{I_{r+s-2}}+p_{i_{r+s-1}})} 
\nonumber \\ &
\hspace{50pt}  \text{for} \quad k\in \{1,2,\dots, n_{I_{r+s-2}}\} \quad \text{if} \quad i_{r+s} \in \Bm ,
\end{align}
{\bf for $(r+s-1)$-th node of type D:}
\begin{align}
- 1&= \frac{p_{i_{r+s-1}}z_{i_{r+s-1}}}{z_{i_{r+s}}} 
\frac{\Qb_{I_{r+s-2}}(v_{k}^{\breve{I}_{r+s}}-p_{i_{r+s-1}})  \Qf_{\breve{I}_{r+s}}(v_{k}^{\breve{I}_{r+s}}+2p_{i_{r+s-1}}) 
\Qf_{I_{r+s}}(v_{k}^{\breve{I}_{r+s}}+2p_{i_{r+s-1}})} 
{\Qb_{I_{r+s-2}}(v_{k}^{\breve{I}_{r+s}}+p_{i_{r+s-1}})  \Qf_{\breve{I}_{r+s}}(v_{k}^{\breve{I}_{r+s}}-2)
\Qf_{I_{r+s}}(v_{k}^{\breve{I}_{r+s}}+2)} 
\nonumber  \\
&\hspace{50pt}  \text{for} \quad k\in \{1,2,\dots, m_{\breve{I}_{r+s}}\} \quad \text{if} \quad i_{r+s} \in \Bm ,
\end{align}
{\bf for $(r+s)$-th node of type D:}
\begin{align}
- 1&= \frac{z_{i_{r+s-1}}z_{i_{r+s}}}{p_{i_{r+s-1}}} 
\frac{\Qb_{I_{r+s-2}}(v_{k}^{I_{r+s}}-p_{i_{r+s-1}})  \Qf_{\breve{I}_{r+s}}(v_{k}^{I_{r+s}}-2) 
\Qf_{I_{r+s}}(v_{k}^{I_{r+s}}+2)} 
{\Qb_{I_{r+s-2}}(v_{k}^{I_{r+s}}+p_{i_{r+s-1}})  \Qf_{\breve{I}_{r+s}}(v_{k}^{I_{r+s}}-2 p_{i_{r+s-1}} )
\Qf_{I_{r+s}}(v_{k}^{I_{r+s}}-2p_{i_{r+s-1}})} 
\nonumber  \\
&\hspace{50pt}  \text{for} \quad k\in \{1,2,\dots, m_{I_{r+s}}\} \quad \text{if} \quad i_{r+s} \in \Bm .
\label{BAEd6}
\end{align}
The Bethe ansatz equations \eqref{BAEd1}-\eqref{BAEd6}
 for the Bethe roots $\{u_{k}^{I_{a}}\}$ for $1 \le a \le r+s-1$ 
are reductions of \eqref{BAE} on the symmetric nesting path, while the ones for the Bethe roots
 $\{v_{k}^{I_{r+s}}\}$ and  $\{v_{k}^{\breve{I}_{r+s}}\}$ are not.
Eqs. \eqref{tab-fund-d} and \eqref{BAEd1}-\eqref{BAEd6} 
agree with the known results by algebraic Bethe ansatz in case $i_{k} \in \Bm$ for $s+1 \le k \le r+s$  
and $i_{k} \in \Fm$ for $1 \le k \le s$ \cite{GM04}; 
and  in case  $r=1$ 
%, $i_{1} \in \Bm$ and $i_{k} \in \Fm$ for $2 \le k \le s+2$ 
\cite{GM06}. 
Note that the terms ${\mathcal   X}_{I_{r+s+1}}$ and ${\mathcal   X}_{I_{r+s+2}}$ are missing 
in \eqref{tab-fund-d} because of cancellation.
The tableaux sum expression of the T-function \eqref{DVF-tab1} 
(for one row Young diagrams and one column Young diagrams)
reproduces [eq.\ (3.38), \cite{T99}]
\footnote{
The functions $\Qb_{\emptyset}$, 
$\Qb_{I_{b}}$  ($1 \le b \le r+s-2$), $\Qf_{\breve{I}_{r+s}}$, $\Qf_{I_{r+s}}$, 
$\Qb_{\emptyset}^{[2r-2s-2]}\Qb_{\emptyset}{\mathcal   X}_{I_{a}}$ and 
$\Qb_{\emptyset}^{[2r-2s-2]}\Qb_{\emptyset}{\mathcal   X}_{I_{2r+2s+3-a}}$ 
correspond to $\phi(u)$, $Q_{b}(u)$  ($1 \le b \le r+s-2$), $Q_{r+s-1}(u)$, $Q_{r+s}(u)$, 
$\boxed{\overline{a}}_{u}$ and $\boxed{a}_{u}$ in [eqs.\ (3.14), (3.16) \cite{T99}], 
where $1 \le a \le r+s$. 
}
under the reduction. 
As for the case $r=1$, 
 see [eqs.\ (3.18), (3.27) \cite{T99-2}]
\footnote{Set $r=1$. 
The functions $\Qb_{\emptyset}$, 
$\Qb_{I_{b}}$ ($1 \le b \le s$), $\Qf_{I_{s+1}}$, 
$\Qb_{\emptyset}^{[-2s]}\Qb_{\emptyset}{\mathcal   X}_{I_{a}}$ and 
$\Qb_{\emptyset}^{[-2s]}\Qb_{\emptyset}{\mathcal   X}_{I_{2s+5-a}}$ 
correspond to $\phi(u)$, $Q_{b}(u)$ ($1 \le b \le s$), $Q_{s+1}(u)$, 
$\boxed{\overline{a}}_{u}$ and $\boxed{a}_{u}$ in [eqs.\ (3.9), (3.10) \cite{T99-2}], 
where $1 \le a \le s+1$ (one has to set $s \to s+1$ in \cite{T99-2};  
the unit of the shift of the spectral parameter in \cite{T99-2} is half of the one in this paper). 
}.
Moreover,  $\mathsf{T}^{\Bm,\Fm}_{\mu}$ (from \eqref{unnor-t1}) 
and its (super)character limit $\zeta(\mathsf{T}^{\Bm,\Fm}_{\mu})$ 
give a Wronskian expression of the T-function and 
a Weyl-type supercharacter formula respectively after  reduction. 
The Young diagram $\mu$ is related to the labels of the representation through 
\eqref{HW-D}-\eqref{YW-Dm} or \eqref{HW-Cp}-\eqref{YW-Cp}. 
These formulas seem not provide T-functions for irreducible representations 
in the auxiliary space in the general situation. 
 Nevertheless, by investigating the Bethe strap, one can gain some insights 
 on irreducibility (see section \ref{sec:BS}).

The generating functions \eqref{gene1} and \eqref{gene2} reduce to  
\begin{align}
{\mathbf W}_{I_{2r+2s+2}}({\mathbf X})&=
\overrightarrow{\prod_{a=1}^{r+s}} (1-{\mathcal X}_{I_{2r+2s+3-a}}{\mathbf X})^{-p_{i_{a}}} 
 (1-{\mathcal X}_{I_{r+s+3}}{\mathbf X} \, {\mathcal X}_{I_{r+s}}{\mathbf X})
\overleftarrow{\prod_{a=1}^{r+s}} (1-{\mathcal X}_{I_{a}}{\mathbf X})^{-p_{i_{a}}} 
\nonumber \\
&= \sum_{a=0}^{\infty} {\mathcal F}_{(a)}^{I_{2r+2s+2}
[a -1]}{\mathbf X}^{a }, 
\label{gene1D}
\\
{\mathbf W}_{I_{2r+2s+2}}({\mathbf X})^{-1}&=
\overrightarrow{\prod_{a=1}^{r+s}} (1-{\mathcal X}_{I_{a}}{\mathbf X})^{p_{i_{a}}} 
 (1-{\mathcal X}_{I_{r+s+3}}{\mathbf X} \, {\mathcal X}_{I_{r+s}}{\mathbf X})^{-1}
\overleftarrow{\prod_{a=1}^{r+s}} (1-{\mathcal X}_{I_{2r+2s+3-a}}{\mathbf X})^{p_{i_{a}}} 
\nonumber \\
&= \sum_{a=0}^{\infty}(-1)^{a} {\mathcal F}_{(1^{a})}^{I_{2r+2s+2}
[a -1]}{\mathbf X}^{a }, 
\label{gene2D}
\end{align}
where the following relations (follow from \eqref{boxes-zz-d}) are used
\begin{multline}
(1-{\mathcal X}_{I_{r+s+2}}{\mathbf X})(1-{\mathcal X}_{I_{r+s+1}}{\mathbf X})
=
1-({\mathcal X}_{I_{r+s+2}} +{\mathcal X}_{I_{r+s+1}}){\mathbf X}
+{\mathcal X}_{I_{r+s+2}}{\mathbf X} \, {\mathcal X}_{I_{r+s+1}}{\mathbf X}=
\\
=1+{\mathcal X}_{I_{r+s+2}} \, {\mathcal X}^{[2]}_{I_{r+s+1}}{\mathbf X}^{2}
=1-\frac{ \Qf_{I_{r+s}}^{[r-s-3]} \Qf_{I_{r+s}}^{[r-s+3]}   
}{
\Qf_{I_{r+s}}^{[r-s-1]} \Qf_{I_{r+s}}^{[r-s+1]} 
}{\mathbf X}^{2}
%\\
%
=
1-{\mathcal X}_{I_{r+s+3}}{\mathbf X} \, {\mathcal X}_{I_{r+s}}{\mathbf X}.
 \label{can-gen}
\end{multline}
Note that the terms ${\mathcal X}_{I_{r+s+2}}$ and ${\mathcal X}_{I_{r+s+1}}$ disappear from the 
formula irrespective of $i_{r+s}$ because of cancellation. 
In this way, we recover [eqs. (B.4) and (B.3) in \cite{T99}] (see also [eq. (2.9) in \cite{TK96}] for the case $s=0$ case)
\footnote{In \cite{T99}, we considered only the formulas for the distinguished Dynkin diagram, while the formulas here 
are the ones for general Dynkin diagrams. 
For comparison, use an identity \eqref{AB-id}.}. 
Baxter type equations follow from the kernels of  \eqref{gene1D} and \eqref{gene2D}, which are 
reductions of \eqref{Bax3} and \eqref{Bax4}. 
%%%%%%%%%%%%%%

The relation \eqref{re2-con} reduces to 
\begin{align}
\hat{\Ts}^{\Bm,\Fm}_{a,-m-2r+2s+2}&= \hat{\Ts}^{\Bm,\Fm}_{a,m}
\quad \text{for any} \ m \in \mathbb{C} \quad \text{for} \quad \eqref{re2}.
\label{re2-con-d}
\end{align}
We remark that \eqref{re2-con-d} for $a=1$ and $s=0$ (and $q=1$) corresponds to the Yangian $Y(so(2r))$ case of [Proposition 7.58 in \cite{FKT21}]. 
%%%%%%%
Let us write down $a=1$ case of the Wronskian-type formula \eqref{re2}. 
\begin{align}
\Ts^{\Bm,\Fm}_{1,m} &=   
\sum_{i =1}^{2r} 
\frac{ z_{i}^{m-3+2r-2s}
\prod_{f =2r+1 }^{2r+2s+2} (z_{i}-z_{f})}
{\prod_{j=1  \atop j \ne i}^{2r} (z_{i}-z_{j}) }  
\Qb^{[m+2r-2s-2]}_{i}  
\Qb^{[-m]}_{i^{*}}  
\nonumber \\
&=   
\sum_{i =1}^{r}
\left(
\chi_{i}^{+} 
\Qb^{[m+2r-2s-2]}_{i}  
\Qb^{[-m]}_{i^{*}}  
+
\chi_{i^{*}}^{+} 
\Qb^{[m+2r-2s-2]}_{i^{*}}  
\Qb^{[-m]}_{i}  
\right) 
\quad 
\text{for} \quad  2s-2r+3 \le m  , \label{re2a=1d}  
\end{align} 
where the character parts are given by
\begin{align}
\chi_{i}^{+} &=
\frac{ z_{i}^{m-3+2r-2s}
(z_{i}-1)(z_{i}+1)\prod_{f =2r+1 }^{2r+s} (z_{i}-z_{f})(z_{i}-z_{f}^{-1})}
{\prod_{j=1}^{i-1} (z_{i}-z_{j}) \prod_{j=i+1}^{r} (z_{i}-z_{j})
\prod_{j=1}^{r} (z_{i}-z_{j}^{-1})} 
\nonumber
\\
&=
\frac{(-1)^{i-1} z_{i}^{m+i-1}\prod_{j =1 }^{i-1}z_{j}^{-1}
\prod_{f =2r+1 }^{2r+s} (1-\frac{z_{f}}{z_{i}})(1-\frac{1}{z_{i}z_{f}})}
{\prod_{j=1}^{i-1} (1-\frac{z_{i}}{z_{j}}) \prod_{j=i+1}^{r} (1-\frac{z_{j}}{z_{i}})
\prod_{j=1 \atop j \ne i}^{r} (1-\frac{1}{z_{i}z_{j}})} 
\quad \text{for} \quad 1 \le i \le r,
\label{chp1d}
\\
\chi_{i^{*}}^{+} &=
\frac{(-1)^{i-1} z_{i}^{-m+i-2r+2s+1}\prod_{j =1 }^{i-1}z_{j}^{-1}
\prod_{f =2r+1 }^{2r+s} (1-\frac{z_{f}}{z_{i}})(1-\frac{1}{z_{i}z_{f}})}
{\prod_{j=1}^{i-1} (1-\frac{z_{i}}{z_{j}}) \prod_{j=i+1}^{r} (1-\frac{z_{j}}{z_{i}})
\prod_{j=1 \atop j \ne i}^{r} (1-\frac{1}{z_{i}z_{j}})} 
\quad \text{for} \quad 1 \le i \le r.
\label{chp2d}
\end{align}
We remark that \eqref{chp1d} and \eqref{chp2d} for $s=0$ (and $q=1$)  coincide with 
 the Yangian $Y(so(2r))$ case of [eq.\ (9.25) in \cite{FKT21}] and that \eqref{re2a=1d} for $s=0$  corresponds
\footnote{Compare $\Ts^{\Bm,\Fm [-r]}_{1,m}$ for $s=0$ with [eq. (9.27) in \cite{FKT21}]. 
In our convention, the unit of shift of the spectral parameter is twice as large as theirs. Their parameters $\tau_{j}$ 
correspond to our parameters $z_{j}$. 
The sign factor $(-1)^{i-1}$ is included in the character parts \eqref{chp1d} and \eqref{chp2d}.}
  to [eq. (9.27) in \cite{FKT21}]. 

%%%%%%%%%%%%%%

\paragraph{ $\mathfrak{W}$-symmetry}
We would like to consider a subgroup $\mathfrak{W}= \mathbb{Z}^{r+s}_{2} \rtimes S_{r+s}$ of the 
permutation group $ S(I_{M+N})=S(I_{2r+2s+2})=S({\mathfrak I})$, 
which preserves the set of the entire\footnote{We fix the elements of $\mathfrak{D}$.} 
symmetric nesting paths, and discuss the invariance of
 the T-function ${\mathsf F}_{(1)}^{I_{2r+2s+2}}$ under it. 
$\mathfrak{W}$ is generated by two kinds of operations of the form: 
$\mathfrak{s}=\overline{\tau}_{a,a+1} \circ \overline{\tau}_{2r+2s+2-a,2r+2s+3-a} $, 
$\mathfrak{s} ( I_{2r+2s+2} )=\tau_{i_{a}i_{a+1}} \circ \tau_{i_{a}^{*} i_{a+1}^{*}} ( I_{2r+2s+2} )=(i_{1},i_{2},\dots, i_{a-1},i_{a+1},i_{a},i_{a+2}, \dots, i_{r+s},i_{r+s+1},i_{r+s+1}^{*},  i_{r+s}^{*}, \dots ,
i_{a+2}^{*},i_{a}^{*},i_{a+1}^{*},i_{a-1}^{*},\dots , i_{2}^{*}, i_{1}^{*})$ 
for $a \in \{1,2,\dots , r+s-1 \}$,  
and 
$\mathfrak{k}=\overline{\tau}_{r+s,r+s+3} $,   
$\mathfrak{k} ( I_{2r+2s+2})=\tau_{i_{r+s}, i^{*}_{r+s}} ( I_{2r+2s+2})=(i_{1},i_{2},\dots, i_{r+s-1},i^{*}_{r+s}, i_{r+s+1}, i^{*}_{r+s+1},i_{r+s}, i_{r+s-1}^{*}, \dots , i_{2}^{*}, i_{1}^{*})$.
The condition 
$\mathfrak{s} ( {\mathsf F}_{(1)}^{I_{2r+2s+2}})= {\mathsf F}_{(1)}^{\mathfrak{s} (I_{2r+2s+2})}={\mathsf F}_{(1)}^{I_{2r+2s+2}}$ 
is equivalent to the  8-term QQ-relations
\begin{multline}
p_{i_{a}}{\mathcal   X}_{I_{a}}+
p_{i_{a+1}}{\mathcal   X}_{I_{a+1}} +
p_{i_{a}}{\mathcal   X}_{I_{2r+2s+3-a}}+
p_{i_{a+1}}{\mathcal   X}_{I_{2r+2s+2-a}} =
\\
p_{i_{a+1}}{\mathcal   X}_{\mathfrak{s}(I_{a})}+p_{i_{a}}{\mathcal   X}_{\mathfrak{s}(I_{a+1})} 
+
p_{i_{a+1}}{\mathcal   X}_{\mathfrak{s}(I_{2r+2s+3-a})}+p_{i_{a}}{\mathcal   X}_{\mathfrak{s}(I_{2r+2s+2-a})}.
\label{TinvD0}
\end{multline}
The following set of 4-term QQ-relations is a
 sufficient condition for \eqref{TinvD0}:
\begin{align}
p_{i_{a}}{\mathcal   X}_{I_{a}}+ p_{i_{a+1}} {\mathcal   X}_{I_{a+1}} &=
p_{i_{a+1}} {\mathcal   X}_{\mathfrak{s}(I_{a})}+p_{i_{a}} {\mathcal   X}_{\mathfrak{s}(I_{a+1})}, 
\label{TinvD1}
\\
p_{i_{a}} {\mathcal   X}_{I_{2r+2s+3-a}}+p_{i_{a+1}} {\mathcal   X}_{I_{2r+2s+2-a}} &=
p_{i_{a+1}} {\mathcal   X}_{\mathfrak{s}(I_{2r+2s+3-a})}+p_{i_{a}} {\mathcal   X}_{\mathfrak{s}(I_{2r+2s+2-a})}.
\label{TinvD2} 
\end{align}
The relation \eqref{TinvD2} is equivalent to \eqref{TinvD1}, 
which follows
\footnote{As an example, let us consider the 4-term QQ-relation 
\eqref{TinvD1} for $a=r+s-1$, $i_{r+s-1} \in \Fm$, 
$i_{r+s} \in \Bm$, which is equivalent to 
\begin{align}
\left(
\frac{
 z_{i_{r+s-1}}\Qb_{I_{r+s-2}}^{[1]} \Qf_{I_{r+s}}^{[-2]}
 -
z_{i_{r+s}} \Qb_{I_{r+s-2}}^{[-1]} \Qf_{I_{r+s}}^{[2]}
}{\Qb_{\widetilde{I}_{r+s-1}}\Qf_{\breve{I}_{r+s}} }
\right)^{[2]}
=
\frac{
 z_{i_{r+s-1}}\Qb_{I_{r+s-2}}^{[1]} \Qf_{I_{r+s}}^{[-2]}
 -
z_{i_{r+s}} \Qb_{I_{r+s-2}}^{[-1]} \Qf_{I_{r+s}}^{[2]}
}{\Qb_{\widetilde{I}_{r+s-1}}\Qf_{\breve{I}_{r+s}} }
 \label{QQratD1}
\end{align}
This means that the right hand side of  \eqref{QQratD1} is a periodic function $\phi $ of 
the spectral parameter: $\phi^{[2]}=\phi$.
The 3-term QQ-relation \eqref{QQd8} corresponds to the case that this 
periodic function is a constant $\phi= z_{i_{r+s-1}}-z_{i_{r+s}}$. This comes from the assumption that 
the Q-functions have the form \eqref{Q-poly}, and the deformation parameter is 
generic. Thus 
\eqref{QQd8} can be a sufficient condition for \eqref{QQratD1} in the general situation (it is also a necessary condition under \eqref{Q-poly} for a generic $q$). 
}
 from the 3-term QQ-relations \eqref{QQd1}-\eqref{QQd9} and \eqref{QQd3}-\eqref{QQd10}.
\footnote{To be precise, \eqref{QQd6} and \eqref{QQd10} are for $\mathfrak{k}(I_{2r+2s+2})$.}
%.
The condition 
$\mathfrak{k} ( {\mathsf F}_{(1)}^{I_{2r+2s+2}})= {\mathsf F}_{(1)}^{\mathfrak{k} (I_{2r+2s+2})}= {\mathsf F}_{(1)}^{I_{2r+2s+2}}$ 
is equivalent 
\footnote{Note the relations $\mathfrak{k}(I_{r+s})=\breve{I}_{r+s}$ and $\mathfrak{k}(\breve{I}_{r+s})=I_{r+s}$.}
 to the following 4-term QQ-relations
\begin{align}
{\mathcal   X}_{I_{r+s}} +{\mathcal   X}_{I_{r+s+3}} 
&=
{\mathcal   X}_{\mathfrak{k} (I_{r+s})} + {\mathcal   X}_{\mathfrak{k} (I_{r+s+3})}, 
\label{TinvD3}
\end{align}
which follows from the 3-term QQ-relation \eqref{QQd11} for the case $i_{r+s} \in \mathfrak{F} $, 
and from  
\begin{align}
{\mathcal   X}_{\mathfrak{k} (I_{r+s})} ={\mathcal   X}_{I_{r+s+3}} , \qquad
{\mathcal   X}_{\mathfrak{k} (I_{r+s+3})} ={\mathcal   X}_{I_{r+s}} 
\qquad \text{for} \quad i_{r+s} \in \mathfrak{B} .
\label{chchD}
\end{align}
Eq.\ \eqref{chchD} appears to be related to 
the symmetry that flips the $(r+s-1)$-th and $(r+s)$-th nodes of a Dynkin diagram of type D, 
but produces no QQ-relation. 
The relations  \eqref{TinvD1} and \eqref{TinvD2} 
are reductions of \eqref{Tinv}, while the relation \eqref{TinvD3} is not. Thus the T-function ${\mathsf F}_{(1)}^{I_{2r+2s+2}}$ 
on the symmetric nesting path 
is  $\mathfrak{W}$-invariant under the 3-term QQ-relations \eqref{QQd1}-\eqref{QQd11}. 
%%%%%%%%%

The condition that the generating function ${\mathbf W}_{I_{2r+2s+2}}({\mathbf X})$ 
 is invariant under $\mathfrak{s}$, namely 
${\mathfrak{s}(\mathbf W}_{I_{2r+2s+2}}({\mathbf X}))={\mathbf W}_{\mathfrak{s}(I_{2r+2s+2})}({\mathbf X})={\mathbf W}_{I_{2r+2s+2}}({\mathbf X})$,  
is equivalent to the discrete zero curvature condition (a reduction of \eqref{ZCC}):
\begin{multline}
 (1-{\mathcal X}_{I_{a}}{\mathbf X})^{p_{i_{a}}} (1-{\mathcal X}_{I_{a+1}}{\mathbf X})^{p_{i_{a+1}}}
=
 (1-{\mathcal X}_{\mathfrak{s}(I_{a}) }{\mathbf X})^{p_{i_{a+1}}}
(1-{\mathcal X}_{\mathfrak{s}(I_{a+1})}{\mathbf X})^{p_{i_{a}}},
\\
 (1-{\mathcal X}_{I_{2r+2s+2-a}}{\mathbf X})^{p_{i_{a+1}}} (1-{\mathcal X}_{I_{2r+2s+3-a}}{\mathbf X})^{p_{i_{a}}}
 =
\\
=
 (1-{\mathcal X}_{\mathfrak{s}(I_{2r+2s+2-a}) }{\mathbf X})^{p_{i_{a}}}
(1-{\mathcal X}_{\mathfrak{s}(I_{2r+2s+3-a})}{\mathbf X})^{p_{i_{a+1}}} ,
\label{ZCCD1}
\end{multline}
where $a \in \{1,2,\dots , r+s-1 \}$. 
These relations \eqref{ZCCD1} boil down to \eqref{TinvD1} and \eqref{TinvD2}  and 
a reduction of the identity \eqref{id-A}.  
The condition that the generating function ${\mathbf W}_{I_{2r+2s+2}}({\mathbf X})$ 
 is invariant under $\mathfrak{k}$, namely 
${\mathfrak{k}(\mathbf W}_{I_{2r+2s+2}}({\mathbf X}))={\mathbf W}_{\mathfrak{k}(I_{2r+2s+2})}({\mathbf X})={\mathbf W}_{I_{2r+2s+2}}({\mathbf X})$, 
is equivalent to the following discrete zero curvature condition:
\begin{multline}
(1-{\mathcal X}_{I_{r+s+3}}{\mathbf X})^{-p_{i_{r+s}}} (1-{\mathcal X}_{I_{r+s+3}} \, {\mathcal X}^{[2]}_{I_{r+s}}{\mathbf X}^{2})
(1-{\mathcal X}_{I_{r+s}} {\mathbf X})^{-p_{i_{r+s}}}
=
\\
=
 (1-{\mathcal X}_{\mathfrak{k}(I_{r+s+3})}{\mathbf X})^{-p_{i_{r+s}}}
 (1-{\mathcal X}_{\mathfrak{k}(I_{r+s+3}) } \, {\mathcal X}^{[2]}_{\mathfrak{k}(I_{r+s})}{\mathbf X}^{2})
 (1-{\mathcal X}_{\mathfrak{k}(I_{r+s})}{\mathbf X})^{-p_{i_{r+s}}}.
\label{ZCCD2}
\end{multline}
This relation \eqref{ZCCD2} for the case $p_{i_{r+s}}=1$ reduces to \eqref{chchD} and an identity
\begin{multline}
(1-A{\mathbf X})^{-1}(1-A{\mathbf X} B{\mathbf X})(1-B{\mathbf X})^{-1}=
(1-A{\mathbf X})^{-1}+(1-B{\mathbf X})^{-1}-1=
\\
=(1-B{\mathbf X})^{-1}(1-B{\mathbf X} A{\mathbf X})(1-A{\mathbf X})^{-1}
\label{AB-id}
\end{multline}
for any functions $A$ and $B$.
Consider the expansion of \eqref{ZCCD2} for the case $p_{i_{r+s}}=-1$
with respect to the non-negative powers of $ {\mathbf X}$.  
The coefficients of $ {\mathbf X}$ on both sides of \eqref{ZCCD2} give the relation \eqref{TinvD3}, 
which follows from the QQ-relation \eqref{QQd11}. The relation derived from the coefficients of $ {\mathbf X}^{3}$ 
also  follows from the QQ-relation \eqref{QQd11}. The  relation derived from  the coefficients of 
$ {\mathbf X}^{4}$ 
is trivially valid.  
The other coefficients are $0$ or $1$. 
Therefore the T-functions $ {\mathcal F}_{(b)}^{I_{2r+2s+2}}$ and $ {\mathcal F}_{(1^{b})}^{I_{2r+2s+2}}$ 
on the symmetric nesting paths 
are invariant under ${\mathfrak W}$ if the QQ-relations \eqref{QQd1}-\eqref{QQd10} are imposed.   
%The T-functions for the symmetric nesting paths form a closed system under $\mathfrak{W}$. 
%\paragraph{QQ-relations} 
Thus, it may be possible to exclude the T- and Q-functions on the non-symmetric nesting paths from our consideration, and reformulate the Wronskian-type formulas.
%It is not clear at the moment if they over-constrain the system. 

It is also possible to do the opposite of what we have discussed so far.
Namely, forget about the connection with $U_{q}(gl(2r|2s+2)^{(1)})$ (reduction procedures, especially the non-trivial 
ansatzes \eqref{facd1}-\eqref{facd4}) first.  
Assume the Bethe ansatz equations associated with all 
the simple root systems of $osp(2r|2s)$ \eqref{BAEd1}-\eqref{BAEd6}  
(see also section \ref{sec:QQinv}). 
Construct the T-function \eqref{tab-fund-d} by analytic Bethe ansatz, 
which is free of poles under the Bethe ansatz equations associated with each 
simple root system.  Assume that all these T-functions are equivalent. 
We will obtain the 8-term QQ-relation \eqref{TinvD0} and the 4-term QQ-relation \eqref{TinvD3}. 
The 3-term QQ-relations \eqref{QQd1}-\eqref{QQd10} are a sufficient condition for this equivalence.
\footnote{In other words, if there is a part in which the equivalence breaks down, then the corresponding 
QQ-relation must be excluded. 
It would be possible to restrict our consideration to the orbit of the Weyl reflections and odd reflections 
(starting from the distinguished simple root system) 
via \eqref{wID}-\eqref{wIDC} and exclude QQ-relations outside of the orbit. 
}
In particular, \eqref{TinvD1} and \eqref{TinvD2}, which 
 are a sufficient condition for \eqref{TinvD0},  are 
equivalent to the 3-term QQ-relations \eqref{QQd1}-\eqref{QQd9} 
and \eqref{QQd3}-\eqref{QQd10} under the condition that 
the Q-functions have the form \eqref{Q-poly} and the deformation 
parameter $q$ is generic.
In this context, a question is if  
 there are Q-functions which satisfy \eqref{TinvD0} but do not 
\eqref{TinvD1} and \eqref{TinvD2}, in other words, if there are 
more general QQ-relations that supersede \eqref{QQd1}-\eqref{QQd9} 
and \eqref{QQd3}-\eqref{QQd10}.
Further research is needed on this point, but at present, we expect that they are  enough for \eqref{TinvD0} as they reproduce the standard form of the Bethe ansatz equations.  
 We will be able to reformulate the Wronskian-type formulas 
on T-and Q-functions starting from these. 
The same remark will apply not only to other algebras treated in this paper, but also to other algebras such as 
$U_{q}(D(2,1;\alpha)^{(1)})$, $U_{q}(G(3)^{(1)})$ and $U_{q}(F(4)^{(1)})$, etc. 
As for $s=0$ case, several Wronskian-type formulas of T-functions (or T-operators) 
 are already proposed in \cite{FFK20,ESV20,EV21,FKT21}, which provide alternative expressions of 
 tableau sum  \cite{KS94-1,TK96}, CBR-type determinant or Pfaffian formulas \cite{KNH95,TK96} of T-functions.

 %%%%%%%%%%%%%%%%%%%%%%%%%%%%%%%%%%%%%%%%%%%%%%%%%%%%%%%%%%
\subsection{Bethe ansatz equations associated with root systems}
\label{sec:QQinv}
The QQ-relations and the Bethe ansatz equations discussed in the previous subsections can 
be expressed in terms of root systems of underlying algebras. 
%%%%%
\subsubsection{QQ-relations}
Let $\kappa$ be the order of the map $\sigma$ (see \eqref{sigmaep}). 
Here we consider the case $\kappa=2$, $\sigma^{\kappa}=1$. 
Let  $\{ \alpha_{a} \}_{a=1}^{M+N-1}$ be the  simple roots of $gl(M|N)$ defined in \eqref{rootA} 
for a symmetric nesting path. 
The parameters $(M,N)$ are specified as in the previous subsections.
Set $\rr =r+s$. 
For $a \in \{1,2, \dots , \rr \}$, 
the QQ-relations \eqref{QQb1}-\eqref{QQb4},  \eqref{QQt11}-\eqref{QQt14},  \eqref{QQt21}-\eqref{QQt24} and 
\eqref{QQt31}-\eqref{QQt35} are summarized as 
\begin{multline}
 (e^{-\alpha_{a}(h)}-1)P_{a}\prod_{k=0}^{\kappa-1} \prod_{b=1 \atop (\alpha_{a}|\sigma^{k}(\alpha_{b})) \ne 0, \, 
 \alpha_{a} \ne \sigma^{k}(\alpha_{b})}^{\rr}
 \Qc_{b}^{[k\eta]}
 =
e^{-\alpha_{a}(h)}\prod_{k=0 \atop \alpha_{a}=\sigma^{k}(\alpha_{a})}^{\kappa-1} \Qc_{a}^{[d_{a}+k\eta]} \widetilde{\Qc}_{a}^{[-d_{a}+k\eta]} 
\\
-
\prod_{k=0 \atop \alpha_{a}=\sigma^{k}(\alpha_{a})}^{\kappa-1} \Qc_{a}^{[-d_{a}+k\eta]}
\widetilde{\Qc}_{a}^{[d_{a}+k\eta]}
\quad 
\text{if} \quad (\alpha_{a}|\alpha_{a}) \ne 0, 
\label{QQrb2}
\end{multline}
\begin{multline}
 (e^{-\alpha_{a}(h)}-1)  \Qc_{a} \widetilde{\Qc}_{a}
 =
e^{-\alpha_{a}(h)}P_{a}^{[-d_{a}]}
\prod_{k=0}^{\kappa-1} \prod_{b=1 \atop (\alpha_{a}|\sigma^{k}(\alpha_{b})) \ne 0, \, 
 \alpha_{a} \ne \sigma^{k}(\alpha_{b})}^{\rr}
 \Qc_{b}^{[(\alpha_{a}|\sigma^{k}(\alpha_{b}))+k\eta]}
 \\
 -
P_{a}^{[d_{a}]}
\prod_{k=0}^{\kappa-1} \prod_{b=1 \atop (\alpha_{a}|\sigma^{k}(\alpha_{b})) \ne 0, \, 
 \alpha_{a} \ne \sigma^{k}(\alpha_{b})}^{\rr}
 \Qc_{b}^{[-(\alpha_{a}|\sigma^{k}(\alpha_{b}))+k\eta]}
\quad 
  \text{if} \quad (\alpha_{a}|\alpha_{a}) = 0,
  \label{QQrf2}
\end{multline}
where each element is identified as: \\
for $U_{q}(gl(2r|2s+1)^{(2)})$, $U_{q}(gl(2r+1|2s)^{(2)})$, $U_{q}(osp(2r+1|2s)^{(1)})$,
\begin{align}
\Qc_{a}=\Qb_{I_{a}} , \quad \widetilde{\Qc}_{a}=\Qb_{\widetilde{I}_{a}} , \quad 
 u_{k}^{(a)}=u_{k}^{I_{a}} \quad n_{a}=n_{I_{a}} \quad 
 \text{for} \quad a \in \{1,2,\dots , r+s\},
\end{align}
for $U_{q}(gl(2r|2s)^{(2)})$: 
\begin{multline}
\Qc_{a}=\Qb_{I_{a}} , \quad \widetilde{\Qc}_{a}=\Qb_{\widetilde{I}_{a}} , 
\quad u_{k}^{(a)}=u_{k}^{I_{a}} \quad n_{a}=n_{I_{a}} \quad 
 \text{for} \quad a \in \{1,2,\dots , r+s-1 \}, 
\\
\Qc_{r+s}=\Qf_{I_{r+s}}, \quad \widetilde{\Qc}_{r+s}=\Qf_{\widetilde{I}_{r+s}} , 
\quad u_{k}^{(r+s)}=v_{k}^{I_{r+s}}, \quad n_{r+s}=m_{I_{r+s}} ,
\end{multline}
 $d_{a}=(\alpha_{a}|\alpha_{a})/2$ if $(\alpha_{a}|\alpha_{a}) \ne 0$, 
and $d_{a}=(\alpha_{a}|\alpha_{a^{\prime}}) \ne 0$ for some simple root $\alpha_{a^{\prime}}$ 
 if $(\alpha_{a}|\alpha_{a}) = 0$, 
in particular $d_{1}=p_{i_{1}}$;  $e^{\epsilon_{a}(h)}=z_{a}$ and 
%$\alpha_{0}=\epsilon_{i_{1}}-\epsilon_{i_{M+N}}$,
%
%\footnote{This comes from a simple root of the affine Lie superalgebra $gl(M|N)^{(1)}$, but the null vector (often denoted as $\delta$)
% is formally set to $0$ since it 
%plays no role here.}
%
%$\widetilde{\Qc}_{a}=w_{\alpha_{a}}(\Qc_{a})=\Qb_{\widetilde{I}_{M+N-a}}=\Qb_{(i_{1}.i_{2},\dots, i_{M+N-a-1},i_{M+N-a+1})}$ 
%and 
\begin{align}
P_{a}=
\begin{cases}
\Qc_{0}=\Qb_{\emptyset} & \text{if} \quad  a=1, \\
1 & \text{otherwise} .
\end{cases}
\label{vacQQ2}
\end{align}
In addition to the above, we formally set $\eta=0$ for $U_{q}(osp(2r+1|2s)^{(1)})$. 
We remark that \eqref{QQrb2} and \eqref{QQrf2} reduce to \eqref{QQrb0} and \eqref{QQrf0} 
if we set $\kappa=1$, $\rr=M+N-1$ and replace \eqref{vacQQ2}  with \eqref{vacQQ0}. 
In this case, the corresponding simple root system is not necessary on a symmetric nesting path. 
In order to compare  \eqref{QQrb2} for $s=0$ with the QQ-relations for the twisted quantum affine (non-super) algebras 
discussed in \cite{FH16,W22}, one should set:  
$P_{a} \to 1$, $e^{\pm \alpha_{a}(h)/2} \to [\pm \alpha_{a}/2]$, 
  $\widetilde{\Qc}_{a}/(e^{-\alpha_{a}(h)/2}-e^{\alpha_{a}(h)/2}) \to \widetilde{\Qc}_{a}$.

%%%%%%%%%%%%%%%
Let $\{ \beta_{a} \}_{a=1}^{\rr }$ be the simple roots of the orthosymplectic Lie superalgebras defined in
 \eqref{rootB}, \eqref{rootD} and \eqref{rootDC}. 
For $a \in \{1,2, \dots , \rr \}$, 
the QQ-relations \eqref{QQsp1}-\eqref{QQsp3}, \eqref{QQd1}-\eqref{QQd11}, \eqref{QQb1}-\eqref{QQb3} and  \eqref{QQb5}
are summarized as
\begin{multline}
 (e^{-\beta_{a}(h)}-1)\varphi_{a} \prod_{b=1 \atop (\beta_{a}|\beta_{b}) \ne 0, \, b \ne a }^{\rr} \prod_{k=0}^{-C_{ab}-1}
 \Qc_{b}^{[-(\beta_{a}|\beta_{b})-d_{a}(1+2k)]}=
e^{-\beta_{a}(h)}\Qc_{a}^{[d_{a}]} \widetilde{\Qc}_{a}^{[-d_{a}]} -
\Qc_{a}^{[-d_{a}]} \widetilde{\Qc}_{a}^{[d_{a}]}
\\
\text{if} \quad (\beta_{a}|\beta_{a}) \ne 0, \quad p_{\beta_{a}}=1,
\label{QQrb3}
\end{multline}
\begin{multline}
 (e^{-\beta_{a}(h)}+1)\varphi_{a} \prod_{b=1 \atop (\beta_{a}|\beta_{b}) \ne 0}^{\rr} 
 \Qc_{b}=
e^{-\beta_{a}(h)}\Qc_{a}^{[2d_{a}]} \widetilde{\Qc}_{a}^{[-2d_{a}]} +
\Qc_{a}^{[-2d_{a}]} \widetilde{\Qc}_{a}^{[2d_{a}]}
\\
\text{if} \quad (\beta_{a}|\beta_{a}) \ne 0, \quad p_{\beta_{a}}=-1,
\label{QQrbb3}
\end{multline}
\begin{multline}
 (e^{-\beta_{a}(h)}-1)  \Qc_{a} \widetilde{\Qc}_{a}
 =
e^{-\beta_{a}(h)}
 \varphi_{a}^{-} \prod_{b=1 \atop (\beta_{a}|\beta_{b}) \ne 0, a \ne b}^{\rr}
 \Qc_{b}^{[(\beta_{a}|\beta_{b})]}
 -
\varphi_{a}^{+} \prod_{b=1 \atop (\beta_{a}|\beta_{b}) \ne 0, a \ne b}^{\rr}
 \Qc_{b}^{[-(\beta_{a}|\beta_{b})]}\\
  \text{if} \quad (\beta_{a}|\beta_{a}) = 0,
  \label{QQrf3}
\end{multline}
where each element is identified as: \\
for $U_{q}(sp(2r)^{(1)})$: 
\begin{multline}
\beta_{a}=\epsilon_{i^{*}_{a}}-\epsilon_{i^{*}_{a+1}}, \quad 
\Qc_{a}=\Qb_{I_{a}} , \quad \widetilde{\Qc}_{a}=\Qb_{\widetilde{I}_{a}} , 
\quad u_{k}^{(a)}=u_{k}^{I_{a}} \quad n_{a}=n_{I_{a}} 
\\
 \text{for} \quad a \in \{1,2,\dots , r-1 \}, 
\\
\beta_{r}=2\epsilon_{i^{*}_{r}}, \quad 
\Qc_{r}=\Qf_{I_{r}}, \quad 
\widetilde{\Qc}_{r}=\Qb_{\breve{I}_{r}}, \quad 
u_{k}^{(r)}=v_{k}^{I_{r}}, \quad n_{r}=m_{I_{r}} ,\quad s=0,
\end{multline}
for $U_{q}(osp(2r|2s)^{(1)})$, $i_{r+s} \in \Bm $: 
\begin{multline}
\beta_{a}=\epsilon_{i^{*}_{a}}-\epsilon_{i^{*}_{a+1}}, \quad 
\Qc_{a}=\Qb_{I_{a}} , \quad \widetilde{\Qc}_{a}=\Qb_{\widetilde{I}_{a}},
\quad 
u_{k}^{(a)}=u_{k}^{I_{a}} \quad n_{a}=n_{I_{a}} 
\\
 \text{for} \quad a \in \{1,2,\dots , r+s-2 \}, 
 \\
\beta_{r+s-1}=\epsilon_{i^{*}_{r+s-1}}-\epsilon_{i^{*}_{r+s}}, \quad 
\Qc_{r+s-1}=\Qf_{\breve{I}_{r+s}},  
\\
\widetilde{\Qc}_{r+s-1}=\Qb_{\widetilde{I}_{r+s-1}} \quad \text{if} \quad i_{r+s-1} \in \Fm,  
\quad 
\widetilde{\Qc}_{r+s-1}=\Qf_{\acute{I}_{r+s}} \quad \text{if} \quad i_{r+s-1} \in \Bm,  
\\
 u_{k}^{(r+s-1)}=v_{k}^{\breve{I}_{r+s}}, \quad n_{r+s-1}=m_{\breve{I}_{r+s}},
\\
\beta_{r+s}=\epsilon_{i^{*}_{r+s-1}}+\epsilon_{i^{*}_{r+s}}, \quad 
\Qc_{r+s}=\Qf_{I_{r+s}}, 
\\
\widetilde{\Qc}_{r+s}=\Qb_{\dot{I}_{r+s-1}} \quad \text{if} \quad i_{r+s-1} \in \Fm,  
\quad 
\widetilde{\Qc}_{r+s}=\Qf_{\grave{I}_{r+s}} \quad \text{if} \quad i_{r+s-1} \in \Bm,  
\\
\quad u_{k}^{(r+s)}=v_{k}^{I_{r+s}}, \quad n_{r+s}=m_{I_{r+s}} .
\end{multline}
for $U_{q}(osp(2r|2s)^{(1)})$, $i_{r+s} \in \Fm $: 
\begin{multline}
\beta_{a}=\epsilon_{i^{*}_{a}}-\epsilon_{i^{*}_{a+1}}, \quad 
\Qc_{a}=\Qb_{I_{a}} , \quad u_{k}^{(a)}=u_{k}^{I_{a}}, \quad n_{a}=n_{I_{a}} \quad 
 \text{for} \quad a \in \{1,2,\dots , r+s-1 \}, 
\\
\beta_{r+s}=2\epsilon_{i^{*}_{r+s}}, \quad 
\Qc_{r+s}=\Qf_{I_{r+s}}, \quad u_{k}^{(r+s)}=v_{k}^{I_{r+s}}, \quad n_{r+s}=m_{I_{r+s}} , 
\\
\widetilde{\Qc}_{r+s-1}=\Qb_{\widetilde{I}_{r+s-1}} \quad \text{if} \quad i_{r+s-1} \in \Fm,  
\quad 
\widetilde{\Qc}_{r+s-1}=\Qf_{\acute{I}_{r+s}} \quad \text{if} \quad i_{r+s-1} \in \Bm,  
\\
\widetilde{\Qc}_{r+s}=\Qf_{\breve{I}_{r+s}},
\end{multline}
for $U_{q}(osp(2r+1|2s)^{(1)})$: 
\begin{multline}
\beta_{a}=\epsilon_{i^{*}_{a}}-\epsilon_{i^{*}_{a+1}} \quad
 \text{for} \quad a \in \{1,2,\dots , r+s-1 \}, 
 \quad \beta_{r+s}=\epsilon_{i^{*}_{r+s}},
\\
\Qc_{a}=\Qb_{I_{a}} ,  
\quad u_{k}^{(a)}=u_{k}^{I_{a}}, \quad n_{a}=n_{I_{a}} \quad 
 \text{for} \quad a \in \{1,2,\dots , r+s \} ,
 \\
 \widetilde{\Qc}_{a}=\Qb_{\widetilde{I}_{a}} \quad 
 \text{for} \quad a \in \{1,2,\dots , r+s-1 \}, 
 \\
 \widetilde{\Qc}_{r+s}=\Qb_{\breve{I}_{r+s}} \quad \text{if} \quad i_{r+s} \in \Bm ,
  \quad 
 \widetilde{\Qc}_{r+s}=\Qb_{\widetilde{I}_{r+s}} \quad \text{if} \quad i_{r+s} \in \Fm ,
\end{multline}
 and $C_{ab}=2(\beta_{a}|\beta_{b})/(\beta_{a}|\beta_{a})$, 
%, 
%
%$\widetilde{\Qc}_{a}=w_{\alpha_{a}}(\Qc_{a})=\Qb_{\widetilde{I}_{M+N-a}}=\Qb_{(i_{1}.i_{2},\dots, i_{M+N-a-1},i_{M+N-a+1})}$; 
 $d_{a}=(\beta_{a}|\beta_{a})/2$ for $(\beta_{a}|\beta_{a}) \ne 0$.
% $d_{1}=p_{i_{1}}=-p_{i_{2}}$ if $(\beta_{1}|\beta_{1})=0$ [thus $d_{1}=(\beta_{1}|\beta_{2})/2 $ for  
%  $osp(2|2)$, $d_{1}=(\beta_{1}|\beta_{2}) $ for the other];  
% $d_{a}$ for $a \ge 2$,  $(\beta_{a}|\beta_{a}) = 0$ will not be used here
% (but in most cases, $d_{a}=(\beta_{a}|\beta_{a^{\prime}}) \ne 0$ for some simple root $\beta_{a^{\prime}}$).
%  $P_{1}=P_{2}=\Qb_{\emptyset}$  for $U_{q}(osp(2|2)^{(1)})$ 
% with $-p_{i_{1}}=p_{i_{2}}=1$ ($p_{\beta_{1}}=p_{\beta_{2}}=-1$), 
%and for the other case $P_{a}$ is defined by \eqref{vacQQ2}. 
In our examples, the vacuum parts are defined as follows: for $U_{q}(osp(3|0)^{(1)})$,
\begin{align}
\varphi_{1}&=\varphi_{1}(u)=\Qb_{\emptyset}^{[-\frac{1}{2}]} \Qb_{\emptyset}^{[\frac{1}{2}]}, 
\end{align}
and for the other case
\begin{align}
\varphi_{a}&=\varphi_{a}(u)=
\begin{cases}
\Qb_{\emptyset}  & \text{if} \quad (\epsilon_{i_{1}^{*}} |\beta_{a}) \ne 0
\\
1 & \text{if} \quad  (\epsilon_{i_{1}^{*}} |\beta_{a}) = 0 ,
\end{cases}
\\
\varphi_{a}^{\pm}&=\varphi_{a}(u \pm  (\epsilon_{i_{1}^{*}} |\beta_{a}) ) .
\end{align}
%\begin{align}
%P_{a}=
%\begin{cases}
%\Qc_{0}=\Qb_{\emptyset} & \text{if} \quad  a=1, \\
%1 & \text{otherwise}.
%\end{cases}
%\end{align}
%
The vacuum parts depend on the Hilbert space on which transfer matrices act. 
In the theory of q-characters, the vacuum parts should be formally set to $1$. 
We remark that \eqref{QQrb3} and \eqref{QQrf3} reduce to \eqref{QQrb0} and \eqref{QQrf0} 
if we set $\rr=M+N-1$ and replace $\{\beta_{b}\}$ with $\{\alpha_{b}\}$ 
(not necessary on a symmetric nesting path), 
$\varphi_{a}$ with $P_{a}$ in \eqref{vacQQ0}, and $\varphi_{a}^{\pm}$ with $P_{a}^{[\pm d_{a}]}$ 
($d_{a}$ is the one in \eqref{BAEr2}). 

As already remarked, 
 QQ-relations for non-super algebras are expressed in terms of root systems 
of underlying (non-super) Lie algebras \cite{MV04,MRV15,MRV15-2}.  
Our formulation is different from theirs in 
that we start from the QQ-relations associated with root systems of the superalgebra $gl(M|N)$ even 
for the non-superalgebra  $U_{q}(so(2r+1)^{(1)}) \simeq U_{q}(osp(2r+1|0)^{(1)})$ case 
(\eqref{QQrb2} for $s=0$). 

The ODE/IM correspondence is an efficient tool to derive QQ-relations. 
In particular, the ODE/IM correspondence for supersymmetric integrable models 
was discussed in \cite{Su00,BT08} for $U_{q}(gl(2|1)^{(1)})$ (or $U_{q}(sl(2|1)^{(1)})$), 
and in \cite{IZ22} for supersymmetric affine Toda field equations associated to  
affine Lie superalgebras with purely fermionic simple root systems, including $osp(2|2)^{(2)}$. 
It is desirable to reconsider  the QQ-relations discussed here 
in connection with the ODE/IM correspondence, and generalize them further. 
In this context,  it is important to clarify the object corresponding to 
the relations 
\eqref{reduction-sigma}  in  the ODE/IM correspondence for $U_{q}(gl(M|N)^{(1)})$. 
%%%%%%%%%%%
\subsubsection{Bethe ansatz equations}
The Bethe ansatz equations for
$U_{q}(gl(2r|2s+1)^{(2)})$ \eqref{BAEt1}, $U_{q}(gl(2r+1|2s)^{(2)})$ \eqref{BAEt2}, 
$U_{q}(gl(2r|2s)^{(2)})$ \eqref{BAEt3}, $U_{q}(osp(2r+1|2s)^{(1)})$ \eqref{BAEb} 
 are expressed
 \footnote{Note that $\Qb_{I_{b}}(u^{I_{a}}_{k}+\xi)=\Qb_{I_{M+N-b}}(u^{I_{M+N-a}}_{k}+\xi)$, $\xi \in {\mathbb C}$, 
 $0 \le a,b \le M+N$ holds for any 
 symmetric nesting path.}
  in terms of a part of a symmetric simple root system of $gl(M|N)$: 
\begin{multline}
 - \frac{P_{a}(u_{k}^{(a)}+d_{a})}{P_{a}(u_{k}^{(a)}-d_{a})}
 =
 p_{\alpha_{a}}
 e^{-\alpha_{a}(h)}
 \prod_{t=0}^{\kappa -1}
 \prod_{b=1}^{\rr}
\frac{\Qc_{b}(u_{k}^{(a)}+(\alpha_{a}| \sigma^{t}(\alpha_{b}) ) + \eta t)}{\Qc_{b}(u_{k}^{(a)}-(\alpha_{a}|\sigma^{t}(\alpha_{b}) ) + \eta  t)}
\\
\text{for} \quad k\in \{1,2,\dots, n_{a} \} \quad \text{and} \quad a \in \{1,2,\dots, \rr \} .
\label{BAEr3}
\end{multline}
For $N=0$, this fits into the form of the 
Bethe ansatz equations for the twisted quantum affine (non-super) algebras in \cite{RW87}. 
However, the formulation of the $U_{q}(so(2r)^{(1)})$ case in \cite{RW87} is different from ours 
in that we use a simple root system of $gl(2r|1)$. 
Substituting
$u=u^{(a)}_{k}\pm d_{a}$ into \eqref{QQrb2} and eliminating $\widetilde{\Qc}_{a}(u^{(a)}_{k})$, 
we obtain \eqref{BAEr3} for $(\alpha_{a}|\alpha_{a}) \ne 0$. 
Substituting
$u=u^{(a)}_{k}$ into \eqref{QQrf2}, 
we obtain \eqref{BAEr3} for $(\alpha_{a}|\alpha_{a}) = 0$.
Here we assume that the roots of the Q-functions are sufficiently generically distributed (thus $\widetilde{\Qc}_{a}(u^{(a)}_{k}) \ne 0$). 

%%%%%%%%%%%
The Bethe ansatz equations for
$U_{q}(sp(2r)^{(1)})$ \eqref{BAEc01}-\eqref{BAEc03}, $U_{q}(osp(2r|2s)^{(1)})$ 
\eqref{BAEd1}-\eqref{BAEd6}, and 
 $U_{q}(osp(2r+1|2s)^{(1)})$ \eqref{BAEb}  
 are expressed in terms of a simple root system of each finite algebra ($\mathfrak{g}$ for $U_{q}(\mathfrak{g}^{(1)})$):
\begin{multline}
 - \frac{\psi_{a}^{+}(u_{k}^{(a)})}{\psi_{a}^{-}(u_{k}^{(a)} )}=
 p_{\beta_{a}}
 e^{-\beta_{a}(h)}
 \prod_{b=1 \atop b \ne a}^{\rr}
\frac{\Qc_{b}(u_{k}^{(a)}+(\beta_{a}| \beta_{b})) }{\Qc_{b}(u_{k}^{(a)}-(\beta_{a}|\beta_{b} ) )}
 \prod_{l =1}^{\kappa_{a}}
\frac{\Qc_{a}(u_{k}^{(a)}+p_{l \beta_{a}} l (\beta_{a}| \beta_{a})) }{\Qc_{a}(u_{k}^{(a)}-p_{l \beta_{a}} l(\beta_{a}|\beta_{a} ) )}
\\
\text{for} \quad k\in \{1,2,\dots, n_{a} \} \quad \text{and} \quad a \in \{1,2,\dots, \rr \} ,
\label{BAEr4}
\end{multline}
where $\kappa_{a}=2$ if $p_{\beta_{a}}=-1$ and  $(\beta_{a}|\beta_{a} )\ne 0$ (black dot), 
$\kappa_{a}=1$ if $p_{\beta_{a}}=1$ and  $(\beta_{a}|\beta_{a} )\ne 0$ (white dot), 
$\kappa_{a}=0$ if $p_{\beta_{a}}=-1$ and  $(\beta_{a}|\beta_{a} )= 0$ (gray dot); 
%$g_{a}=2$ if $p_{\beta_{a}}=-1$ and  $(\beta_{a}|\beta_{a} )\ne 0$, otherwise  $g_{a}=1$ 
%(but we need only $g_{1}d_{1}=p_{i_{1}}$); 
$p_{l \beta_{a}}=(-1)^{\kappa_{a}-l}$; $\prod_{l=1}^{0}(\cdots )=1$. 
$\psi_{a}^{\pm}$ are vacuum eigenvalues of diagonal elements of a monodromy matrix
\footnote{Depending on the normalization, a shift in the spectral parameter is needed  (cf. \eqref{BAEr}).
}.
In general, cancellation of a common factor 
occurs between the numerator and denominator of the left-hand side of \eqref{BAEr4}. 
Thus in our example, we may set 
\begin{align}
\psi_{a}^{\pm}(u)=\Qb_{\emptyset}(u \pm (\epsilon_{i_{1}^{*}} |\beta_{a})). 
\label{vacbae}
\end{align}
Substituting
$u=u^{(a)}_{k}\pm d_{a}$ into \eqref{QQrb3} and eliminating $\widetilde{\Qc}_{a}(u^{(a)}_{k})$, 
we obtain \eqref{BAEr4} for $(\beta_{a}|\beta_{a}) \ne 0$, $p_{\beta_{a}}=1$.  
Substituting
$u=u^{(a)}_{k}\pm 2d_{a}$ into \eqref{QQrbb3} and eliminating $\widetilde{\Qc}_{a}(u^{(a)}_{k})$, 
we obtain \eqref{BAEr4} for $(\beta_{a}|\beta_{a}) \ne 0$, $p_{\beta_{a}}=-1$. 
Substituting
$u=u^{(a)}_{k}$ into \eqref{QQrf3}, 
we obtain \eqref{BAEr4} for $(\beta_{a}|\beta_{a}) = 0$. 
In general, the vacuum parts of the Bethe ansatz equation \eqref{BAEr4}  
and those of the QQ-relations \eqref{QQrb3}-\eqref{QQrf3}
are related as 
\begin{align}
\frac{\psi_{a}^{+}(u_{k}^{(a)})}{\psi_{a}^{-}(u_{k}^{(a)} )}=
\begin{cases}
\frac{\varphi_{a}(u_{k}^{(a)}+d_{a})}{\varphi_{a}(u_{k}^{(a)} -d_{a})} 
& \text{for} \quad (\beta_{a}| \beta_{a}) \ne 0, \quad p_{\beta_{a}}=1
\\
\frac{\varphi_{a}(u_{k}^{(a)}+2d_{a})}{\varphi_{a}(u_{k}^{(a)} -2d_{a})} 
& \text{for} \quad (\beta_{a}| \beta_{a}) \ne 0, \quad p_{\beta_{a}}=-1
\\
\frac{\varphi_{a}^{+}(u_{k}^{(a)})}{\varphi_{a}^{-}(u_{k}^{(a)} )} 
& \text{for} \quad (\beta_{a}| \beta_{a})=0 .
\end{cases}
\label{vac=vac}
\end{align}
In our examples, \eqref{vac=vac} reduces to $\Qb_{\emptyset}(u+(\epsilon_{i_{1}^{*}}| \beta_{a}))/\Qb_{\emptyset}(u-(\epsilon_{i_{1}^{*}}| \beta_{a}))$.
In order to describe a two-body self-interaction
\footnote{Ref.\ \cite{MR97} pointed out that the formulation of the Bethe ansatz equations in terms of their 
corresponding Lie algebras \cite{ORW87} 
can be extended to the case of superalgebras (for rational vertex models), but the black dot of $osp(1|2s)$ 
 is an exception for this.  Then in \cite{T01}, we tried to make use of the correspondence between $osp(1|2s)^{(1)}$ 
 and $sl(2s+1)^{(2)}$ for the descripution of the black dot in the Bethe ansatz equation. 
 This is incorporated into \eqref{BAEr3}. Eq.\ \eqref{BAEr4} is an alternative expression 
 for this.}
 at $b=a$, we need the even root $2\beta_{a}$ in addition to 
the odd simple root $\beta_{a}$ for the black dot. 
Substituting
$u=u^{(a)}_{k}$ into \eqref{QQrf3}, 
we obtain \eqref{BAEr4} for $(\beta_{a}|\beta_{a}) = 0$.
%%%%%%%%%%%%%%%%%%%%%%%
\subsubsection{Extended Weyl group symmetry}
Eq. \eqref{QQrb3} is related to the Weyl reflection \eqref{Weven} by the simple even root $\beta_{a}$, 
and \eqref{QQrf3} is related to the odd reflection \eqref{Wodd} by 
the simple odd root $\beta_{a}$ (for the case $(\beta_{a} |\beta_{a} ) = 0$): 
$w_{\beta_{a}} (\Qc_{a}) =\widetilde{\Qc}_{a}$, $w_{\beta_{a}} (\Qc_{b})=\Qc_{b}$ 
 for $b \ne a$. 
In addition,  \eqref{QQrbb3} (for $U_{q}(osp(2r+1|2s)^{(1)})$) is related to the Weyl reflection \eqref{Weven} 
by the simple even root $\alpha_{a}$ of $gl(2r|2s+1)$ under reduction: 
$w_{\alpha_{a}} (\Qc_{a}) =\widetilde{\Qc}_{a}$, $w_{\alpha_{a}} (\Qc_{b})=\Qc_{b}$ 
 for $b \ne a$. 
 As far as we could see, \eqref{QQrbb3} is not enough to realize the odd reflection by the 
 odd root $\beta_{r+s}$ of $osp(2r+1|2s)$ with $(\beta_{r+s}|\beta_{r+s}) \ne 0$, $p_{\beta_{r+s}}=-1$. 
 In order to realize this, we have to consider a composition of at least three 3-term QQ-relations, 
 which corresponds to the 6-term QQ-relation \eqref{TinvB3} for the case $p_{i_{r+s}}=-1$. 
 Then the action of this odd reflection becomes 
  $w_{\beta_{r+s}} (\Qc_{r+s}) =\Qb_{\breve{I}_{r+s}}$, $w_{\beta_{r+s}} (\Qc_{b})=\Qc_{b}$ 
 for $b \ne r+s$. 
We show these by a method similar to the one discussed in \cite{T98}, 
noting that the Bethe ansatz equations  \eqref{BAEr4} are expressed in terms of  root systems of underlying 
superalgebras. In \cite{T98}, we adopted an argument in \cite{Wo83} 
(amplified in \cite{BCFH92}) based on the residue theorem associated with the particle-hole transformation 
(a preliminary form of the QQ-relation). Instead we extend  
  a more simplified version \cite{GS03} (cf.\ \cite{PS00}) of it.
%Instead of using the residue theorem associated with
% the particle-hole transformation \cite{Wo83}, we use a more simplified argument (cf.\ \cite{GS03,PS00}).
%We use ``polynomial form'' of QQ-relations (cf.\ \cite{PS00,GS03}) 
%instead of particle-hold transformations \cite{Wo83}, 
%that is, the ``residue theorem version'' of QQ-relations.
 
Let $\{ \widetilde{u}_{k}^{(a)} \}_{k=1}^{\widetilde{n}_{a}}$ be the zeros of $\widetilde{\Qc}_{a}$. 
One can show the following 
for a fixed $a \in \{1,2,\dots , \rr \}$ ($a$ corresponds to a vertex of the Dynkin diagram associated with 
a simple root system $\{ \beta_{c} \}_{c=1}^{\rr}$). 
%%%
\paragraph{White or gray dots [other than the case $(\beta_{a}|\beta_{a}) \ne 0$, $p_{\beta_{a}}=-1$]}
Under the QQ-relations \eqref{QQrb3} and \eqref{QQrf3}, 
the Bethe ansatz equation \eqref{BAEr4},  namely 
\begin{multline}
 - \frac{\psi_{c}^{+}(u_{k}^{(c)})}{\psi_{c}^{-}(u_{k}^{(c)})}=
 p_{\beta_{c}}
 e^{-\beta_{c}(h)}
 \prod_{b=1 \atop b \ne c}^{\rr}
\frac{\Qc_{b}(u_{k}^{(c)}+(\beta_{c}| \beta_{b})) }{\Qc_{b}(u_{k}^{(c)}-(\beta_{c}|\beta_{b} ) )}
\times
\\
\times 
\begin{cases}
\frac{\Qc_{c}(u_{k}^{(c)}+2(\beta_{c}| \beta_{c}))\Qc_{c}(u_{k}^{(c)}-(\beta_{c}|\beta_{c} ) ) }{\Qc_{c}(u_{k}^{(c)}-2(\beta_{c}|\beta_{c} ) )\Qc_{c}(u_{k}^{(c)}+(\beta_{c}|\beta_{c} ) )} & 
\text{if} \quad p_{\beta_{c}}=-1, \quad  (\beta_{c}|\beta_{c} )\ne 0 \\
1 & \text{if}  \quad (\alpha_{c}|\alpha_{c}) =0 \\
 \frac{\Qc_{c}(u_{k}^{(c)}+(\beta_{c}| \beta_{c})) }{\Qc_{c}(u_{k}^{(c)}-(\beta_{c}|\beta_{c} ) )} & \text{otherwise}
\end{cases}
\\
\text{for} \quad k\in \{1,2,\dots, n_{c} \} \quad \text{and} \quad c \in \{1,2,\dots, \rr \} .
\label{BAEr4p} 
\end{multline}
is equivalent
%
%\footnote{In \eqref{BAEr4w}, we will have to consider the action of $w_{\beta_{a}}$ on $P_{c}$ in the general situation. Here %we consider only the case $P_{c}=1$ for $c  \ne 1$, for simplicity. 
%The discussion on the Gaudin model \cite{LM21} would be useful for generalization.
%}
%
 to  
\begin{multline}
 - \frac{w_{\beta_{a}}(\psi_{c}^{+})(\hat{u}_{k}^{(c)} )}{w_{\beta_{a}}(\psi_{c}^{-})(\hat{u}_{k}^{(c)} )}=
 p_{w_{\beta_{a}}(\beta_{c})}
 e^{- w_{\beta_{a}}(\beta_{c})(h)}
 \prod_{b=1 \atop b \ne c}^{\rr}
\frac{w_{\beta_{a}}(\Qc_{b})(\hat{u}_{k}^{(c)}+
(w_{\beta_{a}}(\beta_{c})| w_{\beta_{a}}(\beta_{b}))) }{w_{\beta_{a}}(\Qc_{b})(\hat{u}_{k}^{(c)}-
(w_{\beta_{a}}(\beta_{c}) | w_{\beta_{a}}(\beta_{b}) ) )} 
\times
\\
\times 
\begin{cases}
\frac{w_{\beta_{a}}(\Qc_{c})(\hat{u}_{k}^{(c)}+2(w_{\beta_{a}}(\beta_{c})| w_{\beta_{a}}(\beta_{c})))w_{\beta_{a}}(\Qc_{c})(\hat{u}_{k}^{(c)}-(w_{\beta_{a}}(\beta_{c})|w_{\beta_{a}}(\beta_{c}) ) ) }{w_{\beta_{a}}(\Qc_{c})(\hat{u}_{k}^{(c)}-2(w_{\beta_{a}}(\beta_{c})|w_{\beta_{a}}(\beta_{c}) ) )w_{\beta_{a}}(\Qc_{c})(\hat{u}_{k}^{(c)}+(w_{\beta_{a}}(\beta_{c})|w_{\beta_{a}}(\beta_{c}) ) )} & 
\text{if} \quad p_{w_{\beta_{a}}(\beta_{c})}=-1, \\
&
\quad (w_{\beta_{a}}(\beta_{c})|w_{\beta_{a}}(\beta_{c}) )\ne 0 \\
1 & \text{if}  \quad (w_{\beta_{a}}(\beta_{c})|w_{\beta_{a}}(\beta_{c}) ) =0 \\
 \frac{w_{\beta_{a}}(\Qc_{c})(\hat{u}_{k}^{(c)}+(w_{\beta_{a}}(\beta_{c})| w_{\beta_{a}}(\beta_{c}))) }{w_{\beta_{a}}(\Qc_{c})(\hat{u}_{k}^{(c)}-(w_{\beta_{a}}(\beta_{c})|w_{\beta_{a}}(\beta_{c}) ) )} & \text{otherwise}
\end{cases}
\\
%%%
\text{for} \quad k\in \{1,2,\dots, w_{\beta_{a}}(n_{c}) \} \quad \text{and} \quad c  \in \{1,2,\dots , \rr \},
\label{BAEr4w}
\end{multline}
where $w_{\beta_{a}} (\Qc_{a}) =\widetilde{\Qc}_{a}$, $\hat{u}_{k}^{(a)} =\widetilde{u}_{k}^{(a)}$, 
$w_{\beta_{a}}(n_{a})=\widetilde{n}_{a}$;
$w_{\beta_{a}} (\Qc_{b})=\Qc_{b}$, $\hat{u}_{k}^{(b)} =u_{k}^{(b)}$, $w_{\beta_{a}}(n_{b})=n_{b}$
 for $b \ne a$;
%$w_{\beta_{a}}(g_{a}d_{a})=- g_{a}d_{a}$ if $ (\beta_{a}|\beta_{a}) = 0$, for the other case  
%$w_{\beta_{a}}(g_{b}d_{b})=g_{b}d_{b}$ (but we need only $b=1$ case), 
$w_{\beta_{a}}(\psi_{c}^{\pm})(u)=
\Qb_{\emptyset}(u \pm (w_{\beta_{a}}(\epsilon_{i_{1}^{*}}) |w_{\beta_{a}}(\beta_{c})))$.
%$w_{\beta_{a}}(g_{c})=2$ if $ (w_{\beta_{a}}(\beta_{c})|w_{\beta_{a}}(\beta_{c})) \ne 0$ and 
%$p_{w_{\beta_{a}}(\beta_{c})}=-1$, otherwise $w_{\beta_{a}}(g_{c})=1$. 
Here $w_{\beta_{a}}(\epsilon_{i_{1}^{*}})$ is defined by  
replacing the index $i_{1}^{*}$ of $\epsilon_{i_{1}^{*}}$ with the last  element of the corresponding tuple $w_{\beta_{a}}(I_{\cdots})$ 
 in \eqref{wIA}-\eqref{wIDC}. 
 In particular, $w_{\beta_{a}}(\epsilon_{i_{1}^{*}})=\epsilon_{i_{1}^{*}}$ if $(\epsilon_{i_{1}^{*}}|\beta_{a})=0$. 
 For $osp(2r+1|2s)$ and $osp(2r|2s)$, we have 
 $w_{\beta_{1}}(\epsilon_{i_{1}^{*}})=\epsilon_{i_{2}^{*}}$, 
 $w_{\beta_{a}}(\epsilon_{i_{1}^{*}})=\epsilon_{i_{i}^{*}}$ for $2 \le a \le r+s$ if $r+s \ge 3$, or  
 $r+s=2$ for type B  ($p_{i_{2}}=\pm 1$) or C  ($p_{i_{2}}=-1$); 
 $w_{\beta_{1}}(\epsilon_{i_{1}^{*}})=\epsilon_{i_{2}^{*}}$,  $w_{\beta_{2}}(\epsilon_{i_{1}^{*}})=\epsilon_{i_{2}}=-\epsilon_{i_{2}^{*}}$ if $r+s=2$ for type D ($p_{i_{2}}=1$);  
 $w_{\beta_{1}}(\epsilon_{i_{1}^{*}})=\epsilon_{i_{1}}=-\epsilon_{i_{1}^{*}}$ if $r+s=1$.

%%%%
\paragraph{Black dot [the case $(\beta_{a}|\beta_{a}) \ne 0$, $p_{\beta_{a}}=-1$]}
The only case in which this is realized (for the algebras in question in this paper) is when 
$a$ corresponds to the black dot of a Dynkin diagram of $osp(2r+1|2s)$ ($a=r+s$).  
In this case, the above statement holds if we assume the QQ-relations  \eqref{QQb6} and \eqref{QQb7} 
in addition to \eqref{QQb3} (namely, \eqref{QQrbb3}), and 
replace the above $w_{\beta_{a}} (\Qc_{a}) =\widetilde{\Qc}_{a}$ with 
$w_{\beta_{r+s}} (\Qc_{r+s}) =\Qb_{\breve{I}_{r+s}}$.

%%%%%%%%%%%
Let us explain these, case by case. 
%%%%%%%%%%%%%%%%%
\paragraph{Bosonic QQ-relation \eqref{QQrb3}: Weyl reflection (white dot)}
%%%
\subparagraph{The case $(\beta_{a}| \beta_{a}) \ne 0$, $p_{\beta_{a}}=1$, $c=a$}
Substituting
$u=\widetilde{u}^{(a)}_{k}\pm d_{a}$ into \eqref{QQrb3} and eliminating $\Qc_{a}(\widetilde{u}^{(a)}_{k})$, 
we obtain \eqref{BAEr4w} for $c=a$.
%%%
\subparagraph{The case $(\beta_{a}| \beta_{a}) \ne 0$, $p_{\beta_{a}}=1$, $(\beta_{a}| \beta_{c}) \ne 0$, $c \ne a$}
Substituting
$u=u^{(c)}_{k} +(\beta_{a}|\beta_{c})+d_{a}(1+2j)$ for $j \in \{0,1,\dots, -C_{ac}-1\}$ into \eqref{QQrb3}, 
we obtain  
\begin{multline}
 \frac{\Qc_{a}(u_{k}^{(c)}+(\beta_{a}|\beta_{c})+2d_{a}j) }{\Qc_{a}(u_{k}^{(c)}+(\beta_{a}|\beta_{c})+2d_{a}(1+j) )}=
 e^{-\beta_{a}(h)}
\frac{\widetilde{\Qc}_{a}(u_{k}^{(c)}+(\beta_{a}|\beta_{c})+2d_{a}j) }{\widetilde{\Qc}_{a}(u_{k}^{(c)} +(\beta_{a}|\beta_{c})+2d_{a}(1+j) )}
\\
\text{for} \quad k\in \{1,2,\dots, n_{c} \} .
\label{BAEp3pre}
\end{multline}
Taking the product over $j \in \{0,1,\dots, -C_{ac}-1\}$ on both sides of \eqref{BAEp3pre}, 
we obtain
\begin{align}
 \frac{\Qc_{a}(u_{k}^{(c)}+(\beta_{a}|\beta_{c})) }{\Qc_{a}(u_{k}^{(c)}-(\beta_{a}|\beta_{c}) )}=
 e^{C_{ac}\beta_{a}(h)}
\frac{\widetilde{\Qc}_{a}(u_{k}^{(c)}+(\beta_{a}|\beta_{c})) }{\widetilde{\Qc}_{a}(u_{k}^{(c)} -(\beta_{a}|\beta_{c}) )}
\quad
\text{for} \quad k\in \{1,2,\dots, n_{c} \} .
\label{BAEp3}
\end{align}
Substituting \eqref{BAEp3} into the right hand side of \eqref{BAEr4p} (the part $b=a$), 
we arrive at \eqref{BAEr4w}. Here we use the relations $w_{\beta_{a}}(\beta_{c})=\beta_{c}-C_{ac}\beta_{a}$, 
\eqref{IPWb}. 
%%%
\subparagraph{The case $(\beta_{a}| \beta_{a}) \ne 0$, $p_{\beta_{a}}=1$, $(\beta_{a}| \beta_{c}) =0$, $c \ne a$}
This case is trivial. 
%%%%%%%%%%%%%%%%%%
\paragraph{Fermionic QQ-relation \eqref{QQrf3}: odd reflection (gray dot)}
\subparagraph{The case $(\beta_{a}| \beta_{a}) = 0$, $c=a$}
Substituting
$u=\widetilde{u}^{(a)}_{k}$ into \eqref{QQrf3}, 
we obtain 
\begin{align}
  \frac{\varphi_{a}^{-}(\widetilde{u}_{k}^{(a)})}{\varphi_{a}^{+}(\widetilde{u}_{k}^{(a)})}=
 e^{\beta_{a}(h)}
 \prod_{b=1 \atop b \ne a}^{\rr}
\frac{\Qc_{b}(\widetilde{u}_{k}^{(a)}-(\beta_{a}| \beta_{b})) }{\Qc_{b}(\widetilde{u}_{k}^{(a)}+(\beta_{a}|\beta_{b} ) )}
\quad 
\text{for} \quad k\in \{1,2,\dots, \widetilde{n}_{a} \}  .
 \label{BAEf1}
\end{align}
Because of the condition 
$(w_{\beta_{a}}(\beta_{a})|w_{\beta_{a}}(\beta_{a}) ) =(-\beta_{a}|-\beta_{a} ) =0$, and 
the relations \eqref{Wodd} and \eqref{IPWf}, 
 \eqref{BAEr4w} for $c=a$ has the form: 
\begin{align}
 \frac{w_{\beta_{a}}(\psi_{a}^{+})(\widetilde{u}_{k}^{(a)} )}{w_{\beta_{a}}(\psi_{a}^{-})(\widetilde{u}_{k}^{(a)} )}=
 e^{\beta_{a}(h)}
 \prod_{b=1 \atop b \ne a}^{\rr}
\frac{\Qc_{b} (\widetilde{u}_{k}^{(a)}-
(\beta_{a}| \beta_{b})) }{\Qc_{b}(\widetilde{u}_{k}^{(a)}+
(\beta_{a} | \beta_{b} ) )} 
\quad 
\text{for} \quad k\in \{1,2,\dots, \widetilde{n}_{a} \} .
\label{BAEfp1}
\end{align}
In order to show that \eqref{BAEf1} and \eqref{BAEfp1} coincide, we have to check
\begin{align}
  \frac{\varphi_{a}^{-}(\widetilde{u}_{k}^{(a)})}{\varphi_{a}^{+}(\widetilde{u}_{k}^{(a)})}=
  \frac{w_{\beta_{a}}(\psi_{a}^{+})(\widetilde{u}_{k}^{(a)} )}{w_{\beta_{a}}(\psi_{a}^{-})(\widetilde{u}_{k}^{(a)} )}
  \label{va=va0}
\end{align}
In our examples, \eqref{va=va0} reduces to
\begin{align}
\frac{\Qb_{\emptyset}(\widetilde{u}_{k}^{(a)} -(\epsilon_{i_{1}^{*}}|\beta_{a}) ) }{\Qb_{\emptyset}(\widetilde{u}_{k}^{(a)} +(\epsilon_{i_{1}^{*}}|\beta_{a}) )  } 
=\frac{
\Qb_{\emptyset}(\widetilde{u}_{k}^{(a)}-(w_{\beta_{a}}(\epsilon_{i_{1}^{*}})|\beta_{a} ) )}{
\Qb_{\emptyset}(\widetilde{u}_{k}^{(a)}+(w_{\beta_{a}}(\epsilon_{i_{1}^{*}})|\beta_{a} ) )} .
\label{va=va02}
\end{align}
This relation follows from 
\begin{align}
(\epsilon_{i_{1}^{*}}|\beta_{a})=(w_{\beta_{a}}(\epsilon_{i_{1}^{*}})|\beta_{a} ) .
\label{ipw1}
\end{align}
Let us check \eqref{ipw1} for 
(i)  $osp(2r|2s)$ for $r+s=2$ of type D ($p_{i_{2}}=1$), and 
(ii) the other case for  $osp(2r|2s)$ and $osp(2r+1|2s)$ for $r+s \ge 2$. 
(i) \eqref{ipw1} reduces to  the condition $(\beta_{a}|\beta_{a})=0$ for $a=1$ or $2$ because of 
$\beta_{1}=\epsilon_{i_{1}^{*}}-\epsilon_{i_{2}^{*}}$, $\beta_{2}=\epsilon_{i_{1}^{*}}+\epsilon_{i_{2}^{*}}$, 
$w_{\beta_{1}}(\epsilon_{i_{1}^{*}})=\epsilon_{i_{2}^{*}}$, 
 $w_{\beta_{2}}(\epsilon_{i_{1}^{*}})=-\epsilon_{i_{2}^{*}}$. 
 (ii)  \eqref{ipw1}  follows from $\beta_{1}=\epsilon_{i_{1}^{*}}-\epsilon_{i_{2}^{*}}$, 
 $w_{\beta_{1}}(\epsilon_{i_{1}^{*}})=\epsilon_{i_{2}^{*}}$, 
$(\beta_{a}|\beta_{a})=0$ 
for $a=1$; and 
 $(\epsilon_{i_{1}^{*}}|\beta_{a}) = 0$, $w_{\beta_{a}}(\epsilon_{i_{1}^{*}})=\epsilon_{i_{1}^{*}}$  for   $a \ge 2$. 

%%%%%%%%
\subparagraph{The case $(\beta_{a}| \beta_{a}) = 0$, $(\beta_{a}| \beta_{c}) \ne 0$, $c \ne a$}
Substituting 
$u=u^{(c)}_{k} \pm (\beta_{a}|\beta_{c})$ into \eqref{QQrf3} and taking the ratio of the resultant equations on 
both sides, we obtain
\begin{multline}
 \frac{\Qc_{a}(u_{k}^{(c)}+(\beta_{a}|\beta_{c})) }{\Qc_{a}(u_{k}^{(c)}-(\beta_{a}|\beta_{c}) )}=
- e^{- \beta_{a}(h)}
  \frac{\varphi_{a}^{-}(u_{k}^{(c)}+(\beta_{a} | \beta_{c})) \widetilde{\Qc}_{a}(u_{k}^{(c)}-(\beta_{a}| \beta_{c})) 
}{ \varphi_{a}^{+}(u_{k}^{(c)} -(\beta_{a} | \beta_{c} )) \widetilde{\Qc}_{a} (u_{k}^{(c)} + (\beta_{a} | \beta_{c} ) )} 
\times 
\\
\times 
 \prod_{b=1 \atop (\beta_{a}|\beta_{b}) \ne 0,b \ne a  }^{\rr}
\frac{\Qc_{b}(u_{k}^{(c)}+(\beta_{a} | (\beta_{b}+\beta_{c} )) }{\Qc_{b}(u_{k}^{(c)}-( \beta_{a} | (\beta_{b}+\beta_{c}) )} 
\quad
\text{for} \quad k\in \{1,2,\dots, n_{c} \} ,
\label{BAEp4}
\end{multline}
Substituting \eqref{BAEp4} into the right hand side of \eqref{BAEr4p} (the part $b=a$), we obtain
\begin{multline}
 - \frac{\psi_{c}^{+}(u_{k}^{(c)}) \varphi_{a}^{+}(u_{k}^{(c)}-(\beta_{a} | \beta_{c}) ) }{\psi_{c}^{-}(u_{k}^{(c)} ) \varphi_{a}^{-}(u_{k}^{(c)}+(\beta_{a} | \beta_{c}) )}=
 p_{\beta_{a}} p_{\beta_{c}}
 e^{-(\beta_{a} +\beta_{c} )(h)}
  \frac{ \widetilde{\Qc}_{a}(u_{k}^{(c)}-(\beta_{a}| \beta_{c})) 
}{  \widetilde{\Qc}_{a} (u_{k}^{(c)} + (\beta_{a} | \beta_{c} ) )}  
\times 
\\
\times 
 \prod_{b=1 \atop (\beta_{a}|\beta_{b}) \ne 0,b \ne a  }^{\rr}
\frac{\Qc_{b}(u_{k}^{(c)}+(\beta_{a} | (\beta_{b}+\beta_{c} ))) }{\Qc_{b}(u_{k}^{(c)}-( \beta_{a} | (\beta_{b}+\beta_{c})))} 
 \prod_{b=1 \atop  b \ne a,c}^{\rr}
\frac{\Qc_{b}(u_{k}^{(c)}+(\beta_{c}| \beta_{b})) }{\Qc_{b}(u_{k}^{(c)}-(\beta_{c}|\beta_{b} ) )} 
\times
\\
\times 
\begin{cases}
\frac{\Qc_{c}(u_{k}^{(c)}+2(\beta_{c}| \beta_{c}))\Qc_{c}(u_{k}^{(c)}-(\beta_{c}|\beta_{c} ) ) }{\Qc_{c}(u_{k}^{(c)}-2(\beta_{c}|\beta_{c} ) )\Qc_{c}(u_{k}^{(c)}+(\beta_{c}|\beta_{c} ) )} & 
\text{if} \quad p_{\beta_{c}}=-1, \quad  (\beta_{c}|\beta_{c} )\ne 0 \\[5pt]
1 & \text{if} \quad (\beta_{c}|\beta_{c} )= 0
\\
 \frac{\Qc_{c}(u_{k}^{(c)}+(\beta_{c}| \beta_{c})) }{\Qc_{c}(u_{k}^{(c)}-(\beta_{c}|\beta_{c} ) )} & \text{otherwise}
\end{cases}
\\
\text{for} \quad k\in \{1,2,\dots, n_{c} \} .
\label{BAE5p} 
\end{multline}
Let us write down \eqref{BAEr4w} in this case.
\begin{multline}
 -  \frac{w_{\beta_{a}}(\psi_{c}^{+})(u_{k}^{(c)} )}{w_{\beta_{a}}(\psi_{c}^{-})(u_{k}^{(c)} )}=
 p_{\beta_{a} + \beta_{c}}
 e^{-(\beta_{a} +\beta_{c} )(h)}
  \frac{ \widetilde{\Qc}_{a}(u_{k}^{(c)}-(\beta_{c}| \beta_{a})) 
}{  \widetilde{\Qc}_{a} (u_{k}^{(c)} + (\beta_{c} | \beta_{a} ) )}  
\times 
\\
\times 
 \prod_{b=1 \atop (\beta_{a}|\beta_{b}) \ne 0,b \ne a,c  }^{\rr}
\frac{\Qc_{b}(u_{k}^{(c)}+(\beta_{c}|\beta_{b} )+(\beta_{a} | (\beta_{b}+\beta_{c} ))) }{\Qc_{b}(u_{k}^{(c)}-(\beta_{c}|\beta_{b} )-( \beta_{a} | (\beta_{b}+\beta_{c}) ))} 
 \prod_{b=1 \atop  (\beta_{a}|\beta_{b}) = 0, b \ne a}^{\rr}
\frac{\Qc_{b}(u_{k}^{(c)}+(\beta_{c}| \beta_{b})) }{\Qc_{b}(u_{k}^{(c)}-(\beta_{c}|\beta_{b} ) )}
\times
\\
\times 
\begin{cases}
\frac{\Qc_{c}(u_{k}^{(c)}+2(\beta_{c}| \beta_{c})+4(\beta_{a}|\beta_{c} ))\Qc_{c}(u_{k}^{(c)}-(\beta_{c}|\beta_{c} )-2(\beta_{a}|\beta_{c} ) ) }{\Qc_{c}(u_{k}^{(c)}-2(\beta_{c}|\beta_{c} )-4(\beta_{a}|\beta_{c} ) )\Qc_{c}(u_{k}^{(c)}+(\beta_{c}|\beta_{c} )+2(\beta_{a}|\beta_{c} ) )} & 
\text{if} \quad p_{\beta_{c}}=1, \quad  (\beta_{c}|\beta_{c} )+2(\beta_{a}|\beta_{c} ) \ne 0 \\
1 & \text{if} \quad  (\beta_{c}|\beta_{c} )+2(\beta_{a}|\beta_{c} )= 0
\\
 \frac{\Qc_{c}(u_{k}^{(c)}+(\beta_{c}| \beta_{c})+2(\beta_{a}|\beta_{c} )) }{\Qc_{c}(u_{k}^{(c)}-(\beta_{c}|\beta_{c} )-2(\beta_{a}|\beta_{c} ) )} & \text{otherwise}
\end{cases}
\\
\text{for} \quad k\in \{1,2,\dots, n_{c} \} .
\label{BAE6p} 
\end{multline}
In order to show that \eqref{BAE5p} and  \eqref{BAE6p} coincide, we have to check 
\begin{align}
\frac{\psi_{c}^{+}(u_{k}^{(c)}) \varphi_{a}^{+}(u_{k}^{(c)}-(\beta_{a} | \beta_{c}) ) }{\psi_{c}^{-}(u_{k}^{(c)} ) \varphi_{a}^{-}(u_{k}^{(c)}+(\beta_{a} | \beta_{c}) )}
=\frac{w_{\beta_{a}}(\psi_{c}^{+})(u_{k}^{(c)} )}{w_{\beta_{a}}(\psi_{c}^{-})(u_{k}^{(c)} )}
\label{va=va1}
\end{align}
and
\begin{multline}
\frac{
\Qc_{c}(u_{k}^{(c)}+2(\beta_{a}| \beta_{c})) }{
\Qc_{c}(u_{k}^{(c)}-2(\beta_{a}|\beta_{c} ) ) } 
 \prod_{b=1 \atop (\beta_{a}|\beta_{b}) \ne 0, (\beta_{c}|\beta_{b}) \ne 0, b \ne a,c}^{\rr}
\frac{\Qc_{b}(u_{k}^{(c)}+(\beta_{a} | (\beta_{b}+\beta_{c} ))) 
\Qc_{b}(u_{k}^{(c)}+(\beta_{c}| \beta_{b})) }{\Qc_{b}(u_{k}^{(c)}-( \beta_{a} | (\beta_{b}+\beta_{c}))) 
\Qc_{b}(u_{k}^{(c)}-(\beta_{c}|\beta_{b} ) ) } 
\times
\\
\times 
\begin{cases}
\frac{\Qc_{c}(u_{k}^{(c)}+2(\beta_{c}| \beta_{c}))\Qc_{c}(u_{k}^{(c)}-(\beta_{c}|\beta_{c} ) ) }{\Qc_{c}(u_{k}^{(c)}-2(\beta_{c}|\beta_{c} ) )\Qc_{c}(u_{k}^{(c)}+(\beta_{c}|\beta_{c} ) )} & 
\text{if} \quad p_{\beta_{c}}=-1, \quad  (\beta_{c}|\beta_{c} )\ne 0 \\
1 & \text{if} \quad  (\beta_{c}|\beta_{c} )= 0
\\
 \frac{\Qc_{c}(u_{k}^{(c)}+(\beta_{c}| \beta_{c})) }{\Qc_{c}(u_{k}^{(c)}-(\beta_{c}|\beta_{c} ) )} & \text{otherwise}
\end{cases}
\\
=
 \prod_{b=1 \atop (\beta_{a}|\beta_{b}) \ne 0, (\beta_{c}|\beta_{b}) \ne 0, b \ne a  }^{\rr}
\frac{\Qc_{b}(u_{k}^{(c)}+(\beta_{c}|\beta_{b} )+(\beta_{a} | (\beta_{b}+\beta_{c} ))) }{\Qc_{b}(u_{k}^{(c)}-(\beta_{c}|\beta_{b} )-( \beta_{a} | (\beta_{b}+\beta_{c}) ))} 
\times
\\
\times 
\begin{cases}
\frac{\Qc_{c}(u_{k}^{(c)}+2(\beta_{c}| \beta_{c})+4(\beta_{a}|\beta_{c} ))\Qc_{c}(u_{k}^{(c)}-(\beta_{c}|\beta_{c} )-2(\beta_{a}|\beta_{c} ) ) }{\Qc_{c}(u_{k}^{(c)}-2(\beta_{c}|\beta_{c} )-4(\beta_{a}|\beta_{c} ) )\Qc_{c}(u_{k}^{(c)}+(\beta_{c}|\beta_{c} )+2(\beta_{a}|\beta_{c} ) )} & 
\text{if} \quad p_{\beta_{c}}=1, \quad  (\beta_{c}|\beta_{c} )+2(\beta_{a}|\beta_{c} ) \ne 0 \\
1 & \text{if} \quad  (\beta_{c}|\beta_{c} )+2(\beta_{a}|\beta_{c} )= 0
\\
 \frac{\Qc_{c}(u_{k}^{(c)}+(\beta_{c}| \beta_{c})+2(\beta_{a}|\beta_{c} )) }{\Qc_{c}(u_{k}^{(c)}-(\beta_{c}|\beta_{c} )-2(\beta_{a}|\beta_{c} ) )} & \text{otherwise}
\end{cases}
\\
\quad
\text{for} \quad k\in \{1,2,\dots, n_{c} \} .
\label{BAE7p} 
\end{multline}
In our examples, \eqref{va=va1} reduces to
\begin{multline}
\frac{\Qb_{\emptyset}(u_{k}^{(c)} +(\epsilon_{i_{1}^{*}}|\beta_{c}) ) \Qb_{\emptyset}(u_{k}^{(c)}-(\beta_{a} | \beta_{c})+(\epsilon_{i_{1}^{*}}|\beta_{a}) ) }{\Qb_{\emptyset}(u_{k}^{(c)} -(\epsilon_{i_{1}^{*}}|\beta_{c}) ) \Qb_{\emptyset}(u_{k}^{(c)}+(\beta_{a} | \beta_{c})-(\epsilon_{i_{1}^{*}}|\beta_{a}) ) } 
=\frac{
\Qb_{\emptyset}(u_{k}^{(c)}+(w_{\beta_{a}}(\epsilon_{i_{1}^{*}})|(\beta_{a} + \beta_{c})) )}{
\Qb_{\emptyset}(u_{k}^{(c)}-(w_{\beta_{a}}(\epsilon_{i_{1}^{*}})|(\beta_{a} + \beta_{c})) )}
\\
 \text{if} \quad (\epsilon_{i_{1}^{*}}|\beta_{a}) \ne 0,
\label{va=va2}
\end{multline}
and a trivial identity if $ (\epsilon_{i_{1}^{*}}|\beta_{a}) = 0$. 
Let us prove \eqref{va=va2} for 
(i)  $osp(2r|2s)$ for $r+s=2$ of type D ($p_{i_{2}}=1$), and 
(ii) the other case for  $osp(2r|2s)$ and $osp(2r+1|2s)$ for $r+s \ge 2$. 
(i) In this case, $(a,c)=(1,2)$ or $(2,1)$, and the simple roots have the form 
$\beta_{1}=\epsilon_{i_{1}^{*}}-\epsilon_{i_{2}^{*}}$, $\beta_{2}=\epsilon_{i_{1}^{*}}+\epsilon_{i_{2}^{*}}$. 
Thus the condition $(\beta_{a}|\beta_{a})=0$ leads to $p_{i_{1}}=-1$. 
 Taking note on the fact $w_{\beta_{1}}(\epsilon_{i_{1}^{*}})=\epsilon_{i_{2}^{*}}$, 
 $w_{\beta_{2}}(\epsilon_{i_{1}^{*}})=-\epsilon_{i_{2}^{*}}$, one can show the relations 
 $(\epsilon_{i_{1}^{*}}|\beta_{c}) =(\beta_{a} | \beta_{c})-(\epsilon_{i_{1}^{*}}|\beta_{a})$ and 
 $(w_{\beta_{a}}(\epsilon_{i_{1}^{*}})|(\beta_{a} + \beta_{c}))=0$, from which \eqref{va=va2} follows. 
 (ii) The condition $(\epsilon_{i_{1}^{*}}|\beta_{a}) \ne 0$ leads to $a=1$. Thus  
 $\beta_{a}=\beta_{1}=\epsilon_{i_{1}^{*}}-\epsilon_{i_{2}^{*}}$. We also have $(\epsilon_{i_{1}^{*}}|\beta_{c})=0$ 
 since $c \ne a=1$. 
 It suffices to show the relation  
 $(\beta_{a} | \beta_{c})-(\epsilon_{i_{1}^{*}}|\beta_{a})=-(w_{\beta_{a}}(\epsilon_{i_{1}^{*}})|(\beta_{a} + \beta_{c}))$. 
 This reduces to $(\beta_{a} | \beta_{c})=(\beta_{a}|\beta_{a})-(\epsilon_{i_{2}^{*}}| \beta_{c})$ 
since $w_{\beta_{1}}(\epsilon_{i_{1}^{*}})=\epsilon_{i_{2}^{*}}$. One can show this based on 
$(\beta_{a}|\beta_{a})=0$.

%%%%%%%%%%%%%
The relation \eqref{BAE7p} holds true if the following two relations are valid in the product: 
\\
for the part $ b \ne c,a$, $(\beta_{a}|\beta_{b}) \ne 0$,  
$(\beta_{c}|\beta_{b}) \ne 0$, 
\begin{multline}
\frac{\Qc_{b}(u_{k}^{(c)}+(\beta_{a} | (\beta_{b}+\beta_{c} ))) 
\Qc_{b}(u_{k}^{(c)}+(\beta_{c}| \beta_{b})) }{\Qc_{b}(u_{k}^{(c)}-( \beta_{a} | (\beta_{b}+\beta_{c}))) 
\Qc_{b}(u_{k}^{(c)}-(\beta_{c}|\beta_{b} ) ) } 
=
\frac{\Qc_{b}(u_{k}^{(c)}+(\beta_{c}|\beta_{b} )+(\beta_{a} | (\beta_{b}+\beta_{c} ))) }{\Qc_{b}(u_{k}^{(c)}-(\beta_{c}|\beta_{b} )-( \beta_{a} | (\beta_{b}+\beta_{c}) ))} 
\\
\text{for } \quad 
k\in \{1,2,\dots, n_{c} \} ;
\label{BAE9p} 
\end{multline}
and for the rest,
\begin{multline}
\frac{
\Qc_{c}(u_{k}^{(c)}+2(\beta_{a}| \beta_{c})) }{
\Qc_{c}(u_{k}^{(c)}-2(\beta_{a}|\beta_{c} ) ) } 
\times
\begin{cases}
\frac{\Qc_{c}(u_{k}^{(c)}+2(\beta_{c}| \beta_{c}))\Qc_{c}(u_{k}^{(c)}-(\beta_{c}|\beta_{c} ) ) }{\Qc_{c}(u_{k}^{(c)}-2(\beta_{c}|\beta_{c} ) )\Qc_{c}(u_{k}^{(c)}+(\beta_{c}|\beta_{c} ) )} & 
\text{if} \quad p_{\beta_{c}}=-1, \quad  (\beta_{c}|\beta_{c} )\ne 0 \\
1 & \text{if} \quad  (\beta_{c}|\beta_{c} )= 0 \\
 \frac{\Qc_{c}(u_{k}^{(c)}+(\beta_{c}| \beta_{c})) }{\Qc_{c}(u_{k}^{(c)}-(\beta_{c}|\beta_{c} ) )} & \text{otherwise}
\end{cases}
\\
=
\begin{cases}
\frac{\Qc_{c}(u_{k}^{(c)}+2(\beta_{c}| \beta_{c})+4(\beta_{a}|\beta_{c} ))\Qc_{c}(u_{k}^{(c)}-(\beta_{c}|\beta_{c} )-2(\beta_{a}|\beta_{c} ) ) }{\Qc_{c}(u_{k}^{(c)}-2(\beta_{c}|\beta_{c} )-4(\beta_{a}|\beta_{c} ) )\Qc_{c}(u_{k}^{(c)}+(\beta_{c}|\beta_{c} )+2(\beta_{a}|\beta_{c} ) )} & 
\text{if} \quad p_{\beta_{c}}=1, \quad  (\beta_{c}|\beta_{c} )+2(\beta_{a}|\beta_{c} ) \ne 0 \\
1 & \text{if} \quad  (\beta_{c}|\beta_{c} )+2(\beta_{a}|\beta_{c} )= 0 \\
 \frac{\Qc_{c}(u_{k}^{(c)}+(\beta_{c}| \beta_{c})+2(\beta_{a}|\beta_{c} )) }{\Qc_{c}(u_{k}^{(c)}-(\beta_{c}|\beta_{c} )-2(\beta_{a}|\beta_{c} ) )} & \text{otherwise}
\end{cases}
\\
\quad
\text{for} \quad k\in \{1,2,\dots, n_{c} \} .
\label{BAE8p} 
\end{multline}
%%
%
%Eq. \eqref{BAE8p} is equivalent to
%\begin{multline}
%\frac{\Qc_{c}(u_{k}^{(c)}+C_{ca}(\beta_{c} | \beta_{c} )) 
%\Qc_{c}(u_{k}^{(c)}+(\beta_{c}| \beta_{c})) }{\Qc_{c}(u_{k}^{(c)}-C_{ca}(\beta_{c} | \beta_{c})) 
%\Qc_{c}(u_{k}^{(c)}-(\beta_{c}|\beta_{c} ) ) } 
%=
%\frac{\Qc_{c}(u_{k}^{(c)}+(1+C_{ca}(\beta_{c} | \beta_{c} ))) }{\Qc_{c}(u_{k}^{(c)}-(1+C_{ca}( \beta_{a} | \beta_{c})) )} 
%\\
%\text{for} \quad  (\beta_{a}|\beta_{c}) \ne 0, \ (\beta_{c}|\beta_{c}) \ne 0, \ c \ne a  ,\quad 
%k\in \{1,2,\dots, n_{c} \} ,
%\label{BAE10p} 
%\end{multline}
%This becomes trivial in case $C_{ca}=-1$. There is no $(a,c)$ that realizes $C_{ca} \ne -1$ 
%under the conditions imposed on $\beta_{a}$ and $\beta_{c}$  
%\footnote{In case $(a,c)=(r+s-1,r+s)$ for $U_{q}(osp(2r+1|2s)^{(1)})$, we have $C_{ca}=-2$. However, 
% $\beta_{c}^{\prime}=w_{\beta_{a}}(\beta_{\beta_{c}})$ corresponds to 
%an odd root with $(\beta_{c}^{\prime} | \beta_{c}^{\prime}) \ne 0$ (block dot in the Dynkin diagram).
%Thus the resultant Bethe ansatz equation deviates from the form \eqref{BAEr4}
%(we have to use \eqref{BAEr3} instead).}. 
The conditions for \eqref{BAE9p} mean that the three different vertexes $a,b,c$ of the Dynkin 
diagram form a closed loop. This is possible only
\footnote{
We expect that a similar idea can be applicable for the exceptional superalgebras
$G(3)$, $F(4)$, $D(2,1;\alpha)$, which we do not discuss here. 
}
 when $(a,b,c)$ is a permutation of the last 
three vertexes $(r+s-2,r+s-1,r+s)$ of the Dynkin diagram of $osp(2r|2s)$ 
for the simple root \eqref{rootD} with $p_{i_{r+s-1}}=-p_{i_{r+s}}$.  
Thus \eqref{BAE9p} holds true since 
$(\beta_{r+s-2}|\beta_{r+s-1})=(\beta_{r+s-2}|\beta_{r+s})=-p_{i_{r+s-1}}$, 
$(\beta_{r+s-2}|\beta_{r+s})=p_{i_{r+s-1}}-p_{i_{r+s}}=2p_{i_{r+s-1}}$. 
As for \eqref{BAE8p}, we consider the case $(\beta_{c}|\beta_{c} )=0$ first. 
In this case, \eqref{BAE8p} becomes trivial since $p_{\beta_{c}}=-1$. 
Next we consider the following three cases for  $(\beta_{c}|\beta_{c} ) \ne 0$. 
(1) The case $(\beta_{c}|\beta_{c} ) \ne 0$,  $p_{\beta_{c}}=1$, $(\beta_{c}|\beta_{c} )+2(\beta_{a}|\beta_{c} )=0$: 
\eqref{BAE8p} reduces to a trivial identity. 
(2) The case $(\beta_{c}|\beta_{c} ) \ne 0$,  $p_{\beta_{c}}=1$, $(\beta_{c}|\beta_{c} )+2(\beta_{a}|\beta_{c} ) \ne 0$:   
the only possibility is $C_{ca}=2(\beta_{c}|\beta_{a} ) /(\beta_{c}|\beta_{c} )= -2$, from which \eqref{BAE8p} holds
 since the cases $C_{ca}=0$ and $C_{ca}=-1$ contradict the conditions $(\beta_{a}|\beta_{c} ) \ne 0$ 
and $(\beta_{c}|\beta_{c} )+2(\beta_{a}|\beta_{c} ) \ne 0$, respectively. 
(3) The case $(\beta_{c}|\beta_{c} ) \ne 0$,  $p_{\beta_{c}}=-1$: this means that the vertex $c$ of the Dynkin diagram 
 is a black dot. This is possible only when $c$ is the $(r+s)$-th vertex of the Dynkin diagram of
 $osp(2r+1|2s)$ for the simple root \eqref{rootB} with $p_{i_{r+s}}=-1$ ($=p_{\beta_{r+s}}$). 
 Thus \eqref{BAE8p} holds since $(a,c)=(r+s-1,r+s)$, $C_{ca}=-2$.  

%\subparagraph{The case $(\beta_{c}|\beta_{c} ) \ne 0$} 
%only the cases $(a,b,c)=(r+s-2,r+s-1,r+s)$ or  $(a,b,c)=(r+s-2,r+s,r+s-1)$ 
%for $U_{q}(osp(2r|2s)^{(1)})$ with $- p_{i_{r+s-1}}=p_{i_{r+s}}=1$
%satisfy the required conditions on $\beta_{a}$, $\beta_{b}$ and $\beta_{c}$. 
%In both cases, \eqref{BAE9p} becomes trivial since
% $(\beta_{c}|\beta_{b})=-2$, $(\beta_{a}|\beta_{b})=(\beta_{a}|\beta_{c})=1$ hold. 
%%%
\subparagraph{The case $(\beta_{a}| \beta_{a}) = 0$, $(\beta_{a}| \beta_{c}) =0$, $c \ne a$}
This case is trivial. 
% 
%where we use \eqref{IPW}, $w_{\beta_{a}}(\beta_{a})=-\beta_{a}$, $p_{-\beta_{a}}=p_{\beta_{a}}$, 
%$(w_{\beta_{a}}(\beta_{a}) |w_{\beta_{a}}(\beta_{b}) )=(\beta_{a} |\beta_{b} )$ if $(\beta_{a} |\beta_{a} ) \ne 0$, 
%$(w_{\beta_{a}}(\beta_{a}) |w_{\beta_{a}}(\beta_{b}) )=-(\beta_{a} |\beta_{b} )$ if $(\beta_{a} |\beta_{a} ) = 0$, 
%$w_{\beta_{a}}(d_{a})=p_{\beta_{a}}d_{a}$. 
%%%%%%%%%%%%%%%%%
\paragraph{QQ-relation \eqref{QQrbb3} (and \eqref{QQb6} and \eqref{QQb7}): odd reflection 
 (black dot: the case $(\beta_{a}| \beta_{a}) \ne 0$, $p_{\beta_{a}}=-1$)}
Repeating a similar argument as above for the $U_{q}(gl(M|N)^{(1)})$ case 
(for \eqref{BAEr2}-\eqref{QQrf0}), we identity $w_{\alpha_{a}} (\Qc_{a})=\widetilde{\Qc}_{a}$,
$w_{\alpha_{a}} (\Qc_{b})=\Qc_{b}$ 
for $a \ne b$ in \eqref{QQrb0} and \eqref{QQrf0}, where $\alpha_{a}$ is a simple root of $gl(M|N)$.  
The only case in which \eqref{QQrbb3} is realized (for the algebras in question in this paper) is when 
$a$ corresponds to the black dot of a Dynkin diagram of $osp(2r+1|2s)$ ($a=r+s$). 
Taking note on the fact that \eqref{QQrbb3}, namely \eqref{QQb3}
 is a reduction of \eqref{QQb} for $U_{q}(gl(2r|2s+1)^{(1)})$, we identity 
 $w_{\alpha_{r+s}} (\Qc_{r+s})=\widetilde{\Qc}_{r+s}$, $w_{\alpha_{r+s}} (\Qc_{b})=\Qc_{b}$ for $b \ne r+s$ 
 (under the reduction) in \eqref{QQrbb3}. 
 Note however that this does not keep the standard form of the 
 Bethe ansatz equation \eqref{BAEr4p} since $\widetilde{\Qc}_{r+s}=\Qb_{\widetilde{I}_{r+s}}$ is not on the symmetric nesting path. 
 In order to keep the form, we have to consider the odd reflection $w_{\beta_{r+s}}$ of $osp(2r+1|2s)$
  [with $(\beta_{r+s}|\beta_{r+s}) \ne 0$,  $p_{\beta_{r+s}}=-1$],  
 which acts on the Q-functions as 
 $w_{\beta_{r+s}} (\Qc_{r+s})=w_{\alpha_{r+s}^{\prime \prime}}w_{\alpha_{r+s+1}^{\prime}}w_{\alpha_{r+s}} (\Qc_{r+s})=\Qb_{\breve{I}_{r+s}}$, $w_{\beta_{r+s}} (\Qc_{b})=\Qc_{b}$ for $b \ne r+s$ 
 (under the reduction). Here the action of the odd reflection by $\beta_{r+s}$ is realized
\footnote{Another option is  $w_{\beta_{r+s}} (\Qc_{r+s})=w_{\alpha_{r+s+1}^{\prime \prime}}w_{\alpha_{r+s}^{\prime}}w_{\alpha_{r+s+1}} (\Qc_{r+s})=\Qb_{\breve{I}_{r+s}}$, $w_{\beta_{r+s}} (\Qc_{b})=\Qc_{b}$ for $b \ne r+s$ 
 (under the reduction). Here the action of the odd reflection by $\beta_{r+s}$ is realized by the Weyl reflections of 
 $gl(2r|2s+1)$ under the reduction, 
 where $\alpha_{r+s+1}=\epsilon_{i_{r+s+1}}-\epsilon_{i_{r+s}}=\epsilon_{2r+s+1}-\epsilon_{i_{r+s}}$, 
 $\alpha_{r+s}^{\prime}=\epsilon_{i_{r+s+2}}-\epsilon_{i_{r+s}}=\epsilon_{i_{r+s}^{*}}-\epsilon_{i_{r+s}}$, 
 $\alpha_{r+s+1}^{\prime \prime}=\epsilon_{i_{r+s+2}}-\epsilon_{i_{r+s+1}}=\epsilon_{i_{r+s}^{*}}-\epsilon_{2r+s+1}$. 
 These  roots and the roots in the main text reduce to 
 $ \beta_{r+s},2\beta_{r+s}$ 
 by the formal replacement $(\epsilon_{i_{r+s}},\epsilon_{2r+s+1}) \to (- \epsilon_{i_{r+s}^{*}},0) $. 
In order to describe the whole symmetry of the system, we may need BC-like root system. 
This is also the case with $U_{q}(gl(2r|2s+1)^{(2)})$. This point needs further research. 
%It may be an interesting problem to reconsider the QQ-relations for $U_{q}(osp(2r+1|2s)^{(1)})$ and $U_{q}(gl(2r|2s+1)^{(2)})$ 
% with respect to BC-like root systems. 
 }
  by the Weyl reflections of 
 $gl(2r|2s+1)$ under the reduction, 
 where $\alpha_{r+s}=\epsilon_{i_{r+s+2}}-\epsilon_{i_{r+s+1}}=\epsilon_{i_{r+s}^{*}}-\epsilon_{2r+s+1}$, 
 $\alpha_{r+s+1}^{\prime}=\epsilon_{i_{r+s+2}}-\epsilon_{i_{r+s}}=\epsilon_{i_{r+s}^{*}}-\epsilon_{i_{r+s}}$, 
 $\alpha_{r+s}^{\prime \prime}=\epsilon_{i_{r+s+1}}-\epsilon_{i_{r+s}}=\epsilon_{2r+s+1}-\epsilon_{i_{r+s}}$. 
 The Weyl reflections by the even roots $\alpha_{r+s}$, $\alpha_{r+s+1}^{\prime}$ and  $\alpha_{r+s}^{\prime \prime}$ 
 correspond to the bosonic QQ-relations 
 \eqref{QQb3} (namely, \eqref{QQrbb3}), \eqref{QQb6} and \eqref{QQb7}, respectively. 

Starting from the Bethe ansatz equation associated with one of the Dynkin diagrams, 
one can obtain any other Bethe ansatz equation by using QQ-relations repeatedly. 
%
%\footnote{This is not applicable to the black dot in the Dynkin diagrams of 
%$U_{q}(osp(2r+|2s)^{(1)})$ with the condition $i_{r+s} \in \Fm$.}. 
Now that the QQ-relations \eqref{QQrb2}, \eqref{QQrf2},  \eqref{QQrb3} and \eqref{QQrf3},  
and the Bethe ansatz equations \eqref{BAEr3} and \eqref{BAEr4} 
are expressed in terms of the root systems of the underlying algebras, these are 
expected to be valid for other quantum affine superalgebras as they are or with slight modifications.

%%%%%%%%%%%%%%%%%%%
\subsection{Bethe strap}
\label{sec:BS}
In this subsection, we will explain our observation on Bethe straps for orthosymplectic 
superalgebras based on computer experiments with Mathematica (ver.\ 7).

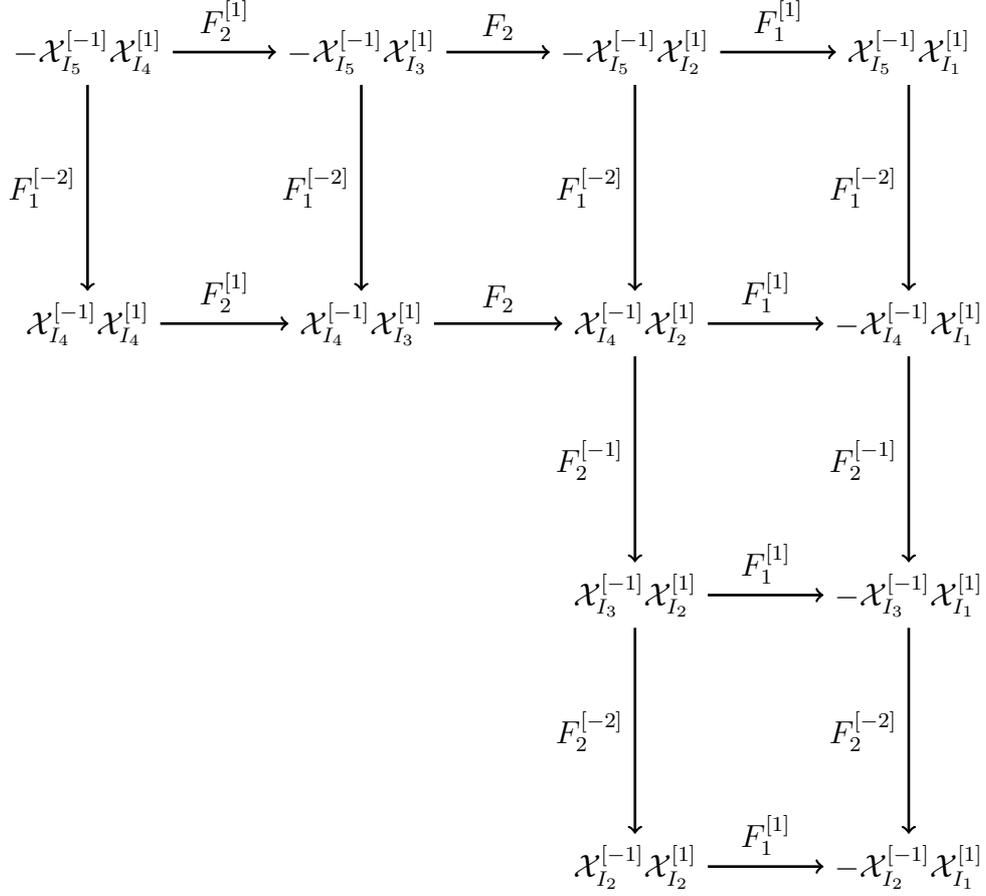
\begin{figure}
 \centering
 \begin{tikzpicture}[x=9mm,y=9mm]
  \node (54) at (-6,6) {$ -{\mathcal  X}_{I_{5}}^{[-1]} {\mathcal  X}_{I_{4}}^{[1]}$};
  \node (53) at (-2,6) {$ -{\mathcal  X}_{I_{5}}^{[-1]} {\mathcal  X}_{I_{3}}^{[1]}$};
  \node (52) at (2,6) {$ -{\mathcal  X}_{I_{5}}^{[-1]} {\mathcal  X}_{I_{2}}^{[1]}$};
  \node (51) at (6,6) {$ {\mathcal  X}_{I_{5}}^{[-1]} {\mathcal  X}_{I_{1}}^{[1]}$};
 \node (44) at (-6,2) {$ {\mathcal  X}_{I_{4}}^{[-1]} {\mathcal  X}_{I_{4}}^{[1]}$};
  \node (43) at (-2,2) {$ {\mathcal  X}_{I_{4}}^{[-1]} {\mathcal  X}_{I_{3}}^{[1]}$};
  \node (42) at (2,2) {$ {\mathcal  X}_{I_{4}}^{[-1]} {\mathcal  X}_{I_{2}}^{[1]}$};
  \node (41) at (6,2) {$- {\mathcal  X}_{I_{4}}^{[-1]} {\mathcal  X}_{I_{1}}^{[1]}$};
 \node (32) at (2,-2) {$ {\mathcal  X}_{I_{3}}^{[-1]} {\mathcal  X}_{I_{2}}^{[1]}$};
 \node (22) at (2,-6) {$ {\mathcal  X}_{I_{2}}^{[-1]} {\mathcal  X}_{I_{2}}^{[1]}$};
 \node (31) at (6,-2) {$ -{\mathcal  X}_{I_{3}}^{[-1]} {\mathcal  X}_{I_{1}}^{[1]}$};
 \node (21) at (6,-6) {$ -{\mathcal  X}_{I_{2}}^{[-1]} {\mathcal  X}_{I_{1}}^{[1]}$};
%%%
  \draw[-to,line width=1pt] (54) -- node[midway,above]{$F_{2}^{[1]}$} (53); 
  \draw[-to,line width=1pt] (53) -- node[midway,above]{$F_{2}$} (52); 
  \draw[-to,line width=1pt] (52) -- node[midway,above]{$F_{1}^{[1]}$} (51); 
 \draw[-to,line width=1pt] (54) -- node[midway,left]{$F_{1}^{[-2]}$} (44); 
  \draw[-to,line width=1pt] (53) -- node[midway,left]{$F_{1}^{[-2]}$} (43); 
   \draw[-to,line width=1pt] (52) -- node[midway,left]{$F_{1}^{[-2]}$} (42); 
    \draw[-to,line width=1pt] (51) -- node[midway,left]{$F_{1}^{[-2]}$} (41); 
 \draw[-to,line width=1pt] (44) -- node[midway,above]{$F_{2}^{[1]}$} (43); 
  \draw[-to,line width=1pt] (43) -- node[midway,above]{$F_{2}$} (42); 
  \draw[-to,line width=1pt] (42) -- node[midway,above]{$F_{1}^{[1]}$} (41); 
 \draw[-to,line width=1pt] (42) -- node[midway,left]{$F_{2}^{[-1]}$} (32); 
 \draw[-to,line width=1pt] (32) -- node[midway,left]{$F_{2}^{[-2]}$} (22); 
 \draw[-to,line width=1pt] (41) -- node[midway,left]{$F_{2}^{[-1]}$} (31); 
 \draw[-to,line width=1pt] (31) -- node[midway,left]{$F_{2}^{[-2]}$} (21); 
  \draw[-to,line width=1pt] (32) -- node[midway,above]{$F_{1}^{[1]}$} (31); 
 \draw[-to,line width=1pt] (22) -- node[midway,above]{$F_{1}^{[1]}$} (21); 
\end{tikzpicture}
\caption{Bethe strap structures of the T-function ${\mathcal F}_{(2)}^{I_{5}}$ for $U_{q}(osp(3|2)^{(1)})$, where 
$\Bm=\{1,2\}$, $\Fm =\{3,4,5 \}$, 
$I_{5}=(5,2,4,1,3)$, $I_{4}=(5,2,4,1)$, $I_{3}=(5,2,4)$, $I_{2}=(5,2)$, $I_{1}=(5)$, $I_{0}=\emptyset $. 
The top term $ -{\mathcal  X}_{I_{5}}^{[-1]} {\mathcal  X}_{I_{4}}^{[1]}$ carries the $osp(3|2)$ highest weight $\epsilon_{3}+\epsilon_{1}$.}
\label{BSb2}
\end{figure}

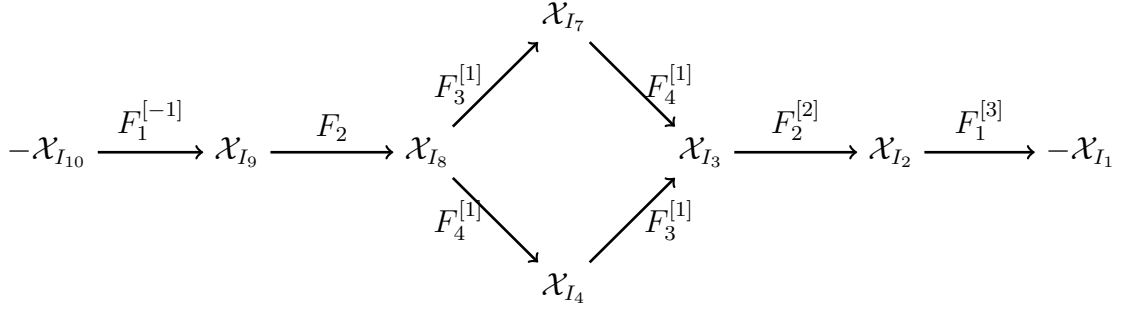
\begin{figure}
 \centering
 \begin{tikzpicture}[x=10mm,y=10mm]
  \node (10) at (-6.8,0) {$- {\mathcal  X}_{I_{10}}$};
  \node (9) at (-4.3,0) {${\mathcal  X}_{I_{9}}$};
  \node (8) at (-1.8,0) {${\mathcal  X}_{I_{8}} $};
  \node (7) at (0,1.8) {${\mathcal  X}_{I_{7}} $};
   \node (4) at (0,-1.8) {${\mathcal  X}_{I_{4}} $};
   \node (3) at (1.8,0) {${\mathcal  X}_{I_{3}} $};
   \node (2) at (4.3,0) {${\mathcal  X}_{I_{2}} $};
   \node (1) at (6.8,0) {$-{\mathcal  X}_{I_{1}} $};
 % \node[below of = 4]  {${\mathcal F}_{(1)}^{(10,6,5,4,9,8,3,2,1)}$};
  \draw[-to,line width=1pt] (10) -- node[midway,above]{$F_{1}^{[-1]}$} (9); 
  \draw[-to,line width=1pt] (9) -- node[midway,above]{$F_{2}$} (8); 
  \draw[-to,line width=1pt] (8) -- node[midway,left]{$F_{3}^{[1]}$} (7); 
  \draw[-to,line width=1pt] (8) -- node[midway,left]{$F_{4}^{[1]}$} (4); 
  \draw[-to,line width=1pt] (7) -- node[midway,right]{$F_{4}^{[1]}$} (3); 
  \draw[-to,line width=1pt] (4) -- node[midway,right]{$F_{3}^{[1]}$} (3); 
  \draw[-to,line width=1pt] (3) -- node[midway,above]{$F_{2}^{[2]}$} (2); 
  \draw[-to,line width=1pt] (2) -- node[midway,above]{$F_{1}^{[3]}$} (1); 
\end{tikzpicture}
\caption{Bethe strap structures of the T-function ${\mathcal F}_{(1)}^{I_{10}}$ for $U_{q}(osp(6|2)^{(1)})$, 
where $\Bm =\{1,2,3,4,5,6\}$, $\Fm =\{7,8,9,10\}$, $\mathfrak{D} =\{8,9\}$, 
$I_{10}=(10,6,5,4,9,8,3,2,1,7)$, $I_{9}=(10,6,5,4,9,8,3,2,1)$, $I_{8}=(10,6,5,4,9,8,3,2)$, 
$I_{7}=(10,6,5,4,9,8,3)$, $I_{6}=(10,6,5,4,9,8)$, $I_{5}=(10,6,5,4,9)$, $I_{4}=(10,6,5,4)$, $I_{3}=(10,6,5)$, 
$I_{2}=(10,6)$, $I_{1}=(10)$, $I_{0}=\emptyset$. 
The top term $ -{\mathcal  X}_{I_{10}}$ carries the $osp(6|2)$ highest weight $\epsilon_{7}$.}
\label{BSd}
\end{figure}

%%%%%%%%%%%%%%%%%%
In relation to the Bethe ansatz equation \eqref{BAEr4}, 
we introduce the following function 
\begin{multline}
F_{a}(u)=
 p_{\beta_{a}}
 e^{-\beta_{a}(h)}
%\frac{P_{a}(u-(\epsilon_{i_{1}^{*}}| \beta_{a}) )}{P_{a}(u+(\epsilon_{i_{1}^{*}}| \beta_{a}))}
  \frac{\psi_{a}^{-}(u)}{\psi_{a}^{+}(u)}
 \prod_{b=1\atop b \ne a}^{\rr}
\frac{\Qc_{b}(u+(\beta_{a}| \beta_{b})) }{\Qc_{b}(u-(\beta_{a}|\beta_{b} ) )}
 \prod_{l =1}^{\kappa_{a}}
\frac{\Qc_{a}(u+p_{l \beta_{a}} l (\beta_{a}| \beta_{a})) }{\Qc_{a}(u-p_{l \beta_{a}} l(\beta_{a}|\beta_{a} ) )}
%\\
%
%\times 
%\begin{cases}
%\frac{\Qc_{a}(u+2(\beta_{a}| \beta_{a}))\Qc_{a}(u-(\beta_{a}|\beta_{a} ) ) }{\Qc_{a}(u-2(\beta_{a}|\beta_{a} ) )\Qc_{a}(u+(\beta_{a}|\beta_{a} ) )} & 
%\text{if} \quad p_{\beta_{a}}=-1, \quad  (\beta_{a}|\beta_{a} )\ne 0 \\[5pt]
% \frac{\Qc_{a}(u+(\beta_{a}| \beta_{a})) }{\Qc_{a}(u-(\beta_{a}|\beta_{a} ) )} & \text{otherwise}
%\end{cases}
\\
\text{for}  \quad a \in \{1,2,\dots, \rr \} .
\label{BAEr4F}
\end{multline}
%where $P_{1}=\Qb_{\emptyset}$, $P_{a}=1$ if $a \ne 1$.
%, $g_{1}d_{1}=p_{i_{1}}$. 
In our examples, the vacuum parts $\psi_{a}^{\pm}(u)$ are given by \eqref{vacbae}.  
The Bethe ansatz equation \eqref{BAEr4} is equivalent to  $F_{a}(u_{k}^{(a)})=-1$, $k\in \{1,2,\dots, n_{a} \} $. 
The adjacent terms in \eqref{boxes-b} are related to each other as
\begin{multline}
 {\mathcal   X}_{I_{2r+2s+2-a}} F_{a}^{[\sum_{j \in I_{a}} p_{j} ]}= p_{\beta_{a}} {\mathcal   X}_{I_{2r+2s+1-a}}, 
\qquad 
 {\mathcal   X}_{I_{a+1}} F_{a}^{[2r-2s-1-\sum_{j \in I_{a}} p_{j} ]}= p_{\beta_{a}} {\mathcal   X}_{I_{a}} 
 \\
\text{for} \quad 1 \le a \le r+s.
 \label{acFb}
\end{multline}
The adjacent terms in \eqref{boxes-zz-d} are related to each other as
\begin{align}
\begin{split}
& {\mathcal   X}_{I_{2r+2s+3-a}} F_{a}^{[\sum_{j \in I_{a}} p_{j} ]}= p_{\beta_{a}} {\mathcal   X}_{I_{2r+2s+2-a}}, 
\quad 
 {\mathcal   X}_{I_{a+1}} F_{a}^{[2r-2s-2-\sum_{j \in I_{a}} p_{j} ]}= p_{\beta_{a}} {\mathcal   X}_{I_{a}} 
 \\
& \quad \text{for} \quad 1 \le a \le r+s-1 ;
\\
& {\mathcal   X}_{I_{r+s+4}} F_{r+s}^{[r-s-1]}= p_{\beta_{r+s}} {\mathcal   X}_{I_{r+s}} ,
\quad 
{\mathcal   X}_{I_{r+s+3}} F_{r+s}^{[r-s-1]}= p_{\beta_{r+s}} {\mathcal   X}_{I_{r+s-1}} 
\quad \text{if} \quad i_{r+s} \in \Bm ;
\\
& {\mathcal   X}_{I_{r+s+3}} F_{r+s}^{[r-s-1]}= p_{\beta_{r+s}} {\mathcal   X}_{I_{r+s}} 
\quad \text{if} \quad i_{r+s} \in \Fm .
\end{split}
 \label{acFd}
\end{align}
T-functions form  Bethe strap structures by the relations \eqref{acFb} and \eqref{acFd} (see Figures \ref{BSb1}, \ref{BSb2}, \ref{BSc}, \ref{BSd}).  

\paragraph{$U_{q}(osp(2r+1|2s)^{(1)})$ case}
We consider the tuple 
$I_{2r+2s+1}=(2r+2s+1, 2r+2s, \dots ,2r+s+3, 2 r+s+2,2r,2r-1, 
\dots ,r+2,r+1,2r+s+1 ,r,r-1,\dots, 2,1,2r+s,2r+s-1,\dots ,2r+2,2r+1)$ and 
a partition $\mu$ with the condition $\mu_{r+1} \le s$
 (the Young diagram $\mu$ is on the $[r,s]$-hook in Figure \ref{MN-hookB}).
Let $\mathsf{t}_{\mu}(u)$ be the T-function derived by the Bethe strap procedure with the 
top term (cf.\ eq.\ (3.48) in \cite{T99}) 
\begin{align}
\mathsf{hw}_{\mu}(u)=
\prod_{k=1}^{s }
\prod_{j=1}^{\mu^{\prime}_{k}}
(-1)
{\mathcal  X}_{I_{2r+2s+2-k}}^{[-\mu_{1}+\mu_{1}^{\prime}-2j+2k]} 
\prod_{j=1}^{ r }
\prod_{k=s+1}^{\mu_{j}}
{\mathcal  X}_{I_{2r+s+2-j}}^{[-\mu_{1}+\mu_{1}^{\prime}-2j+2k]} ,
\label{topB}
\end{align}
which carries the $osp(2r+1|2s)$ highest weight \eqref{HW-B} for \eqref{YW-B}. In fact, 
we have 
\begin{align}
\zeta(\mathsf{hw}_{\mu}(u))=(-1)^{\sum_{k=1}^{s} \mu_{k}^{\prime}}e^{\Lambda (h)} ,
\label{topBhw}
\end{align}
where $h$ is a Cartan element such that $e^{\epsilon_{a}(h)}=z_{a}$ ($1 \le a \le r $ or 
$2r+1 \le a \le 2r+s$). 
Here we set $\mu_{j}=0$ if $j>\mu^{\prime}_{1}$, $\mu^{\prime}_{k}=0$ if $k>\mu_{1}$, 
$\prod_{j=a}^{b}(\cdots)=1$ if $a>b$. 
%Eq.\ \eqref{topB} is a reduction of \eqref{topA}. 
 We conjecture that $\mathsf{t}_{\mu}(u)=\mathcal{F}^{I_{2r+2s+1}}_{\mu}$ holds on the $[r,s]$-hook. 

\paragraph{$U_{q}(osp(2r|2s)^{(1)})$ case}
We consider the tuple 
$I_{2r+2s+2}=(2r+2s+2, 2r+2s+1, \dots ,2r+s+4, 2 r+s+3,2r,2r-1, 
\dots ,r+2,r+1,2r+s+2,2r+s+1 ,r,r-1,\dots, 2,1,2r+s,2r+s-1,\dots ,2r+2,2r+1)$ and 
a partition $\mu$ with the condition $\mu_{r+1} \le s$ 
 (the Young diagram $\mu$ is on the $[r,s]$-hook in Figure \ref{MN-hookD}).
Let $\mathsf{t}_{\mu,+}(u)$ be the T-function derived by the Bethe strap procedure with the 
top term (cf.\ eqs.\ (3.49), (3.50) in \cite{T99}) 
\begin{align}
\mathsf{hw}_{\mu,+}(u)=
\prod_{k=1}^{s }
\prod_{j=1}^{\mu^{\prime}_{k}}
(-1)
{\mathcal  X}_{I_{2r+2s+3-k}}^{[-\mu_{1}+\mu_{1}^{\prime}-2j+2k]} 
\prod_{j=1}^{ r }
\prod_{k=s+1}^{\mu_{j}}
{\mathcal  X}_{I_{2r+s+3-j}}^{[-\mu_{1}+\mu_{1}^{\prime}-2j+2k]} ,
\label{topD}
\end{align}
which carries the $osp(2r|2s)$ highest weight \eqref{HW-D} for \eqref{YW-Dp} 
(in the same manner as \eqref{topBhw}). 
% Eq.\ \eqref{topD} is a reduction of \eqref{topA}. 
Let $\mathsf{t}_{\mu,-}(u)$ be the T-function derived by the Bethe strap procedure with the 
top term 
\begin{align}
\mathsf{hw}_{\mu,-}(u)=
\prod_{k=1}^{s }
\prod_{j=1}^{\mu^{\prime}_{k}}
(-1)
{\mathcal  X}_{I_{2r+2s+3-k}}^{[-\mu_{1}+\mu_{1}^{\prime}-2j+2k]} 
\prod_{j=1}^{ r-1 }
\prod_{k=s+1}^{\mu_{j}}
{\mathcal  X}_{I_{2r+s+3-j}}^{[-\mu_{1}+\mu_{1}^{\prime}-2j+2k]}
\prod_{k=s+1}^{\mu_{r}}
{\mathcal  X}_{I_{r+s}}^{[-\mu_{1}+\mu_{1}^{\prime}-2r+2k]} ,
\label{topDm}
\end{align}
which carries the $osp(2r|2s)$ highest weight \eqref{HW-D} for \eqref{YW-Dm} 
(in the same manner as \eqref{topBhw}).  
%We expect that 
%$\mathsf{t}_{\mu,-}(u)=\mathfrak{k}(\mathsf{t}_{\mu,+}(u))$ holds because of the relation 
%$\mathsf{hw}_{\mu,-}(u)=\mathfrak{k}(\mathsf{hw}_{\mu,-}(u))$. 
 As already remarked in the previous paper \cite{T99}, 
 $\mathsf{t}_{\mu,+}(u)=\mathcal{F}^{I_{2r+2s+2}}_{\mu}$ does not always hold, 
 but rather, all the terms of $\mathsf{t}_{\mu,+}(u)$ are expected to be in a subset of those of 
 $\mathcal{F}^{I_{2r+2s+2}}_{\mu}$ 
 since both of them contain the top term \eqref{topD}. 
 In fact, for Young diagrams with one row or column, we observe
\footnote{
 We have confirmed \eqref{FT-D1} and \eqref{FT-D2} for $r=0$, $2 \le s \le 4$, $1 \le a \le 6$; 
 $r=0$, $s=5, 6$, $1 \le a \le 5$; 
 $r=1$, $s=2,3$, $1 \le a \le 6$;  $r=1$, $s=4,5$, $1 \le a \le 5$; 
 $r=2$, $1 \le s \le 4$, $1 \le a \le 5$;   
 $r \ge 3$, $1 \le s \le 6-r$, $1 \le a \le r-2$;
 $3 \le r \le 6$, $s=0$, $1 \le a \le \min(r,5)$;
 and \eqref{FT-D3} and \eqref{FT-D4} for $r \ge 1$, $2 \le r+s \le 6$, $1 \le m \le 6$; 
 $r=0$, $2 \le s \le 6$, $1 \le m \le s$. 
  The Bethe straps seem to have pseudo-top terms at least for
   the cases $\mu=(1^{a})$, $r \ge 3$, $1 \le s \le 6-r$, $r-1 \le a \le 5$. 
   We have to add pseudo-top terms (see \cite{T99}) by hand to make the Bethe straps finite connected graphs. 
   We do not have a systematic way for this at the moment. 
   This is a serious drawback of the Bethe strap procedures, which has to be overcome. 
}
:  
for $a,m \in \mathbb{Z}_{\ge 1}$,  
\begin{align}
\widehat{\mathcal{F}}^{I_{2r+2s+2}}_{(1^{a})} &=
\mathsf{t}_{(1^{a}),+}(u) 
&  \text{for} & \quad s=0, \quad a<r , \quad \text{or} \quad s\ge 1, \quad  r \ge 2,
\nonumber \\
& & & \quad  \text{or} \quad  s \ge 2, \quad  r=0,1, \label{FT-D1} \\ 
\widehat{\mathcal{F}}^{I_{2r+2s+2}}_{(1^{a})} &=
\mathsf{t}_{(1^{r}),+}(u) +\mathsf{t}_{(1^{r}),-}(u)  
& \text{for} & \quad
a=r \ge 3, \quad s=0,
\label{FT-D2} \\
 \widehat{\mathcal{F}}^{I_{2r+2s+2}}_{(m)}&= 
 \mathsf{t}_{(m),+}(u) 
 & \text{for} & \quad r \ge 2, \ r+s \ge 3 \quad \text{or} \quad  r=0,1, \ s \ge 2, \ m \le s, 
 \label{FT-D3} \\
 \widehat{\mathcal{F}}^{I_{2r+2s+2}}_{(m)}&=   \mathsf{t}_{(m),+}(u)  +  \mathsf{t}_{(m),-}(u)
  & \text{for} & \quad
  r=1,\quad  s \ge 2, \quad m \ge s+1,
  \label{FT-D4}
\end{align}
where 
\begin{align}
 \widehat{\mathcal{F}}^{I_{2r+2s+2}}_{(1^{a})} &=
 \begin{cases}
 \mathcal{F}^{I_{2r+2s+2}}_{(1^{a})} - g_{(1^{a})}(u) 
\mathcal{F}^{I_{2r+2s+2}}_{(1^{2(r-s-1)-a})}
 & \text{if} \quad 2 \le r-s \le a \le 2(r-s-1) ,  \\[8pt]
 \mathcal{F}^{I_{2r+2s+2}}_{(1^{a})} & \text{otherwise} ,
 \end{cases} 
 \label{subD1}
 \\
 \widehat{\mathcal{F}}^{I_{2r+2s+2}}_{(m)}  &=
 \begin{cases}
 \mathcal{F}^{I_{2r+2s+2}}_{(m)} - g_{(m)}(u) 
\mathcal{F}^{I_{2r+2s+2}}_{(2(s-r+1)-m)}
 & \text{if} \quad 2 \le  s-r+2 \le m \le 2(s-r+1) , \\[8pt]
 \mathcal{F}^{I_{2r+2s+2}}_{(m)} & \text{otherwise} ,
 \end{cases} 
 \label{subD2}
 \end{align}
 and 
 \footnote{
 In \eqref{gfun1}, 
 we use the following relation repeatedly:
 \begin{align}
\chi_{I_{2r+2s+2}}^{[2r-2s-2 ]} \chi_{I_{1} }
=
 \frac{\Qb^{[2r-2s-4]}_{\emptyset} \Qb^{[2r-2s]}_{\emptyset} }{ ( \Qb^{[2r-2s-2]}_{\emptyset} )^{2} }  . 
\end{align}
There are misprints in \cite{T99}: the condition three lines above eq.\ (3.51), 
``$0\le r-s-1 \le a \le 2(r-s-1)$'' is a misprint of ``$0 < r-s-1 < a \le 2(r-s-1)$''; 
the condition three lines above eq.\ (3.52), 
``$0\le s-r+1 \le m \le 2(s-r+1)$'' is a misprint of ``$0< s-r+1 < a \le 2(s-r+1)$''.
 }
 \begin{align}
 g_{(1^{a})}(u)&=
 \prod_{j=1}^{a-r+s+1}
\chi_{I_{2r+2s+2}}^{[(2r-2s-2) +(2j-a-1)]} \chi_{I_{1} }^{[2j-a-1]}
=
 \frac{\Qb^{[2r-2s-a-3]}_{\emptyset} \Qb^{[a+1]}_{\emptyset} }{ \Qb^{[a-1]}_{\emptyset} \Qb^{[2r-2s-a-1]}_{\emptyset} } ,
 \label{gfun1}
 \\
 g_{(m)}(u)&=
 \prod_{j=1}^{m+r-s-1}
\chi_{I_{2r+2s+3-j}}^{[-m+2j-1]} \chi_{I_{j} }^{[m-2j+1]}
=
 \frac{\Qb^{[m+2r-2s-1]}_{\emptyset} \Qb^{[-m-1]}_{\emptyset} }{ \Qb^{[-m+1]}_{\emptyset} \Qb^{[m+2r-2s-3]}_{\emptyset} }  .
 \end{align}
  Similarly, in order to find the relation between  $\mathsf{t}_{\mu,\pm}(u)$ 
  and $\mathcal{F}^{I_{2r+2s+2}}_{\mu}$,  
 we will have to remove unnecessary terms from $\mathcal{F}^{I_{2r+2s+2}}_{\mu}$. 
 For example, for $\mu=(2,1)$ case, we have checked 
 that   $\mathcal{F}^{I_{2r+2s+2}}_{(2,1)}=\mathsf{t}_{(2,1),\pm}(u)$ holds at least
  for $(r,s)=(1,2),(2,1),(3,0),(4,0)$, while  this is modified as 
$\mathcal{F}^{I_{2r+2s+2}}_{(2,1)}- 
\frac{\Qb^{[-2]}_{\emptyset} \Qb^{[2]}_{\emptyset}}{(\Qb_{\emptyset})^{2}} 
\mathcal{F}^{I_{2r+2s+2}[2(r-s-1)]}_{(1)}=\mathsf{t}_{(2,1),\pm}(u)$
 for $(r,s)=(2,2), (3,1)$. 
 
 %%%%%%%%%%%%%%%%%%%%%%%%
 \paragraph{$U_{q}(osp(2|2s)^{(1)})$ case}
We consider the tuple 
 $I_{2s+4}=(2,2s+4, 2s+3,\dots ,4,3,1)$ and 
a partition $\mu$ with the condition $\mu_{2} \le s$ 
 (the Young diagram $\mu$ is on the $[r,s]$-hook in Figure \ref{MN-hookCp}). 
Note that this is different from the previous case for $r=1$ in that the definition of the 
tuple is different. 
Let $\mathsf{t}_{\mu}(u)$ be the T-function derived by the Bethe strap procedure with the 
top term (cf.\ eqs.\ (3.22), (3.31) in \cite{T99-2}) 
\begin{align}
\mathsf{hw}_{\mu}(u)=
\prod_{k=1}^{\mu_{1}}
{\mathcal  X}_{I_{2s+4}}^{[-\mu_{1}+\mu_{1}^{\prime}-2+2k]}
\prod_{k=1}^{s }
\prod_{j=2}^{\mu^{\prime}_{k}}
(-1)
{\mathcal  X}_{I_{2s+4-k}}^{[-\mu_{1}+\mu_{1}^{\prime}-2j+2k]}  ,
\label{topC}
\end{align}
which carries the $osp(2|2s)$ highest weight \eqref{HW-Cp} for \eqref{YW-Cp}. 
% Eq.\ \eqref{topC} is a reduction of \eqref{topA}. 
In fact, we have 
\begin{align}
\zeta(\mathsf{hw}_{\mu}(u))=(-1)^{\sum_{k=1}^{s} \max\{\mu_{k}^{\prime}-1,0\} }e^{\Lambda (h)} ,
\label{topChw}
\end{align}
where $h$ is a Cartan element such that $e^{\epsilon_{a}(h)}=z_{a}$ 
($ a=1 $ or $3 \le a \le s+2 $). 
In addition to this, we consider 
 the T-function  $\widetilde{\mathsf{t}}_{(m)}(u)$ for $m \ge s+1$
  derived by the Bethe strap procedure with the 
top term 
\begin{align}
\widetilde{\mathsf{hw}}_{(m)}(u)=
\prod_{k=1}^{\min(m-s-1,s)}
{\mathcal  X}_{I_{2s+4}}^{[-m+2k-1]}
\prod_{k=m-s}^{s }
(-1)
{\mathcal  X}_{I_{m+s+3-k}}^{[-m+2k-1]} 
\prod_{k=s+1}^{m }
{\mathcal  X}_{I_{1}}^{[-m+2k-1]}   ,
\label{topCt}
\end{align}
which carries the $osp(2|2s)$ highest weight 
$\widetilde{\Lambda}=-\epsilon_{1}+\sum_{j=3}^{2s+3-m}\epsilon_{j}
=-\varepsilon_{1}+\sum_{j=1}^{2s+1-m}\delta_{j}$ 
for $s+1 \le m \le 2s$, and  $-(m-2s)\epsilon_{1}=-(m-2s)\varepsilon_{1}$ 
for $m \ge 2s+1$. We have 
\begin{align}
\zeta(\widetilde{\mathsf{hw}}_{\mu}(u))=(-1)^{\max\{2s-m+1,0\}}e^{\widetilde{\Lambda} (h)} .
\label{top2Chw}
\end{align} 
 For Young diagrams with one row or column, we observe
\footnote{We have confirmed \eqref{FT-C1} for $s=1$, $1 \le a \le 7$; 
$2 \le s \le3$,  $1 \le a \le 6$; $4 \le s \le 6$, $1 \le a \le 5$;   
and \eqref{FT-C2}-\eqref{defoT} for $1 \le s \le 6$, 
$1 \le m \le 9$.}
 :  
for $a,m \in \mathbb{Z}_{\ge 1}$,  
\begin{align}
\mathcal{F}^{I_{2s+4}}_{(1^{a})} &=
\mathsf{t}_{(1^{a})}(u) 
&  \text{for} &  \quad r \ge 2,  \label{FT-C1} \\ 
 \mathcal{F}^{I_{2s+4}}_{(m)}&= 
 \mathsf{t}_{(m)}(u) 
 & \text{for} & \quad  \ m \le s,  
 \label{FT-C2} \\
 \widehat{\mathcal{F}}^{I_{2s+4}}_{(m)}&=   \mathsf{t}_{(m)}(u)  +  \widetilde{\mathsf{t}}_{(m)}(u)
  & \text{for} &  \quad m \ge s+1,
  \label{FT-C3}
\end{align}
where 
\begin{align}
 \widehat{\mathcal{F}}^{I_{2s+4}}_{(m)}  &=
 \begin{cases}
 \mathcal{F}^{I_{2s+4}}_{(m)} - g_{(m)}(u) 
\mathcal{F}^{I_{2s+4}}_{(2s-m)}
 & \text{if} \quad 2 \le  s+1 \le m \le 2s , \\[8pt]
 \mathcal{F}^{I_{2s+4}}_{(m)} & \text{otherwise} ,
 \end{cases} 
 \label{subC1}
 \end{align}
 and 
 \begin{align}
 g_{(m)}(u)&=
 \prod_{j=1}^{m-s}
\chi_{I_{2s+5-j}}^{[-m+2j-1]} \chi_{I_{j} }^{[m-2j+1]}
=
 \frac{\Qb^{[m-2s+1]}_{\emptyset} \Qb^{[-m-1]}_{\emptyset} }{ \Qb^{[-m+1]}_{\emptyset} \Qb^{[m-2s-1]}_{\emptyset} }  .
 \end{align}
 In addition to the above, we observe: for $m \in \mathbb{Z}_{\ge 1}$, 
 \begin{align}
  \mathsf{t}_{(m)}(u) =
  \begin{cases}
  \mathcal{T}_{m}(u)-
 g_{(m)}(u)  \mathcal{F}^{I_{2s+4}}_{(2s-m)} & \text{if} \quad s+1 \le m \le 2s, \\[8pt]
 \mathcal{T}_{m}(u) & \text{if} \quad 2s+1 \le m,
 \end{cases}
  \label{tsuC99}
 \end{align}
 where 
\begin{align}
\mathcal{T}_{m}(u)= z^{m-s}_{1}
 \frac{\Qb^{[m-2s+1]}_{\emptyset} \Qb^{[-m]}_{I_{1}} }{ \Qb^{[-m+1]}_{\emptyset} \Qb^{[m-2s]}_{I_{1}} } 
  \mathcal{F}^{I_{2s+4}[m-s]}_{(s)} . \label{defoT}
\end{align}
Eqs.\ \eqref{tsuC99} and \eqref{defoT} correspond to [eqs.\ (4.54)-(4.56), \cite{T99-2}]. 
Here we rewrite them in our convention. 
Note that the T-function \eqref{defoT} is well defined for any $m \in \mathbb{C}$, and  is free of poles
\footnote{The trivial poles from $\Qb_{\emptyset}$ are out of the question.} 
under the Bethe ansatz equation. 

 As for the general partition (other than Young diagrams with one rows or columns), 
 we could not find  $\mu$ such that  $\mathsf{t}_{\mu}(u)=\mathcal{F}^{I_{2s+4}}_{\mu}$ holds. 
 We expect that the set of all the terms of $\mathsf{t}_{\mu}(u)$ is a subset of those of 
 $\mathcal{F}^{I_{2s+4}}_{\mu}$. 
 %since they share the same top term \eqref{topC}. 
 For example, for $s=1$, $\mu=(2,1)$ case, 
 $\mathcal{F}^{I_{6}}_{(2,1)}$ has $20$ terms, and $8$ of them constitute  
 $\mathsf{t}_{(2,1)}(u)$. 
 It is desirable to establish the general relation between  $\mathcal{F}^{I_{2s+4}}_{\mu}$ 
 and $\mathsf{t}_{\mu}(u)$.

%%%%%%%%%%%%
 \begin{figure}
  \centering
 \begin{tikzpicture}[x=10mm,y=10mm]
  \node (5) at (-5,0) {$ -{\mathcal  X}_{I_{5}}$};
  \node (4) at (-2.5,0) {${\mathcal  X}_{I_{4}}$};
  \node (3) at (0,0) {${\mathcal  X}_{I_{3}} $};
   \node (2) at (2.5,0) {${\mathcal  X}_{I_{2}} $};
   \node (1) at (5,0) {$-{\mathcal  X}_{I_{1}} $};
 % \node[below of = 6]  {${\mathcal F}_{(1)}^{(2,8,7,6,5,4,3,1)}$};
  \draw[-to,line width=1pt] (5) -- node[midway,above]{$F_{1}^{[-1]}$} (4); 
  \draw[-to,line width=1pt] (4) -- node[midway,above]{$F_{2}$} (3); 
  \draw[-to,line width=1pt] (3) -- node[midway,above]{$F_{2}^{[-1]}$} (2); 
  \draw[-to,line width=1pt] (2) -- node[midway,above]{$F_{1}$} (1); 
\end{tikzpicture}
\caption{Bethe strap structures of the T-function ${\mathcal F}_{(1)}^{I_{5}}$ for $U_{q}(osp(3|2)^{(1)})$, 
where 
$\Bm=\{1,2\}$, $\Fm =\{3,4,5 \}$, 
$I_{5}=(5,2,4,1,3)$, $I_{4}=(5,2,4,1)$, $I_{3}=(5,2,4)$, $I_{2}=(5,2)$, $I_{1}=(5)$, $I_{0}=\emptyset $. 
The top term $ -{\mathcal  X}_{I_{5}}$ carries the $osp(3|2)$ highest weight $\epsilon_{3}$.}
\label{BSb1}
\end{figure}
%%%%%%%%%%%%
%%%%%%%%%%%%
 \begin{figure}
 \centering
 \begin{tikzpicture}[x=10mm,y=10mm]
  \node (8) at (-6.25,0) {$ {\mathcal  X}_{I_{8}}$};
  \node (7) at (-3.75,0) {$-{\mathcal  X}_{I_{7}}$};
  \node (6) at (-1.25,0) {$-{\mathcal  X}_{I_{6}} $};
   \node (3) at (1.25,0) {$-{\mathcal  X}_{I_{3}} $};
   \node (2) at (3.75,0) {$-{\mathcal  X}_{I_{2}} $};
   \node (1) at (6.25,0) {$-{\mathcal  X}_{I_{1}} $};
 % \node[below of = 6]  {${\mathcal F}_{(1)}^{(2,8,7,6,5,4,3,1)}$};
  \draw[-to,line width=1pt] (8) -- node[midway,above]{$F_{1}^{[1]}$} (7); 
  \draw[-to,line width=1pt] (7) -- node[midway,above]{$F_{2}$} (6); 
  \draw[-to,line width=1pt] (6) -- node[midway,above]{$F_{3}^{[-2]}$} (3); 
  \draw[-to,line width=1pt] (3) -- node[midway,above]{$F_{2}^{[-4]}$} (2); 
  \draw[-to,line width=1pt] (2) -- node[midway,above]{$F_{1}^{[-5]}$} (1); 
\end{tikzpicture}
\caption{Bethe strap structures of the T-function ${\mathcal F}_{(1)}^{I_{8}}$ for $U_{q}(osp(2|4)^{(1)})$, 
where $\Bm =\{1,2\}$, $\Fm =\{3,4,5,6,7,8\}$, $\mathfrak{D} =\{5,6\}$, $I_{8}=(2,8,7,6,5,4,3,1)$, $I_{7}=(2,8,7,6,5,4,3)$, $I_{6}=(2,8,7,6,5,4)$, $I_{5}=(2,8,7,6,5)$, 
$I_{4}=(2,8,7,6)$, $I_{3}=(2,8,7)$, $I_{2}=(2,8)$, $I_{1}=(2)$, $I_{0}=\emptyset$. 
The top term $ {\mathcal  X}_{I_{8}}$ carries the $osp(2|4)$ highest weight $\epsilon_{1}$.}
\label{BSc}
\end{figure}
%%%%%%%%%%%%
%%%%%%%%%%%%%%%%%%%
\subsection{T-functions for spinorial representations: $U_{q}(osp(2r+1|2s)^{(1)})$ case} 
\label{sec:SpR}
We introduce a function labeled by a partition $\mu =(\mu_{1},\mu_{2}, \dots ,\mu_{\mu_{1}^{\prime}}) $ 
with $\mu_{1}^{\prime} \le 2r$,  $\mu_{1} \ge \mu_{2} \ge \dots \ge \mu_{\mu^{\prime}_{1}} > 0$,
\begin{multline}
\Ss_{\mu}=
\left(
\prod_{b=1}^{r}(z_{b}^{\frac{1}{2}}+z_{b}^{-\frac{1}{2}}) 
\prod_{b=1}^{r}
\prod_{f=2r+1}^{2r+s} (z_{b}-z_{f}) (1-(z_{b}z_{f})^{-1})
\right)^{-1}
  \Qb^{[2r -\mu_{1}-\mu_{1}^{\prime} +2\mu_{\mu^{\prime}_{1}}]}_{\emptyset }  
  \times 
  \\
  \times
\lim_{c}
\Ts^{\Bm,\Fm [\mu_{1}^{\prime}-c] }_{\mu +((2s+1)^{c})} 
=
\prod_{b=1}^{r}(z_{b}^{\frac{1}{2}}+z_{b}^{-\frac{1}{2}}) 
\prod_{b=1}^{r}
\prod_{f=2r+1}^{2r+s} (z_{b}-z_{f}) (1-(z_{b}z_{f})^{-1})
 \Ts^{\Bm,\emptyset }_{\mu } .
 \label{typ-limB}  
\end{multline} 
This is a reduction of \eqref{typ-lim}. 
The case $s=0$ corresponds to [eq.\  (3.53) in \cite{T21}], which gives T-functions for spinorial representations of $U_{q}(so(2r+1)^{(1)})$.
%%%

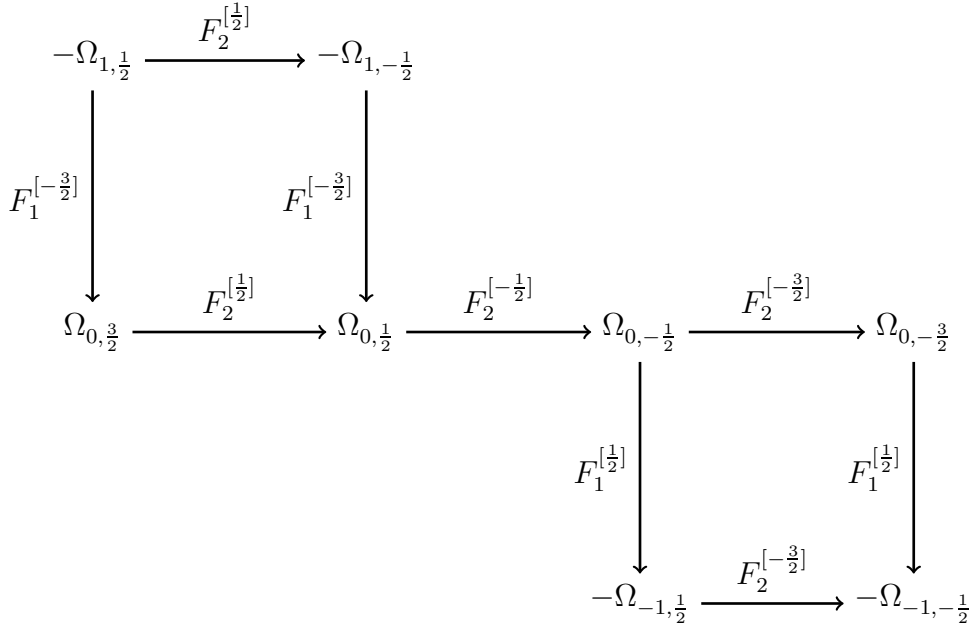
\begin{figure}
 \centering
 \begin{tikzpicture}[x=9mm,y=9mm]
  \node (11) at (-6,4) {$ -\Omega_{1,\frac{1}{2}}$};
  \node (1m1) at (-2,4) {$ -\Omega_{1,-\frac{1}{2}}$};
 \node (03) at (-6,0) {$ \Omega_{0,\frac{3}{2}}$};
  \node (01) at (-2,0) {$ \Omega_{0,\frac{1}{2}}$};
  \node (0m1) at (2,0) {$ \Omega_{0,-\frac{1}{2}}$};
  \node (0m3) at (6,0) {$ \Omega_{0,-\frac{3}{2}}$};
 \node (m11) at (2,-4) {$ -\Omega_{-1,\frac{1}{2}}$};
 \node (m1m1) at (6,-4) {$ -\Omega_{-1,-\frac{1}{2}}$};
%%%
  \draw[-to,line width=1pt] (11) -- node[midway,above]{$F_{2}^{[\frac{1}{2}]}$} (1m1); 
 \draw[-to,line width=1pt] (11) -- node[midway,left]{$F_{1}^{[-\frac{3}{2}]}$} (03); 
 \draw[-to,line width=1pt] (1m1) -- node[midway,left]{$F_{1}^{[-\frac{3}{2}]}$} (01); 
  \draw[-to,line width=1pt] (03) -- node[midway,above]{$F_{2}^{[\frac{1}{2}]}$} (01); 
  \draw[-to,line width=1pt] (01) -- node[midway,above]{$F_{2}^{[-\frac{1}{2}]}$} (0m1); 
  \draw[-to,line width=1pt] (0m1) -- node[midway,above]{$F_{2}^{[-\frac{3}{2}]}$} (0m3); 
\draw[-to,line width=1pt] (0m1) -- node[midway,left]{$F_{1}^{[\frac{1}{2}]}$} (m11); 
\draw[-to,line width=1pt] (0m3) -- node[midway,left]{$F_{1}^{[\frac{1}{2}]}$} (m1m1); 
\draw[-to,line width=1pt] (m11) -- node[midway,above]{$F_{2}^{[-\frac{3}{2}]}$} (m1m1); 
\end{tikzpicture}
\caption{Bethe strap structures of the T-function $\mathsf{t}_{1,\frac{3}{2}}(u)$ for $U_{q}(osp(3|2)^{(1)})$, where 
$\Bm=\{1,2\}$, $\Fm =\{3,4,5 \}$, 
$I_{5}=(5,2,4,1,3)$, $I_{4}=(5,2,4,1)$, $I_{3}=(5,2,4)$, $I_{2}=(5,2)$, $I_{1}=(5)$, $I_{0}=\emptyset $. 
The top term $ -\Omega_{1,\frac{1}{2}}$ carries the $osp(3|2)$ highest weight
 $\epsilon_{3}+\frac{1}{2}\epsilon_{1} =\delta_{1}+\frac{1}{2}\varepsilon_{1}$.}
\label{BSb3}
\end{figure}

Let us consider $U_{q}(osp(3|2)^{(1)})$ case with 
$\Bm=\{1,2\}$, $\Fm =\{3,4,5 \}$, 
$I_{5}=(5,2,4,1,3)$, $I_{4}=(5,2,4,1)$, $I_{3}=(5,2,4)$, $I_{2}=(5,2)$, $I_{1}=(5)$, $I_{0}=\emptyset $. 
We introduce functions of the spectral parameter: 
\begin{align}
\Omega_{1,\frac{1}{2}}&=z_{3}z_{1}^{\frac{1}{2}} \Qb_{\emptyset}^{[-\frac{5}{2}]} 
 \frac{\Qb_{I_{1}}^{[\frac{3}{2}]} \Qb_{I_{2}}^{[-\frac{1}{2}]} }{\Qb_{I_{1}}^{[-\frac{3}{2}]} \Qb_{I_{2}}^{[\frac{1}{2}]} }, 
 &
 \Omega_{1,-\frac{1}{2}}&=z_{3}z_{1}^{-\frac{1}{2}} \Qb_{\emptyset}^{[-\frac{5}{2}]} 
 \frac{\Qb_{I_{1}}^{[-\frac{1}{2}]} \Qb_{I_{2}}^{[\frac{3}{2}]} }{\Qb_{I_{1}}^{[-\frac{3}{2}]} \Qb_{I_{2}}^{[\frac{1}{2}]} }, 
 \notag \\
 \Omega_{0,\frac{3}{2}}&=z_{1}^{\frac{3}{2}} \Qb_{\emptyset}^{[-\frac{1}{2}]} 
 \frac{\Qb_{I_{1}}^{[\frac{3}{2}]} \Qb_{I_{2}}^{[-\frac{5}{2}]} }{\Qb_{I_{1}}^{[-\frac{3}{2}]} \Qb_{I_{2}}^{[\frac{1}{2}]} },
&
  \Omega_{0,\frac{1}{2}}&=z_{1}^{\frac{1}{2}} \Qb_{\emptyset}^{[-\frac{1}{2}]} 
 \frac{\Qb_{I_{1}}^{[-\frac{1}{2}]} \Qb_{I_{2}}^{[-\frac{5}{2}]} \Qb_{I_{2}}^{[\frac{3}{2}]} }{ 
 \Qb_{I_{1}}^{[-\frac{3}{2}]} \Qb_{I_{2}}^{[-\frac{1}{2}]} \Qb_{I_{2}}^{[\frac{1}{2}]} } ,
\notag \\
  \Omega_{0,-\frac{1}{2}}&=z_{1}^{-\frac{1}{2}} \Qb_{\emptyset}^{[-\frac{1}{2}]} 
 \frac{\Qb_{I_{1}}^{[-\frac{1}{2}]} \Qb_{I_{2}}^{[-\frac{5}{2}]} \Qb_{I_{2}}^{[\frac{3}{2}]} }{ 
 \Qb_{I_{1}}^{[\frac{1}{2}]} \Qb_{I_{2}}^{[-\frac{3}{2}]} \Qb_{I_{2}}^{[-\frac{1}{2}]} } ,
 &
  \Omega_{0,-\frac{3}{2}}&=z_{1}^{-\frac{3}{2}} \Qb_{\emptyset}^{[-\frac{1}{2}]} 
 \frac{\Qb_{I_{1}}^{[-\frac{5}{2}]} \Qb_{I_{2}}^{[\frac{3}{2}]} }{\Qb_{I_{1}}^{[\frac{1}{2}]} \Qb_{I_{2}}^{[-\frac{3}{2}]} },
\notag \\
  \Omega_{-1,\frac{1}{2}}&=z_{3}^{-1} z_{1}^{\frac{1}{2}} \Qb_{\emptyset}^{[\frac{3}{2}]} 
 \frac{\Qb_{I_{1}}^{[-\frac{1}{2}]} \Qb_{I_{2}}^{[-\frac{5}{2}]} }{\Qb_{I_{1}}^{[\frac{1}{2}]} \Qb_{I_{2}}^{[-\frac{3}{2}]} },
 &
\Omega_{-1,-\frac{1}{2}}&=z_{3}^{-1} z_{1}^{-\frac{1}{2}} \Qb_{\emptyset}^{[\frac{3}{2}]} 
 \frac{\Qb_{I_{1}}^{[-\frac{5}{2}]} \Qb_{I_{2}}^{[-\frac{1}{2}]} }{\Qb_{I_{1}}^{[\frac{1}{2}]} \Qb_{I_{2}}^{[-\frac{3}{2}]} }.
 \label{spfunb}
\end{align}
The function $\Omega_{j,k}$ carries the $osp(3|2)$ weight $j\epsilon_{3}+k \epsilon_{1}$. 
Then the T-function derived by the Bethe strap procedure with the top term $\Omega_{1,\frac{1}{2}}$ is given by 
the summation over \eqref{spfunb}:
\begin{align}
\mathsf{t}_{1,\frac{3}{2}}(u)=\sum_{j=-1}^{1} \sum_{k=1}^{4-2|j|} (-1)^{j}\Omega_{j,k+|j|-\frac{5}{2}} .
\label{T-bsp}
\end{align}
%%%%
The Bethe strap structures of \eqref{T-bsp} are described in Figure \ref{BSb3}. 
More generally, we conjecture that 
the T-function derived by the Bethe strap procedure with the top term 
\begin{align}
-
\Qb_{\emptyset}^{[\ell-\frac{3}{2}]} \Omega_{1,\frac{1}{2}}^{[1-\ell ]} \prod_{k=1}^{\ell-1 } \chi_{I_{4}}^{[2k-\ell+\frac{3}{2}]} =
-z_{3}z_{1}^{\ell-\frac{1}{2}} \Qb_{\emptyset}^{[-\ell-\frac{3}{2}]} \Qb_{\emptyset}^{[\ell-\frac{3}{2}]}
 \frac{\Qb_{I_{1}}^{[\ell+\frac{1}{2}]} \Qb_{I_{2}}^{[-\ell+\frac{1}{2}]} }{\Qb_{I_{1}}^{[-\ell-\frac{1}{2}]} \Qb_{I_{2}}^{[\ell-\frac{1}{2}]} },
\end{align}
which carries the  $osp(3|2)$ highest weight $\epsilon_{3} + (\ell-\frac{1}{2})\epsilon_{1} =\delta_{1} + (\ell-\frac{1}{2})\varepsilon_{1}$, 
 is given by 
\begin{multline}
\mathsf{t}_{1,\ell+\frac{1}{2} }(u)=(-\Omega_{ 1,\frac{1}{2} }^{ [1-\ell] }+ \Omega_{ 0,\frac{3}{2} }^{ [1-\ell] })
\Qb_{\emptyset}^{[\ell-\frac{3}{2}]} \mathcal{F}_{\ell -1}^{I_{4} [\frac{3}{2}]} +
(-\Omega_{ 1,-\frac{1}{2} }^{ [1-\ell] }+ \Omega_{ 0,\frac{1}{2} }^{ [1-\ell] }+ \Omega_{ 0,-\frac{1}{2} }^{ [1-\ell] }+ \Omega_{ 0,-\frac{3}{2} }^{ [1-\ell] })
\Qb_{\emptyset}^{[\ell-\frac{3}{2}]} \mathcal{F}_{\ell-1}^{I_{2} [\frac{3}{2}]}
 \\
 \text{for} \quad \ell \in \mathbb{Z}_{\ge 2} .
\end{multline}
%%%%%%%%%%%%
%%%%%%%%%%%%
By using QQ-relations, one can transform \eqref{T-bsp} into a Wronskian form:
\begin{align}
\mathsf{t}_{1,\frac{3}{2}}(u)=(z_{1}^{\frac{1}{2}}+z_{1}^{-\frac{1}{2}}) 
(z_{1}-z_{3}) \left(1-\frac{1}{z_{1}z_{3}} \right)
\Qb_{12}^{[-\frac{1}{2}]} .
\end{align}
More generally, we conjecture 
\begin{align}
\mathsf{t}_{1,\ell+\frac{1}{2}}(u)=(z_{1}^{\frac{1}{2}}+z_{1}^{-\frac{1}{2}}) 
(z_{1}-z_{3}) \left(1-\frac{1}{z_{1}z_{3}} \right)
(\Qb_{\emptyset }^{[-\frac{1}{2}]})^{-\delta_{\ell ,1}}
\mathsf{T}^{\Bm,\emptyset [-\frac{3}{2}]}_{1,\ell -1} 
\quad \text{for} \quad \ell \in \mathbb{Z}_{\ge 1} .
\end{align}
Note that the factor $(z_{1}-z_{3}) \left(1-\frac{1}{z_{1}z_{3}} \right)$ coincides with the character 
limit of $\mathsf{T}^{\Bm,\Fm \setminus \{ 4 \} }_{1,1} $.

Let us consider $U_{q}(osp(2r+1|2s)^{(1)})$ case with  
$I_{2r+2s+1}=(2r+2s+1,2r+2s,\dots ,2r+s+2, 2r,2r-1,\dots ,r+1,2r+s+1,r,r-1,\dots,2,1,2r+s,2r+s-1,\dots,2r+2,2r+1)$, 
\dots , 
$I_{r+s}=(2r+2s+1,2r+2s,\dots ,2r+s+2, 2r,2r-1,\dots ,r+1)$, \dots ,
$I_{s}=(2r+2s+1,2r+2s,\dots ,2r+s+2)$, \dots $I_{1}=(2r+2s+1)$, $I_{0}=\emptyset$. 
Based on a computer experiment by Mathematica  (ver.\ 7), 
we conjecture that 
the T-function derived by the Bethe strap procedure with the top term 
\begin{multline}
(-1)^{rs} (z_{1}z_{2}\cdots z_{r})^{\ell -s +\frac{1}{2}} (z_{2r+1}z_{2r+2}\cdots z_{2r+s})^{r} 
\Qb_{\emptyset}^{[-\ell -r -\frac{1}{2}]}
(\Qb_{\emptyset}^{[\ell -2s +r -\frac{1}{2}]})^{1-\delta_{\ell , s}} 
\times 
\\
\times 
\frac{ \Qb_{I_{s}}^{[\ell -s+r + \frac{1}{2}]} \Qb_{I_{r+s}}^{[-\ell +s - \frac{1}{2}]} }{  
 \Qb_{I_{s}}^{[-\ell +s-r -\frac{1}{2}]} \Qb_{I_{r+s}}^{[\ell -s + \frac{1}{2}]} }  
 \qquad \text{for}
 \quad \ell \in \mathbb{Z}_{\ge s}, 
 \label{top-Bsp}
\end{multline}
which carries the  $osp(2r+1|2s)$  highest weight 
$ (\ell -s +\frac{1}{2}) (\epsilon_{1}+\epsilon_{2}+\cdots +\epsilon_{r})+r (\epsilon_{2r+1}+\epsilon_{2r+2} + \cdots + \epsilon_{2r+s})= r \sum_{j=1}^{s}  \delta_{j} + (\ell -s +\frac{1}{2}) \sum_{j=1}^{r}  \varepsilon_{j}$, is given by 
the following Wronskian-type formula
\footnote{
This may be interpreted as a T-function labelled by the Young diagram $((\ell +\frac{1}{2} )^{r})$ with 
the height $r$ and  the 
half integer width $\ell +\frac{1}{2}$. 
%It is also proportional to the T-function of an asymptotic representation labelled by
% the Young diagram  $((\ell -s )^{r})$. 
More generally, we expect that the T-function derived by the Bethe strap procedure with a top term 
which carries the  $osp(2r+1|2s)$  highest weight 
$  r \sum_{j=1}^{s}  \delta_{j} +  \sum_{j=1}^{r} (\mu_{j} +\frac{1}{2}) \varepsilon_{j}$ 
is described by the function $\mathsf{S}_{\mu}$ labelled by 
 a partition $\mu =(\mu_{1},\mu_{2}, \dots ,\mu_{\mu_{1}^{\prime}}) $, where 
 $\mu_{1}^{\prime} \le r$, $\mu_{1} \ge \mu_{2} \ge \dots \ge \mu_{\mu^{\prime}_{1}} > 0$,
  $\mu_{j}=0$ if $j >\mu^{\prime}_{1}$ (see [eq.\ (3.50), \cite{T21}] for $s=0$ case). 
  In order to give a precise description of this, we need further case by case studies. 
}  
\begin{align}
\mathsf{t}_{r,\ell +\frac{1}{2}}(u)=
\left( \Qb_{\emptyset }^{ [3r-s-\frac{1}{2}] } \right)^{ -\delta_{\ell , s} }
\mathsf{S}_{ (\ell -s)^{r} }^{[ r \delta_{ \ell , s}-s-\frac{1}{2}] } 
 \qquad \text{for}
 \quad \ell \in \mathbb{Z}_{\ge s} .
 \label{Wron-Bsp}
\end{align}
The T-function derived from the top term \eqref{top-Bsp} for 
$r=2$, $s=1$, $\ell=1$ corresponds to the 64 term expression mentioned in 
the previous paper [section 5, \cite{T99}]. The expression was too bulky to write down, but now 
the Wronskian formula \eqref{Wron-Bsp} provides an alternative concise expression for it. 
We also expect that reductions of \eqref{typ-lim} for $(M,N)=(2r,2s+2)$ are related to 
certain combinations of T-functions for spinorial representations of $U_{q}(osp(2r|2s)^{(1)})$. 
However, this requires further investigation.
%%%%%%%%%%%%%%%%%%%%%%%%%%%%%%%%%%%%%%%%%
\section{Concluding remarks}
\label{sec:CR}
 In this paper, we continued our trials \cite{T02,BT08,T09,T11,T21} to construct 
various expressions of T-functions, in particular 
 Wronskian-type formulas (analogues of the Weyl character formula)
 associated with any quantum affine (super)algebras or Yangians. 
 The key is an extension of the reduction procedures proposed in \cite{T11}. 
  This also connects our earlier works on the analytic Bethe ansatz for type A
   superalgebras  \cite{T97,T98,T98-2} 
 and the ones for type B, C, D superalgebras \cite{T99,T99-2}, 
 which is one of the motivations for this paper. 
 There remain problems which have to be clarified step by step. 
\begin{itemize}
\item {\bf Refinement of the reduction procedures}
We considered reductions so that the resultant Bethe ansatz equations and 
T-functions (for the fundamental representation)  reproduce those from the algebraic Bethe ansatz 
 on the symmetric nesting paths. 
 There is still a room for generalization or improvement of the reduction procedures. 
 In  \cite{BCFGT17}, QQ-relations with $osp(4|6)$-symmetries were introduced 
 in relation to the quantum spectral curve for $AdS_{4}/CFT_{3}$. 
 In this context, 
 an interesting problem is to modify the reduction procedures and 
 find QQ-relations
 corresponding to [eqs.\ (7.51), (7.52), \cite{BCFGT17}] 
 as substitutes of  the QQ-relations \eqref{QQd9}, \eqref{QQd8}, \eqref{QQd10} and \eqref{QQd9s}. 
 This may fix some unclear points mentioned in subsection \ref{sec:QQD}.

\item {\bf Refinement of T-functions} 
Our discussions on Bethe straps suggest that 
 not all the T-functions obtained by the reduction procedures give T-functions for irreducible representations of underlying algebras in the auxiliary spaces.  
%(the spaces in which the traces of monodromy matrices for transfer matrices are taken). 
Thus it is important to clarify the condition for irreducibility and find the modification to get T-functions for irreducible representations. 

\item {\bf Symmetries of T-functions} 
In this paper, we considered reductions of QQ-relations and T-functions mainly 
along symmetric nesting paths, 
which are related to symmetric Dynkin diagrams of $gl(M|N)$. 
If we had considered non-symmetric nesting paths, we would have come across 
non-standard forms of Bethe ansatz equations. It remains to be seen 
 whether we should exclude
\footnote{in case non-standard Bethe ansatz equations or extra QQ-relations over-constrain the system} them from our consideration, or rather clarify what they mean.  

Let $\mathfrak{g}^{(1)}$ be an affine Lie superalgebra, and $\mathfrak{g}_{k}$ be 
the Lie superalgebra corresponding to the Dynkin diagram derived by  
 removing $k$-th node of a Dynkin diagram of $\mathfrak{g}^{(1)}$. 
In the standard notation,  $\mathfrak{g}_{0}=\mathfrak{g}$. 
The supercharacters of  finite dimensional representations of $U_{q}(\mathfrak{g}^{(1)})$ 
are invariant under the Weyl group
% (and its extension by odd reflections) 
of $\mathfrak{g}_{k}$ and are linear combinations of the supercharacters of finite dimensional 
representations of  $\mathfrak{g}_{k}$ 
\footnote{In \cite{HKOTT01}, the characters of the Kirillov-Reshetikhin modules of $U_{q}(\mathfrak{g}^{(1)})=U_{q}(B_{r}^{(1)})$ were expressed as linear combinations of characters of $\mathfrak{g}_{r}=D_{r}$  
 (cf. $\mathfrak{g}_{0}=B_{r}$). Analogous character formulas for twisted quantum affine algebras 
were also presented. We found a generalization of these results to the case of superalgebras (see Appendix B).}.
A standard way to consider the problem is to set $k=0$. 
We discussed $\mathfrak{W}$-symmetry of T-functions for 
 $U_{q}(\mathfrak{g}^{(1)})$ ($\mathfrak{g}=osp(2r+1|2s), osp(2r|2s)$), which is a part of 
the original $S_{M+N}$-symmetry of T-functions for $U_{q}(gl(M|N)^{(1)})$ and is 
related to the Weyl group (and its extension by odd reflections, 
and a symmetry that flips the $(r+s-1)$-th and $(r+s)$-th nodes of a Dynkin diagram of type D) of $\mathfrak{g}_{0}$. 
It is desirable to consider also the $k\ne 0$ case and to clarify the whole symmetries of 
the T-functions for $U_{q}(\mathfrak{g}^{(1)})$ and their connection with  
the original $S_{M+N}$-symmetry of the T-functions for $U_{q}(gl(M|N)^{(1)})$. 
After this, it is desirable to reformulate the Wronskian-type expressions of T-functions, 
which are invariant under the whole symmetries.

\item {\bf Generalization to other algebras}
One of the interesting superalgebras is $U_{q}(D(2,1;\alpha)^{(1)})$. 
This algebra is unique in that it depends on  an extra parameter $\alpha$. 
There should be some connections to our results at $\alpha=1$ because of the relation $D(2,1;1) \simeq osp(4|2)$. 
 In addition, it is possible to execute the analytic Bethe ansatz based on Bethe ansatz equations 
 with Cartan matrices of $D(2,1;\alpha)$, as in the case of other superalgebras \cite{T97,T98,T98-2,T99,T99-2}. 
It will be possible to consider further reductions of some of the QQ-relations in this paper 
with respect to Dynkin diagram symmetries and derive QQ-relations for
 twisted quantum affine superalgebras including $U_{q}(osp(2r|2s)^{(2)})$ (see Appendix A).  
 
\item 
{\bf Operator realization}
It is important to realize Wronskian-type formulas of T-functions as operators (through Q-operators) and give 
representation theoretical background for them. In \cite{BT08,T12,T19-1}, we constructed  q-oscillator representations of 
$U_{q}(gl(M|N)^{(1)})$ (or $U_{q}(sl(M|N)^{(1)})$) for Q-operators 
(see also \cite{BDKM06,KLT12,FLMS10} for the rational case from various points of view, and \cite{Z14} for representation theoretical background). 
A tentative goal on this topic is to reformulate and generalize the contents of 
\cite{BT08,T12,T19-1} further. 
In particular, it is worthwhile to apply the folding technique described in \cite{FSS89}  
(or a modified version of it) to the results in \cite{T19-1}.
%How much reduction procedures work for these remained to be clarified. 

\item
{\bf Connection to the soliton theory}
As explained in \cite{AKLTZ11} (see also \cite{KLT12}), a generating function of T-operators (master T-operator) for quantum integrable spin chains associated with $Y(gl(M))$ is the $\tau$-function of the modified KP hierarchy. It should be possible to consider reductions of the master T-operator for $U_{q}(gl(M|N)^{(1)})$ and discuss connection to the soliton theory. 
A related issue is the T-system for quantum integrable systems associated with superalgebras. 
Some partial results have already been obtained in the previous papers \cite{T97,T98,T99,T99-2}, 
but the whole picture is still unclear, which contrasts with the well-understood non-superalgebra cases \cite{KNS93,KS94-2,KNS10}. 

\item
{\bf Grassmannian formalism}
In \cite{KLV15}, 
determinant formulas of T-and Q-functions in \cite{T11,T09} 
were reformulated in terms of exterior forms of Q-functions.  
In light of this, it will be possible to 
reformulate the reduction procedures in terms of the Grassmannian formalism. 
The recent papers \cite{ESV20,EV21} on QQ-relations for $so(2r)$, which use 
pure spinors,  would be clues for this. 
\end{itemize}

In addition to these, it would be possible to extend the reduction procedures and the above topics 
to the case of open super spin chains (at least for diagonal K-matrices). The Bethe ansatz equations for the open super 
spin chain based on $Y(osp(M|2s))$ are formulated in \cite{AACDFR03}.  

The T-functions are not the only generalization of (super)characters. 
Although it is not a subject of our series of papers, it might be mathematically 
meaningful to consider reductions 
similar to \eqref{reduction-sigma}   
for any series in $\{z_{j}\}_{j=1}^{M+N}$ (supersymmetric functions or polynomials) 
as well as q-(super)characters (and their extensions)
 that generalize supercharacters of $gl(M|N)$. 
 
% All in all, this paper is based on ansatzes, empirical methods, and computer experiments that are not necessarily %mathematically rigorous, but we hope that it will serve as a stepping stone for further research in the future.

%%%%%%%%%%%%%%%
\section*{Acknowledgments} 
The work is supported by Grant No. 0657-2020-0015 of the Ministry of Science and Higher Education of Russia. 
The author thanks the anonymous referee for useful comments. 

\paragraph{Note added}
%On the same daythat we submitted this paper to arXiv, 
When we had almost finished writing this paper, two papers \cite{BSLD23,FR23} 
appeared on arXiv. 
%[\cite{BSLD23} on 11 September 2023, \cite{FR23} on 26 September 2023]. 
They studied rational L-operators related to $osp(M|2s)$, 
which might be useful for advancing the research on the operator realization 
of the functional relations discussed in this paper.
%%%%%%%%%%%%%%%%%%%%%%%%%%%%%%%%%%%%%%%%%%%%%%%%%
\section*{Appendix A: Regular reductions in a singular reduction: $U_{q}(osp(2r|2s)^{(2)})$ case}
\label{sec:SRR}
\addcontentsline{toc}{section}{Appendix A}
\def\theequation{A\arabic{equation}}
\setcounter{equation}{0}

One can consider more reductions to some of the formulas on T-and Q-functions derived by reductions in the 
 main text. 
Let us consider reductions of the $U_{q}(osp(2r|2s)^{(1)})$ case (subsection \ref{sec:QQD}) with respect to the symmetry of 
exchanging the $(r+s-1)$-th and $(r+s)$-th nodes of a Dynkin diagram of type D. 
In the tuple $I_{2r+2s+2}=(i_{1},i_{2},\dots , i_{r+s+1},i_{r+s+1}^{*},\dots , i_{2}^{*},i_{1}^{*})$, $i_{r+s+1} \in \mathfrak{D}$, 
we fix $(i_{r+s},i_{r+s}^{*})=(r,r+1)$, or $(r+1,r)$, thus $i_{r+s},i_{r+s}^{*} \in \Bm $. 
We are interested in the action of $\sigma^{\prime}=\tau_{i_{r+s},i_{r+s}^{*}} \in \mathfrak{W}$:
\begin{align}
\begin{split}
& \sigma^{\prime}(\Qb_{I})=\Qb_{\breve{I}} \quad \text{for} \quad   I \subset \mathfrak{I} ,
\\
&
\sigma^{\prime}(\Qf_{I_{r+s}})=\Qf_{\breve{I}_{r+s}}, 
\quad 
\sigma^{\prime}(\Qf_{\breve{I}_{r+s}})=\Qf_{I_{r+s}},
\\
& \sigma^{\prime}(z_{a})=z_{a} \quad \text{for} \quad a \in \mathfrak{I} \setminus \{i_{r+s}, i_{r+s}^{*}\},
\quad  \sigma^{\prime}(z_{i_{r+s}})=z_{i_{r+s}^{*}}, 
\quad  \sigma^{\prime}(z_{i_{r+s}^{*}})=z_{i_{r+s}}, 
\end{split}
\end{align}
where  $\breve{I}=\sigma^{\prime}(I)$, namely $\breve{I}=I$ if $i_{r+s},i_{r+s}^{*} \in I$ or $i_{r+s}, i_{r+s}^{*} \notin I$; 
$\breve{I}=(I \setminus \{i_{r+s} \}) \sqcup \{i_{r+s}^{*}\}$ if $i_{r+s} \in I$ and 
$i_{r+s}^{*} \notin I$; 
$\breve{I}=(I \setminus \{i_{r+s}^{*} \}) \sqcup \{i_{r+s}\}$ if $i_{r+s}^{*} \in I$ and 
$i_{r+s} \notin I$. 
Then we consider the following reduction:
\begin{align}
& \Qb_{\breve{I}}=\Qb_{I}^{[\eta]} \quad \text{for} \quad I \subset \mathfrak{I}, 
\label{Q=Q1}
\\
&
\Qf_{\breve{I}_{r+s}} =\Qf_{I_{r+s}}^{[\eta]} , 
\label{Q=Q2}
\\
& z_{i_{r+s}^{*}}=z_{i_{r+s}} .
\label{z=zd}
\end{align}
The condition \eqref{z=zd} means $z_{i_{r+s}}=\pm 1$. In case $z_{i_{r+s}}= 1$, we assume that 
$\eta$ is the half the period of the Q-functions
\footnote{In case $z_{i_{r+s}}= -1$, we assume $\eta =0$
% is parallel to the case  $z_{i_{r+s}}= 1$, $\eta \ne 0$. 
}.
\eqref{Q=Q1} suggest a factorization of the form 
$\Qb_{I}=\Qf_{I} \Qf_{I}^{[\eta]}$ if  $i_{r+s},i_{r+s}^{*} \in I$ or $i_{r+s}, i_{r+s}^{*} \notin I$, where 
$\Qf_{I}^{[2\eta]}=\Qf_{I}$. Thus on the symmetric nesting path defined by the aforementioned tuple $I_{2r+2s+2}$, 
we have $\Qb_{I_{a}}=\Qb_{I_{2r+2s+2-a}}=\Qf_{I_{a}} \Qf_{I_{a}}^{[\eta]}$ for $0 \le a \le r+s-1$. 
Combining this for $a=r+s-1$ and the relation 
 $\Qb_{I_{r+s-1}}=\Qf_{I_{r+s}} \Qf_{I_{r+s}}^{[\eta]}$ derived from \eqref{facd2} and \eqref{Q=Q2}, 
 we find $\Qf_{I_{r+s-1}} \Qf_{I_{r+s-1}}^{[\eta]}=\Qf_{I_{r+s}} \Qf_{I_{r+s}}^{[\eta]} $. 
 In the following we consider the case 
 %
%\footnote{The other choice $\Qf_{I_{r+s}} =\Qf_{I_{r+s-1}}^{[\eta]}$ , 
%$\Qf_{\breve{I}_{r+s}} =\Qf_{I_{r+s-1}}$ is also possible.}
 % 
\begin{align}
&
\Qf_{I_{r+s}} =\Qf_{I_{r+s-1}} , 
\qquad 
\Qf_{\breve{I}_{r+s}} =\Qf_{I_{r+s-1}}^{[\eta]} .
\label{Q=Q3}
\end{align}
The QQ-relations \eqref{QQd1} and \eqref{QQd10} reduce to the following functional relations:
\\ \noindent
{\bf for $a$-th node ($1 \le a \le r+s-2$):}
\begin{align}
& (z_{i_{a}}-z_{i_{a+1}})\Qf^{2}_{I_{a-1}}\Qf^{2}_{I_{a+1}}
=z_{i_{a}}\Qf_{I_{a}}^{2[p_{i_{a}}]}
\Qf_{\widetilde{I}_{a}}^{2[-p_{i_{a}}]}-
z_{i_{a+1}}\Qf_{I_{a}}^{2[-p_{i_{a}}]}
\Qf_{\widetilde{I}_{a}}^{2[p_{i_{a}}]}
\quad \text{if} \quad p_{i_{a}}=p_{i_{a+1}} ,  
\label{QQd1h}  \\
& (z_{i_{a}}-z_{i_{a+1}})\Qf^{2}_{I_{a}}\Qf^{2}_{\widetilde{I}_{a}}
=z_{i_{a}}\Qf_{I_{a-1}}^{2[-p_{i_{a}}]}
\Qf_{I_{a+1}}^{2[p_{i_{a}}]}-
z_{i_{a+1}}\Qf_{I_{a-1}}^{2[p_{i_{a}}]}
\Qf_{I_{a+1}}^{2[-p_{i_{a}}]}
\quad \text{if} \quad p_{i_{a}}=-p_{i_{a+1}},  
\label{QQd2h}  
\end{align}
%%%%%%%%%%%
{\bf for $(r+s-1)$-th node (from simply laced):}
\begin{align}
& (z_{i_{r+s-1}}-1)\Qf^{2}_{I_{r+s-2}}
=
 z_{i_{r+s-1}}\Qf_{I_{r+s-1}}^{[\eta +1]} \Qf_{\acute{I}_{r+s-1}}^{[-1]}
 -
\Qf_{I_{r+s-1}}^{[\eta-1]} \Qf_{\acute{I}_{r+s-1}}^{[1]}
\quad \text{if} \quad  i_{r+s-1} \in \Bm,  
\label{QQd5h}  
\end{align}
%%%%%%%%
{\bf for $(r+s-1)$-th node (from non-simply laced):}
\begin{align}
& (z_{i_{r+s-1}}-1)\Qb_{\widetilde{I}_{r+s-1}}\Qf_{I_{r+s-1}}^{[\eta]}
=
 z_{i_{r+s-1}}\Qf_{I_{r+s-2}}^{2[1]} \Qf_{I_{r+s-1}}^{[-2]}
 -
 \Qf_{I_{r+s-2}}^{2[-1]} \Qf_{I_{r+s-1}}^{[2]}
\quad  \text{if} \quad  i_{r+s-1} \in \Fm,  
\label{QQd8h}   
\end{align} 
where $\Qf_{I_{a}}^{2}=\Qf_{I_{a}} \Qf^{[\eta]}_{I_{a}}$, $\Qf_{\widetilde{I}_{a}}^{2}=\Qf_{\widetilde{I}_{a}} \Qf^{[\eta]}_{\widetilde{I}_{a}}$ for $0 \le a \le r+s-1$, 
$\acute{I}_{r+s-1}=(i_{1},i_{2},\dots , i_{r+s-2},i^{*}_{r+s-1})$.
%%%%%%%%%%%%%%%
%%%%%%
\paragraph{T-functions and Bethe ansatz equations}
Under the reduction, \eqref{boxes-zz-d} reduces to 
\begin{align}
%\begin{split}
{\mathcal   X}_{I_{a}}&=
z_{i_{a}}
\frac{\Qf_{I_{a-1}}^{2[2r-2s-2-\sum_{j \in I_{a}}p_{j}-p_{i_{a}}]}
\Qf_{I_{a}}^{2[2r-2s-2-\sum_{j \in I_{a}}p_{j}+2p_{i_{a}}]}
}{
\Qf_{I_{a-1}}^{2[2r-2s-2-\sum_{j \in I_{a}}p_{j}+p_{i_{a}}]}
\Qf_{I_{a}}^{2[2r-2s-2-\sum_{j \in I_{a}}p_{j}]}
}
% 
%\nonumber \\ &
 \quad \text{for} \quad 
 1 \le a \le r+s-1
\nonumber  \\
{\mathcal   X}_{I_{r+s}}&=
\frac{ \Qf_{I_{r+s-1}}^{[r-s-3+\eta]}
  \Qf_{I_{r+s-1}}^{[r-s+1]}   
}{
\Qf_{I_{r+s-1}}^{2[r-s-1]} 
} ,
\nonumber  \\
{\mathcal   X}_{I_{r+s+1}}&=-{\mathcal   X}_{I_{r+s+2}}=
z_{i_{r+s+1}}
\frac{ \Qf_{I_{r+s-1}}^{[r-s+1]} \Qf_{I_{r+s-1}}^{[r-s-3]}  
}{
(\Qf_{I_{r+s-1}}^{[r-s-1]} )^{2}
} ,
\nonumber \\
{\mathcal   X}_{I_{r+s+3}}&=
\frac{ \Qf_{I_{r+s-1}}^{[r-s+1+\eta]}
  \Qf_{I_{r+s-1}}^{[r-s-3]}   
}{
\Qf_{I_{r+s-1}}^{2[r-s-1]} 
} ,  
\nonumber \\
{\mathcal   X}_{I_{2r+2s+3-a}}&=
z_{i_{a}}^{-1}
\frac{\Qf_{I_{a-1}}^{2[\sum_{j \in I_{a}}p_{j}+p_{i_{a}}]}
\Qf_{I_{a}}^{2[\sum_{j \in I_{a}}p_{j}-2p_{i_{a}}]}
}{
\Qf_{I_{a-1}}^{2[\sum_{j \in I_{a}}p_{j}-p_{i_{a}}]}
\Qf_{I_{a}}^{2[\sum_{j \in I_{a}}p_{j} ]}
} 
 \quad \text{for} \quad 
 1 \le a \le r+s-1 .
%\end{split}
\label{boxes-zz-dh} 
\end{align}
The T-function \eqref{tab-fund-d} reduces to 
\begin{align}
{\mathsf F}_{(1)}^{I_{2r+2s+2}}=
\Qf_{\emptyset}^{2[2r-2s-2]}
\Qf_{\emptyset}^{2}
\sum_{a=1}^{r+s}p_{i_{a}}({\mathcal   X}_{I_{a}}+{\mathcal   X}_{I_{2r+2s+3-a}}) .
 \label{tab-fund-dh}
\end{align}
Note that the terms ${\mathcal   X}_{I_{r+s+1}}$ and ${\mathcal   X}_{I_{r+s+2}}$ are missing 
in \eqref{tab-fund-dh} because of cancellation. 
The pole-free condition of the T-function \eqref{tab-fund-dh} 
produces the following Bethe ansatz equations: 
\\
\noindent
{\bf for $a$-th node ($1 \le a \le r+s-2$):}
\begin{multline}
 -1=\frac{p_{i_{a}}z_{i_{a}}}{p_{i_{a+1}}z_{i_{a+1}}}
\frac{\Qf^{2}_{I_{a-1}}(u_{k}^{I_{a}}-p_{i_{a}})
\Qf^{2}_{I_{a}}(u_{k}^{I_{a}}+2p_{i_{a}})
\Qf^{2}_{I_{a+1}}(u_{k}^{I_{a}}-p_{i_{a+1}})} 
{\Qf^{2}_{I_{a-1}}(u_{k}^{I_{a}}+p_{i_{a}})
\Qf^{2}_{I_{a}}(u_{k}^{I_{a}}-2p_{i_{a+1}})
\Qf^{2}_{I_{a+1}}(u_{k}^{I_{a}}+p_{i_{a+1}})}
 \\
 \text{for} \quad k\in \{1,2,\dots, n_{I_{a}}\}  ,
\label{BAEd1h}
\end{multline}
{\bf for $(r+s-1)$-th node:}
\begin{multline}
- 1= \frac{z_{i_{r+s-1}}}{p_{i_{r+s-1}}} 
\frac{\Qf^{2}_{I_{r+s-2}}(u_{k}^{I_{r+s-1}}-p_{i_{r+s-1}})  \Qf_{I_{r+s-1}}(u_{k}^{I_{r+s-1}}-2+\eta) 
} 
{\Qf^{2}_{I_{r+s-2}}(u_{k}^{I_{r+s-1}}+p_{i_{r+s-1}})  \Qf_{I_{r+s-1}}(u_{k}^{I_{r+s-1}}-2 p_{i_{r+s-1}} +\eta )
} 
\times 
 \\
\times 
 \frac{
\Qf_{I_{r+s-1}}(u_{k}^{I_{r+s-1}}+2)} 
{
\Qf_{I_{r+s-1}}(u_{k}^{I_{r+s-1}}-2p_{i_{r+s-1}})} 
\qquad 
 \text{for} \quad k\in \{1,2,\dots, n_{I_{r+s-1}}\}  ,
\label{BAEd6h}
\end{multline}
where $\Qb_{I_{a}}(u_{k}^{I_{a}})=0$, $\Qf_{I_{a}}(v_{k}^{I_{a}})=0$, 
$\{ u_{k}^{I_{a}} \}_{k=1}^{n_{I_{a}}}=\{ v_{k}^{I_{a}} \}_{k=1}^{m_{I_{a}}} \sqcup \{ v_{k}^{I_{a}} + \eta \}_{k=1}^{m_{I_{a}}}$, 
$n_{I_{a}}=2m_{I_{a}}$, $0 \le a \le r+s-1$.
The Bethe ansatz equations \eqref{BAEd1h}-\eqref{BAEd6h} 
are reductions of  \eqref{BAEd1}-\eqref{BAEd6}.
Eqs. \eqref{boxes-zz-dh}-\eqref{BAEd6h} for $s=0$ 
agree
%
%\footnote{One has to correct misprints in \cite{R87} beforehand.}
%
 with the known results \cite{R87} by analytic Bethe ansatz. 
As for $s>0$ case, we could not find  appropriate references
\footnote{As remarked in \cite{GM06}, the result in \cite{GM04} denoted as $U_{q}(osp(2m|2n)^{(2)})$ 
is something different. We also remark that 
Q-operators for $r=s=1$ case were studied in \cite{KZ05}.} to compare. 
The generating functions of the T-functions have the same form as \eqref{gene1D} and \eqref{gene2D}, 
but the functions \eqref{boxes-zz-d} have to be replaced with \eqref{boxes-zz-dh}. 
The tableaux sum and  Wronskian expressions of the T-functions are given by 
 \eqref{DVF-tab1} and \eqref{unnor-t1} under the reductions. 
 However we have not identified the conditions that the auxiliary spaces of the corresponding transfer matrices 
 become irreducible representations of $U_{q}(osp(2r|2s)^{(2)})$. 

%%%%%%%%%%%%%%%%%%%%%%%%%%%%%%%%%%%%%%%%%
\section*{Appendix B: Decomposition of supercharacters}
\label{ApB}
\addcontentsline{toc}{section}{Appendix B}
\def\theequation{B\arabic{equation}}
\setcounter{equation}{0}
%%%
In this appendix, we will consider the (super)character limit of T-functions in sections 
\ref{sec:RR} and \ref{sec:SR}, and compare special cases of them with character 
formulas for Kirillov-Reshetikhin modules of quantum affine algebras (or their Yangian counterparts). 
We assume that the characters of $U_{q}(\mathfrak{g})$ for generic $q$ are the same as those of 
 the corresponding Lie superalgebra  $\mathfrak{g}$, and use the same symbols for representations of $U_{q}(\mathfrak{g})$  
 and those of $\mathfrak{g}$. 

The (super)character limit of the generating function \eqref{gene1} at $K=M+N$, namely 
$\zeta(\mathbf{W}_{I_{M+N}}(\mathbf{X})) =w(t)$, is given by 
\begin{align}
w(t)=
\prod_{j=1}^{M} (1-z_{j} t)^{-1} \prod_{j=1}^{N} (1-z_{j+M} t)
=\sum_{m=0}^{\infty} \chi_{m} (\{z_{b} \}_{b=1}^{M}  | \{z_{b+M}\}_{b=1}^{N} ) t^{m} ,
\label{gench-A}
\end{align}
where 
$ \chi_{0} (\{z_{b} \}_{b=1}^{M}  | \{z_{b+M}\}_{b=1}^{N} )=1$, $ \chi_{m} (\{z_{b} \}_{b=1}^{M}  | \{z_{b+M}\}_{b=1}^{N} )=0$ if $m<0$, 
 $\zeta(\mathcal{F}^{I_{M+N}}_{(m)})=\chi_{m}$, $\zeta(\mathbf{X})=t$. 
Then the (super)character limit of  \eqref{superJT2} at $K=M+N$, $\lambda=\emptyset$ is 
the supersymmetric Jacobi-Trudi determinant:
\begin{align}
\zeta(\mathcal{F}^{I_{M+N}}_{\mu})=
\chi_{\mu} (\{z_{b} \}_{b=1}^{M}  | \{z_{b+M}\}_{b=1}^{N} ) =
|(\chi_{\mu_{i}-i+j} )_{1 \le i,j \le \mu_{1}^{\prime}} | ,
\label{JTdet}
\end{align}
where the arguments are omitted on the right hand side: $ \chi_{m}=\chi_{m}(\{z_{b} \}_{b=1}^{M}  | \{z_{b+M}\}_{b=1}^{N} )$.
Note that $\zeta(\mathcal{F}^{I_{M+N}}_{\mu})=\zeta(\mathsf{T}^{\Bm,\Fm}_{\mu})$ always holds, 
from which a Weyl-type formula for supercharacters is given. 
The determinant \eqref{JTdet}  in the $[M,N]$-hook 
gives the supercharacter of the irreducible representation 
$V(\Lambda)$ with the highest weight \eqref{HW-A}, \eqref{YW-A}.

In addition to \eqref{JTdet}, we need the following determinants
\footnote{The right hand side of \eqref{JT-c}  is the determinant of a block matrix consisting of 
a $\mu_{1}^{\prime} \times 1$ matrix and a $\mu_{1}^{\prime} \times (\mu_{1}^{\prime}-1)$ matrix. 
One may rewrite \eqref{JT-c} as
$\chi_{\langle \mu \rangle }  (\{z_{b} \}_{b=1}^{M}  | \{z_{b+M}\}_{b=1}^{N} )= \frac{1}{2} |
 (\chi_{\mu_{i}-i+j} +\chi_{\mu_{i}-i-j+2})_{1  \le i, j \le \mu_{1}^{\prime} } | 
$ if $\mu \ne \emptyset$,  
 and $\chi_{\langle \emptyset \rangle }  (\{z_{b} \}_{b=1}^{M}  | \{z_{b+M}\}_{b=1}^{N} )=1$. 
One can also rewrite \eqref{JT-o} as 
$\chi_{[ \mu ] }  (\{z_{b} \}_{b=1}^{M}  | \{z_{b+M}\}_{b=1}^{N} )
=|
(h_{\mu_{i}-i+1})_{1 \le i \le \mu_{1}^{\prime}} \
 (h_{\mu_{i}-i+j} +h_{\mu_{i}-i-j+2})_{1 \le i \le \mu_{1}^{\prime} \atop 2 \le j \le \mu_{1}^{\prime} } |$, where 
 $h_{m}=\chi_{m}-\chi_{m-2}$. 
 This is equivalent to 
$\chi_{[ \mu ] }  (\{z_{b} \}_{b=1}^{M}  | \{z_{b+M}\}_{b=1}^{N} )= \frac{1}{2} |
 (h_{\mu_{i}-i+j} +h_{\mu_{i}-i-j+2})_{1  \le i, j \le \mu_{1}^{\prime} } | 
$ if $\mu \ne \emptyset$,  
 and $\chi_{[ \emptyset ] }  (\{z_{b} \}_{b=1}^{M}  | \{z_{b+M}\}_{b=1}^{N} )=1$. }
:
\begin{align}
\chi_{[\mu]}  (\{z_{b} \}_{b=1}^{M}  | \{z_{b+M}\}_{b=1}^{N} )&= |(\chi_{\mu_{i}-i+j} -\chi_{\mu_{i}-i-j})_{1 \le i,j \le \mu_{1}^{\prime}} | ,
\label{JT-o}
\\
\chi_{\langle \mu \rangle }  (\{z_{b} \}_{b=1}^{M}  | \{z_{b+M}\}_{b=1}^{N} )&=|
(\chi_{\mu_{i}-i+1})_{1 \le i \le \mu_{1}^{\prime}} \
 (\chi_{\mu_{i}-i+j} +\chi_{\mu_{i}-i-j+2})_{1 \le i \le \mu_{1}^{\prime} \atop 2 \le j \le \mu_{1}^{\prime} } |,
\label{JT-c}
\end{align}
where the arguments are omitted on the right hand sides: $ \chi_{m}=\chi_{m}(\{z_{b} \}_{b=1}^{M}  | \{z_{b+M}\}_{b=1}^{N} )$. 
From now on, we will use the following shorthand notations for the arguments of the above 
determinants: $\mathbf{x}=\{x_{b}\}_{b=1}^{r}$, $\mathbf{x}^{-1}=\{x_{b}^{-1}\}_{b=1}^{r}$, 
$\mathbf{y}=\{y_{b}\}_{b=1}^{s}$, $\mathbf{y}^{-1}=\{y_{b}^{-1}\}_{b=1}^{s}$, 
$(\mathbf{x},\mathbf{x}^{-1}|\mathbf{y},\mathbf{y}^{-1})=
(\mathbf{x} \sqcup \mathbf{x}^{-1}|\mathbf{y} \sqcup \mathbf{y}^{-1})$, 
$(\mathbf{x},\mathbf{x}^{-1}|\mathbf{y}, \pm 1, \mathbf{y}^{-1})=
(\mathbf{x} \sqcup \mathbf{x}^{-1}|\mathbf{y} \sqcup \{\pm 1\} \sqcup \mathbf{y}^{-1})$, 
$(\mathbf{x}, \pm 1, \mathbf{x}^{-1}|\mathbf{y},  \mathbf{y}^{-1})=
(\mathbf{x} \sqcup \{\pm 1\} \sqcup \mathbf{x}^{-1}|\mathbf{y} \sqcup \mathbf{y}^{-1})$. 
We will consider the following specializations of the determinants \eqref{JT-o} and \eqref{JT-c} 
(cf.\ \cite{BB81,BSR96}). 
\begin{align}
 &\chi_{[\mu]} (\mathbf{x} ,1,\mathbf{x}^{-1}|\mathbf{y},  \mathbf{y}^{-1})
& & \text{[type B]},
\label{ch-fin-B}
\\
&  \chi_{[\mu]} (\mathbf{x}  , \mathbf{x}^{-1}|\mathbf{y},  \mathbf{y}^{-1})
& & \text{[type D]},
\label{ch-fin-D}
\\
& \chi_{\langle \mu \rangle }  (\mathbf{x}  , \mathbf{x}^{-1}|\mathbf{y},  \mathbf{y}^{-1}) 
& &  \text{[type C]} .
\label{ch-fin-C}
\end{align}
In case the Young diagram $\mu$ is defined on the $[r,s]$-hook, 
\eqref{ch-fin-B} and \eqref{ch-fin-D} are expected to give supercharacters of 
representations of $osp(2r+1|2s)$ and $osp(2r|2s)$ with the highest weights \eqref{HW-B}, \eqref{YW-B} and 
 \eqref{HW-D}, \eqref{YW-Dp}, \eqref{YW-Dm}, respectively.
 \footnote{We do not know whether these are 
 irreducible supercharacters in the general situation. One may have to subtract unnecessary supercharacters of invariant subspaces 
 from them  to get irreducible ones. 
}
In particular at $s=0$, 
$\chi_{[\mu]} (\mathbf{x} , 1, \mathbf{x}^{-1}|\emptyset)$ 
$\chi_{[\mu]} (\mathbf{x} ,\mathbf{x}^{-1}|\emptyset)$,  
$\chi_{\langle \mu \rangle }  (\mathbf{x}  , \emptyset)$  
 give  irreducible characters of $so(2r+1)$, $O(2r)$ and $sp(2r)$, respectively. 
 In case $\mu_{r}=0$, $\chi_{[\mu]} (\mathbf{x} ,\mathbf{x}^{-1}|\emptyset)$ gives 
 irreducible characters of $so(2r)$, while in case $\mu_{r} \ne 0$, 
 it  becomes summation of the characters of 
 irreducible representations of $so(2r)$ with the highest weights  \eqref{HW-D}, \eqref{YW-Dp} 
 and \eqref{HW-D}, \eqref{YW-Dm} (at $s=0$). 
 We observe similar phenomena for the case $s>0$ on the level of T-functions 
 (thus in relation to representations of $U_{q}(osp(2r|2s)^{(1)})$): see \eqref{FT-D2} and 
 \eqref{FT-D4}. 

Let us introduce the set of all the partitions 
$ \lambda =(\lambda_{1},\lambda_{2}, \dots , \lambda_{a})$ 
in a rectangular Young diagram $(m^{a})$ in the $[M,N]$-hook ($a,m \in \mathbb{Z}_{\ge 1}$):
\begin{align}
\ytableausetup{boxsize=0.5em}
\mathcal{S}_{\tiny \ydiagram{1}}=
\bigl\{ \lambda \ \bigl| \ 
&m \ge \lambda_{1} \ge \lambda_{2}  \ge \dots\ge \lambda_{a} \ge 0
\bigr\} ,
\label{req-dec}
\end{align}
and two kinds of subsets of this set:
\begin{align}
\mathcal{S}_{\tiny \ydiagram{1,1}}=
\Biggl\{ \lambda \ \Biggl| \ 
\begin{split}
&m = \lambda_{1} \ge \lambda_{2} =\lambda_{3} \ge \lambda_{4}=\lambda_{5} 
\ge \dots\ge \lambda_{a-1}=\lambda_{a} \ge 0 \quad \text{if} \quad a \in 2\mathbb{Z}+1,
\\
&m \ge \lambda_{1} = \lambda_{2} \ge \lambda_{3}=\lambda_{4} 
\ge \dots\ge \lambda_{a-1}=\lambda_{a} \ge 0 
\quad \text{if} \quad a \in 2\mathbb{Z}
\end{split}
\Biggr\} ,
\label{req-dect}
\\
\mathcal{S}_{\tiny \ydiagram{2}}=
\Biggl\{ \lambda \ \Biggl| \ 
\begin{split}
&m \ge \lambda_{1} \ge \lambda_{2}  \ge \dots\ge \lambda_{a} \ge 1, \quad \lambda_{j} \in 
2\mathbb{Z}+1
 \quad \text{if} \quad m \in 2\mathbb{Z}+1,
\\
&m \ge \lambda_{1} \ge \lambda_{2} \ge \dots\ge \lambda_{a} \ge 0 , \quad \lambda_{j} \in 
2\mathbb{Z} 
\quad \text{if} \quad m \in 2\mathbb{Z}
\end{split}
\Biggr\} .
\label{req-decy}
\end{align}

%%%%%%%%%%%%%%%%%%%%%%%%%%%%%%%%%%%%%%%%
\paragraph{$U_{q}(osp(2r+1|2s)^{(1)})$ case:}
The (super)character limit of the generating function \eqref{gene1B} is 
%$\zeta(\mathbf{W}_{I_{2r+2s+1}}(\mathbf{X})) =w(t)$
  given by 
\begin{align}
%w(t)&=
\prod_{j=1}^{r} (1-x_{j} t)^{-1} (1-x_{j}^{-1} t)^{-1} \prod_{j=1}^{s} (1-y_{j} t)(1-y_{j}^{-1} t)(1+t)
% \nonumber \\
%
%&
=\sum_{m=0}^{\infty} \chi_{m}(\mathbf{x},\mathbf{x}^{-1}|\mathbf{y},-1,\mathbf{y}^{-1}) t^{m} ,
\label{gench-B}
\end{align}
where  
%$\zeta(\mathcal{F}^{I_{2r+2s+1}}_{(m)})=\chi_{m}$, $\zeta(\mathbf{X})=t$, 
$x_{j}=z_{j}$ for $1 \le j \le r$, $y_{j}=z_{2r+j}$ for $1 \le j \le s$.

We conjecture that the following decompositions hold 
for any rectangular Young diagram $(m^{a})$ in the $[2r,2s+1]$-hook ($a,m \in \mathbb{Z}_{\ge 1}$):
\begin{align}
\ytableausetup{boxsize=0.5em}
\chi_{(m^{a})}(\mathbf{x},\mathbf{x}^{-1}|\mathbf{y},-1,\mathbf{y}^{-1})
&=\sum_{\lambda \in \mathcal{S}_{\tiny \ydiagram{1,1}}} \chi_{[\lambda]} (\mathbf{x},1,\mathbf{x}^{-1}|\mathbf{y},\mathbf{y}^{-1}) 
&& \text{[on type B]}
\label{de-BB}
\\
&=\sum_{\lambda \in \mathcal{S}_{\tiny \ydiagram{1}}} \chi_{[\lambda]} (\mathbf{x},\mathbf{x}^{-1}|\mathbf{y},\mathbf{y}^{-1}) 
&& \text{[on type D]}.
\label{de-BD}
\end{align}
To relate this formula to the labels of representations, we  restrict it to the $[r,s]$-hook. 
%In this case, our observation on the Bethe strap suggest that  it gives an
% irreducible character of $U_{q}(osp(2r+1|2s)^{(1)})$. 
 Eq.\ \eqref{de-BB} suggests a decomposition of a
 representation $W_{a,m}$ of $U_{q}(osp(2r+1|2s)^{(1)})$ (or $Y(osp(2r+1|2s))$) 
into representations $\{V^{\prime}(\lambda )\}$ of $U_{q}(osp(2r+1|2s))$ (or $osp(2r+1|2s)$): 
$W_{a,m} \simeq
\oplus_{\lambda \in \mathcal{S}_{\tiny \ydiagram{1,1}}} V^{\prime}(\lambda)  $.  
Here we use the same symbol $\lambda$ for a Young diagram 
and the highest weight specified by it (via \eqref{HW-B}, \eqref{YW-B}). We do not know whether 
 $V^{\prime}(\lambda)$ coincides with the irreducible representation  $V(\lambda)$ 
 in the general situation, but at least for $s=0$ it does. 
 In this case ($s=0$), \eqref{de-BB} coincides with the 
character formula of the Kirillov-Reshetikhin modules \cite{KR90} 
for $U_{q}(so(2r+1)^{(1)})$ or $Y(so(2r+1))$ (see also \cite{HKOTY98}; 
use \eqref{KD-B} for comparison). Note that the case $a=r$ (with $s=0$) 
corresponds to a spin-even 
(tensor-like) representation, and the spin-odd (spinor-like) representations have 
to be treated separately.
The second equality \eqref{de-BD} for $s=0$ corresponds to [eq.\ (C.1), \cite{HKOTT01}], 
which suggests another decomposition of $W_{a,m}$. 

%$\chi_{a,m}=\chi_{(m^{a})}(\mathbf{x},\mathbf{x}^{-1}|\mathbf{y},-1,\mathbf{y}^{-1})$ 
%for $1 \le a \le r-1$, 
%$\chi_{r,2m}=\chi_{(m^{r})}(\mathbf{x},\mathbf{x}^{-1}|\mathbf{y},-1,\mathbf{y}^{-1})$.  

For more general Young diagram $\mu$ in the $[2r,2s+1]$-hook, 
we expect a decomposition of the form: 
\begin{align}
\chi_{\mu} (\mathbf{x},\mathbf{x}^{-1}|\mathbf{y},-1,\mathbf{y}^{-1}) 
&=\sum_{\kappa,\lambda } LR^{ \mu}_{(2\kappa)^{\prime},\lambda}
\chi_{[ \lambda ]} 
(\mathbf{x},1, \mathbf{x}^{-1}|\mathbf{y},\mathbf{y}^{-1})
&&  \text{[on type B] },
\label{gen-dec}
\end{align}
where $LR^{ \mu}_{(2\kappa)^{\prime},\lambda}$ is a  
 Littlewood-Richardson coefficient for $gl(M|N)$, 
 and the sum is taken over all the partitions $\kappa, \lambda$ 
such that
\footnote{Do not confuse $\lambda \subset \mu $ with a skew-Young diagram. 
$2\kappa=(2\kappa_{1},2\kappa_{2},\dots )$} 
$(2\kappa)^{\prime},\lambda \subset \mu $. 
A proof of \eqref{gen-dec} restricted to the $[r,0]$-hook for the case $s=0$
 is available in  [Lemma 7.3, \cite{HKOTY98}] (see also [eq.\ (3.14), \cite{KOS95}]).

%%%%%%%%%%%%%%%%%%%%%%%%%%%%%%%%%%%%%%%%
\paragraph{$U_{q}(gl(2r|2s+1)^{(2)})$ case:}
The (super)character limit of the generating function \eqref{gene1t1} is 
  given by 
\begin{align}
%w(t)&=
\prod_{j=1}^{r} (1-x_{j} t)^{-1} (1-x_{j}^{-1} t)^{-1} \prod_{j=1}^{s} (1-y_{j} t)(1-y_{j}^{-1} t)(1-t)
% \nonumber \\
%
%&
=\sum_{m=0}^{\infty} \chi_{m}(\mathbf{x},\mathbf{x}^{-1}|\mathbf{y},1,\mathbf{y}^{-1}) t^{m} ,
\label{gench-t1}
\end{align}
where  
%$\zeta(\mathcal{F}^{I_{2r+2s+1}}_{(m)})=\chi_{m}$, $\zeta(\mathbf{X})=t$, 
$x_{j}=z_{j}$ for $1 \le j \le r$, $y_{j}=z_{2r+j}$ for $1 \le j \le s$. 

We conjecture that the following decompositions hold  
for any rectangular Young diagram $(m^{a})$ in the $[2r,2s+1]$-hook ($a,m \in \mathbb{Z}_{\ge 1}$):
\begin{align}
\ytableausetup{boxsize=0.5em}
\chi_{(m^{a})}(\mathbf{x},\mathbf{x}^{-1}|\mathbf{y},1,\mathbf{y}^{-1})
&=\sum_{\lambda \in \mathcal{S}_{\tiny \ydiagram{1,1}}} \chi_{[\lambda]} (\mathbf{x},-1,\mathbf{x}^{-1}|\mathbf{y},\mathbf{y}^{-1}) 
&&  \text{[on type B$^{\prime}$] }
\label{de-t1Bp}
\\
&=\sum_{\lambda \in \mathcal{S}_{\tiny \ydiagram{1}}} (-1)^{ma+|\lambda |}
 \chi_{[\lambda]} (\mathbf{x},\mathbf{x}^{-1}|\mathbf{y},\mathbf{y}^{-1}) 
&& \text{[on type D]},
\label{de-t1Dp}
\end{align}
where $|\lambda|=\sum_{j=1}^{r}\lambda_{j}$ is the size of the Young diagram $\lambda$. 
This case is parallel to the case $U_{q}(osp(2r+1|2s)^{(1)})$. However, 
because of the difference between the factors $1+t$ in
 \eqref{gench-B} and $1-t$ in \eqref{gench-t1}, 
we have to modify \eqref{ch-fin-B} as in \eqref{de-t1Bp}, or add a sign factor as in \eqref{de-t1Dp}. 

%%%%%%%%%%%%%%%%%%%%%%%%%%%%%%%%%%%%%%%%
\paragraph{$U_{q}(gl(2r+1|2s)^{(2)})$ case:}
The (super)character limit of the generating function \eqref{gene1t2} is 
%$\zeta(\mathbf{W}_{I_{2r+2s+1}}(\mathbf{X})) =w(t)$
  given by 
\begin{align}
%w(t)&=
\prod_{j=1}^{r} (1-x_{j} t)^{-1} (1-x_{j}^{-1} t)^{-1} \prod_{j=1}^{s} (1-y_{j} t)(1-y_{j}^{-1} t)(1-t)^{-1}
% \nonumber \\
%
%&
=\sum_{m=0}^{\infty} \chi_{m}(\mathbf{x},1,\mathbf{x}^{-1}|\mathbf{y},\mathbf{y}^{-1}) t^{m} ,
\label{gench-t2}
\end{align}
where  
%$\zeta(\mathcal{F}^{I_{2r+2s+1}}_{(m)})=\chi_{m}$, $\zeta(\mathbf{X})=t$, 
$x_{j}=z_{j}$ for $1 \le j \le r$, $y_{j}=z_{2r+1+j}$ for $1 \le j \le s$.

We conjecture that the following decompositions hold  
for any rectangular Young diagram $(m^{a})$ in the $[2r+1,2s]$-hook ($a,m \in \mathbb{Z}_{\ge 1}$):
\begin{align}
\ytableausetup{boxsize=0.5em}
\chi_{(m^{a})}(\mathbf{x},1,\mathbf{x}^{-1}|\mathbf{y},\mathbf{y}^{-1})
&=\sum_{\lambda \in \mathcal{S}_{\tiny \ydiagram{2}}} \chi_{[\lambda]} (\mathbf{x},1,\mathbf{x}^{-1}|\mathbf{y},\mathbf{y}^{-1}) 
&& \text{[on type B]}
\label{de-t2B}
\\
&=\sum_{\lambda \in \mathcal{S}_{\tiny \ydiagram{1}}} \chi_{\langle \lambda \rangle} (\mathbf{x},\mathbf{x}^{-1}|\mathbf{y},\mathbf{y}^{-1}) 
&& \text{[on type C]}.
\label{de-t2C}
\end{align}
To relate this formula to the labels of representations, we  restrict it to the $[r,s]$-hook. 
%In this case, our observation on the Bethe strap suggests that  it gives an
% irreducible character of $U_{q}(gl(2r+1|2s)^{(2)})$. 
 Eq.\  \eqref{de-t2B} suggests a decomposition of a
  representation $W_{a,m}$ of $U_{q}(gl(2r+1|2s)^{(2)})$ 
into representations $\{V^{\prime}(\lambda )\}$ of $U_{q}(osp(2r+1|2s))$: $W_{a,m} \simeq
\oplus_{\lambda \in \mathcal{S}_{\tiny \ydiagram{2}}} V^{\prime}(\lambda)  $.  
In particular, 
 $V^{\prime}(\lambda)$ coincides with the irreducible representation  $V(\lambda)$ 
 at least for $s=0$, and then \eqref{de-t2B} coincides with the 
character formula of the Kirillov-Reshetikhin modules for 
$U_{q}(gl(2r+1)^{(2)})$ [eq.\ (6.7), \cite{HKOTT01}].  
The second equality \eqref{de-t2C} for $s=0$ corresponds to [eq.\ (6.6), \cite{HKOTT01}], 
which suggests another decomposition of $W_{a,m}$. 

%%%%%%%%%%%%%%%%%%%%%%%%%%%%%%%%%%%%%%%%
\paragraph{$U_{q}(gl(2r|2s)^{(2)})$ case:}
The (super)character limit of the generating function \eqref{gene1t3} is 
%$\zeta(\mathbf{W}_{I_{2r+2s+1}}(\mathbf{X})) =w(t)$
  given by 
\begin{align}
%w(t)&=
\prod_{j=1}^{r} (1-x_{j} t)^{-1} (1-x_{j}^{-1} t)^{-1} \prod_{j=1}^{s} (1-y_{j} t)(1-y_{j}^{-1} t)
% \nonumber \\
%
%&
=\sum_{m=0}^{\infty} \chi_{m}(\mathbf{x},\mathbf{x}^{-1}|\mathbf{y},\mathbf{y}^{-1}) t^{m} ,
\label{gench-t3}
\end{align}
where  
%$\zeta(\mathcal{F}^{I_{2r+2s+1}}_{(m)})=\chi_{m}$, $\zeta(\mathbf{X})=t$, 
$x_{j}=z_{j}$ for $1 \le j \le r$, $y_{j}=z_{2r+j}$ for $1 \le j \le s$.

We conjecture that the following decompositions hold  
for any rectangular Young diagram $(m^{a})$ in the $[2r,2s]$-hook ($a,m \in \mathbb{Z}_{\ge 1}$):
\begin{align}
\ytableausetup{boxsize=0.5em}
\chi_{(m^{a})}(\mathbf{x},\mathbf{x}^{-1}|\mathbf{y},\mathbf{y}^{-1})
&=\sum_{\lambda \in \mathcal{S}_{\tiny \ydiagram{2}}} \chi_{[\lambda]} (\mathbf{x},\mathbf{x}^{-1}|\mathbf{y},\mathbf{y}^{-1}) 
&& \text{[on type D]}
\label{de-t3D}
\\
&=\sum_{\lambda \in \mathcal{S}_{\tiny \ydiagram{1,1}}} \chi_{\langle \lambda \rangle} (\mathbf{x},\mathbf{x}^{-1}|\mathbf{y},\mathbf{y}^{-1}) 
&& \text{[on type C]}.
\label{de-t3C}
\end{align}
To relate this formula to the labels of representations, we  restrict it to the $[r,s]$-hook. 
%In this case, our observation on the Bethe strap suggests that  it gives an
% irreducible character of $U_{q}(gl(2r|2s)^{(2)})$.
Eq.\  \eqref{de-t3D} suggests a decomposition of a representation $W_{a,m}$ of $U_{q}(gl(2r|2s)^{(2)})$ 
into representations $\{V^{\prime}(\lambda )\}$ of $U_{q}(osp(2r|2s))$: $W_{a,m} \simeq
\oplus_{\lambda \in \mathcal{S}_{\tiny \ydiagram{2}}} V^{\prime}(\lambda)  $.  
In particular, 
 $V^{\prime}(\lambda)$ coincides with the irreducible representation  $V(\lambda)$ 
 at least for $s=0$, $\lambda_{r}=0$, and
  $V^{\prime}(\lambda) \simeq V(\lambda) \oplus V(\lambda)|_{\lambda_{r} \to -\lambda_{r}}$ 
   at least for $s=0$, $\lambda_{r} \ne 0$, and then \eqref{de-t3D} coincides with the 
character formula of the Kirillov-Reshetikhin modules for $U_{q}(gl(2r)^{(2)})$
 [eq.\ (6.9), \cite{HKOTT01}].  
The second equality \eqref{de-t3C} for $s=0$ corresponds to [eq.\ (6.8), \cite{HKOTT01}], 
which suggests another decomposition of $W_{a,m}$. 

%%%%%%%%%%%%%%%%%%%%%%%%%%%%%%%%%%%%%%%%
\paragraph{$U_{q}(osp(2r|2s)^{(1)})$ case:}
The (super)character limit of the generating function \eqref{gene1D} is 
%$\zeta(\mathbf{W}_{I_{2r+2s+1}}(\mathbf{X})) =w(t)$
  given by 
\begin{multline}
\prod_{j=1}^{r} (1-x_{j} t)^{-1} (1-x_{j}^{-1} t)^{-1} \prod_{j=1}^{s} (1-y_{j} t)(1-y_{j}^{-1} t)(1-t^{2})
=
\\
=\sum_{m=0}^{\infty} \chi_{m}(\mathbf{x},\mathbf{x}^{-1}|\mathbf{y},1,-1,\mathbf{y}^{-1}) t^{m} ,
\label{gench-D}
\end{multline}
where  
%$\zeta(\mathcal{F}^{I_{2r+2s+1}}_{(m)})=\chi_{m}$, $\zeta(\mathbf{X})=t$, 
$x_{j}=z_{j}$ for $1 \le j \le r$, $y_{j}=z_{2r+j}$ for $1 \le j \le s$.

We conjecture that the following decomposition holds  
for any rectangular Young diagram $(m^{a})$ in the $[2r,2s+2]$-hook ($a,m \in \mathbb{Z}_{\ge 1}$):
\begin{align}
\ytableausetup{boxsize=0.5em}
\chi_{(m^{a})}(\mathbf{x},\mathbf{x}^{-1}|\mathbf{y},1,-1,\mathbf{y}^{-1})
&=\sum_{\lambda \in \mathcal{S}_{\tiny \ydiagram{1,1}}} \chi_{[\lambda]} (\mathbf{x},\mathbf{x}^{-1}|\mathbf{y},\mathbf{y}^{-1}) 
&& \text{[on type D]} .
\label{de-D}
\end{align}
To relate this formula to the labels of representations, we  restrict it to the $[r,s]$-hook. 
In this case, our observation on the Bethe strap suggests that  it does not always give an
 irreducible character of $U_{q}(osp(2r|2s)^{(1)})$ in the general situation. In addition, 
 \eqref{de-D} suggests a decomposition of a representation $W_{a,m}$ of $U_{q}(osp(2r|2s)^{(1)})$ 
  (or $Y(osp(2r|2s))$)  
into representations $\{V^{\prime}(\lambda )\}$ of $U_{q}(osp(2r|2s))$ (or $osp(2r|2s)$): $W_{a,m} \simeq
\oplus_{\lambda \in \mathcal{S}_{\tiny \ydiagram{1,1}}} V^{\prime}(\lambda)  $.  
In particular, 
 $V^{\prime}(\lambda)$ coincides with the irreducible representation  $V(\lambda)$ 
 at least for $s=0$, $\lambda_{r}=0$, and
  $V^{\prime}(\lambda) \simeq V(\lambda) \oplus V(\lambda)|_{\lambda_{r} \to -\lambda_{r}}$ 
   at least for $s=0$, $\lambda_{r} \ne 0$, and then \eqref{de-D} 
   for $1 \le a \le r-2$ coincides with the 
character formula of the Kirillov-Reshetikhin modules for $U_{q}(so(2r)^{(1)})$ or $Y(so(2r))$
 \cite{KR90} (see also [eq.\ (7.4), \cite{HKOTY98}]). The cases $a=r-1,r$ for $s=0$ have to be treated separately. 
 
 %%%%%%%%%%%%%%%%%%%%%%%%%%%%%%%%%%%%%%%%
\paragraph{$U_{q}(osp(2r|2s)^{(2)})$, $r \ge 1$, $s\ge 0$ case:}
The (super)character limit of the reduction of the generating function \eqref{gene1D} is 
%$\zeta(\mathbf{W}_{I_{2r+2s+1}}(\mathbf{X})) =w(t)$
  given by 
\begin{multline}
\prod_{j=1}^{r-1} (1-x_{j} t)^{-1} (1-x_{j}^{-1} t)^{-1} \prod_{j=1}^{s} (1-y_{j} t)(1-y_{j}^{-1} t)(1- t)^{-1}(1+t)
=
\\
=\sum_{m=0}^{\infty} \chi_{m}(\tilde{\mathbf{x}},1,1,\tilde{\mathbf{x}}^{-1}|\mathbf{y},1,-1,\mathbf{y}^{-1}) t^{m} ,
\label{gench-D2}
\end{multline}
where  
%$\zeta(\mathcal{F}^{I_{2r+2s+1}}_{(m)})=\chi_{m}$, $\zeta(\mathbf{X})=t$, 
$\tilde{\mathbf{x}}=\{x_{j} \}_{j=1}^{r-1}$, $\tilde{\mathbf{x}}^{-1}=\{x_{j}^{-1} \}_{j=1}^{r-1}$, 
$x_{j}=z_{j}$ for $1 \le j \le r-1$, $y_{j}=z_{2r+j}$ for $1 \le j \le s$.

We conjecture that the following decomposition holds  
for any rectangular Young diagram $(m^{a})$ in the $[2r,2s+2]$-hook ($a,m \in \mathbb{Z}_{\ge 1}$):
\begin{align}
\ytableausetup{boxsize=0.5em}
\chi_{(m^{a})}(\tilde{\mathbf{x}},1,1,\tilde{\mathbf{x}}^{-1}|\mathbf{y},1,-1,\mathbf{y}^{-1})
&=\sum_{\lambda \in \mathcal{S}_{\tiny \ydiagram{1}}} \chi_{[\lambda]} (\tilde{\mathbf{x}},1,\tilde{\mathbf{x}}^{-1}|\mathbf{y},\mathbf{y}^{-1}) 
&& \text{[on type B]} .
\label{de-D2}
\end{align}
To relate this formula to the labels of representations, we  restrict it to the $[r-1,s]$-hook. 
%In this case, our observation on the Bethe strap suggests that  it does not always give an
% irreducible character of $U_{q}(osp(2r|2s)^{(2)})$ in the general situation. 
% In addition, 
 \eqref{de-D2} suggests a decomposition of a representation $W_{a,m}$ of $U_{q}(osp(2r|2s)^{(2)})$ 
into representations $\{V^{\prime}(\lambda )\}$ of $U_{q}(osp(2r-1|2s))$: $W_{a,m} \simeq
\oplus_{\lambda \in \mathcal{S}_{\tiny \ydiagram{1}}} V^{\prime}(\lambda)  $.  
In particular, 
 $V^{\prime}(\lambda)$ coincides with the irreducible representation  $V(\lambda)$ 
 at least for $s=0$, $\lambda_{r-1}=0$, and then \eqref{de-D2} 
   for $1 \le a \le r-2$ coincides with the 
character formula of the Kirillov-Reshetikhin modules for $U_{q}(so(2r)^{(2)})$ 
 [eq.\ (6.10), \cite{HKOTT01}]. The case $a=r-1$ for $s=0$ has to be treated separately. 

 We have not tried to prove \eqref{de-BB}, \eqref{de-BD}, \eqref{de-t1Bp}, \eqref{de-t1Dp}, \eqref{de-t2B}, \eqref{de-t2C}, \eqref{de-t3D}, \eqref{de-t3C}, \eqref{de-D} and \eqref{de-D2} yet, but checked them by Mathematica (ver.\ 7) for the case $0 \le r+s \le 6$. 
 Apparently, all these formulas hold independent of representation theory 
 and could be proved by a combinatorial method. 
 Nevertheless, it is desirable to study them from the perspective of representation theory and clarify their relationship with irreducible (super)characters of superalgebras.
 
 %%%%%%%%
\paragraph{Declaration of Generative AI and AI-assisted technologies in the writing process}
During the preparation of this work the author used Microsoft Copilot in order to ensure grammatical accuracy for several English sentences. After using this tool/service, the author reviewed and edited the content as needed and takes full responsibility for the content of the publication.
%%%%%%%%%%%%

\end{document}